\documentclass[11pt]{article}
\usepackage{amsfonts}
\usepackage{amssymb}
\usepackage{graphicx}
\usepackage{amsmath}
\usepackage{geometry}
\usepackage{appendix}

\input{tcilatex}
\DeclareMathSizes{10.95}{10}{7}{6}

\begin{document}

\title{Statistical Field Theory and Neural Structures Dynamics IV:
Field-Theoretic Formalism for Interacting Collective States}
\author{Pierre Gosselin\thanks{%
Pierre Gosselin : Institut Fourier, UMR 5582 CNRS-UGA, Universit\'{e}
Grenoble Alpes, BP 74, 38402 St Martin d'H\`{e}res, France.\ E-Mail:
Pierre.Gosselin@univ-grenoble-alpes.fr} \and A\"{\i}leen Lotz\thanks{%
A\"{\i}leen Lotz: Cerca Trova, BP 114, 38001 Grenoble Cedex 1, France.\
E-mail: a.lotz@cercatrova.eu}}
\maketitle

\begin{abstract}
Building upon the findings presented in the first three papers of this
series, we formulate an effective field theory for interacting collective
states. These states consist of a large number of interconnected neurons and
are distinguished by their intrinsic activity. The field theory encompasses
an infinite set of fields, each of which characterizes the dynamics of a
specific type of collective state. Interaction terms within the theory drive
transitions between various collective states, allowing us to describe
processes such as activation, association, and deactivation of these states.
\end{abstract}

\section{Introduction}

In this series of papers, we present a field-theoretic approach to the
dynamics of connectivities within a set of interacting spiking neurons. In (%
\cite{GLr}), we introduced a two-field model that characterizes both the
dynamics of neural activity and the connectivity between points along the
thread. The first field describes the assembly of neurons and is akin to the
one introduced in (\cite{GL}). The second field, on the other hand, captures
the dynamics of connectivity between cells. Both fields interact with
themselves, representing interactions within the thread, and also interact
with each other, encapsulating the mutual influences between neural
activities and connectivities. This field-based framework encompasses both
collective and individual aspects of the system. The system composed of
these two fields is described by a field action functional, which
encapsulates the interactions within the system at the microscopic level.
This action functional provides insights into the dynamics of the entire
system.

In (\cite{GLr}), by focusing on the field for connectivity functions, our
framework enabled the derivation of background fields for both neural
interactions and connectivities, which minimize the action functional. These
background fields encompass the collective configurations of the system and
determine the potential static equilibria for neural activities and
connectivities. These equilibria serve as the foundational framework for
conditioning and shaping fluctuations and signal propagation within the
system. Their existence is contingent on internal system parameters and
external stimuli. We demonstrated the feasibility of several background
states, each associated with specific connectivities, with the thread
primarily organized into groups of interconnected points.

In (\cite{GLs}), we demonstrated how repeated activations at specific points
can propagate throughout the thread, progressively altering the connectivity
functions. We discussed interference phenomena and their influence on
connectivities. At points of constructive interference, the background state
for connectivities and average connectivities is modified, leading to the
emergence of states with enhanced connectivities between certain points.
These states gradually fade over time but can be reactivated by external
perturbations. Furthermore, the association of these emerging states is
possible when their activation occurs at closed times. Activating any of
these states rekindles their combination. These enhanced connectivity states
exhibit characteristics typical of interacting partial neuronal assemblies.

In Article (\cite{GLt}), we expanded this approach by focusing on the system
of connectivities itself. Expressing the individual cells background field
as an effective quantity depending on the connectivity field, we described
the effective dynamics of the connectivity fields. The ntegration of the
cell field degrees of freedom led to self-interactions for the connectivity
field. This introduced internal dynamics that could modify the static
background state we initially started with at some points of the thread. The
self-interactions, possibly induced by specific perturbations, initiated
internal patterns of connections between cells. Depending on various
internal parameters, permanent shifts in the background states of
connectivity in specific areas of the thread arise while others remain
unaffected. This effective theory also allowed us to elucidate the
mechanisms behind the reinforcement of connectivities between multiple cells
and the emergence of groups with altered connectivities. These collective
shifts can be seen as additional structures that emerge above the background
field, but whose possibility of emergence depends on these background fields.

The emergence of such states brings about a shift in perspective. In our
previous works, collective states emerged from interactions within the
entire self-interacting system of neurons and consisted of a large number of
interacting individual states. Here, conversely, the dynamics of individual
neurons take on a secondary role as they contribute to one or several
connected states. Cells may exhibit varying activation patterns depending on
their involvement in collective patterns. The collective states become
prominent and determine the individual neurons'activity.

Further investigation of the implications of our formalism should prompt us
to explicitly consider families of collective states, including the
possibility for multiple activations or deactivations of such states. In
other words, we should contemplate an effective field theory for emerging
and interacting states in its own right. This constitutes the focus of the
present study.

Building upon our prior research, this paper establishes a formalism for
interacting collective sets of connectivities. By directly considering these
emerging states as our starting point, we introduce an effective field
formalism to describe their dynamics and interactions. This process unfolds
in two stages. Initially, assuming the existence of such states, we
elucidate their principal attributes as assemblies of individual states.
These structures are characterized by various parameters, including their
spatial extent, the average connectivities between their constituent
elements, and the activities of each element. The activities exhibit
potential dynamic oscillations whose frequencies play a role in interactions
with other structures. Importantly, these parameters are not unique. In
fact, the same structure can be described by a discrete, and potentially
infinite, set of parameters, each corresponding to different possible
average connectivities and activity frequencies. As a result, a structure
may exist in different states and may undergo transitions between these
states.

To account for the existence of multiple states and transitions between
them, we then develop a field theoretic model of collective states. We
introduce a field for each possible structure, so that the states of this
field representing activations or multiple activations of the structure. The
action functional describes not only the primary characteristics of these
states but also their dynamic interactions. The field formalism enables the
activation or deactivation of these structures, as well as transitions of
the same structure between states induced by interactions or external
perturbations. Consequently, a single structure may experience activation or
deactivation, which, in turn, may activate other structures. In this
context, state transitions of a given structure and the associated
modifications in terms of connectivities and activity frequencies can
facilitate or inhibit interactions with other collective states, resulting
in synchronization or desynchronization with other structures.

The interactions between collective states also involve mechanisms that
describe the assembly of some states to create a larger set. An effective
formalism, based on the integration of intermediate degrees of freedom,
characterizes the binding of several structures through indirect
interactions mediated by intermediate collective states. This effective
formalism is presented in two alternative ways: a field formalism in which
several structures are projected in their background state, depending on
their interactions with others, and a perturbation formalism that enables
the extraction of the interaction content of the bound structures.

This work is organized as follows. The first part describes the emergence of
collective states from the formalism developed in our previous works and
their properties. Starting from the field formalism for neurons and
connectivities in Section 2, we recall the concept of a background field for
the system. We provide conditions for the emergence of enhanced activation
between large sets of cells and describe the characteristics of such states.
Section 3 reviews the definition and the principles of deriving the
background field, as presented in (\cite{GLr}). Section 4 outlines the
effective formalism for emerging collective states, which was developed in (%
\cite{GLs}) and (\cite{GLt}). In Section 5, we derive the characteristics of
these states, including enhanced connectivities and multiple activity
frequencies. Section 6 anticipates the formalism of the second part by
describing the characteristics of composed collective states involving
different types of cell fields. This allows in Section 7 to consider
transitions of several collective states to form a composite one. This
description serves as a guide for the next part, which significantly expands
upon this approach through the cell states formalism.

The second part of this article develops the field formalism for a large
number of interacting collective states. Section 8 elaborates on this field
formalism, beginning with collective states as basic elements and then
constructing the associated field approach. Section 9 incorporates
interactions between structures.

The third part of this work describes the different mechanisms of dynamical
transitions implied by the formalism, providing several examples of these
mechanisms. In Section 10, we present three approaches to transitions
between structures. The first approach is based on a perturbative expansion,
allowing for a straightforward description of transition mechanisms induced
by interactions but misses non-perturbative effects. The second approach is
an effective formalism that helps understand how some structures, considered
as backgrounds, can initiate activations of other structures. Two examples
are presented. Then, an operator formalism that elucidates how indirect
interactions can bind several structures is developed. Section 11 presents
the application of the effective and perturbative formalism to a system with
three structures. Section 12 extends the formalism by introducing an
extension considering non-localized structures. Section 13 is a conclusion.

\part*{I Field theory, emergence and description of collective states}

In this first part, we start by revisiting the foundational model developed
in (\cite{GLr}), (\cite{GLs}) and (\cite{GLt}), along with the pertinent
results relevant to our present study. In (\cite{GLt}), we derived the
possibility of emerging collective states above some given background. We
conducted an examination of the associated modification to the background
state, interpreting these modifications as distinct states themselves. While
we partially described their interactions, a comprehensive elucidation of
their properties in terms of activities or connectivity frequencies remained
incomplete. The primary objective of this first part, therefore, is to
further elucidate the characteristics of these emerging states. Ultimately,
by qualitatively exploring the transitions between various states, we will
pave the way for the field formalism in the subsequent part of \ this paper.

\section{Field theoretic description of the system}

In the three next sections, we sum-up some results of the previous papers of
this series. Then, we focus on the effective action for connectivity field.

\subsection{Fields and action functional}

Based on \cite{GL1}\cite{GL2}\cite{GL3}\cite{GL4}, we gave in ((\cite{GLr}),
resume in (\cite{GLs})) a statistical field formalism to describe both cells
and connectivities dynamics. This decription relies on two fields, $\Psi $
for cells, and $\Gamma $ for connectivities. The field action for the system
is:

\begin{eqnarray}
S_{full} &=&-\frac{1}{2}\Psi ^{\dagger }\left( \theta ,Z,\omega \right)
\nabla \left( \frac{\sigma _{\theta }^{2}}{2}\nabla -\omega ^{-1}\left(
J,\theta ,Z,\left\vert \Psi \right\vert ^{2}\right) \right) \Psi \left(
\theta ,Z\right) +V\left( \Psi \right)   \label{flt} \\
&&+\frac{1}{2\eta ^{2}}\left( S_{\Gamma }^{\left( 0\right) }+S_{\Gamma
}^{\left( 1\right) }+S_{\Gamma }^{\left( 2\right) }+S_{\Gamma }^{\left(
3\right) }+S_{\Gamma }^{\left( 4\right) }\right) +U\left( \left\{ \left\vert
\Gamma \left( \theta ,Z,Z^{\prime },C,D\right) \right\vert ^{2}\right\}
\right)   \notag
\end{eqnarray}%
where the activities $\omega \left( J,\theta ,Z,\left\vert \Psi \right\vert
^{2}\right) $ satisfy:%
\begin{equation}
\omega \left( J,\theta ,Z,\left\vert \Psi \right\vert ^{2}\right) =G\left(
J\left( \theta \right) +\frac{\kappa }{N}\int T\left( Z,Z_{1},\theta \right) 
\frac{\omega \left( \theta -\frac{\left\vert Z-Z_{1}\right\vert }{c}%
,Z_{1}\right) }{\omega \left( \theta ,Z\right) }\left\vert \Psi \left(
\theta -\frac{\left\vert Z-Z_{1}\right\vert }{c},Z_{1}\right) \right\vert
^{2}dZ_{1}\right)   \label{qf}
\end{equation}%
and $S_{\Gamma }^{\left( 1\right) }$, $S_{\Gamma }^{\left( 2\right) }$, $%
S_{\Gamma }^{\left( 3\right) }$, $S_{\Gamma }^{\left( 4\right) }$ are given
in appendix 1. In (\ref{flt}), we added a potential:%
\begin{equation*}
U\left( \left\{ \left\vert \Gamma \left( \theta ,Z,Z^{\prime },C,D\right)
\right\vert ^{2}\right\} \right) =U\left( \int T\left\vert \Gamma \left( T,%
\hat{T},\theta ,Z,Z^{\prime },C,D\right) \right\vert ^{2}dTd\hat{T}\right) 
\end{equation*}%
that models the constraint about the number of active connections in the
system.

\subsection{Interpretation of the various field}

The action functional depends on two distinct fields: $\Psi \left( \theta
,Z\right) $ and $\Gamma \left( T,\hat{T},\theta ,Z,Z^{\prime },C,D\right) $.
These abstract quantities play a crucial role in deriving the comprehensive
dynamics of the entire system and subsequently, in the analysis of
transitions between distinct states. However, the squared modulus of these
two functions can be interpreted in terms of statistical distribution,
depending on the chosen framework. If we consider a system comprised of
simple cells distributed along a thread, the function $\left\vert \Psi
\left( \theta ,Z\right) \right\vert ^{2}$ measures at time $\theta $, the
density of active cells at point $Z$. In the context of complex cells with
multiple axons and dendrites, we can regard each cell as residing at point $%
Z $, and $\left\vert \Psi \left( \theta ,Z\right) \right\vert ^{2}$ measures
the density of axons for that particular cell. A similar interpretation can
be applied to $\Gamma \left( T,\hat{T},\theta ,Z,Z^{\prime },C,D\right) $.
Within the framework of a system composed of simple cells 'accumulated' in
the vicinity of $Z$, $\left\vert \Gamma \left( T,\hat{T},\theta ,Z,Z^{\prime
},C,D\right) \right\vert ^{2}$ quantifies the density of connections with
the value $T$ \ (and auxiliary variables $\hat{T}$, $C,D$ between the set of
cells plying at points $Z$ and $Z^{\prime }$ respectivly. In the context of
complex cells, it characterizes the density of connections with strength $T$
between sets of axons and dendrites of the cells.

\section{Derivation of background fields of the model}

In the previous parts, we derived the background fields of the system
described above. We considered first that the evolution for activity is at a
faster pace than the dynamics of connectivits described by $\sum S_{\Gamma
}^{\left( i\right) }$.

Starting by saddle point equations for:%
\begin{equation*}
-\frac{1}{2}\Psi ^{\dagger }\left( \theta ,Z\right) \nabla \left( \frac{%
\sigma _{\theta }^{2}}{2}\nabla -\omega ^{-1}\left( J,\theta ,Z,\left\vert
\Psi \right\vert ^{2}\right) \right) \Psi \left( \theta ,Z\right) +V\left(
\Psi \right) 
\end{equation*}%
allowed to describe the background states for $\Psi \left( \theta ,Z\right) $
and activity $\omega ^{-1}\left( J,\theta ,Z,\left\vert \Psi \right\vert
^{2}\right) $ as a function of connectvity $\Gamma \left( T,\hat{T},\theta
,Z,Z^{\prime },C,D\right) $ (see appendix 1) for more details.

We then minimized the action for connectivties:%
\begin{equation*}
\frac{\delta \sum S_{\Gamma }^{\left( i\right) }}{\delta \Gamma ^{\dag
}\left( T,\hat{T},\theta ,Z,Z^{\prime }\right) }=\frac{\delta \sum S_{\Gamma
}^{\left( i\right) }}{\delta \Gamma \left( T,\hat{T},\theta ,Z,Z^{\prime
}\right) }=0
\end{equation*}%
and obtained the background field for this field along with the associated
average values of connectivities and activities. The relevant formulas are
given in appendix 1. It is important to note that various background fields
are conceivable, contingent upon specific external conditions. This, in
turn, results in several potential configurations of average connectivities.
When subjected to external perturbations, we observed that configurations
featuring localized clusters of heightened connectivities were favored, with
these clusters partially connected\footnote{%
Formulas for the averages are detailed in the appendix.}. In other words
perturbations around the background state induce emergence of collective
states, but this emergence is conditioned by the background states.

\section{ Effective formalism and emerging connectivity collective states}

\subsection{Effective action for variation around the background field.}

Pertubations and internal dynamics in a given background obtained by
expansion of action around this background. We where led to define:%
\begin{eqnarray*}
\Gamma \left( T,\hat{T},\theta ,Z,Z^{\prime }\right) &=&\Gamma _{0}\left( T,%
\hat{T},\theta ,Z,Z^{\prime }\right) +\Delta \Gamma \left( T,\hat{T},\theta
,Z,Z^{\prime }\right) \\
\Gamma ^{\dag }\left( T,\hat{T},\theta ,Z,Z^{\prime }\right) &=&\Gamma
_{0}^{\dag }\left( T,\hat{T},\theta ,Z,Z^{\prime }\right) +\Delta \Gamma
^{\dag }\left( T,\hat{T},\theta ,Z,Z^{\prime }\right)
\end{eqnarray*}%
where $\Gamma _{0}$ and $\Gamma _{0}^{\dag }$ are the background fields.

A second order expansion around the background fields $\Gamma _{0}$ and $%
\Gamma _{0}^{\dag }$ which minimize $\sum S_{\Gamma }^{\left( i\right) }$
led to the effective action for the connectivity field. Formally, this
effective action is given by:%
\begin{equation*}
S\left( \Delta \Gamma \left( T,\hat{T},\theta ,Z,Z^{\prime }\right) \right)
=\Delta \Gamma ^{\dag }\left( T,\hat{T},\theta ,Z,Z^{\prime }\right) \frac{%
\delta ^{2}\sum S_{\Gamma }^{\left( i\right) }}{\delta \Delta \Gamma \left(
T,\hat{T},\theta ,Z,Z^{\prime }\right) \delta \Delta \Gamma ^{\dag }\left( T,%
\hat{T},\theta ,Z,Z^{\prime }\right) }\Delta \Gamma \left( T,\hat{T},\theta
,Z,Z^{\prime }\right) 
\end{equation*}%
This effective action above the background field will be the starting point
of a formalism of collective states. Its expanded form is given by\footnote{%
See \cite{GLt} and appendix 1 for a summary of the derivation}: 
\begin{eqnarray}
&&S\left( \Delta \Gamma \left( T,\hat{T},\theta ,Z,Z^{\prime }\right)
\right)   \label{fcp} \\
&=&-\Delta \Gamma ^{\dag }\left( T,\hat{T},\theta ,Z,Z^{\prime }\right)
\left( \nabla _{T}\left( \nabla _{T}+\frac{\Delta T-\lambda \Delta \hat{T}}{%
\tau \omega _{0}\left( Z\right) }\right) \right) \Delta \Gamma \left( T,\hat{%
T},\theta ,Z,Z^{\prime }\right)   \notag \\
&&-\Delta \Gamma ^{\dag }\left( T,\hat{T},\theta ,Z,Z^{\prime }\right)
\nabla _{\hat{T}}\left( \nabla _{\hat{T}}+\left\vert \bar{\Psi}_{0}\left(
Z,Z^{\prime }\right) \right\vert ^{2}\Delta \hat{T}\right) \Delta \Gamma
\left( T,\hat{T},\theta ,Z,Z^{\prime }\right) -\hat{V}\left( \Delta \Gamma
,\Delta \Gamma ^{\dag }\right)   \notag \\
&&+U_{\Delta \Gamma }\left( \left\Vert \Delta \Gamma \left( Z,Z^{\prime
}\right) \right\Vert ^{2}\right)   \notag
\end{eqnarray}%
where the average field $\left\vert \bar{\Psi}_{0}\left( Z,Z^{\prime
}\right) \right\vert ^{2}$ is defined by:%
\begin{equation*}
\left\vert \bar{\Psi}_{0}\left( Z,Z^{\prime }\right) \right\vert ^{2}=\frac{%
\rho \left( C\left( \theta \right) \left\vert \Psi _{0}\left( Z\right)
\right\vert ^{2}\omega _{0}\left( Z\right) +D\left( \theta \right) \hat{T}%
\left\vert \Psi _{0}\left( Z^{\prime }\right) \right\vert ^{2}\omega
_{0}\left( Z^{\prime }\right) \right) }{\omega _{0}\left( Z\right) }
\end{equation*}%
and the potential $\hat{V}\left( \Delta \Gamma ,\Delta \Gamma ^{\dag
}\right) $ is:%
\begin{eqnarray}
&&\hat{V}\left( \Delta \Gamma ,\Delta \Gamma ^{\dag }\right) =\Delta \Gamma
^{\dag }\left( T,\hat{T},\theta ,Z,Z^{\prime }\right)   \label{pnl} \\
&&\times \left( \nabla _{\hat{T}}\left( \frac{\rho D\left( \theta \right)
\left\langle \hat{T}\right\rangle \left\vert \Psi _{0}\left( Z^{\prime
}\right) \right\vert ^{2}}{\omega _{0}\left( Z\right) }\check{T}\left(
1-\left( 1+\left\langle \left\vert \Psi _{\Gamma }\right\vert
^{2}\right\rangle \right) \check{T}\right) ^{-1}\left[ O\frac{\Delta
T\left\vert \Delta \Gamma \left( \theta _{1},Z_{1},Z_{1}^{\prime }\right)
\right\vert ^{2}}{T\Lambda ^{2}}\right] \right) \right) \Delta \Gamma \left(
T,\hat{T},\theta ,Z,Z^{\prime }\right)   \notag
\end{eqnarray}%
The expression for the kernel of the operator $O$ is given by:%
\begin{equation}
O\left( Z,Z^{\prime },Z_{1}\right) =-\frac{\left\vert Z-Z^{\prime
}\right\vert }{c}\nabla _{\theta _{1}}+\frac{\left( Z^{\prime }-Z\right) ^{2}%
}{2}\left( \frac{\nabla _{Z_{1}}^{2}}{2}+\frac{\nabla _{\theta _{1}}^{2}}{%
2c^{2}}-\frac{\nabla _{Z}^{2}\omega _{0}\left( Z\right) }{2}\right) 
\label{fms}
\end{equation}%
The variables $\Delta T$ and $\Delta \hat{T}$ are the deviations of
connectivities with respect to the averages: 
\begin{eqnarray*}
\Delta T &=&T-\left\langle T\right\rangle  \\
\Delta \hat{T} &=&\hat{T}-\left\langle \hat{T}\right\rangle 
\end{eqnarray*}%
with $\left\langle T\right\rangle $ and $\left\langle \hat{T}\right\rangle $
are averages of $T$ and $\hat{T}$\ in the background field. The potential%
\begin{equation*}
U_{\Delta \Gamma }\left( \left\Vert \Delta \Gamma \left( Z,Z^{\prime
}\right) \right\Vert ^{2}\right) 
\end{equation*}%
is the second order expansion of $U\left( \left\{ \left\vert \Gamma \left(
\theta ,Z,Z^{\prime },C,D\right) \right\vert ^{2}\right\} \right) $ around $%
\Gamma _{0}$ and $\Gamma _{0}^{\dag }$.

\subsection{Static collective states}

In (\cite{GLt}), we derived the conditions of existence for collective
states of the effective action (\ref{fcp}). Such state are described as a
collection of shifted states. The values of these shifts depend both on the
characteristics of the background field and the potential $U_{\Delta \Gamma }
$. For shifted states, the average connectivities are modified: $%
\left\langle \hat{T}\right\rangle \rightarrow \left\langle \hat{T}%
\right\rangle +\underline{\left\langle \Delta \hat{T}\right\rangle }$ and $%
\left\langle T\right\rangle \rightarrow \left\langle T\right\rangle +%
\underline{\left\langle \Delta T\right\rangle }$, where the values of the
shift $\underline{\left\langle \Delta \hat{T}\right\rangle }$ and $%
\underline{\left\langle \Delta T\right\rangle }$\ are computed in (\cite{GLt}%
).\ \ For later purposes, in appendix 1 we recall the derivation of these
states and the associated average shifts along with the description of
collective states.

\subsubsection{Formula for shifted states}

The static collective states are described by a set $W$ of doublet, such
that if $\left( Z,Z^{\prime }\right) \in W$ \ The solutions to the saddle
point equation of (\ref{fcp}) at these points become:%
\begin{eqnarray}
&&\Delta \Gamma _{\delta }\left( T,\hat{T},\theta ,Z,Z^{\prime }\right)
\label{SF} \\
&=&\exp \left( -\frac{1}{2}\left( \mathbf{\Delta T-}\underline{\left\langle 
\mathbf{\Delta T}\right\rangle }\right) ^{t}\hat{U}\left( \mathbf{\Delta T-}%
\underline{\left\langle \mathbf{\Delta T}\right\rangle }\right) \right) 
\notag \\
&&\times H_{p}\left( \left( \mathbf{\Delta T}^{\prime }\mathbf{-}\underline{%
\left\langle \mathbf{\Delta T}\right\rangle }^{\prime }\right) _{2}\frac{%
\sigma _{T}\lambda _{+}}{2\sqrt{2}}\left( \mathbf{\Delta T}^{\prime }\mathbf{%
-}\underline{\left\langle \mathbf{\Delta T}\right\rangle }^{\prime }\right)
_{2}\right) H_{p-\delta }\left( \left( \mathbf{\Delta T}^{\prime }\mathbf{-}%
\underline{\left\langle \mathbf{\Delta T}\right\rangle }^{\prime }\right)
_{2}\frac{\sigma _{\hat{T}}\lambda _{-}}{2\sqrt{2}}\left( \mathbf{\Delta T}%
^{\prime }\mathbf{-}\underline{\left\langle \mathbf{\Delta T}\right\rangle }%
^{\prime }\right) _{2}\right)  \notag
\end{eqnarray}%
and:%
\begin{eqnarray}
&&\Delta \Gamma _{\delta }^{\dagger }\left( T,\hat{T},\theta ,Z,Z^{\prime
}\right)  \label{CJ} \\
&=&H_{p}\left( \left( \mathbf{\Delta T}^{\prime }\mathbf{-}\underline{%
\left\langle \mathbf{\Delta T}\right\rangle }^{\prime }\right) _{2}\frac{%
\sigma _{T}\lambda _{+}}{2\sqrt{2}}\left( \mathbf{\Delta T}^{\prime }\mathbf{%
-}\underline{\left\langle \mathbf{\Delta T}\right\rangle }^{\prime }\right)
_{2}\right) H_{p-\delta }\left( \left( \mathbf{\Delta T}^{\prime }\mathbf{-}%
\underline{\left\langle \mathbf{\Delta T}\right\rangle }^{\prime }\right)
_{2}\frac{\sigma _{\hat{T}}\lambda _{-}}{2\sqrt{2}}\left( \mathbf{\Delta T}%
^{\prime }\mathbf{-}\underline{\left\langle \mathbf{\Delta T}\right\rangle }%
^{\prime }\right) _{2}\right)  \notag
\end{eqnarray}%
where $H_{p}$ and $H_{p-\delta }$ are Hermite polynomials.

The variables involved are:%
\begin{eqnarray}
\mathbf{\Delta T-}\underline{\left\langle \mathbf{\Delta T}\right\rangle }
&=&\left( 
\begin{array}{c}
\Delta T-\underline{\left\langle \Delta T\right\rangle } \\ 
\Delta \hat{T}-\underline{\left\langle \Delta \hat{T}\right\rangle }%
\end{array}%
\right) \\
\mathbf{\Delta T}^{\prime }\mathbf{-}\underline{\left\langle \mathbf{\Delta T%
}\right\rangle }^{\prime } &=&P^{-1}\left( \mathbf{\Delta T-}\underline{%
\left\langle \mathbf{\Delta T}\right\rangle }\right)  \notag
\end{eqnarray}%
The matrices and parameters involved are provided in the appendices, along
with the formula for $\underline{\left\langle \Delta T\right\rangle }$, $%
\underline{\left\langle \Delta \hat{T}\right\rangle }$.

The density of connectivities between $Z$ and $Z^{\prime }$ is given by $%
\left\vert \Delta \Gamma _{\delta }\left( T,\hat{T},\theta ,Z,Z^{\prime
}\right) \right\vert ^{2}$and can be understood as follows: regardless of
how the system is interpreted, whether as a set of groups of simple cells or
single complex cells at each point, the stable backgrounds are not defined
with a specific connectivity value. On the contrary, the background states
are described by a distribution around some average value. In other words,
the cells or groups of axons/dendrites are connected with srength of
connectivities that are distributed around this average.

\subsubsection{Average shifts}

When a shifted state exists, the average shifts are given by:%
\begin{equation}
\underline{\left\langle \Delta T\right\rangle }\simeq \frac{\omega
_{0}\left( Z\right) \left\langle T\right\rangle }{\rho D\left( \theta
\right) \left\langle \hat{T}\right\rangle \left\vert \Psi _{0}\left(
Z^{\prime }\right) \right\vert ^{2}k\underline{A_{1}}\left\Vert \Delta
\Gamma \right\Vert ^{6}}\left\langle \rho \frac{\left\vert \bar{\Psi}%
_{0}\left( Z,Z^{\prime }\right) \right\vert ^{2}}{A}\right\rangle
^{2}\left\langle T\right\rangle
\end{equation}%
and:%
\begin{equation}
\left\langle \Delta \right\rangle \hat{T}=\hat{A}\underline{\left\langle
\Delta T\right\rangle }
\end{equation}%
with:%
\begin{equation*}
A_{1}\left( Z,Z^{\prime }\right) =\frac{\rho D\left( \theta \right)
\left\langle \hat{T}\right\rangle \left\vert \Psi _{0}\left( Z^{\prime
}\right) \right\vert ^{2}}{\omega _{0}\left( Z\right) }\underline{A_{1}}%
\left( Z,Z^{\prime }\right)
\end{equation*}%
and:%
\begin{equation*}
\underline{A_{1}}\left( Z,Z^{\prime }\right) =\left\langle \left[ F\left(
Z_{2},Z_{2}^{\prime }\right) \left[ \check{T}\left( 1-\left\langle
\left\vert \Psi _{0}\right\vert ^{2}\right\rangle \frac{\underline{%
\left\langle \Delta T\right\rangle }}{T}\left\Vert \Delta \Gamma \right\Vert
^{2}\right) ^{-1}O\right] \right] _{\left( T,\hat{T},\theta ,Z,Z^{\prime
}\right) }\right\rangle
\end{equation*}%
\begin{equation}
\hat{A}\simeq -\frac{1}{v}A\frac{\left\Vert \Delta \Gamma \right\Vert ^{2}}{%
\left\langle T\right\rangle }
\end{equation}

\subsubsection{Conditions for shifted state}

Shifted connectivities emerge at certain points under stability conditions.
Depending on the potential and the activity of cells at these points,
connectivity may either be enhanced or reduced. We established the
conditions for the existence of a shift in a previous work (\cite{GLt}). The
existence conditions for such a set depend on the background $\left\vert
\Psi _{0}\left( Z\right) \right\vert ^{2}$ and the potential $U_{\Delta
\Gamma }^{\prime \prime }$:%
\begin{equation}
\left\vert \left( u+v\right) -\left\langle u+v\right\rangle \right\vert <%
\sqrt{-8U_{\Delta \Gamma }\left( \left\Vert \Delta \Gamma \left( Z,Z^{\prime
}\right) \right\Vert _{\min }^{2}\right) U_{\Delta \Gamma }^{\prime \prime
}\left( \left\Vert \Delta \Gamma \left( Z,Z^{\prime }\right) \right\Vert
_{\min }^{2}\right) }  \label{thr}
\end{equation}%
where $\left\Vert \Delta \Gamma \left( Z,Z^{\prime }\right) \right\Vert
_{\min }^{2}$ is the minimum of the potential $U_{\Delta \Gamma }$ at $%
\left( Z,Z^{\prime }\right) $ and:

\begin{eqnarray*}
u &=&\frac{\left\vert \Psi _{0}\left( Z\right) \right\vert ^{2}}{\tau \omega
_{0}\left( Z\right) } \\
v &=&\rho C\frac{\left\vert \Psi _{0}\left( Z\right) \right\vert
^{2}h_{C}\left( \omega _{0}\left( Z\right) \right) }{\omega _{0}\left(
Z\right) }+\rho D\frac{\left\vert \Psi _{0}\left( Z^{\prime }\right)
\right\vert ^{2}h_{D}\left( \omega _{0}\left( Z^{\prime }\right) \right) }{%
\omega _{0}\left( Z\right) } \\
s &=&-\frac{\lambda \left\vert \Psi _{0}\left( Z\right) \right\vert ^{2}}{%
\omega _{0}\left( Z\right) }
\end{eqnarray*}

The bracket $\left\langle u+v\right\rangle $ represents the average of $u+v$
over the entire space. Therefore, the condition (\ref{thr}) for a shift is
relative. The main factor for allowing the emergence of modified
connectivities is the relative level of activity and frequencies with
respect to the entire system.

This condition means that, in first approximation, $\left\vert \Psi
_{0}\left( Z\right) \right\vert ^{2}$ must be below a threshold provided by
the right-hand-side of (\ref{thr}) for a state with enhanced connectivities
to exist.

\section{Description of dynamic emergent collective states}

Our earlier results were obtained by averaging over the entire system to
identify the elements that would become activated when connectivity was
modified. However, the resulting states themselves were not extensively
studied, especially in terms of their group interactions or the interactions
between different potential collective states. Additionally, there was no
mention of the activities associated with these possible states.

In the current context, when a group of states experiences a shift in
connectivity, these states are expected to interact collectively due to
these changes. Furthermore, several group of states should interact with
each other. These interactions are contingent upon the activities of each
constituent element within the group. The primary objective is to describe
the characteristic activity patterns for such independent groups of
interconnected cells and investigate the role of these characteristics in
their interactions.

To achieve this, we need to introduce dynamic aspects into the description
of collective states and expand the formalism to encompass interacting
groups of collective states. This begins with a reworking of (\ref{fcp}) to
incorporate dynamic aspects of interactions between elements within the
activated group. In (\cite{GLt}), the effective action (\ref{fcp}) was
initially derived to ascertain conditions for the emergence of states, and
it was sufficient to work with the neurons background activity.
Consequently, we will assume that neuronal activity is given by:%
\begin{equation*}
\omega \left( \theta ,Z,\left\vert \Psi \right\vert ^{2}\right) =\omega
_{0}\left( Z\right) +\Delta \omega \left( \theta ,Z,\left\vert \Psi
\right\vert ^{2}\right)
\end{equation*}%
where $\Delta \omega \left( \theta ,Z,\left\vert \Psi \right\vert
^{2}\right) $ is the internal activity of the group. We will find the
formula for the possible values $\Delta \omega \left( \theta ,Z,\left\vert
\Psi \right\vert ^{2}\right) $ and this will describe the set of possible
activities for the emerging states. In fact we will look specifically for
stable\ oscillating forms for $\Delta \omega \left( \theta ,Z,\left\vert
\Psi \right\vert ^{2}\right) $. This choice is justified in detail in
appendices 2 and 3 based on (\cite{GL}), (\cite{GLr}), (\cite{GLs}). The
reason relies on a field-theoretic perturbative argument in (\cite{GL}): a
perturbation in activities modifies the background state $\Psi $ which
compensate any dampening or enhancement in activity oscillations, resulting
in stable patterns.

\subsection{Effective action}

We assume the existence of states with finite\ set $S^{2}$ $=\left\{ \left(
Z,Z^{\prime }\right) \right\} $, with $\left\Vert \Delta \Gamma \left( T,%
\hat{T},\theta ,Z,Z^{\prime }\right) \right\Vert ^{2}\neq 0$\footnote{%
The condition of existence for such collective state in a dynamic context is
discussed below.}. Introducing the averages:%
\begin{equation*}
\left( \underline{\left\langle \Delta T\right\rangle },\underline{%
\left\langle \Delta \hat{T}\right\rangle }\right)
\end{equation*}%
that are both implicitely dependent on doublet of points $\left( Z,Z^{\prime
}\right) $, we can rewrite the effective action (\ref{fcp}) for the elements
of the group of activated states as:

\begin{eqnarray}
&&S\left( \Delta \Gamma \left( T,\hat{T},\theta ,Z,Z^{\prime }\right) \right)
\label{TRM} \\
&=&-\Delta \Gamma ^{\dag }\left( T,\hat{T},\theta ,Z,Z^{\prime }\right)
\left( \nabla _{T}\left( \nabla _{T}+\frac{\left( \Delta T-\underline{%
\left\langle \Delta T\right\rangle }\right) -\lambda \left( \Delta \hat{T}-%
\underline{\left\langle \Delta \hat{T}\right\rangle }\right) }{\tau \left(
\omega _{0}\left( Z\right) +\Delta \omega \left( \theta ,Z,\left\vert \Psi
\right\vert ^{2}\right) \right) }\right) \right) \Delta \Gamma \left( T,\hat{%
T},\theta ,Z,Z^{\prime }\right)  \notag \\
&&-\Delta \Gamma ^{\dag }\left( T,\hat{T},\theta ,Z,Z^{\prime }\right)
\nabla _{\hat{T}}\left( \nabla _{\hat{T}}+\left\vert \bar{\Psi}_{0}\left(
Z,Z^{\prime }\right) \right\vert ^{2}\left( \Delta \hat{T}-\underline{%
\left\langle \Delta \hat{T}\right\rangle }\right) \right) \Delta \Gamma
\left( T,\hat{T},\theta ,Z,Z^{\prime }\right) +U_{\Delta \Gamma }\left(
\left\Vert \Delta \Gamma \left( Z,Z^{\prime }\right) \right\Vert ^{2}\right)
\notag
\end{eqnarray}

Where the sum over $Z\in S$ and $Z^{\prime }\in S$ is implicit. However,
this formula relies on the averages background activities $\omega _{0}\left(
Z\right) $, since it was designed to find average conditions for emergence
of modified states. Once a set of cells are activated, their interaction
implies some additional activity frequency $\Delta \omega \left( \theta
,Z,\left\vert \Psi \right\vert ^{2}\right) $ inducing a modification of
action that becomes:%
\begin{equation}
\hat{S}\left( \Delta \Gamma \left( T,\hat{T},\theta ,Z,Z^{\prime }\right)
\right) =S\left( \Delta \Gamma \left( T,\hat{T},\theta ,Z,Z^{\prime }\right)
\right) -\Delta V\left( \Delta \Gamma ,\Delta \Gamma ^{\dag }\right)
\label{MFD}
\end{equation}%
where:%
\begin{eqnarray}
&&\Delta V\left( \Delta \Gamma ,\Delta \Gamma ^{\dag }\right) =\Delta \Gamma
^{\dag }\left( T,\hat{T},\theta ,Z,Z^{\prime }\right)  \label{TSN} \\
&&\times \left( \nabla _{\hat{T}}\left( \frac{\rho \left( D\left( \theta
\right) \left\langle \hat{T}\right\rangle \left\vert \Psi _{0}\left(
Z^{\prime }\right) \right\vert ^{2}\left( \left( \left( Z-Z^{\prime }\right)
\left( \nabla _{Z}+\nabla _{Z}\omega _{0}\left( Z\right) \right) +\frac{%
\left\vert Z-Z^{\prime }\right\vert }{c}\right) \Delta \omega \left( \theta
,Z,\left\vert \Psi \right\vert ^{2}\right) \right) \right) }{\omega
_{0}\left( Z\right) +\Delta \omega \left( \theta ,Z,\left\vert \Psi
\right\vert ^{2}\right) }\right) \right)  \notag \\
&&\times \Delta \Gamma \left( T,\hat{T},\theta ,Z,Z^{\prime }\right)  \notag
\end{eqnarray}%
This additional term is similar to the potential (\ref{pnl}) but accounts
for an additional interaction between the activated elements as implied by
the terms $\Delta \omega \left( \theta ,Z,\left\vert \Psi \right\vert
^{2}\right) $. Eventhough similar to (\ref{pnl}), this term was not present
in (\ref{fcp}) since this action describes the possibility of emerging
groups, not the internal dynamics of such states.

To consider \ the possible dynamics of the group as a whole, we first
consider its possible dynamic activities given its shape and connectivity
magnitude.

\subsection{Dynamic activities $\Delta \protect\omega \left( \protect\theta %
,Z,\left\vert \Psi \right\vert ^{2}\right) $ of collective stats}

In a state with $S^{2}$ $=\left\{ \left( Z,Z^{\prime }\right) \right\} $, we
aim at finding the possible activity frequencies $\omega \left( \theta
,Z,\left\vert \Psi \right\vert ^{2}\right) $ associated to the state $\Delta
T$. \ We consider the state $\left\{ \Delta T\left( Z,Z^{\prime }\right)
\right\} $ as a system with its own associated activity $\Delta \omega
\left( \theta ,Z,\left\vert \Psi \right\vert ^{2}\right) $ in the given
background field $\left\vert \Psi \right\vert ^{2}$.

To find $\omega \left( \theta ,Z,\left\vert \Psi \right\vert ^{2}\right) $,
we start with the defining equation: 
\begin{eqnarray*}
&&\left( \omega _{0}+\Delta \omega \right) \left( \theta ,Z,\left\vert \Psi
\right\vert ^{2}\right) \\
&=&G\left( \int \frac{\kappa }{N}\frac{\left( \omega _{0}+\Delta \omega
\right) \left( J,\theta -\frac{\left\vert Z-Z_{1}\right\vert }{c},Z_{1},\Psi
\right) \left( \left\langle T\right\rangle +\left\langle \Delta
T\right\rangle \right) \left( Z,\theta ,Z_{1},\theta -\frac{\left\vert
Z-Z_{1}\right\vert }{c}\right) }{\left( \omega _{0}+\Delta \omega \right)
\left( J,\theta ,Z,\left\vert \Psi \right\vert ^{2}\right) }\left\vert \Psi
\left( \theta -\frac{\left\vert Z-Z_{1}\right\vert }{c},Z_{1}\right)
\right\vert ^{2}dZ_{1}\right)
\end{eqnarray*}

The notation $\omega _{0}+\Delta \omega $ stands for the background plus
modified activity and $\left\langle T\right\rangle +\left\langle \Delta
T\right\rangle $ is the sum of the average connectivty in the background
state plus the average modification when the shift occurs.

We show in appendix 3 that $\Delta \omega ^{-1}\left( \theta ,Z,\left\vert
\Psi \right\vert ^{2}\right) $ decomposes into a static part and a variable
part.

\subsubsection{Static part of the activities}

We decompose the additional activity into a static and a dynamic
contribution:%
\begin{equation*}
\Delta \omega \left( \theta ,Z,\left\vert \Psi \right\vert ^{2}\right) =%
\overline{\Delta \omega }\left( Z,\left\vert \Psi \right\vert ^{2}\right)
+\Delta \omega _{D}\left( \theta ,Z,\left\vert \Psi \right\vert ^{2}\right)
\end{equation*}%
so that the overall activity is:%
\begin{equation*}
\omega \left( \theta ,Z,\left\vert \Psi \right\vert ^{2}\right) =\omega
_{0}\left( Z\right) +\overline{\Delta \omega }\left( Z,\left\vert \Psi
\right\vert ^{2}\right) +\Delta \omega _{D}\left( \theta ,Z,\left\vert \Psi
\right\vert ^{2}\right)
\end{equation*}%
The static part of the activity:%
\begin{equation*}
\omega _{S}\left( Z,\left\vert \Psi \right\vert ^{2}\right) =\omega
_{0}\left( Z\right) +\overline{\Delta \omega }\left( Z,\left\vert \Psi
\right\vert ^{2}\right)
\end{equation*}%
satisfies:%
\begin{eqnarray*}
&&\omega _{S}\left( Z,\left\vert \Psi \right\vert ^{2}\right) \\
&=&G\left( \int \frac{\kappa }{N}\frac{\omega _{S}\left( J,Z_{1},\Psi
\right) \left( \left\langle T\right\rangle +\left\langle \Delta
T\right\rangle \right) \left( Z,\theta ,Z_{1}\right) }{\omega _{S}\left(
J,Z,\left\vert \Psi \right\vert ^{2}\right) }\left\vert \Psi \left(
Z_{1}\right) \right\vert ^{2}dZ_{1}\right)
\end{eqnarray*}%
The notation $\omega _{0}+\Delta \omega $ stands for the background plus
modified activity.

Using that the modifications states is a group over a bounded domain and
involve some finite number of points we find the equation for the static
modification at each point $Z_{i}$ of this group:%
\begin{equation*}
\omega _{S}\left( Z_{i},\left\vert \Psi \right\vert ^{2}\right) =G\left(
\sum_{j}\frac{\kappa }{N}\frac{\omega _{S}\left( Z_{j},\Psi \right) \left(
\left\langle T\right\rangle +\left\langle \Delta T\right\rangle \right)
\left( Z_{i},Z_{j}\right) }{\omega _{S}\left( Z_{i},\left\vert \Psi
\right\vert ^{2}\right) }\left\vert \Psi \left( Z_{j}\right) \right\vert
^{2}\right)
\end{equation*}%
with solution:%
\begin{equation*}
\omega _{S}\left( \mathbf{Z},\mathbf{T},\left\vert \Psi \right\vert
^{2}\right) =\omega _{0}\left( Z\right) +\overline{\Delta \omega }\left( 
\mathbf{Z},\mathbf{T},\left\vert \Psi \right\vert ^{2}\right)
\end{equation*}%
where $\omega _{S}\left( \mathbf{Z},\mathbf{T},\left\vert \Psi \right\vert
^{2}\right) $ is the vector with coordinates $\omega _{S}\left(
Z_{i},\left\vert \Psi \right\vert ^{2}\right) $.

\subsubsection{Variable oscillatory part of the activities}

The variable part is a first order variation:%
\begin{equation*}
\Delta \omega _{D}\left( J,\theta ,Z,\left\vert \Psi \right\vert ^{2}\right)
\end{equation*}%
In fact this variable part should induce some fluctuations in $\left\vert
\Psi \right\vert ^{2}$. We show in appendix 2 that we can equivalently
consider $\left\vert \Psi \right\vert ^{2}$ to remain constant while the
variable part exhibits stable oscillations. Therefore, in a first
approximation, we can investigate stable oscillations for $\Delta \omega
\left( J,\theta ,Z,\left\vert \Psi \right\vert ^{2}\right) $.

Using the fact that the modified states form a group over a bounded domain
and involve a finite number of points we find in appendix 3 the equation for
this modification at each point $Z_{i}$ of this group: 
\begin{equation}
\Delta \omega _{D}\left( \theta ,Z_{i},\left\vert \Psi \right\vert
^{2}\right) =\sum_{j}\check{T}\left( Z_{i},Z_{j}\right) \Delta \omega
_{D}\left( \theta -\frac{\left\vert Z_{i}-Z_{j}\right\vert }{c},Z_{j},\Psi
\right)   \label{FRQ}
\end{equation}%
where:%
\begin{equation*}
\check{T}\left( Z_{i},Z_{j}\right) =\frac{\kappa }{N}G^{\prime }\left(
G^{-1}\left( \omega _{S}^{-1}\left( J,Z,\left\vert \Psi \right\vert
^{2}\right) \right) \right) \frac{T\left( Z_{i},Z_{j}\right) \omega
_{S}\left( J,Z_{j},\Psi \right) \left\vert \Psi _{0}\left( Z_{j}\right)
\right\vert ^{2}}{G^{\prime }\left( G^{-1}\left( \omega _{S}^{-1}\left(
J,Z,\left\vert \Psi \right\vert ^{2}\right) \right) \right) \omega
_{S}\left( J,Z_{1},\Psi \right) +\omega _{S}^{2}\left( J,Z,\left\vert \Psi
\right\vert ^{2}\right) }
\end{equation*}

Appendix 3 shows the existence of stable oscillatory solutions:%
\begin{equation}
\Delta \omega _{D}\left( \theta ,\mathbf{Z},\left\vert \Psi \right\vert
^{2}\right) =A\left( Z_{1}\right) \left( 1,\left( 1-\mathbf{\check{T}}\exp
\left( -i\Upsilon _{p}\frac{\left\vert \mathbf{\Delta Z}\right\vert }{c}%
\right) \right) ^{-1}\check{T}_{1}\left( \mathbf{Z}\right) \exp \left(
-i\Upsilon _{p}\frac{\left\vert \mathbf{\Delta Z}_{1}\right\vert }{c}\right)
\right) ^{t}\exp \left( i\Upsilon _{p}\left( \mathbf{\check{T}}\right)
\theta \right)   \label{VR}
\end{equation}%
where $\Delta \omega _{D}\left( \theta ,\mathbf{Z},\left\vert \Psi
\right\vert ^{2}\right) $ and $\check{T}_{1}\left( \mathbf{Z}\right) $\ are
vectors with coordinates $\Delta \omega _{D}\left( J,\theta
,Z_{i},\left\vert \Psi \right\vert ^{2}\right) $ and $\check{T}\left(
Z_{1},Z_{j}\right) $ respectively. The point $Z_{1}$ is an arbitrary point
chosen in the group and $A\left( Z_{1}\right) $ is the amplitude of $\Delta
\omega _{D}$ at some given $Z_{1}$.

The matrix $\mathbf{\check{T}}$ has elements $\check{T}\left(
Z_{i},Z_{j}\right) $ and the frequencies $\Upsilon _{p}\left( \mathbf{\check{%
T}}\right) $ belong to a discrete set satisfying the equation:%
\begin{equation}
\det \left( 1-\check{T}\left( Z_{i},Z_{j}\right) \exp \left( -i\Upsilon _{p}%
\frac{\left\vert Z_{i}-Z_{j}\right\vert }{c}\right) \right) =0  \label{Frc}
\end{equation}%
The possible oscillatory activities associated to the assembly is thus given
by the sets:%
\begin{equation*}
\left\{ \left\{ A\left( Z_{i}\right) \right\} _{i=1,...n},\Upsilon
_{p}\left( \left\{ \check{T}\left( Z_{i},Z_{j}\right) \right\} \right)
\right\} _{p}
\end{equation*}%
where $p$ refers to the frequencies $\Upsilon $, solutions of (\ref{FRQ}):%
\begin{equation*}
A\left( Z_{i}\right) =\sum_{j\neq i}A\left( Z_{j}\right) \check{T}\left(
Z_{i},Z_{j}\right) \exp \left( -i\Upsilon _{p}\left( \left\{ \check{T}\left(
Z_{i},Z_{j}\right) \right\} \right) \frac{\left\vert Z_{i}-Z_{j}\right\vert 
}{c}\right)
\end{equation*}%
the $\Upsilon _{p}\left( \check{T}\left( Z_{i},Z_{j}\right) \right) $ are
solutions of (\ref{Frc}).

\subsubsection{Overall activity of the collective state}

Gathering (\ref{FRQ}) and (\ref{VR}), the solution for the specific
frequencies of the group is:

\begin{eqnarray*}
&&\Delta \omega \left( \theta ,Z_{i},\mathbf{T},\left\vert \Psi \right\vert
^{2}\right) \\
&=&\overline{\Delta \omega }\left( \mathbf{Z},\mathbf{T},\left\vert \Psi
\right\vert ^{2}\right) \\
&&+A\left( Z_{1}\right) \left( 1,\left( 1-\mathbf{\check{T}}\exp \left(
-i\Upsilon _{p}\left( \mathbf{\check{T}}\right) \frac{\left\vert \mathbf{%
\Delta Z}\right\vert }{c}\right) \right) _{1}^{-1}\check{T}_{1}\left( 
\mathbf{Z}\right) \exp \left( -i\Upsilon _{p}\left( \mathbf{\check{T}}%
\right) \frac{\left\vert \mathbf{\Delta Z}_{1}\right\vert }{c}\right)
\right) ^{t}\exp \left( i\Upsilon _{p}\left( \mathbf{\check{T}}\right)
\theta \right)
\end{eqnarray*}%
where:%
\begin{eqnarray*}
\left( \check{T}_{1}\left( \mathbf{Z}\right) \right) _{i} &=&\check{T}\left(
Z_{i},Z_{1}\right) \\
\Upsilon _{p}\left( \mathbf{\check{T}}\right) &=&\Upsilon _{p}\left( \left\{ 
\check{T}\left( Z_{i},Z_{j}\right) \right\} \right)
\end{eqnarray*}%
The amplitde $A\left( Z_{1}\right) $ for oscillations is not arbitrary.
Actually, we show in Part $0$ that $A\left( Z_{1}\right) $ depends on the
parameters of the system, the field $\Psi _{0}$, th bckgr ctv $\omega _{0}$
and the potential for $\Psi _{0}$.

\subsection{averages computations for activated states}

Now, considering (\ref{MFD}) for the group of shifted states, we rewrite the
action taking into account their particular interactions. Having obtained
the frequencies, we can come back to the action minimization and compute the
connectivity states along with average connectivities. To do so, we replace
in (\ref{TSN}) (see (\cite{GLt})):%
\begin{eqnarray}
&&\left( \left( \left( Z-Z^{\prime }\right) \left( \nabla _{Z}+\nabla
_{Z}\left( \omega _{0}+\overline{\Delta \omega }\right) \left( Z,\left\vert
\Psi \right\vert ^{2}\right) \right) +\frac{\left\vert Z-Z^{\prime
}\right\vert }{c}\right) \Delta \omega \left( \theta ,Z,\left\vert \Psi
\right\vert ^{2}\right) \right)   \label{dtn} \\
&\rightarrow &\frac{\left\vert Z-Z^{\prime }\right\vert }{c}\Delta \omega
\left( \theta ,Z,\left\vert \Psi \right\vert ^{2}\right)   \notag \\
&\rightarrow &g\left\vert Z-Z^{\prime }\right\vert \Delta \omega \left(
\theta ,Z,\left\vert \Psi \right\vert ^{2}\right)   \notag
\end{eqnarray}%
At the connectivity time scale we replace $\Delta \omega \left( \theta
,Z,\left\vert \Psi \right\vert ^{2}\right) $ by its static part $\overline{%
\Delta \omega }\left( Z,\left\vert \Psi \right\vert ^{2}\right) $. \
Replacing this formula in (\ref{TSN}), using (\ref{MFD}) and (\ref{TRM}),
leads to the following action:%
\begin{eqnarray}
&&\hat{S}\left( \Delta \Gamma \left( T,\hat{T},\theta ,Z,Z^{\prime }\right)
\right)   \label{ctv} \\
&=&-\Delta \Gamma ^{\dag }\left( T,\hat{T},\theta ,Z,Z^{\prime }\right)
\left( \nabla _{T}\left( \nabla _{T}+\frac{\left( \Delta T-\underline{%
\left\langle \Delta T\right\rangle }\right) -\lambda \left( \Delta \hat{T}-%
\underline{\left\langle \Delta \hat{T}\right\rangle }\right) }{\tau \omega
_{S}\left( Z,\left\vert \Psi \right\vert ^{2}\right) }\left\vert \Psi \left(
\theta ,Z\right) \right\vert ^{2}\right) \right) \Delta \Gamma \left( T,\hat{%
T},\theta ,Z,Z^{\prime }\right)   \notag \\
&&-\Delta \Gamma ^{\dag }\left( T,\hat{T},\theta ,Z,Z^{\prime }\right)
\nabla _{\hat{T}}\left( \nabla _{\hat{T}}+\left\vert \bar{\Psi}_{0}\left(
Z,Z^{\prime }\right) \right\vert ^{2}\left( \Delta \hat{T}-\left( \underline{%
\left\langle \Delta \hat{T}\right\rangle }+\Delta ^{\omega }\left\langle 
\hat{T}\right\rangle \right) \right) \right) \Delta \Gamma \left( T,\hat{T}%
,\theta ,Z,Z^{\prime }\right)   \notag \\
&&+U_{\Delta \Gamma }\left( \left\Vert \Delta \Gamma \left( Z,Z^{\prime
}\right) \right\Vert ^{2}\right)   \notag
\end{eqnarray}%
with:%
\begin{equation*}
\omega _{S}=\omega _{0}+\overline{\Delta \omega }
\end{equation*}%
and:%
\begin{equation}
\Delta ^{\omega }\hat{T}=\frac{\left( D\left( \theta \right) \left\langle 
\hat{T}\right\rangle \left\vert \Psi _{0}\left( Z^{\prime }\right)
\right\vert ^{2}\left\vert Z-Z^{\prime }\right\vert g\overline{\Delta \omega 
}\left( Z,\mathbf{T},\left\vert \Psi \right\vert ^{2}\right) \right) }{%
\left( C\left( \theta \right) \left\vert \Psi _{0}\left( Z\right)
\right\vert ^{2}\omega _{S}\left( Z,\left\vert \Psi \right\vert ^{2}\right)
+D\left( \theta \right) \left\langle \hat{T}\right\rangle \left\vert \Psi
_{0}\left( Z^{\prime }\right) \right\vert ^{2}\left( \omega _{0}+\overline{%
\Delta \omega }\right) \left( Z^{\prime },\left\vert \Psi \right\vert
^{2}\right) \right) }
\end{equation}%
Then we define $\left\langle \Delta T\right\rangle $ and $\left\langle
\Delta \hat{T}\right\rangle $ as solutions of:%
\begin{eqnarray*}
\left\langle \Delta T\right\rangle  &=&\underline{\left\langle \Delta
T\right\rangle }+\lambda \left\langle \Delta ^{\omega }\hat{T}\right\rangle 
\\
\left\langle \Delta \hat{T}\right\rangle  &=&\underline{\left\langle \Delta 
\hat{T}\right\rangle }+\left\langle \Delta ^{\omega }\hat{T}\right\rangle 
\end{eqnarray*}%
The first equation defines the set $\left\langle \Delta T\right\rangle $ and
the second one yields $\left\langle \Delta \hat{T}\right\rangle $. Writing: $%
\left\langle \Delta ^{\omega }\hat{T}\right\rangle =\frac{\left\langle
\Delta T\right\rangle -\underline{\left\langle \Delta T\right\rangle }}{%
\lambda }$, the averaged equation becomes:%
\begin{eqnarray}
&&\frac{\left\langle \Delta T\right\rangle -\underline{\left\langle \Delta
T\right\rangle }}{\lambda } \\
&\simeq &\frac{D\left( \theta \right) \left\langle \hat{T}\right\rangle
\left\vert \Psi _{0}\left( Z^{\prime }\right) \right\vert ^{2}\left\vert
Z-Z^{\prime }\right\vert g\overline{\Delta \omega }\left( Z,\left\langle 
\mathbf{T}\right\rangle +\left\langle \Delta \mathbf{T}\right\rangle
,\left\vert \Psi \right\vert ^{2}\right) }{\left( C\left( \theta \right)
\left\vert \Psi _{0}\left( Z\right) \right\vert ^{2}\omega _{S}\left(
Z,\left\langle \mathbf{T}\right\rangle +\left\langle \Delta \mathbf{T}%
\right\rangle ,\left\vert \Psi \right\vert ^{2}\right) +D\left( \theta
\right) \left\langle \hat{T}\right\rangle \left\vert \Psi _{0}\left(
Z^{\prime }\right) \right\vert ^{2}\omega _{S}\left( Z^{\prime
},\left\langle \mathbf{T}\right\rangle +\left\langle \Delta \mathbf{T}%
\right\rangle ,\left\vert \Psi \right\vert ^{2}\right) \right) }  \notag
\end{eqnarray}

There are several solutions:%
\begin{equation}
\left( \left\langle \Delta \mathbf{T}\right\rangle ^{\alpha },\left\langle
\Delta \mathbf{\hat{T}}\right\rangle ^{\alpha }\right)  \label{VRS}
\end{equation}%
For each of these solutions, a sequence of frequencies $\left( \Upsilon
_{p}^{\alpha }\right) $ satisfying (\ref{Frc}) are compatible. The variable
contribution of activities is given by:%
\begin{equation*}
\Delta \omega _{p}^{\alpha }\left( \theta ,Z,\mathbf{\Delta T}\right) =%
\overline{\Delta \omega }\left( Z,\mathbf{T},\left\vert \Psi \right\vert
^{2}\right) +\left( \mathbf{N}_{p}^{\alpha }\right) ^{-1}\Delta \mathbf{%
\omega }_{0}\exp \left( -i\Upsilon _{p}^{\alpha }\frac{\left\vert \Delta 
\mathbf{Z}_{i}\right\vert }{c}\right)
\end{equation*}%
where:%
\begin{equation*}
\left[ \mathbf{N}_{p}^{\alpha }\right] _{\left( Z_{i},Z_{j}\right) }=\left(
\delta _{ij}-\left[ \mathbf{\Delta T}\right] _{\left( Z_{i},Z_{j}\right)
}\exp \left( -i\Upsilon _{p}\frac{\left\vert Z_{i}-Z_{j}\right\vert }{c}%
\right) \right) ^{-1}
\end{equation*}

\subsection{Form of the activated state}

We will now derive the formula for $\Delta \Gamma $ and its conjugate $%
\Delta \Gamma ^{\dag }$ in the activated state. In appendix 1, we show that
after a change of variables, as (\ref{ctv}), the effective action rewrites :%
\begin{eqnarray}
&&\hat{S}\left( \Delta \Gamma \left( T,\hat{T},\theta ,Z,Z^{\prime }\right)
\right)  \label{CFR} \\
&=&-\Delta \Gamma ^{\dag }\left( T,\hat{T},\theta ,Z,Z^{\prime }\right)
\left( \nabla _{T}^{2}+\nabla _{\hat{T}}^{2}-\frac{1}{2}\left( \mathbf{%
\Delta T-}\left\langle \mathbf{\Delta T}\right\rangle _{p}^{\alpha }\right)
^{t}\mathbf{A}_{p}^{\alpha }\left( \mathbf{\Delta T-}\left\langle \mathbf{%
\Delta T}\right\rangle _{p}^{\alpha }\right) \right) \Delta \Gamma \left( T,%
\hat{T},\theta ,Z,Z^{\prime }\right)  \notag \\
&&+C\left( Z,Z^{\prime }\right) \left\Vert \Delta \Gamma \left( T,\hat{T}%
,\theta ,Z,Z^{\prime }\right) \right\Vert ^{2}  \notag
\end{eqnarray}%
and that the minimization equations leads to the activated state:%
\begin{equation}
\Delta \Gamma =\prod\limits_{Z,Z^{\prime }}\left\vert \left\langle \Delta
T\right\rangle \left( Z,Z^{\prime }\right) ,\left\langle \Delta \hat{T}%
\right\rangle \left( Z,Z^{\prime }\right) ,\alpha \left( Z,Z^{\prime
}\right) ,p\left( Z,Z^{\prime }\right) \right\rangle \equiv \left\vert 
\mathbf{\alpha },\mathbf{p},S^{2}\right\rangle  \label{CST}
\end{equation}%
with $S^{2}=\left\{ \left( Z,Z^{\prime }\right) \right\} $ where the states
are activated. In a developped form, this state is given by formula similar
to (\ref{SF}) and (\ref{CJ}):%
\begin{eqnarray}
&&\left\vert \mathbf{\alpha },\mathbf{p},S^{2}\right\rangle  \label{CL} \\
&=&\exp \left( -\frac{1}{2}\left( \mathbf{\Delta T-}\left\langle \mathbf{%
\Delta T}\right\rangle _{p}^{\alpha }\right) ^{t}\mathbf{A}_{p}^{\alpha
}\left( \mathbf{\Delta T-}\left\langle \mathbf{\Delta T}\right\rangle
_{p}^{\alpha }\right) \right) H_{p}\left( \frac{1}{2}\left( \mathbf{\Delta T-%
}\left\langle \mathbf{\Delta T}\right\rangle _{p}^{\alpha }\right) ^{t}%
\mathbf{A}_{p}^{\alpha }\left( \mathbf{\Delta T-}\left\langle \mathbf{\Delta
T}\right\rangle _{p}^{\alpha }\right) \right)  \notag \\
&&\times H_{p}\left( \left( \mathbf{\Delta T}^{\prime }\mathbf{-}%
\left\langle \mathbf{\Delta T}\right\rangle _{p}^{\alpha }\right)
_{1}^{t}\left( \mathbf{D}_{p}^{\alpha }\right) _{1}\left( \mathbf{\Delta T}%
^{\prime }\mathbf{-}\left\langle \mathbf{\Delta T}\right\rangle _{p}^{\alpha
}\right) _{1}^{t}\right) H_{p-\delta }\left( \left( \mathbf{\Delta T}%
^{\prime }\mathbf{-}\left\langle \mathbf{\Delta T}\right\rangle _{p}^{\alpha
}\right) _{2}^{t}\left( \mathbf{D}_{p}^{\alpha }\right) _{2}\left( \mathbf{%
\Delta T}^{\prime }\mathbf{-}\left\langle \mathbf{\Delta T}\right\rangle
_{p}^{\alpha }\right) _{2}^{t}\right)  \notag
\end{eqnarray}%
with the conjugate:%
\begin{eqnarray}
\Delta \Gamma ^{\dag } &=&\left\langle \mathbf{\alpha },\mathbf{p}%
,S^{2}\right\vert  \label{CT} \\
&=&H_{p}\left( \left( \mathbf{\Delta T}^{\prime }\mathbf{-}\left\langle 
\mathbf{\Delta T}\right\rangle _{p}^{\alpha }\right) _{1}^{t}\left( \mathbf{D%
}_{p}^{\alpha }\right) _{1}\left( \mathbf{\Delta T}^{\prime }\mathbf{-}%
\left\langle \mathbf{\Delta T}\right\rangle _{p}^{\alpha }\right)
_{1}^{t}\right) H_{p-\delta }\left( \left( \mathbf{\Delta T}^{\prime }%
\mathbf{-}\left\langle \mathbf{\Delta T}\right\rangle _{p}^{\alpha }\right)
_{2}^{t}\left( \mathbf{D}_{p}^{\alpha }\right) _{2}\left( \mathbf{\Delta T}%
^{\prime }\mathbf{-}\left\langle \mathbf{\Delta T}\right\rangle _{p}^{\alpha
}\right) _{2}^{t}\right)  \notag
\end{eqnarray}%
where $H_{p}$ and $H_{p-\delta }$ are Hermite polynomials. The coordinates $%
\left( \mathbf{\Delta T}^{\prime }\mathbf{-}\left\langle \mathbf{\Delta T}%
\right\rangle _{p}^{\alpha }\right) _{1}^{t}$ as well as elements $\left( 
\mathbf{D}_{p}^{\alpha }\right) _{i}$ are obtained through the
diagonalization of $\mathbf{A}_{p}^{\alpha }$.

\section{Collective state for $n$ interacting fields}

In this section, we explore collective states that emerge from the
interactions of various types of fields. This section follows a similar
structure to the previous one. We begin by revisiting the findings from (%
\cite{GLt}) and then delve into the dynamic collective states that result
from the interactions within the system. This generalization will be used in
the following section when we discuss the transition from a system with
multiple collective states (one for each field type) to a state that
combines each type of cells. Consequently, introducing the extended field
formalism for collective states developed in (\cite{GLs}) and (\cite{GLt})
becomes useful. Describing the formalism for collective states involving $n$%
\ different fields will allow us to compare multiple collective states, each
associated with a specific field, to a single state resulting from
interactions and mergers. This, in turn, will provide insights for
developing a formalism for transitions between collective states.

\subsection{General set up}

We saw in (\cite{GLr}) how to describe $n$ interacting types of cells, with
arbitrary interactions. Each type of cells is caracterized by its activity $%
i=1,...,n$, and interacts either positively or negatively with each other.
Each type is defined by a field $\Psi _{i}$ and activities $\omega
_{i}\left( \theta ,Z\right) $. \ The system is described by a sum of terms
similar to (\ref{flt}): 
\begin{eqnarray}
S_{full} &=&-\frac{1}{2}\Psi _{i}^{\dagger }\left( \theta ,Z\right) \nabla
\left( \frac{\sigma _{\theta }^{2}}{2}\nabla -\omega _{i}^{-1}\left(
J,\theta ,Z,\left( \left\vert \Psi _{k}\right\vert ^{2}\right) _{k\leqslant
n}\right) \right) \Psi _{i}\left( \theta ,Z\right) +V\left( \Psi _{i}\right)
\\
&&+\frac{1}{2\eta ^{2}}\left( S_{\Gamma _{ij}}^{\left( 0\right) }+S_{\Gamma
_{ij}}^{\left( 1\right) }+S_{\Gamma _{ij}}^{\left( 2\right) }+S_{\Gamma
_{ij}}^{\left( 3\right) }+S_{\Gamma _{ij}}^{\left( 4\right) }\right)
+U\left( \left( \left\vert \Gamma _{ij}\left( \theta ,Z,Z^{\prime
},C,D\right) \right\vert ^{2}\right) _{i\leqslant n,j\leqslant n}\right) 
\notag
\end{eqnarray}%
The field $\Gamma _{ij}$ describes the connectivities between types $i$ and $%
j$. The functionals $S_{\Gamma _{ij}}^{\left( c\right) }$ involve $\Gamma
_{ij}$ and $\Psi _{i}\left( \theta ,Z\right) $ and $\Psi _{j}\left( \theta
,Z\right) $. The formula are given in appendix 4. As in the one field case,
we replace $\Psi _{i}\left( \theta ,Z\right) $ and $\omega _{i}^{-1}$ by
functionals of $\Gamma _{ij}\left( \theta ,Z,Z^{\prime },C,D\right) $.

Then the background fields are derived by minimizing:%
\begin{equation*}
\frac{\delta \sum S_{\Gamma _{ij}}^{\left( a\right) }}{\delta \Delta \Gamma
_{ij}^{\dag }\left( T,\hat{T},\theta ,Z,Z^{\prime }\right) }=\frac{\delta
\sum S_{\Gamma _{ij}}^{\left( a\right) }}{\delta \Delta \Gamma _{ij}\left( T,%
\hat{T},\theta ,Z,Z^{\prime }\right) }=0
\end{equation*}%
We obtain the background for this field along with the associated average
values of connectivities and activities. The average connectivity functions $%
\left\langle T_{ij}\right\rangle $ is given by:%
\begin{equation*}
\left\langle T_{ij}\right\rangle \left( Z,Z_{1}\right) =\int
T_{ij}\left\vert \Gamma _{ij}\left( T_{ij},\hat{T}_{ij},\theta ,Z,Z^{\prime
},C_{ij},D_{ij}\right) \right\vert ^{2}
\end{equation*}%
and the equilibrium static activities are derived as functions of the non
interacting activities, under the assumption:%
\begin{equation*}
G^{ij}T_{ij}<<T_{ii}\text{ for }i\neq j
\end{equation*}%
that is in limit of weak interaction between different fields\footnote{%
The matrix $G^{ij}$ describes the interaction between these different types
of fields (see appendix 4).}. The formula are recalled in appendix 4.

\subsection{Effective action for collective states of several type of
interacting structures}

\subsubsection{Effective action without collective internal dynamics}

As in (\cite{GLt}), the effective action involving $n$ field is obtained
directly by a second order expansion around the background field minimizing $%
\sum S_{\Gamma }^{\left( i\right) }$. This leads to the effective action for
connectivities. Formally it is given by:%
\begin{equation*}
S\left( \Delta \Gamma \left( T,\hat{T},\theta ,Z,Z^{\prime }\right) \right)
=\Delta \Gamma ^{\dag }\left( T,\hat{T},\theta ,Z,Z^{\prime }\right) \frac{%
\delta ^{2}\sum S_{\Gamma }^{\left( i\right) }}{\delta \Delta \Gamma \left(
T,\hat{T},\theta ,Z,Z^{\prime }\right) \delta \Delta \Gamma ^{\dag }\left( T,%
\hat{T},\theta ,Z,Z^{\prime }\right) }\Delta \Gamma \left( T,\hat{T},\theta
,Z,Z^{\prime }\right)
\end{equation*}%
Modifying (\ref{fcp}) and (\ref{pnl}) to the case of several fields leads to:%
\begin{eqnarray}
&&S\left( \Delta \Gamma _{ij}\left( T,\hat{T},\theta ,Z,Z^{\prime }\right)
\right)  \label{fcg} \\
&=&\Delta \Gamma _{ij}^{\dag }\left( T,\hat{T},\theta ,Z,Z^{\prime }\right)
\left( \nabla _{T}\left( \nabla _{T}+\frac{\left( T_{ij}-\left\langle
T_{ij}\right\rangle \right) }{\tau \omega _{i}\left( Z\right) }\left\vert
\Psi \left( \theta ,Z\right) \right\vert ^{2}\right) \right) \Delta \Gamma
_{ij}\left( T,\hat{T},\theta ,Z,Z^{\prime }\right)  \notag \\
&&+\Delta \Gamma _{ij}^{\dag }\left( T,\hat{T},\theta ,Z,Z^{\prime }\right)
\nabla _{\hat{T}}\left( \nabla _{\hat{T}}+\left\vert \bar{\Psi}_{0ij}\left(
Z,Z^{\prime }\right) \right\vert ^{2}\left( \hat{T}_{ij}-\left\langle \hat{T}%
_{ij}\right\rangle \right) \right) \Delta \Gamma _{ij}\left( T,\hat{T}%
,\theta ,Z,Z^{\prime }\right) +\hat{V}\left( \Delta \Gamma _{ij},\Delta
\Gamma _{ij}^{\dag }\right)  \notag \\
&&+U_{\Delta \Gamma _{ij}}\left( \left( \left\vert \Delta \Gamma _{ij}\left(
\theta ,Z,Z^{\prime },C,D\right) \right\vert ^{2}\right) _{i\leqslant
n,j\leqslant n}\right)  \notag
\end{eqnarray}%
with:%
\begin{equation*}
\left\vert \bar{\Psi}_{0ij}\left( Z,Z^{\prime }\right) \right\vert ^{2}=%
\frac{\rho \left( C\left( \theta \right) \left\vert \Psi _{0i}\left(
Z\right) \right\vert ^{2}\omega _{i}\left( Z\right) +D\left( \theta \right) 
\hat{T}_{ij}\left\vert \Psi _{0j}\left( Z^{\prime }\right) \right\vert
^{2}\omega _{j}\left( Z^{\prime }\right) \right) }{\omega _{i}\left(
Z\right) }
\end{equation*}%
\begin{eqnarray}
&&\hat{V}\left( \Delta \Gamma _{ij},\Delta \Gamma _{ij}^{\dag }\right)
=\Delta \Gamma _{ij}^{\dag }\left( T_{ij},\hat{T}_{ij},\theta ,Z,Z^{\prime
}\right) \\
&&\times \left( \nabla _{\hat{T}}\left( \frac{\rho D\left( \theta \right)
\left\langle \hat{T}_{ij}\right\rangle \left\vert \Psi _{0j}\left( Z^{\prime
}\right) \right\vert ^{2}}{\omega _{0i}\left( Z\right) }\check{T}_{ij}\left(
1-\left( 1+\left\langle \left\vert \Psi _{\Gamma _{ij}}\right\vert
^{2}\right\rangle \right) \check{T}_{ij}\right) ^{-1}\left[ O_{i}\frac{%
\Delta T_{ij}\left\vert \Delta \Gamma _{ij}\left( \theta
_{1},Z_{1},Z_{1}^{\prime }\right) \right\vert ^{2}}{T_{ij}\Lambda ^{2}}%
\right] \right) \right)  \notag \\
&&\times \Delta \Gamma _{ij}\left( T_{ij},\hat{T}_{ij},\theta ,Z,Z^{\prime
}\right)  \notag
\end{eqnarray}%
and:%
\begin{equation}
O_{i}\left( Z,Z^{\prime },Z_{1}\right) =-\frac{\left\vert Z-Z^{\prime
}\right\vert }{c}\nabla _{\theta _{1}}+\frac{\left( Z^{\prime }-Z\right) ^{2}%
}{2}\left( \frac{\nabla _{Z_{1}}^{2}}{2}+\frac{\nabla _{\theta _{1}}^{2}}{%
2c^{2}}-\frac{\nabla _{Z}^{2}\omega _{0i}\left( Z\right) }{2}\right)
\end{equation}%
\begin{eqnarray*}
\Delta T_{ij} &=&T_{ij}-\left\langle T_{ij}\right\rangle \\
\Delta \hat{T} &=&\hat{T}_{ij}-\left\langle \hat{T}\right\rangle _{ij}
\end{eqnarray*}%
with $\left\langle T_{ij}\right\rangle $ and $\left\langle \hat{T}%
_{ij}\right\rangle $ are the connectivities averages in the background
field. The potential:%
\begin{equation*}
U_{\Delta \Gamma _{ij}}\left( \left( \left\vert \Delta \Gamma _{ij}\left(
\theta ,Z,Z^{\prime },C,D\right) \right\vert ^{2}\right) _{i\leqslant
n,j\leqslant n}\right)
\end{equation*}%
is the second order expansion of $U\left( \left( \left\vert \Gamma
_{ij}\left( \theta ,Z,Z^{\prime },C,D\right) \right\vert ^{2}\right)
_{i\leqslant n,j\leqslant n}\right) $ around the background field.

\subsubsection{Effective action including internal dynamics}

Introducing the average modifications $\left( \underline{\left\langle \Delta
T_{ij}\right\rangle }\underline{,\left\langle \Delta \hat{T}%
_{ij}\right\rangle }\right) $ of $\left( \Delta T_{ij},\Delta \hat{T}%
_{ij}\right) $ and including internal dynamics leads to consider the
generalization of (\ref{TRM}) and (\ref{MFD}):%
\begin{equation}
\hat{S}\left( \Delta \Gamma _{ij}\left( T_{ij},\hat{T}_{ij},\theta
,Z,Z^{\prime }\right) \right) =S\left( \Delta \Gamma _{ij}\left( T_{ij},\hat{%
T}_{ij},\theta ,Z,Z^{\prime }\right) \right) -\Delta V\left( \Delta \Gamma
_{ij},\Delta \Gamma _{ij}^{\dag }\right)
\end{equation}%
wth:%
\begin{eqnarray}
&&S\left( \Delta \Gamma _{ij}\left( T_{ij},\hat{T}_{ij},\theta ,Z,Z^{\prime
}\right) \right) \\
&=&-\Delta \Gamma _{ij}^{\dag }\left( T_{ij},\hat{T}_{ij},\theta
,Z,Z^{\prime }\right) \left( \nabla _{T_{ij}}\left( \nabla _{T_{ij}}+\frac{%
\left( \Delta T_{ij}-\underline{\left\langle \Delta T_{ij}\right\rangle }%
\right) -\lambda \left( \Delta \hat{T}_{ij}-\underline{\left\langle \Delta 
\hat{T}_{ij}\right\rangle }\right) }{\tau \left( \omega _{0i}\left( Z\right)
+\Delta \omega _{i}\right) \left( \theta ,Z,\left\vert \Psi _{i}\right\vert
^{2}\right) }\right) \right) \Delta \Gamma _{ij}\left( T_{ij},\hat{T}%
_{ij},\theta ,Z,Z^{\prime }\right)  \notag \\
&&-\Delta \Gamma _{ij}^{\dag }\left( T_{ij},\hat{T}_{ij},\theta ,Z,Z^{\prime
}\right) \nabla _{\hat{T}_{ij}}\left( \nabla \hat{T}_{ij}+\left\vert \bar{%
\Psi}_{0ij}\left( Z,Z^{\prime }\right) \right\vert ^{2}\left( \Delta \hat{T}%
_{ij}-\underline{\left\langle \Delta \hat{T}_{ij}\right\rangle }\right)
\right) \Delta \Gamma _{ij}\left( T_{ij},\hat{T}_{ij},\theta ,Z,Z^{\prime
}\right)  \notag \\
&&+U_{\Delta \Gamma _{ij}}\left( \left( \left\vert \Delta \Gamma _{ij}\left(
\theta ,Z,Z^{\prime },C,D\right) \right\vert ^{2}\right) _{i\leqslant
n,j\leqslant n}\right)  \notag
\end{eqnarray}

As before $\Delta \omega _{i}\left( \theta ,Z,\left\vert \Psi
_{i}\right\vert ^{2}\right) $ is some additional activity inducing a
modification of action. In the sequel, we write:%
\begin{equation*}
\omega _{i}\left( \theta ,Z,\left\vert \Psi _{i}\right\vert ^{2}\right)
=\omega _{i0}\left( Z,\left\vert \Psi _{i}\right\vert ^{2}\right) +\Delta
\omega _{i}\left( \theta ,Z,\left\vert \Psi _{i}\right\vert ^{2}\right)
\end{equation*}%
where the static part is wrttn:%
\begin{equation*}
\left( \omega _{S}\right) _{i}\left( Z,\left\vert \Psi _{i}\right\vert
^{2}\right) =\omega _{i0}\left( Z,\left\vert \Psi _{i}\right\vert
^{2}\right) +\overline{\Delta \omega _{i}}\left( Z,\left\vert \Psi
_{i}\right\vert ^{2}\right)
\end{equation*}%
and the dynamic part is $\left( \Delta \omega _{D}\right) _{i}\left( \theta
,Z,\left\vert \Psi _{i}\right\vert ^{2}\right) $ so that the decomposition
between static and dynamic part of activity is:%
\begin{equation*}
\omega _{i}\left( \theta ,Z,\left\vert \Psi _{i}\right\vert ^{2}\right)
=\left( \omega _{S}\right) _{i}\left( Z,\left\vert \Psi _{i}\right\vert
^{2}\right) +\left( \Delta \omega _{D}\right) _{i}\left( \theta
,Z,\left\vert \Psi _{i}\right\vert ^{2}\right)
\end{equation*}

The potential $V\left( \Delta \Gamma _{ij},\Delta \Gamma _{ij}^{\dag
}\right) $ is:%
\begin{eqnarray}
&&V\left( \Delta \Gamma _{ij},\Delta \Gamma _{ij}^{\dag }\right) =-\Delta
\Gamma _{ij}^{\dag }\left( T_{ij},\hat{T}_{ij},\theta ,Z,Z^{\prime }\right)
\label{Tlr} \\
&&\times \nabla _{\hat{T}}\left( \frac{\rho \left( D\left( \theta \right)
\left\langle \hat{T}\right\rangle \left\vert \Psi _{0j}\left( Z^{\prime
}\right) \right\vert ^{2}\left( \left( \omega _{S}\right) _{i}\left(
Z\right) \left( \Delta \omega _{D}\right) _{j}\left( \theta -\frac{%
\left\vert Z-Z^{\prime }\right\vert }{c},Z^{\prime },\left\vert \Psi
_{0j}\right\vert ^{2}\right) -\left( \omega _{S}\right) _{j}\left( Z^{\prime
}\right) \left( \Delta \omega _{D}\right) _{i}\left( \theta ,Z,\left\vert
\Psi _{0i}\right\vert ^{2}\right) \right) \right) }{\left( \left( \omega
_{S}\right) _{i}\left( Z\right) \right) ^{2}}\right)  \notag \\
&&\times \Delta \Gamma _{ij}\left( T_{ij},\hat{T}_{ij},\theta ,Z,Z^{\prime
}\right)  \notag
\end{eqnarray}

Given our assumptions, the interaction $V$ is of relatively small magnitude
between two different type of fields. Thus, describing the full collective
state relies on mixing non interacting states.

\subsection{Activities for collective several types of fields}

The previous results generalize for a state composed of several type of
fields. We consider both the static and variable components of activity. In
this section indices $i$ and $j$ denote the distinct types of fields. We
consider several groups of points $\left\{ a_{i}\right\} $, $\left\{
b_{j}\right\} $,... Summing over the entire group will be denoted by $%
\left\{ \left\{ b_{j}\right\} \right\} $. We define $\left\vert \Psi
\right\vert ^{2}=\left\{ \left\vert \Psi _{i}\right\vert ^{2}\right\} $, $%
\Psi =\left\{ \Psi _{i}\right\} $. The derivation is similar to the case of
a single type of field.

\subsubsection{Static part of activity}

The static part of activity for the $i$-th group at $Z_{a_{i}}$ is denoted $%
\omega _{i}^{-1}\left( Z_{a_{i}},\left\vert \Psi \right\vert ^{2}\right) $
and solves: 
\begin{equation}
\left( \omega _{S}\right) _{i}\left( Z_{a_{i}},\left\vert \Psi \right\vert
^{2}\right) =G\left( \sum_{\left\{ \left\{ b_{j}\right\} \right\} }\frac{%
\kappa }{N}G_{ij}\frac{\left( \omega _{S}\right) _{j}\left( Z_{j},\Psi
_{j}\right) \left( \left\langle T\right\rangle \left( Z_{i},Z_{j}\right)
+\left\langle \Delta T\right\rangle \left( Z_{i},Z_{j}\right) \right) }{%
\left( \omega _{S}\right) _{j}\left( Z_{a_{i}},\left\vert \Psi
_{i}\right\vert ^{2}\right) }\left\vert \Psi _{j}\left( Z_{b_{j}}\right)
\right\vert ^{2}\right)  \label{STPT}
\end{equation}

\subsubsection{Variable oscillarory part of activity}

To find the dynamic part of activities in intrction, we expand (\ref{frQ})
around (\ref{bsc}) in appendix 5. We find:%
\begin{eqnarray*}
&&\left( \Delta \omega _{D}\right) _{i}\left( Z_{a_{i}},\left\vert \Psi
\right\vert ^{2}\right)  \\
&=&\Delta G\left( \sum_{\left\{ \left\{ b_{j}\right\} \right\} }\frac{\kappa 
}{N}g^{ij}\frac{\omega _{j}\left( \theta -\frac{\left\vert
Z_{a_{i}}-Z_{b_{j}}\right\vert }{c},Z_{b_{j}}\right) T_{ij}\left(
Z_{a_{i}},Z_{b_{j}},\theta -\frac{\left\vert Z_{a_{i}}-Z_{b_{j}}\right\vert 
}{c}\right) }{\omega _{i}\left( Z_{a_{i}},\theta \right) }\left\vert \Psi
_{j}\left( \theta -\frac{\left\vert Z_{a_{i}}-Z_{b_{j}}\right\vert }{c}%
,Z_{b_{j}}\right) \right\vert ^{2}\right) 
\end{eqnarray*}%
As for one field case, we can neglect:%
\begin{equation*}
\Delta \left\vert \Psi _{j}\left( \theta -\frac{\left\vert
Z_{a_{i}}-Z_{b_{j}}\right\vert }{c},Z_{b_{j}}\right) \right\vert ^{2}
\end{equation*}%
and we obtain the following relation:%
\begin{equation}
\left( \Delta \omega _{D}\right) _{i}\left( \theta ,Z_{a_{i}},\left\vert
\Psi \right\vert ^{2}\right) =\sum_{\left\{ \left\{ b_{j}\right\} \right\} }%
\check{T}_{ij}\left( Z_{a_{i}},Z_{b_{j}}\right) \left( \Delta \omega
_{D}\right) _{b_{j}}\left( J,\theta -\frac{\left\vert
Z_{a_{i}}-Z_{b_{j}}\right\vert }{c},Z_{b_{j}},\Psi \right)   \label{SR}
\end{equation}%
where:%
\begin{eqnarray}
\check{T}_{ij}\left( Z_{a_{i}},Z_{b_{j}}\right)  &=&G^{\prime }\left(
G^{-1}\left( \left( \omega _{S}\right) _{i}^{-1}\left( Z_{a_{i}},\left\vert
\Psi \right\vert ^{2}\right) \right) \right)   \label{TNH} \\
&&\times \frac{\kappa }{N}g^{ij}\frac{\left( \omega _{S}\right) _{j}\left(
Z_{b_{j}}\right) \left\langle T_{ij}\right\rangle \left(
Z_{a_{i}},Z_{b_{j}}\right) }{G^{\prime }\left( G^{-1}\left( \left( \omega
_{S}\right) _{i}^{-1}\left( Z_{a_{i}},\left\vert \Psi \right\vert
^{2}\right) \right) \right) +\left( \omega _{S}\right) _{i}^{2}\left(
Z_{a_{i}}\right) }\left\vert \Psi _{j}\left( Z_{b_{j}}\right) \right\vert
^{2}  \notag
\end{eqnarray}%
The solution of (\ref{SR}) was given before. Defining:%
\begin{equation*}
A_{i}\left( Z_{a_{i}}\right) =\sum_{\left\{ \left\{ b_{j}\right\} \right\}
}A_{j}\left( Z_{b_{j}}\right) \check{T}_{ij}\left(
Z_{a_{i}},Z_{b_{j}}\right) \exp \left( -i\Upsilon _{p}\left( \left\{ \check{T%
}_{ij}\left( Z_{a_{i}},Z_{b_{j}}\right) \right\} \right) \frac{\left\vert
Z_{a_{i}}-Z_{b_{j}}\right\vert }{c}\right) 
\end{equation*}%
the solution is:%
\begin{eqnarray*}
&&\Delta \omega _{i}\left( \theta ,Z_{a_{i}},\mathbf{T},\left\vert \Psi
\right\vert ^{2}\right)  \\
&=&\overline{\Delta \omega _{i}}\left( Z_{a_{i}},\mathbf{\check{T}}\right) 
\\
&&+\sum_{\left\{ \left\{ b_{j}\right\} \right\} }A\left( Z_{b_{j}}\right)
\left( 1,\left( 1-\mathbf{\check{T}}\exp \left( -i\Upsilon _{p}\left( 
\mathbf{\check{T}}\right) \frac{\left\vert \mathbf{\Delta Z}\right\vert }{c}%
\right) \right) _{1}^{-1}\check{T}\left( \mathbf{Z}\right) \exp \left(
-i\Upsilon _{p}\left( \mathbf{\check{T}}\right) \frac{\left\vert \mathbf{%
\Delta Z}_{1}\right\vert }{c}\right) \right) ^{t}\exp \left( i\Upsilon
_{p}\left( \mathbf{\check{T}}\right) \theta \right) 
\end{eqnarray*}%
where we defined:%
\begin{equation*}
\mathbf{\check{T}=}\left\{ \check{T}_{ij}\left( Z_{a_{i}},Z_{b_{j}}\right)
\right\} 
\end{equation*}%
However, to describe interactions between different collective states and
transition between them, it is useful to describe the equilibrium
frequencies of oscillations in the activity of a mixed state as a function
of the frequencies of collective states of different types.

\subsection{Activities and frequencies for interacting collective states as
function of non interacting states}

We formulate the activity equations for a collective state by considering
separately the different type of collectives states involved in this state.
The activity equations for each type are influenced by the activities of
other types. The equations for activities of the $i$-th group as:%
\begin{eqnarray}
&&\omega _{i}\left( Z_{a_{i}},\theta \right)  \label{frQ} \\
&=&G\left( \sum_{j}\sum_{\left\{ b_{j}\right\} }\frac{\kappa }{N}g^{ij}\frac{%
\omega _{j}\left( \theta -\frac{\left\vert Z_{a_{i}}-Z_{b_{j}}\right\vert }{c%
},Z_{b_{j}}\right) \left( \left( T_{ij}+\Delta T_{ij}\right) \left(
Z_{a_{i}},Z_{b_{j}},\theta -\frac{\left\vert Z_{a_{i}}-Z_{b_{j}}\right\vert 
}{c}\right) \right) }{\Delta \omega _{i}\left( Z_{a_{i}},\theta \right) }%
\left\vert \Psi _{j}\left( \theta -\frac{\left\vert
Z_{a_{i}}-Z_{b_{j}}\right\vert }{c},Z_{b_{j}}\right) \right\vert ^{2}\right)
\notag
\end{eqnarray}%
where as in the previous section:%
\begin{equation*}
\omega _{i}^{-1}\left( Z_{a_{i}},\theta \right) =\omega _{i0}^{-1}\left(
Z_{a_{i}}\right) +\Delta \omega _{i}^{-1}\left( Z_{a_{i}},\theta \right)
\end{equation*}%
Said differently, this is the equation for the activities of the $i$-th
component of the whole state.

Assuming a weak interactions between the components:%
\begin{equation*}
g^{ij}\Delta T_{ij}<<\Delta T_{ii}
\end{equation*}%
we expand this equation (\ref{frQ}) to the first order in $g^{ij}\Delta
T_{ij}$. It leads to writing the static and dynamic parts of activities as
functions of the non-interacting ones. Details are provided in appendix 5.
This description will be useful for depicting transitions between
non-interacting and interacting states.

\subsubsection{Static part}

Equilibrium\ static activities satify:%
\begin{eqnarray}
&&\overline{\Delta \omega }_{i}\left( Z_{a_{i}}\right)  \label{bsc} \\
&=&\sum_{j,b_{j}}\left( \delta _{\left( i,a_{i}\right) \left( j,b_{j}\right)
}-G^{\prime }\left( G^{-1}\left( \left( \omega _{Sf}\right) _{i}\right)
\right) \right.  \notag \\
&&\times \left. \left( \frac{\kappa }{N}\frac{G^{ij}\left( \left( \omega
_{Sf}\right) _{j}\right) \left( Z_{b_{j}}\right) T_{ij}\left(
Z_{a_{i}},Z_{b_{j}}\right) }{\left( \left( \omega _{Sf}\right) _{j}\right)
\left( Z_{a_{i}}\right) }\left( \mathcal{G}_{j0}+\left\vert \Psi _{j}\left(
Z_{b_{j}}\right) \right\vert ^{2}\right) \right) _{j\neq i}\right) _{ij}^{-1}%
\overline{\left( \Delta \omega _{f}\right) }_{j}\left( Z_{b_{j}}\right) 
\notag
\end{eqnarray}%
where:%
\begin{equation*}
\left( \omega _{Sf}\right) _{i}=\omega _{0i}+\overline{\left( \Delta \omega
_{f}\right) }_{i}
\end{equation*}
are the static part of the activities without interaction. Here, the $%
\overline{\left( \Delta \omega _{f}\right) }_{i}\left( Z_{a_{i}}\right) $
are the additional activities in absence of interactions and $\omega _{0i}$
is the background activity for field $\Psi _{i}$.

The $\overline{\left( \Delta \omega _{f}\right) }_{i}\left( Z_{a_{i}}\right) 
$ satisfy an equation similar to (\ref{STPT}):%
\begin{equation}
\left( \omega _{Sf}\right) _{i}\left( Z_{a_{i}}\right) =G\left( \sum_{\beta
_{i}}\frac{\kappa }{N}\frac{\left( \left( \omega _{Sf}\right) _{i}\right)
\left( Z_{\beta _{i}}\right) T_{ij}\left( Z_{a_{i}},Z_{b_{j}}\right) }{%
\left( \left( \omega _{Sf}\right) _{i}\right) \left( Z_{a_{i}},\theta
\right) }\left\vert \Psi _{j}\left( Z_{b_{j}}\right) \right\vert ^{2}\right) 
\label{STN}
\end{equation}%
where we assume $G^{ii}=1$. Formula for $\overline{\Delta \omega }_{i}\left(
Z_{a_{i}}\right) $ are derived in appendix 5 at the first order in $G^{ij}$:%
\begin{eqnarray}
&&\overline{\Delta \omega }_{i}\left( Z_{a_{i}}\right)   \label{STP} \\
&=&\sum_{j,b_{j}}\left( \delta _{\left( i,a_{i}\right) \left( j,b_{j}\right)
}-G^{\prime }\left( G^{-1}\left( \left( \omega _{Sf}\right) _{i}\left(
Z\right) \right) \right) \right.   \notag \\
&&\times \left. \left( \frac{\kappa }{N}\frac{G^{ij}\left( \omega
_{Sf}\right) _{j}\left( Z_{b_{j}}\right) T_{ij}\left(
Z_{a_{i}},Z_{b_{j}}\right) }{\left( \omega _{Sf}\right) _{i}\left(
Z_{a_{i}}\right) }\left( \mathcal{G}_{j0}+\left\vert \Psi _{j}\left(
Z_{b_{j}}\right) \right\vert ^{2}\right) \right) _{j\neq i}\right) _{ij}^{-1}%
\overline{\left( \Delta \omega _{f}\right) }_{j}\left( Z_{b_{j}}\right)  
\notag
\end{eqnarray}

\subsubsection{Non-static part}

\paragraph{\protect\bigskip Activities without interactions}

As before, the non static part of $n$ non-interacting activities are
solutions of:%
\begin{equation}
\left( \Delta \omega _{Df}\right) _{i}\left( Z_{a_{i}},\theta \right)
=\sum_{a_{j}}\check{T}_{ii}\left( Z_{a_{i}},Z_{a_{j}}\right) \left( \Delta
\omega _{Df}\right) _{i}\left( \theta -\frac{\left\vert Z_{\alpha
_{i}}-Z_{a_{j}}\right\vert }{c},Z_{a_{j}}\right)   \label{Sr}
\end{equation}%
where:%
\begin{eqnarray}
&&\check{T}_{ij}\left( Z_{a_{i}},Z_{b_{j}}\right)   \label{St} \\
&=&G^{\prime }\left( G^{-1}\left( \left( \omega _{S}\right) _{i}^{-1}\left(
Z_{a_{i}},\left\vert \Psi \right\vert ^{2}\right) \right) \right)   \notag \\
&&\times \frac{\kappa }{N}g^{ij}\frac{\left( \omega _{S}\right) _{j}\left(
Z_{b_{j}}\right) T_{ij}\left( Z_{a_{i}},Z_{b_{j}}\right) }{\sum_{b_{j}}\frac{%
\kappa }{N}g^{ij}\left( \omega _{S}\right) _{j}\left( Z_{b_{j}}\right)
G^{\prime }\left( G^{-1}\left( \left( \omega _{S}\right) _{i}^{-1}\left(
Z_{a_{i}},\left\vert \Psi \right\vert ^{2}\right) \right) \right) +\left(
\omega _{S}\right) _{i}^{2}\left( Z_{a_{i}}\right) }\left\vert \Psi
_{j}\left( Z_{b_{j}}\right) \right\vert ^{2}  \notag
\end{eqnarray}%
Rewriting the solutions as a vector:%
\begin{equation*}
\left( \left( \Delta \omega _{Df}\right) ^{-1}\left( Z_{a_{i}},\theta
\right) \right) _{a_{i}}\equiv \Delta \mathbf{\omega }_{Df}^{-1}\left( Z_{%
\mathbf{a}},\theta \right) 
\end{equation*}%
and looking for oscillatory solutions:%
\begin{equation*}
\Delta \mathbf{\omega }_{Df}^{-1}\left( Z_{\mathbf{a}},\theta \right)
=\Delta \mathbf{\omega }_{f}^{-1}\left( Z_{\mathbf{a}}\right) \exp \left(
i\Upsilon \theta \right) 
\end{equation*}%
yields the following formula, similar to the single type of cells case:

\begin{eqnarray*}
&&\left( \Delta \omega _{Df}\right) _{i}\left( \theta ,\mathbf{Z},\left\vert
\Psi \right\vert ^{2}\right)  \\
&=&A_{i}\left( \left( Z_{1}\right) _{i}\right) \left( 1,\left( 1-\mathbf{%
\check{T}}_{ii}\exp \left( -i\Upsilon _{ip}\frac{\left\vert \mathbf{\Delta Z}%
\right\vert }{c}\right) \right) ^{-1}\check{T}_{1ii}\left( \mathbf{Z}\right)
\exp \left( -i\Upsilon _{ip}\frac{\left\vert \mathbf{\Delta Z}%
_{1}\right\vert }{c}\right) \right) ^{t}\exp \left( i\Upsilon _{ip}\left( 
\mathbf{\check{T}}_{ii}\right) \theta \right) 
\end{eqnarray*}%
As in the $1$-field case, $A_{i}\left( \left( Z_{1}\right) _{i}\right) $ is
the amplitude of activity at one given point $\left( Z_{1}\right) _{i}$ of
the group $i$ and $\check{T}_{1ii}\left( \mathbf{Z}\right) $ is a vector
with components $\left( \check{T}_{1ii}\left( \mathbf{Z}\right) \right)
_{a_{i}}=\check{T}_{ii}\left( Z_{a_{i}},\left( Z_{1}\right) _{i}\right) $.

The $\Upsilon _{ip}\left( \mathbf{\check{T}}_{ii}\right) $ are equilbrium
frequencies. They are solutions of:%
\begin{equation}
\det \left( 1-\check{T}_{ii}\left( Z,Z_{1}\right) \exp \left( -i\Upsilon _{p}%
\frac{\left\vert Z-Z_{j}\right\vert }{c}\right) \right) =0  \label{MTV}
\end{equation}%
We write the solutions $\Upsilon _{p}\left( \mathbf{\check{T}}_{ii}\right) $%
. By diagonalization of the matrix involved in (\ref{MTV}), these
frequencies satisfy:%
\begin{equation*}
\prod\limits_{k}\left( 1-f_{i,k}\left( \Upsilon _{p}\right) \right) =0
\end{equation*}%
for some functions $f_{i,k}$. Thus, the solutions form a set:%
\begin{equation*}
\Upsilon _{p}^{i}=\left\{ \gamma _{i,k}\right\} _{k}
\end{equation*}

\paragraph{Activities with interactions}

Comparing (\ref{SR}) and (\ref{Sr}) leads to the lowest order:%
\begin{equation*}
\left( \Delta \omega _{D}\right) _{i}\left( Z_{a_{i}},\theta \right) =\left(
\Delta \omega _{Df}\right) _{i}\left( Z_{a_{i}},\theta \right) +\sum_{j\neq
i,\left\{ \left\{ b_{j}\right\} \right\} }\check{T}_{ij}\left(
Z_{a_{i}},Z_{b_{j}}\right) \left( \Delta \omega _{Df}\right) _{j}\left(
\theta -\frac{\left\vert Z_{a_{i}}-Z_{b_{j}}\right\vert }{c},Z_{b_{j}}\right)
\end{equation*}%
with $\check{T}_{ij}\left( Z_{a_{i}},Z_{b_{j}}\right) $ given by (\ref{TNH}).

\subsubsection{Frequencies for sets n interactions}

\paragraph{Solutions for similar groups}

Rewriting the solutions of (\ref{SR}) as a vector:%
\begin{equation*}
\left( \left( \Delta \omega _{D}\right) _{i}^{-1}\left( Z_{\alpha
_{i}},\theta \right) \right) _{i}\equiv \left( \Delta \mathbf{\omega }%
_{D}\right) ^{-1}\left( Z_{\mathbf{\alpha }},\theta \right)
\end{equation*}%
we look for oscillatory solutions. For a composed state to exist, we have to
consider non-destructive interactions and this imply that we have to look
for similar frequencies between the various groups. As a consequence, we
assume that the solutions have the following form:%
\begin{equation*}
\left( \Delta \mathbf{\omega }_{D}\right) \left( Z_{\mathbf{a}},\theta
\right) =\left( \Delta \mathbf{\omega }\right) \left( Z_{\mathbf{a}}\right)
\exp \left( i\Upsilon \theta \right)
\end{equation*}%
which implies a solution for:%
\begin{equation*}
\Delta \mathbf{\omega }\left( Z_{\mathbf{a}}\right) =M\Delta \mathbf{\omega }%
\left( Z_{\mathbf{a}}\right)
\end{equation*}%
with:%
\begin{equation*}
M_{\left( ia_{i}\right) ,\left( jb_{j}\right) }=\hat{T}_{ij}\left(
Z_{a_{i}},Z_{b_{j}}\right) \exp \left( -i\Upsilon \frac{\left\vert
Z_{a_{i}}-Z_{b_{j}}\right\vert }{c}\right)
\end{equation*}%
\begin{equation*}
M=\left( M_{\left( ia_{i}\right) ,\left( jb_{j}\right) }\right) +\left(
M_{\left( ia_{i}\right) ,\left( jb_{j}\right) }\right) _{i\neq j}
\end{equation*}

\paragraph{Composed frequencies as functions of witht-interaction frequencies%
}

Without interactions, the frequencies are in some state $\gamma _{i,l_{i}}$
satisfying: 
\begin{equation*}
f_{i,l_{i}}\left( \gamma _{i,l_{i}}\right) =1
\end{equation*}%
and the solution for the vector $\Upsilon $ depends on these quantities:%
\begin{equation}
\Upsilon _{\left( i,l_{i}\right) }=\sum_{i,j\neq i}\left[ \left[ M\right] %
\right] _{i,j}\frac{\gamma _{i,k_{i}}+\gamma _{j,l_{j}}}{2}\pm \sqrt{\left(
\sum_{i,j\neq i}\frac{\left[ \left[ M\right] \right] _{i,j}}{\sum_{i,j\neq i}%
\left[ \left[ M\right] \right] _{i,j}}\frac{\gamma _{i,k_{i}}-\gamma
_{j,l_{j}}}{2}\right) ^{2}+\sum_{i,j\neq i}\left[ \left[ M\right] \right]
_{i,j}}  \label{ntr}
\end{equation}%
The notation $\Upsilon _{\left( i,l_{i}\right) }$ encapsulates that the
resulting frequency for the new structure depends on the states $\left(
\gamma _{i,l_{i}}\right) $ of initial ones. The expression $\left[ \left[ M%
\right] \right] _{i,j}$: 
\begin{equation*}
\left[ \left[ M\right] \right] _{i,j}=\frac{\left[ \hat{M}_{j,i}\right]
_{l_{j},k_{i}}\left[ \hat{M}_{i,j}\right] _{k_{i},l_{j}}}{\left( \frac{%
\partial }{\partial \gamma }f_{j,l_{j}}\left( \gamma \right) \right)
_{\gamma _{j,l_{j}}}\left( \frac{\partial }{\partial \gamma }%
f_{i,k_{i}}\left( \gamma \right) \right) _{\gamma _{i,k_{i}}}}
\end{equation*}%
depends on the matrix $\hat{M}_{j,i}$ obtained from $M_{ij}$ by a change of
basis (see appendix 5).

\subsubsection{Full activity for composed collective states}

Ultimately, the possible activities for a composed state are obtained by
gathering static and non-static parts. The modified activity for component $%
i $ 
\begin{equation*}
\overline{\Delta \omega }_{i}\left( Z_{a_{i}}\right) +\left( \Delta \omega
_{D}\right) _{i}\left( Z_{a_{i}}\right) \exp \left( i\Upsilon _{\left\{
i,l_{i}\right\} }\theta \right)
\end{equation*}%
wher $\Upsilon _{\left( i,l_{i}\right) }$ is defined in (\ref{ntr}).

\section{Interactions and transitions between collective states}

We can now explore the possibilities for several collective states to
potentially undergo transitions into other collective states. We have given
the descriptions of individual and mixed collective states, defined by
connectivities and their associated activities. We may, therefore, consider
transitions in which two states can merge to form a third state with its own
activities. This transition should depend on the characteristics of each
state. Dynamically, the interaction modifies the activities of the initial
states, inducing a dynamic process that may converge towards the merged
state. The initial states are minima of some action $S_{i}$, meaning the
collection of initial states minimizes $\sum $ $S_{i}$ without including
interactions. The final states minimize the action, including certain
interactions $\sum $ $S_{i}+\sum $ $S_{i,j}$.

This section develops the mechanism of transitions and provides some
elements that we will incorporate into a more general field formalism where
transitions will result from interaction terms between fields of structures.

\subsection{Dynamic evolution of connectivity}

We want to describe the transition between states wth $S_{i}$, $T_{ij}$,
towards states with $T_{ij}$. We first describe the dynamic of the
connectivity $T_{ij}$ of a state. Assuming that starting with $\Delta
T_{ij}\left( Z,Z^{\prime }\right) =\Delta \hat{T}_{ij}\left( Z,Z^{\prime
}\right) =0$, some connectivity above the threshold arises. The term inside
the gradient in (\ref{Tlr}) yields the evolution of a single connectivity.
This corresponds to a modification:%
\begin{eqnarray*}
&&\frac{d}{dt}\Delta \hat{T}_{ij}\left( Z,Z^{\prime }\right) \\
&=&\frac{\rho }{\omega _{0i}^{2}\left( Z\right) }\frac{D\left( \theta
\right) \left\langle \Delta \hat{T}_{ij}\left( Z,Z^{\prime }\right)
\right\rangle \left\vert \Psi _{0j}\left( Z^{\prime }\right) \right\vert
^{2}\left( \omega _{0i}\left( Z\right) \Delta \omega _{j}\left( \theta -%
\frac{\left\vert Z-Z^{\prime }\right\vert }{c},Z^{\prime },\left\vert \Psi
\right\vert ^{2}\right) -\omega _{0j}\left( Z^{\prime }\right) \Delta \omega
_{i}\left( \theta ,Z,\left\vert \Psi \right\vert ^{2}\right) \right) }{%
C\left( \theta \right) \left\vert \Psi _{0i}\left( Z\right) \right\vert
^{2}\omega _{0i}\left( Z\right) +D\left( \theta \right) \left\langle \Delta 
\hat{T}_{ij}\right\rangle \left\vert \Psi _{0j}\left( Z^{\prime }\right)
\right\vert ^{2}\omega _{0j}\left( Z^{\prime }\right) }
\end{eqnarray*}%
Given that the connectivities fluctuate arounfd their average value, we can
focus on the dynamics of: 
\begin{equation*}
\left\langle \Delta \hat{T}_{ij}\left( Z,Z^{\prime }\right) \right\rangle
\end{equation*}%
which becomes:%
\begin{eqnarray*}
&&\frac{d}{dt}\left\langle \Delta \hat{T}_{ij}\left( Z,Z^{\prime }\right)
\right\rangle \\
&=&\frac{\rho }{\omega _{0i}^{2}\left( Z\right) }\frac{D\left( \theta
\right) \left\langle \Delta \hat{T}_{ij}\left( Z,Z^{\prime }\right)
\right\rangle \left\vert \Psi _{0j}\left( Z^{\prime }\right) \right\vert
^{2}\left( \omega _{0i}\left( Z\right) \Delta \omega _{j}\left( \theta -%
\frac{\left\vert Z-Z^{\prime }\right\vert }{c},Z^{\prime },\left\vert \Psi
\right\vert ^{2}\right) -\omega _{0j}\left( Z^{\prime }\right) \Delta \omega
_{i}\left( \theta ,Z,\left\vert \Psi \right\vert ^{2}\right) \right) }{%
C\left( \theta \right) \left\vert \Psi _{0i}\left( Z\right) \right\vert
^{2}\omega _{0i}\left( Z\right) +D\left( \theta \right) \left\langle \Delta 
\hat{T}_{ij}\right\rangle \left\vert \Psi _{0j}\left( Z^{\prime }\right)
\right\vert ^{2}\omega _{0j}\left( Z^{\prime }\right) }
\end{eqnarray*}%
using:%
\begin{equation*}
\left\langle \Delta \hat{T}_{ij}\left( Z,Z^{\prime }\right) \right\rangle
\simeq \frac{\left\langle \Delta T_{ij}\left( Z,Z^{\prime }\right)
\right\rangle }{\lambda }
\end{equation*}%
this becomes a dynamics for:%
\begin{eqnarray*}
&&\frac{d}{dt}\left\langle \Delta T_{ij}\left( Z,Z^{\prime }\right)
\right\rangle \\
&=&\frac{\lambda \rho }{\omega _{0i}^{2}\left( Z\right) }\frac{D\left(
\theta \right) \left\langle \Delta T_{ij}\left( Z,Z^{\prime }\right)
\right\rangle \left\vert \Psi _{0j}\left( Z^{\prime }\right) \right\vert
^{2}\left( \omega _{0i}\left( Z\right) \Delta \omega _{j}\left( \theta -%
\frac{\left\vert Z-Z^{\prime }\right\vert }{c},Z^{\prime },\left\vert \Psi
\right\vert ^{2}\right) -\omega _{0j}\left( Z^{\prime }\right) \Delta \omega
_{i}\left( \theta ,Z,\left\vert \Psi \right\vert ^{2}\right) \right) }{%
\lambda \left( C\left( \theta \right) \left\vert \Psi _{0i}\left( Z\right)
\right\vert ^{2}\omega _{0i}\left( Z\right) +D\left( \theta \right)
\left\langle \Delta T_{ij}\left( Z,Z^{\prime }\right) \right\rangle
\left\vert \Psi _{0j}\left( Z^{\prime }\right) \right\vert ^{2}\omega
_{0j}\left( Z^{\prime }\right) \right) }
\end{eqnarray*}%
This is associated to the dynamics for activities, obtained through mdfctn (%
\ref{mdf}) at the lowest order for $Z$ and its closest point $Z^{\prime }$
(assuming that $Z$ is also the closest point of $Z^{\prime }$):%
\begin{equation*}
\frac{d}{dt}\omega _{i}\left( Z\right) =\sum_{j,b_{j}}G^{\prime }\left(
G^{-1}\left( \omega _{0i}\left( Z\right) \right) \right) \left( \frac{\kappa 
}{N}\frac{G^{ij}\omega _{0j}\left( Z^{\prime }\right) \Delta T_{ij}\left(
Z,Z^{\prime }\right) }{\omega _{0i}\left( Z\right) }\left( \mathcal{G}%
_{j0}+\left\vert \Psi _{j}\left( Z^{\prime }\right) \right\vert ^{2}\right)
\right) _{j\neq i}\omega _{0j}\left( Z^{\prime }\right)
\end{equation*}%
This yields system with four dynamic variables:%
\begin{equation*}
\left\langle \Delta T_{ij}\left( Z,Z^{\prime }\right) \right\rangle
,\left\langle \Delta T_{ji}\left( Z,Z^{\prime }\right) \right\rangle ,\omega
_{i}\left( Z\right) ,\omega _{j}\left( Z^{\prime }\right)
\end{equation*}%
whose matrix writes:%
\begin{equation}
A=\left( 
\begin{array}{cccc}
0 & 0 & A_{i,j} & B_{i,j} \\ 
0 & 0 & B_{j,i} & A_{j,i} \\ 
C_{i,j} & 0 & 0 & 0 \\ 
0 & C_{j,i} & 0 & 0%
\end{array}%
\right)  \label{MT}
\end{equation}%
the eigenvalues of (\ref{MT}) are:%
\begin{equation*}
\pm \frac{1}{\sqrt{2}}\sqrt{A_{i,j}C_{i,j}+A_{j,i}C_{j,i}+\sqrt{\left(
A_{i,j}C_{i,j}-A_{j,i}C_{j,i}\right) ^{2}+4C_{i,j}C_{j,i}B_{j,i}B_{i,j}}}
\end{equation*}%
and:%
\begin{equation*}
\pm \frac{1}{\sqrt{2}}\sqrt{A_{i,j}C_{i,j}+A_{j,i}C_{j,i}-\sqrt{\left(
A_{i,j}C_{i,j}-A_{j,i}C_{j,i}\right) ^{2}+4C_{i,j}C_{j,i}B_{j,i}B_{i,j}}}
\end{equation*}%
In the case of enhancing interaction, the dominant eigenvalue:%
\begin{equation*}
\frac{1}{\sqrt{2}}\sqrt{A_{i,j}C_{i,j}+A_{j,i}C_{j,i}+\sqrt{\left(
A_{i,j}C_{i,j}-A_{j,i}C_{j,i}\right) ^{2}+4C_{i,j}C_{j,i}B_{j,i}B_{i,j}}}
\end{equation*}%
is real. As a consequence, the system depart from its initial value:%
\begin{equation*}
\left\langle \Delta T_{ij}\left( Z,Z^{\prime }\right) \right\rangle
=0,\left\langle \Delta T_{ji}\left( Z,Z^{\prime }\right) \right\rangle
=0,\omega _{i0}\left( Z\right) ,\omega _{j0}\left( Z^{\prime }\right)
\end{equation*}%
and move towards an other equilibrium, given by the previous derivations. In
the case of inhibitory interactions, the system oscillates, but starts from:%
\begin{equation*}
\left\langle \Delta T_{ij}\left( Z,Z^{\prime }\right) \right\rangle
=0,\left\langle \Delta T_{ji}\left( Z,Z^{\prime }\right) \right\rangle =0
\end{equation*}%
and grows slowly toward a new equilibrium if the amplitude of oscillations
is large enough.

\subsection{Dynamic transition for activities}

While the $\Delta T_{ij}\left( Z,Z^{\prime }\right) $ are evolving towards
their new values, the activities depending on these values evolve also
towards their new equilibrium values. Temporarily, the term:%
\begin{equation*}
\omega _{0i}\left( Z\right) \Delta \omega _{j}\left( \theta -\frac{%
\left\vert Z-Z^{\prime }\right\vert }{c},Z^{\prime },\left\vert \Psi
\right\vert ^{2}\right) -\omega _{0j}\left( Z^{\prime }\right) \Delta \omega
_{i}\left( \theta ,Z,\left\vert \Psi \right\vert ^{2}\right)
\end{equation*}%
presents some interferences, but these ones may initiate the dynamics for $%
\Delta T_{ij}\left( Z,Z^{\prime }\right) $ by allowing $\Delta \hat{T}_{ij}$
to overcome the threshold for the connectivity dynamics when $T_{ij}\left(
Z,Z^{\prime }\right) <\delta $:%
\begin{eqnarray*}
&&\frac{d}{dt}\Delta \hat{T}_{ij}\left( Z,Z^{\prime }\right) \\
&=&\frac{\rho }{\omega _{0i}^{2}\left( Z\right) }\frac{D\left( \theta
\right) \left\langle \Delta \hat{T}_{ij}\left( Z,Z^{\prime }\right)
\right\rangle \left\vert \Psi _{0j}\left( Z^{\prime }\right) \right\vert
^{2}\left( \omega _{0i}\left( Z\right) \Delta \omega _{j}\left( \theta -%
\frac{\left\vert Z-Z^{\prime }\right\vert }{c},Z^{\prime },\left\vert \Psi
\right\vert ^{2}\right) -\omega _{0j}\left( Z^{\prime }\right) \Delta \omega
_{i}\left( \theta ,Z,\left\vert \Psi \right\vert ^{2}\right) \right) }{%
C\left( \theta \right) \left\vert \Psi _{0i}\left( Z\right) \right\vert
^{2}\omega _{0i}\left( Z\right) +D\left( \theta \right) \left\langle \Delta 
\hat{T}_{ij}\right\rangle \left\vert \Psi _{0j}\left( Z^{\prime }\right)
\right\vert ^{2}\omega _{0j}\left( Z^{\prime }\right) } \\
&&-\eta H\left( \delta -T_{ij}\left( Z,Z^{\prime }\right) \right)
\end{eqnarray*}%
The convergence towards the new equilibrium should be ensured by the fact
that the system's backgrnd minimize an action functional.

\subsection{Transitions betweenn states}

In dynmics perspective, we can formally describe the previous transition as
a states transition between several states of different types:%
\begin{equation*}
\prod\limits_{i}\left\vert \mathbf{a}_{i},\mathbf{p}_{i},S_{i}^{2}\right%
\rangle
\end{equation*}%
where, as before, the states $\left\vert \mathbf{a}_{i},\mathbf{p}%
_{i},S_{i}^{2}\right\rangle $ are defined by the state of each of its
components:%
\begin{equation*}
\left\vert \mathbf{a}_{i},\mathbf{p}_{i},S_{i}^{2}\right\rangle
=\prod\limits_{Z_{a_{i}},Z_{a_{i}^{\prime }}}\left\vert \Delta T_{ii}\left(
Z_{a_{i}},Z_{a_{i}^{\prime }}\right) ,\Delta \hat{T}_{ii}\left(
Z_{a_{i}},Z_{a_{i}^{\prime }}\right) ,\alpha \left(
Z_{a_{i}},Z_{a_{i}^{\prime }}\right) ,p\left( Z_{a_{i}},Z_{a_{i}^{\prime
}}\right) \right\rangle
\end{equation*}%
and a composed state:%
\begin{eqnarray*}
&&\left\vert \mathbf{a}_{i^{\prime }},\mathbf{p}_{i^{\prime }},\left( \cup
S_{i}\right) \times \left( \cup S_{i}\right) \right\rangle \\
&=&\prod\limits_{a_{i},b_{j}}\prod\limits_{Z_{a_{i}},Z_{b_{j}}}\left\vert
\Delta T_{ij}\left( Z_{a_{i}},Z_{b_{j}}\right) ,\Delta \hat{T}_{ij}\left(
Z_{a_{i}},Z_{b_{j}}\right) ,\alpha \left( Z_{a_{i}},Z_{b_{j}}\right)
,p\left( Z_{a_{i}},Z_{b_{j}}\right) \right\rangle
\end{eqnarray*}%
where indices $i^{\prime }\in \left( \cup S_{i}\right) \times \left( \cup
S_{i}\right) $. The previous sections show that has to happen so that (\ref%
{ntr}) linking activities between non interacting stats and composed one is
satisfied. We thus may expect some transitions of the form:%
\begin{equation}
\prod\limits_{i}\left\vert \mathbf{a}_{i},\mathbf{p}_{i},S_{i}^{2}\right%
\rangle \rightarrow f\left( \left( \Upsilon _{\mathbf{p}_{i}}^{\mathbf{a}%
_{i}}\right) ,\Upsilon _{\mathbf{p}_{i}^{\prime }}^{\mathbf{a}_{i}^{\prime
}}\right) \left\vert \mathbf{a}_{i^{\prime }},\mathbf{p}_{i^{\prime
}},\left( \cup S_{i}\right) \times \left( \cup S_{i}\right) \right\rangle
\label{TRC}
\end{equation}%
where $f\left( \left( \Upsilon _{\mathbf{p}_{i}}^{\mathbf{a}_{i}}\right)
,\Upsilon _{\mathbf{p}_{i}^{\prime }}^{\mathbf{a}_{i}^{\prime }}\right) $ is
a function close to the Dirac function:%
\begin{equation*}
\delta \left( \Upsilon _{\left( i,l_{i}\right) }-\left\{ \sum_{i,j\neq i} 
\left[ \left[ M\right] \right] _{i,j}\frac{\gamma _{i,k_{i}}+\gamma
_{j,l_{j}}}{2}\pm \sqrt{\left( \sum_{i,j\neq i}\frac{\left[ \left[ M\right] %
\right] _{i,j}}{\sum_{i,j\neq i}\left[ \left[ M\right] \right] _{i,j}}\frac{%
\gamma _{i,k_{i}}-\gamma _{j,l_{j}}}{2}\right) ^{2}+\sum_{i,j\neq i}\left[ %
\left[ M\right] \right] _{i,j}}\right\} \right)
\end{equation*}%
to impose (\ref{ntr}) up to some fluctuations. This will be used in the
field formalism: transition may occur when the frequencies of the initial
states activities and those of the final ones satisfy some consistency
conditions.

\part*{II Field formalism for multiple interacting collective states}

So far, we have described interactions and activity within a given
collective state and considered the transition mechanism between two states
and a mixed one resulting from combined signals between the two initial
ones. This chapter aims to systematize the interactions and transitions
between large sets of collective states.

To do so, we develop a formalism for assemblies of collective states. This
formalism extends the previous description developed in the first three
parts of this work, as the variables involved in the model will directly
represent sets of interacting cells rather than individual cells themselves.
The number of degrees of freedom are thus much larger than in our previous
field formalism. After rewriting the collective stats in a suitable form for
our formalism, we develop the field description of an assembly of collective
stat. Interactions are introduced to account for the mechanisms of
transition described in the previous section.

\section{Effective field formalism for assemblies}

Developing a field formalism for assemblies of collectiv states $\left\vert 
\mathbf{\alpha },\mathbf{p},S^{2}\right\rangle $ amounts to consider
arbitrary numbers of such states along with the possibilities of transitions
between them. Thus, we first need to consider products of such states. To
explore the dynamic transitions between these states, we also need to
consider any "intermediate states" in the transition between stable states.
In other words, we must account for any possible collective states
characterized by activated connectivity between an arbitrary set of cells.
This is performed in several steps.

First of all, as demonstrated in equations (\ref{CL}) and (\ref{CT}), the
states $\left\vert \mathbf{\alpha },\mathbf{p},S^{2}\right\rangle $ are
non-local states characterized by probability density functions for the
connectivities of the activated state. Therefore, we first define states $%
\left\vert \Delta \mathbf{T}_{p}^{\alpha },\mathbf{\alpha },\mathbf{p}%
,S^{2}\right\rangle $ with given values of connectivities $\Delta \mathbf{T}%
_{p}^{\alpha }$. These states serve as a basis for defining arbitrary
collective states through linear combinations:%
\begin{equation}
\left\vert \underline{\gamma }\left( \mathbf{T,}\boldsymbol{\alpha },\mathbf{%
p},S^{2}\right) \right\rangle =\int \underline{\gamma }\left( \mathbf{T},%
\mathbf{\alpha },\mathbf{p},S^{2}\right) \left\vert \Delta \mathbf{T}%
_{p}^{\alpha },\mathbf{\alpha },\mathbf{p},S^{2}\right\rangle d\mathbf{T}
\label{lc}
\end{equation}

These states encompass all possible collective states, including unstable
ones. However, each of these states describes a single collective state. To
transition to a representation where multiple states are involved and
interact, we replace the components $\underline{\gamma }\left( \mathbf{T},%
\mathbf{\alpha },\mathbf{p},S^{2}\right) $ with fields $\underline{\Gamma }%
\left( \mathbf{T},\mathbf{\alpha },\mathbf{p},S^{2}\right) $ that account
for the possibility of multiple states.

These fields encompass all possible realizations of the components $%
\underline{\gamma }\left( \mathbf{T},\mathbf{\alpha },\mathbf{p}%
,S^{2}\right) $. The dynamic aspects of the states are governed by the field
action functional $S\left( \underline{\Gamma }\left( \mathbf{T},\mathbf{%
\alpha },\mathbf{p},S^{2}\right) \right) $. This functional is derived by
initially formulating the action for the individual states $\left\vert 
\mathbf{\alpha },\mathbf{p},S^{2}\right\rangle $, and then extending it to
encompass their combinations (\ref{lc}), leading to an action $S\left( 
\underline{\gamma }\left( \mathbf{T},\mathbf{\alpha },\mathbf{p}%
,S^{2}\right) \right) $ for the components. By replacing these components $%
\underline{\gamma }\left( \mathbf{T},\mathbf{\alpha },\mathbf{p}%
,S^{2}\right) $.with the field of which they are realizations, we obtain a
field action functional $S\left( \underline{\Gamma }\left( \mathbf{T},%
\mathbf{\alpha },\mathbf{p},S^{2}\right) \right) $ which accounts for the
possibility of multiple states. Note that doing so, the system as a whole is
described by an infinite number of fields $\underline{\Gamma }\left( \mathbf{%
T},\mathbf{\alpha },\mathbf{p},S^{2}\right) $, each of them characerized by
a set $\left( \mathbf{\alpha },\mathbf{p},S^{2}\right) $ describing average
connectivities, activations and their frequencies, along with its action.

This substitution of components by fields $\underline{\Gamma }\left( \mathbf{%
T},\mathbf{\alpha },\mathbf{p},S^{2}\right) $, which are random variables,
is equivalent to describing the system's dynamics through a partition
function for the fields $\underline{\Gamma }\left( \mathbf{T},\mathbf{\alpha 
},\mathbf{p},S^{2}\right) $. This partition function is determined by the
exponentiated sum of action functionals, with integration over the infinite
number of fields $\underline{\Gamma }\left( \mathbf{T},\mathbf{\alpha },%
\mathbf{p},S^{2}\right) $. In this context, interactions between states are
modeled through modifications of this action functional. These modifications
involve various different fields, and dynamically induce transitions between
states associated with these fields. The amplitudes of these transitions are
derived by computing Green functions between such states.

Alternatively, the field description of the system is equivalent to an
operator-based perspective. In this approach, the fields $\underline{\Gamma }%
\left( \mathbf{T},\mathbf{\alpha },\mathbf{p},S^{2}\right) $ can be
interpreted as operators:%
\begin{equation}
\underline{\Gamma }\left( \mathbf{T},\mathbf{\alpha },\mathbf{p}%
,S^{2}\right) =\underline{\Gamma }^{+}\left( \mathbf{T},\mathbf{\alpha },%
\mathbf{p},S^{2}\right) +\underline{\Gamma }^{-}\left( \mathbf{T},\mathbf{%
\alpha },\mathbf{p},S^{2}\right)  \label{FT}
\end{equation}%
The field $\underline{\Gamma }\left( \mathbf{T},\mathbf{\alpha },\mathbf{p}%
,S^{2}\right) $ is decomposed into a sum of creation operator $\underline{%
\Gamma }^{+}\left( \mathbf{T},\mathbf{\alpha },\mathbf{p},S^{2}\right) $ and
destruction operator $\underline{\Gamma }^{-}\left( \mathbf{T},\mathbf{%
\alpha },\mathbf{p},S^{2}\right) $ which act on an abstract multiple-state
space, constructed by iteratively applying creation operators on a vacuum
state, representing the absence of any activated state. This extension of
the formalism allows us to consider tensor products of individual states
through the action of field operators on the vacuum state.\ In this
approach, acting on such products, operator ,$\underline{\Gamma }^{+}\left( 
\mathbf{T},\mathbf{\alpha },\mathbf{p},S^{2}\right) $ creates an additional
structures with characteristics $\left( \mathbf{\alpha },\mathbf{p}%
,S^{2}\right) $ while $\underline{\Gamma }^{-}\left( \mathbf{T},\mathbf{%
\alpha },\mathbf{p},S^{2}\right) $ deactivates states with such
characteristics, enabling transitions between different states.

The same description can be obtained directly through a dual representation,
which involves considering the creation and destruction operators $\mathbf{A}%
^{+}\left( \alpha ,p,S^{2}\right) $ and $\mathbf{A}^{-}\left( \alpha
,p,S^{2}\right) $ associated with non-local states $\left\vert \mathbf{%
\alpha },\mathbf{p},S^{2}\right\rangle $. These operators are directly
derived from the saddle point equations governing the states $\left\vert 
\mathbf{\alpha },\mathbf{p},S^{2}\right\rangle $. Furthermore, the field
operators $\underline{\Gamma }\left( \mathbf{T},\mathbf{\alpha },\mathbf{p}%
,S^{2}\right) $ (\ref{FT}) are themselve derived from $\mathbf{A}^{+}\left(
\alpha ,p,S^{2}\right) $ and $\mathbf{A}^{-}\left( \alpha ,p,S^{2}\right) $.
The space of multiple states is generated by the successive application of
this family of operators on the vacuum state. For instance, a specific state
with given connectivity values $\left\vert \Delta \mathbf{T}_{p}^{\alpha },%
\mathbf{\alpha },\mathbf{p},S^{2}\right\rangle $ emerges as an eigenstate of
a position operator $\Delta \mathbf{T}_{p}^{\alpha }$\footnote{%
This operator will be defined in the next section.}, which combines the
creation and annihilation operators.

In this approach, the interaction terms in the action functional depend on
the $\underline{\Gamma }\left( \mathbf{T},\mathbf{\alpha },\mathbf{p}%
,S^{2}\right) $ and are expressed as series of products of creation and
destruction operators Thus, interactions can modify collective states or
replace one state with another. As a consequence, they can induce
transitions between states

At this point, it's important to note that the field description utilized
here operates within a much broader space compared to the formalism
presented in the first three parts of this work.\textbf{\ }In those previous
articles, the field defining the connectivity functions $\Delta \Gamma
\left( T,\hat{T},\theta ,Z,Z^{\prime }\right) $ described the state of
connectivity between two cells at locations $\left( Z,Z^{\prime }\right) $.
However, in the current approach, there exists a separate field for each
type of potential collective states.

Each of these fields depends on a substantial number of variables, as the
collective states involve a multitude of points along with potential values
for their associated connectivities, activities frequencies and amplitudes.
Our description encompasses an infinite number of these variables, although,
due to the action functional under consideration and initial conditions,
only certain types of fields may become effective. Nevertheless, this
formalism prevents imposing overly restrictive constraints on the emergence
of collective states. From this perspective, there shouldn't be a fixed
"repertoire" of possible states, even if it may appear that way in practice.

Ultimately, it should be noted that the field formalism considered here
implies that several collective states may be activated multiple times or in
various different configurations. This pertains to states with enhanced
connectivity and their involvement in several interaction processes. This
description aligns with our previous works in which individual cells can be
perceived as complex entities participating in multiple interactions and
assemblies.

\subsection{Collective states as basic elements}

This paragraph summarizes and systematizes the results of the previous
chapter. The description of cell states is formulated in a manner that
allows for the modeling of multiple states. Each state is characterized by
its spatial extent, connectivities, and activities. These variables are
encapsulated in several vector parameters.

\subsubsection{Description of collective states}

Consider assemblies defined as set of points obtained by shifts in
background connectivities. To each assembly:%
\begin{equation*}
S=\left\{ Z_{1},...,Z_{n}\right\}
\end{equation*}%
we associate the field:%
\begin{equation*}
\Delta \Gamma \left( \left( T_{ij},\hat{T}_{ij},Z_{i},Z_{j}\right)
_{Z_{i},Z_{j}\in S},\theta \right)
\end{equation*}%
To account for activation of possible subpattern independently from an
initial collective state, we consider a sequence of independent fields:%
\begin{equation*}
\left\{ \Delta \Gamma _{\beta }\left( \left( T_{ij},\hat{T}%
_{ij},Z_{i}^{\beta },Z_{j}^{\beta }\right) _{Z_{i}^{\beta },Z_{j}^{\beta
}\in S_{\beta }},\theta \right) \right\} _{S_{\beta }\subset S}
\end{equation*}%
associated to $\Delta \Gamma $. The link between a pattern and the
subpattern will be described by some interactions terms. Our choice implies
that both a pattern and several subpatterns may be activated simltaneously,
implying the existence, for one neuron, of a multiplicity of connections
involving a cell in several patterns.

In the sequel, we will write:

\begin{equation*}
S^{2}=S\times S
\end{equation*}%
with points:%
\begin{equation*}
Z_{ij}=\left( Z_{i},Z_{j}\right)
\end{equation*}%
to account for the fact that connectivities in a set $S$ are described by
set of doublets $Z_{ij}$.

We will also introduce a change of notation compared to the previous
sections. In those sections, collective states were described with an
"extra" connectivity vector $\left( \mathbf{\Delta T},\Delta \mathbf{\hat{T}}%
\right) $ compared to background field values. To simplify notations and
emphasize our focus on the structures themselves, we will use the notation $%
\mathbf{T}$ and $\mathbf{\hat{T}}$ for the connectivity variables, or $%
\mathbf{T}_{S^{2}}$ and $\mathbf{\hat{T}}_{S^{2}}$ when the spatial
extension needs to be specified. The notation $\Delta \mathbf{T}$, with
superscripts or subscripts, will be retained to indicate deviations in
connectivity from their average values: $\Delta \mathbf{T}=\mathbf{T-}%
\left\langle \mathbf{T}\right\rangle $. Similarly, we rewrite the "extra"
activity along the structure $\Delta \omega \left( Z,\mathbf{\Delta T}%
\right) $ as $\omega \left( Z,\mathbf{\Delta T}\right) $.

Given our previous results a state will be parametrized by:%
\begin{equation}
\left( S^{2},\left\langle \mathbf{T}\left( \mathbf{Z}\right) \right\rangle
^{\alpha },\Upsilon _{p}^{\alpha },\omega \left( Z,\mathbf{T}\right) \right)
\label{PRS}
\end{equation}%
where $\left\langle \mathbf{T}\left( \mathbf{Z}\right) \right\rangle
^{\alpha }$ is one of the possible solution of (\ref{VRS}) for the average
connectivities, and $\Upsilon _{p}^{\alpha }$ one of the associated
frequencies for the activities:%
\begin{equation*}
\omega _{p}^{\alpha }\left( \theta ,Z,\mathbf{\Delta T}\right) =\omega
\left( Z,\mathbf{\Delta T}\right) +\left( \mathbf{N}_{p}^{\alpha }\right)
^{-1}\mathbf{\omega }_{0}\exp \left( -i\Upsilon _{p}\frac{\left\vert \Delta 
\mathbf{Z}_{i}\right\vert }{c}\right)
\end{equation*}%
with:%
\begin{equation*}
\left[ \mathbf{N}_{p}^{\alpha }\right] _{\left( Z_{i},Z_{j}\right) }=\left(
\delta _{ij}-\left[ \mathbf{\Delta T}\right] _{\left( Z_{i},Z_{j}\right)
}\exp \left( -i\Upsilon _{p}\frac{\left\vert Z_{i}-Z_{j}\right\vert }{c}%
\right) \right) ^{-1}
\end{equation*}

The states we are considering have the form:

\begin{equation}
\prod\limits_{Z,Z^{\prime }}\left\vert \Delta T\left( Z,Z^{\prime }\right)
,\Delta \hat{T}\left( Z,Z^{\prime }\right) ,\alpha \left( Z,Z^{\prime
}\right) ,p\left( Z,Z^{\prime }\right) \right\rangle \equiv \left\vert 
\mathbf{\alpha },\mathbf{p},S^{2}\right\rangle
\end{equation}%
The states $\left\vert \mathbf{\alpha },\mathbf{p},S^{2}\right\rangle $ and
their conjugates $\left\langle \mathbf{\alpha },\mathbf{p},S^{2}\right\vert $
have the form (\ref{CL}) and (\ref{CT}) in absence of interactions.and, or,
transitions.

\subsubsection{Field action functional for states}

Having determined the averages for connectivities and activities, we can
reformulate the action for the collective states. We consider the same
action as (\ref{TRM}), but summing only over points $\left( Z,Z^{\prime
}\right) \in S^{2}$ belonging to the collective state with spatial extension 
$S^{2}$:%
\begin{equation}
S\left( \Delta \Gamma \right) =\sum_{\left( Z,Z^{\prime }\right) \in
S^{2}}S\left( \Delta \Gamma \left( T,\hat{T},\theta ,Z,Z^{\prime }\right)
\right)  \label{FC}
\end{equation}%
While considering directly collective states on their own, we saw that the
correspondent action for such state is:%
\begin{eqnarray}
&&\hat{S}\left( \left\vert \mathbf{\alpha },\mathbf{p},S^{2}\right\rangle
\right)  \label{SC} \\
&=&\left\langle \mathbf{\alpha },\mathbf{p},S^{2}\right\vert \left( -\nabla
_{T}^{2}-\nabla _{\hat{T}}^{2}+\frac{1}{2}\left( \mathbf{\Delta T-}%
\left\langle \mathbf{\Delta T}\right\rangle _{p}^{\alpha }\right) ^{t}%
\mathbf{A}_{p}^{\alpha }\left( \mathbf{\Delta T-}\left\langle \mathbf{\Delta
T}\right\rangle _{p}^{\alpha }\right) +\mathbf{C}\right) \left\vert \mathbf{%
\alpha },\mathbf{p},S^{2}\right\rangle  \notag
\end{eqnarray}%
wth:%
\begin{equation}
\prod\limits_{Z,Z^{\prime }}\left\vert \Delta T\left( Z,Z^{\prime }\right)
,\Delta \hat{T}\left( Z,Z^{\prime }\right) ,\alpha \left( Z,Z^{\prime
}\right) ,p\left( Z,Z^{\prime }\right) \right\rangle \equiv \left\vert
\Delta \mathbf{T},\Delta \mathbf{\hat{T},\alpha },\mathbf{p}%
,S^{2}\right\rangle
\end{equation}%
and:%
\begin{equation*}
\left\langle \mathbf{\alpha },\mathbf{p},S^{2}\right\vert =\left\vert 
\mathbf{\alpha },\mathbf{p},S^{2}\right\rangle ^{\dag }
\end{equation*}%
In (\ref{SC}), we have defined:%
\begin{equation}
\mathbf{\Delta T-}\left\langle \mathbf{\Delta T}\right\rangle _{p}^{\alpha
}=\left( 
\begin{array}{c}
\Delta T-\left\langle \Delta T\right\rangle \\ 
\Delta \hat{T}-\left\langle \Delta \hat{T}\right\rangle%
\end{array}%
\right)
\end{equation}%
and the matrix $\mathbf{A}_{p}^{\alpha }$ is derived in appendix 3:%
\begin{equation*}
\mathbf{A}_{p}^{\alpha }=\left( 
\begin{array}{cc}
\left( \frac{1}{\tau \omega _{0}\left( Z\right) }\right) ^{2}+\left( \mathbf{%
M}^{\alpha }\left( Z,Z^{\prime }\right) \right) ^{2} & -\lambda \left( \frac{%
1}{\tau \omega _{0}\left( Z\right) }\right) ^{2}+D\left( Z,Z^{\prime
}\right) \mathbf{M}^{\alpha }\left( Z,Z^{\prime }\right) \\ 
-\lambda \left( \frac{1}{\tau \omega _{0}\left( Z\right) }\right)
^{2}+D\left( Z,Z^{\prime }\right) \mathbf{M}^{\alpha }\left( Z,Z^{\prime
}\right) & \left( \frac{\lambda }{\tau \omega _{0}\left( Z\right) }\right)
^{2}+D^{2}\left( Z,Z^{\prime }\right)%
\end{array}%
\right)
\end{equation*}%
The matrix $\mathbf{D}$ is diagonal with element:%
\begin{equation*}
\mathbf{D}\left( Z,Z^{\prime }\right) =D\left[ \frac{\rho \left( C\left(
\theta \right) \left\vert \Psi _{0}\left( Z\right) \right\vert ^{2}\omega
_{0}\left( Z\right) +D\left( \theta \right) \hat{T}\left\vert \Psi
_{0}\left( Z^{\prime }\right) \right\vert ^{2}\omega _{0}\left( Z^{\prime
}\right) \right) }{\omega _{0}\left( Z\right) }\right]
\end{equation*}%
and:%
\begin{equation*}
\mathbf{M}^{\alpha }\left( Z,Z^{\prime }\right) =\frac{\rho }{\omega
_{0}\left( Z\right) }\left( D\left( \theta \right) \left\langle \hat{T}%
\right\rangle \left\vert \Psi _{0}\left( Z^{\prime }\right) \right\vert
^{2}A\left\vert Z-Z^{\prime }\right\vert \left( \nabla _{\mathbf{\Delta T}%
_{\left( Z_{1},Z_{1}^{\prime }\right) }}\left( \Delta \omega \left(
Z,\left\langle \mathbf{\Delta T}\right\rangle \right) \right) _{\left(
\left\langle \Delta \mathbf{T}_{\left( Z_{1},Z_{1}^{\prime }\right)
}\right\rangle ^{\alpha }\right) }\right) \right)
\end{equation*}%
The vector $\mathbf{C}\left( Z,Z^{\prime }\right) $ is defined by:%
\begin{equation}
\mathbf{C}\left( Z,Z^{\prime }\right) =\frac{\tau \omega _{0}\left( Z\right) 
}{2}+\frac{\rho \left( C\left( \theta \right) \left\vert \Psi _{0}\left(
Z\right) \right\vert ^{2}\omega _{0}\left( Z\right) +D\left( \theta \right)
\left\vert \Psi _{0}\left( Z^{\prime }\right) \right\vert ^{2}\omega
_{0}\left( Z^{\prime }\right) \right) }{2\omega _{0}\left( Z\right) }
\label{LC}
\end{equation}%
Ultimately, given our hypothesis that:%
\begin{equation*}
\left\Vert \Delta \mathbf{\hat{T}}-\left\langle \Delta \mathbf{\hat{T}}%
\right\rangle \right\Vert <<\left\Vert \Delta \mathbf{T}-\left\langle \Delta 
\mathbf{T}\right\rangle ^{\alpha }\right\Vert
\end{equation*}%
the field action simplifies: 
\begin{eqnarray}
&&\hat{S}\left( \left\vert \mathbf{\alpha },\mathbf{p},S^{2}\right\rangle
\right)  \label{FCT} \\
&=&\left\langle \mathbf{\alpha },\mathbf{p},S^{2}\right\vert \left( -\nabla
_{T}^{2}+\frac{1}{2}\left( \mathbf{\Delta T-}\left\langle \mathbf{\Delta T}%
\right\rangle _{p}^{\alpha }\right) ^{t}\mathbf{A}_{p}^{\alpha }\left( 
\mathbf{\Delta T-}\left\langle \mathbf{\Delta T}\right\rangle _{p}^{\alpha
}\right) +\mathbf{C}\right) \left\vert \mathbf{\alpha },\mathbf{p}%
,S^{2}\right\rangle +U\left( \left\vert \mathbf{\alpha },\mathbf{p}%
,S^{2}\right\rangle \right)  \notag
\end{eqnarray}%
The matrix $\mathbf{A}^{\alpha }$ is defined by:%
\begin{equation*}
\mathbf{A}^{\alpha }=\sqrt{\mathbf{D}^{2}+\left( \mathbf{M}^{\alpha }\right)
^{t}\mathbf{M}^{\alpha }}
\end{equation*}

\subsection{From states to fields}

Starting from the collective states described in the previous paragraph, we
describe general collective states, i.e., states that are not inherently
stable. Considering such states is necessary to study transitions.\ 

To do so, we rewrite the states $\left\vert \mathbf{\alpha },\mathbf{p}%
,S^{2}\right\rangle $ as eigenstates of a certain differential operator.
This operator is itself built from some creation and annihilation operators.
Creation and annihilation operatrs are the fundamental components used to
describe the assembly of states. The creation operators, acting on a vacuum
state, provide the product of states we are looking for.

Moreover, a combination of these creation and annihilation operators allows
to define collective states that are peaked at some fixed values of
connectivty function. These localized states are a suitable basis to define
general collective states. The state space of the system is thus generated
by tensor products of such spaces. The system is then described by a field,
a random variable whose realization are the arbitrary states defined before.
We derive a field action functionl that encompass the dynamics and
transition between states.\ 

\subsubsection{Collective states and operators}

To introduce products of states, we use that the states $\left\vert \mathbf{%
\alpha },\mathbf{p},S^{2}\right\rangle $ are eigenstates of some operators.
Actually, let us consider the saddle point equation for $\left\vert \mathbf{%
\alpha },\mathbf{p},S^{2}\right\rangle $:%
\begin{equation*}
0=\left( -\nabla _{T}^{2}+\frac{1}{2}\left( \mathbf{\Delta T-}\left\langle 
\mathbf{\Delta T}\right\rangle _{p}^{\alpha }\right) ^{t}\mathbf{A}%
_{p}^{\alpha }\left( \mathbf{\Delta T-}\left\langle \mathbf{\Delta T}%
\right\rangle _{p}^{\alpha }\right) +\mathbf{C}\right) \left\vert \mathbf{%
\alpha },\mathbf{p},S^{2}\right\rangle +U^{\prime }\left( \left\vert \mathbf{%
\alpha },\mathbf{p},S^{2}\right\rangle \right) \left\vert \mathbf{\alpha },%
\mathbf{p},S^{2}\right\rangle
\end{equation*}%
In first approximation, for a slowly varying $U^{\prime }\left( \left\vert 
\mathbf{\alpha },\mathbf{p},S^{2}\right\rangle \right) $, we can replace%
\footnote{%
the constant $C$ has been defined in (\ref{LC}).}:%
\begin{equation*}
\mathbf{C}+U^{\prime }\left( \left\vert \mathbf{\alpha },\mathbf{p}%
,S^{2}\right\rangle \right) \rightarrow \mathbf{C}
\end{equation*}%
so that $\left\vert \mathbf{\alpha },\mathbf{p},S^{2}\right\rangle $ are
eigenvectors of:%
\begin{equation}
-\frac{1}{2}\nabla _{\left( \mathbf{T}\right) _{S^{2}}}^{2}+\frac{1}{2}%
\left( \mathbf{T}_{S^{2}}-\left\langle \mathbf{T}_{p}^{\alpha }\right\rangle
_{S^{2}}\right) ^{t}\mathbf{A}_{S^{2}}^{\alpha }\left( \mathbf{T}%
_{S^{2}}-\left\langle \mathbf{T}_{p}^{\alpha }\right\rangle _{S^{2}}\right)
=-\frac{1}{2}\nabla _{\left( \mathbf{T}\right) _{S^{2}}}^{2}+\frac{1}{2}%
\left( \Delta \mathbf{T}_{p}^{\alpha }\right) _{S^{2}}^{t}\mathbf{A}%
_{S^{2}}^{\alpha }\left( \Delta \mathbf{T}_{p}^{\alpha }\right)  \label{PR}
\end{equation}%
where:%
\begin{equation*}
\left( \Delta \mathbf{T}_{p}^{\alpha }\right) _{S^{2}}=\mathbf{T}%
_{S^{2}}-\left\langle \mathbf{T}_{p}^{\alpha }\right\rangle _{S^{2}}
\end{equation*}%
To describe further the collective states, we assume that the operator:%
\begin{equation*}
-\frac{1}{2}\nabla _{\left( \mathbf{T}\right) _{S^{2}}}^{2}+\frac{1}{2}%
\left( \Delta \mathbf{T}_{p}^{\alpha }\right) _{S^{2}}^{t}\mathbf{A}%
_{S^{2}}^{\alpha }\left( \Delta \mathbf{T}_{p}^{\alpha }\right)
\end{equation*}%
can be diagonalized as:%
\begin{equation}
-\frac{1}{2}\nabla _{\left( \mathbf{\bar{T}}\right) _{S^{2}}}^{2}+\frac{1}{2}%
\left( \Delta \mathbf{\bar{T}}_{p}^{\alpha }\right) _{S^{2}}^{t}\mathbf{\bar{%
D}}_{S^{2}}^{\alpha }\left( \Delta \mathbf{\bar{T}}_{p}^{\alpha }\right)
_{S^{2}}  \label{DPR}
\end{equation}%
with:%
\begin{eqnarray*}
\mathbf{\bar{D}}_{S^{2}}^{\alpha } &=&O^{-1}\left( \mathbf{A}%
_{S^{2}}^{\alpha }\right) O \\
\left( \Delta \mathbf{\bar{T}}_{p}^{\alpha }\right) _{S^{2}} &=&O^{-1}\left(
\Delta \mathbf{T}_{p}^{\alpha }\right) _{S^{2}}
\end{eqnarray*}

The \ possible eigenvalues of (\ref{DPR}) associated to $\left\vert \mathbf{%
\alpha },\mathbf{p},S^{2}\right\rangle $ are determined simulaneously with
the norm of $\left\vert \mathbf{\alpha },\mathbf{p},S^{2}\right\rangle $
(see (\cite{GLt})). Actually, in first approximation, the norm of $%
\left\vert \mathbf{\alpha },\mathbf{p},S^{2}\right\rangle $ is given by the
minimum of $U\left( \left\vert \mathbf{\alpha },\mathbf{p}%
,S^{2}\right\rangle \right) $. However, imposing that the norm of $%
\left\vert \mathbf{\alpha },\mathbf{p},S^{2}\right\rangle $ is bounded,
implies a condition on the eigenvalue (\ref{DPR}). These eigenvalues have
the form:%
\begin{equation*}
\sum_{i}\left( \sqrt{\mathbf{\bar{D}}_{S^{2}}^{\alpha }}\right) _{i}\left(
l_{i}+\frac{1}{2}\right)
\end{equation*}%
where the $\left( \sqrt{\mathbf{\bar{D}}_{S^{2}}^{\alpha }}\right) _{i}$ are
the components of $\sqrt{\mathbf{\bar{D}}_{S^{2}}^{\alpha }}$ and $l_{i}$ is
a sequence of integer. This imposes a condition yielding the norm of $%
\left\vert \mathbf{\alpha },\mathbf{p},S^{2}\right\rangle $ (see also below
the condition for collective states).

\subsubsection{Creation and annihilation operators}

We can now introduce the creation and annihilation operators relevant to the
system. To do so, we use that operator (\ref{DPR}) writes also:%
\begin{eqnarray}
-\frac{1}{2}\nabla _{\left( \mathbf{\bar{T}}\right) _{S^{2}}}^{2}+\frac{1}{2}%
\left( \Delta \mathbf{\bar{T}}_{p}^{\alpha }\right) _{S^{2}}^{t}\mathbf{\bar{%
D}}_{S^{2}}^{\alpha }\left( \Delta \mathbf{\bar{T}}_{p}^{\alpha }\right)
_{S^{2}} &=&\mathbf{\bar{D}}_{S^{2}}^{\alpha }\left( \mathbf{A}^{+}\left(
\alpha ,p,S^{2}\right) \mathbf{A}^{-}\left( \alpha ,p,S^{2}\right) +\frac{1}{%
2}\right)  \label{DRT} \\
&=&\left( \mathbf{\bar{D}}_{S^{2}}^{\alpha }\right) _{i}\left( \mathbf{A}%
_{i}^{+}\left( \alpha ,p,S^{2}\right) \mathbf{A}_{i}^{-}\left( \alpha
,p,S^{2}\right) +\frac{1}{2}\right)  \notag
\end{eqnarray}%
where:%
\begin{eqnarray}
\mathbf{A}^{-}\left( \alpha ,p,S^{2}\right) &=&\frac{1}{2}\left( \sqrt{%
\mathbf{\bar{D}}_{S^{2}}^{\alpha }}\Delta \mathbf{\bar{T}}_{p}^{\alpha }-%
\frac{1}{\sqrt{\mathbf{\bar{D}}_{S^{2}}^{\alpha }}}\nabla _{\left( \mathbf{%
\bar{T}}\right) _{S^{2}}}^{2}\right)  \label{CRN} \\
\mathbf{A}^{+}\left( \alpha ,p,S^{2}\right) &=&\frac{1}{2}\left( \sqrt{%
\mathbf{\bar{D}}_{S^{2}}^{\alpha }}\Delta \mathbf{\bar{T}}_{p}^{\alpha }+%
\frac{1}{\sqrt{\mathbf{\bar{D}}_{S^{2}}^{\alpha }}}\nabla _{\left( \mathbf{%
\bar{T}}\right) _{S^{2}}}^{2}\right)  \notag
\end{eqnarray}%
and $\mathbf{A}_{i}^{\pm }\left( \alpha ,p,S^{2}\right) $ and $\left( 
\mathbf{\bar{D}}_{S^{2}}^{\alpha }\right) _{i}$ are cmpnnts of $\mathbf{A}%
^{\pm }\left( \alpha ,p,S^{2}\right) $ and $\mathbf{\bar{D}}_{S^{2}}^{\alpha
}$ respectively.

The operators $\mathbf{A}^{\pm }\left( \alpha ,p,S^{2}\right) $\ satisfy the
following commutation relation:%
\begin{equation*}
\left[ \mathbf{A}_{i}^{-}\left( \mathbf{\alpha },\mathbf{p},S^{2}\right) ,%
\mathbf{A}_{j}^{+}\left( \mathbf{\alpha }^{\prime },\mathbf{p}^{\prime
},S^{\prime 2}\right) \right] =\delta \left( \mathbf{p}-\mathbf{p}^{\prime
}\right) \delta \left( \mathbf{\alpha }-\mathbf{\alpha }^{\prime }\right)
\delta \left( S^{2}-S^{\prime 2}\right) \delta _{i,j}
\end{equation*}%
The states $\left\vert \mathbf{\alpha },\mathbf{p},S^{2}\right\rangle $ are
obtained by successive applications of the components $\mathbf{A}%
_{i}^{+}\left( \alpha ,p,S^{2}\right) $ of $\mathbf{A}^{+}\left( \alpha
,p,S^{2}\right) $ on the minimum:%
\begin{equation}
\prod\limits_{i}\left( \mathbf{A}_{i}^{+}\left( \alpha ,p,S^{2}\right)
\right) ^{l_{i}}\left\vert vac\right\rangle =\left\vert \mathbf{\alpha },%
\mathbf{p},S^{2}\right\rangle  \label{PTS}
\end{equation}%
The eigenvalue of (\ref{DPR}), or which is equivalent, of (\ref{DRT})
associated to such state is $\sum_{i}\left( \sqrt{\mathbf{\bar{D}}%
_{S^{2}}^{\alpha }}\right) _{i}\left( l_{i}+\frac{1}{2}\right) $

For large number of elementsnote that states $\left\vert \mathbf{\alpha },%
\mathbf{p},S^{2}\right\rangle $\ are mutually orthogonl:%
\begin{equation*}
\left\langle \mathbf{\alpha },\mathbf{p},S^{2}\right\vert \left\vert \left( 
\mathbf{\alpha }^{\prime },\mathbf{p}^{\prime },S^{2}\right) \right\rangle
\simeq \exp \left( -\sum \left( \left\langle \Delta \mathbf{T}\right\rangle
_{p}^{\alpha }-\left\langle \Delta \mathbf{T}\right\rangle _{p^{\prime
}}^{\alpha ^{\prime }}\right) ^{2}\right) \simeq 0
\end{equation*}

\subsubsection{Connectivity states and operator}

Recall that states $\left\vert \mathbf{\alpha },\mathbf{p}%
,S^{2}\right\rangle $ are characterized by connectivities at each point $%
\left( Z,Z^{\prime }\right) $ being distributed around an average value (see
(\ref{SF})), modeling that connections between axons and dendrites are
defined by a range of values.

However, to calculate transitions between states, it is necessary to
consider states with specific values $\Delta \mathbf{T}_{p}^{\alpha }$ or
its diagonalized form $\Delta \mathbf{\bar{T}}_{p}^{\alpha }$. These states
differ from $\left\vert \mathbf{\alpha },\mathbf{p},S^{2}\right\rangle $ and
are not stable, instead, they represent all possible connectivity states at
a given time. These form the basis for transient states that enable the
description of dynamic transitions. Indeed, connectivity values fluctuate
due to interactions, and these fluctuations drive state changes.

To derive the form of these states, we assume that for a large number of
states most of them are in fundamental states and we consider:%
\begin{equation*}
\left\vert \Delta \mathbf{\bar{T}}_{p}^{\alpha },\mathbf{\alpha },\mathbf{p}%
,S^{2}\right\rangle =\prod \left\vert \Delta \bar{T}_{p}^{\alpha },\alpha
,p,S^{2}\right\rangle
\end{equation*}%
that are eigenstate of operator:%
\begin{equation}
\Delta \mathbf{\bar{T}}_{p}^{\alpha }=\sqrt{\mathbf{\bar{D}}_{S^{2}}}\left( 
\mathbf{A}^{-}\left( \mathbf{\alpha },\mathbf{p},S^{2}\right) +\mathbf{A}%
^{+}\left( \mathbf{\alpha },\mathbf{p},S^{2}\right) \right)  \label{CP}
\end{equation}%
We also define::%
\begin{equation*}
\left\vert \mathbf{\bar{T}}_{p}^{\alpha },\mathbf{\alpha },\mathbf{p}%
,S^{2}\right\rangle \equiv \left\vert \Delta \mathbf{\bar{T}}_{p}^{\alpha },%
\mathbf{\alpha },\mathbf{p},S^{2}\right\rangle
\end{equation*}%
that are the same states, seen as eigenstate of:%
\begin{equation}
\mathbf{\mathbf{\bar{T}}_{p}^{\alpha }\equiv }\sqrt{\mathbf{\bar{D}}_{S^{2}}}%
\left( \mathbf{A}^{-}\left( \mathbf{\alpha },\mathbf{p},S^{2}\right) +%
\mathbf{A}^{+}\left( \mathbf{\alpha },\mathbf{p},S^{2}\right) \right)
+\left\langle \mathbf{\bar{T}}_{p}^{\alpha }\right\rangle  \label{PN}
\end{equation}%
States $\left\vert \Delta \mathbf{\bar{T}}_{p}^{\alpha },\mathbf{\alpha },%
\mathbf{p},S^{2}\right\rangle $ also satisfy:%
\begin{equation}
\left\langle \left( \Delta \mathbf{\bar{T}}_{p}^{\alpha }\right) ^{\prime },%
\mathbf{\alpha },\mathbf{p},S^{2}\right\vert \nabla _{\Delta \mathbf{\bar{T}}%
}\left\vert \Delta \mathbf{\bar{T}}_{p}^{\alpha },\mathbf{\alpha },\mathbf{p}%
,S^{2}\right\rangle =\nabla _{\Delta \mathbf{\bar{T}}}\delta \left( \left(
\Delta \mathbf{\bar{T}}_{p}^{\alpha }\right) ^{\prime }-\Delta \mathbf{\bar{T%
}}_{p}^{\alpha }\right)  \label{DRV}
\end{equation}%
Moreover, coming back to orignal variables:%
\begin{equation*}
\left\vert \Delta \mathbf{T}_{p}^{\alpha },\mathbf{\alpha },\mathbf{p}%
,S^{2}\right\rangle =O\left\vert \Delta \mathbf{\bar{T}}_{p}^{\alpha },%
\mathbf{\alpha },\mathbf{p},S^{2}\right\rangle
\end{equation*}%
this state is eigenstate of operator $\Delta \mathbf{T}_{p}^{\alpha }$ and
satisfy:%
\begin{equation}
\left\langle \left( \Delta \mathbf{T}_{p}^{\alpha }\right) ^{\prime },%
\mathbf{\alpha },\mathbf{p},S^{2}\right\vert \nabla _{\Delta \mathbf{\bar{T}}%
}\left\vert \Delta \mathbf{T}_{p}^{\alpha },\mathbf{\alpha },\mathbf{p}%
,S^{2}\right\rangle =\nabla _{\Delta \mathbf{T}}\delta \left( \left( \Delta 
\mathbf{T}_{p}^{\alpha }\right) ^{\prime }-\Delta \mathbf{T}_{p}^{\alpha
}\right)  \label{DVT}
\end{equation}

\subsubsection{States $\left\vert \Delta \mathbf{\bar{T}}_{p}^{\protect%
\alpha },\mathbf{\protect\alpha },\mathbf{p},S^{2}\right\rangle $ \ from
creation and annihilation operators}

As states $\left\vert \mathbf{\alpha },\mathbf{p},S^{2}\right\rangle $ are
obtained by application of operators on the vacuum, the states $\left\vert
\Delta \mathbf{\bar{T}}_{p}^{\alpha },\mathbf{\alpha },\mathbf{p}%
,S^{2}\right\rangle $ can be obtained through the fundamental states of the
structure. Actually, we use (\ref{CP}):%
\begin{equation*}
\sqrt{\mathbf{\bar{D}}_{S^{2}}}\left( \mathbf{A}^{-}\left( \mathbf{\alpha },%
\mathbf{p},S^{2}\right) +\mathbf{A}^{+}\left( \mathbf{\alpha },\mathbf{p}%
,S^{2}\right) \right) \left\vert \Delta \mathbf{\bar{T}}_{p}^{\alpha },%
\mathbf{\alpha },\mathbf{p},S^{2}\right\rangle =\Delta \mathbf{\bar{T}}%
_{p}^{\alpha }\left\vert \Delta \mathbf{\bar{T}}_{p}^{\alpha },\mathbf{%
\alpha },\mathbf{p},S^{2}\right\rangle
\end{equation*}%
and the solutions of this equation are:

\begin{eqnarray}
&&\left\vert \Delta \mathbf{\bar{T}}_{p}^{\alpha },\mathbf{\alpha },\mathbf{p%
},S^{2}\right\rangle  \label{DCN} \\
&=&\exp \left( -\left( \Delta \mathbf{T}_{p}^{\alpha }\mathbf{\bar{D}}%
_{S^{2}}\Delta \mathbf{T}_{p}^{\alpha }+2\left( \Delta \mathbf{\bar{T}}%
_{p}^{\alpha }\right) ^{t}\sqrt{\mathbf{\bar{D}}_{S^{2}}}\mathbf{A}%
^{+}\left( \mathbf{\alpha },\mathbf{p},S^{2}\right) \right) +\frac{1}{2}%
\mathbf{A}^{+}\left( \mathbf{\alpha },\mathbf{p},S^{2}\right) .\mathbf{A}%
^{+}\left( \mathbf{\alpha },\mathbf{p},S^{2}\right) \right) \left\vert
Vac\right\rangle  \notag
\end{eqnarray}%
Coming back to the original variables, we also write:%
\begin{eqnarray}
&&\left\vert \Delta \mathbf{T}_{p}^{\alpha },\mathbf{\alpha },\mathbf{p}%
,S^{2}\right\rangle  \label{LCN} \\
&=&\exp \left( -\left( \left( \Delta \mathbf{T}_{p}^{\alpha }\right) ^{t}%
\mathbf{A}_{p}^{\alpha }\Delta \mathbf{T}_{p}^{\alpha }+2\left( \Delta 
\mathbf{T}_{p}^{\alpha }\right) ^{t}\sqrt{\mathbf{A}_{p}^{\alpha }}\mathbf{%
\hat{A}}^{+}\left( \mathbf{\alpha },\mathbf{p},S^{2}\right) \right) +\frac{1%
}{2}\mathbf{\hat{A}}^{+}\left( \mathbf{\alpha },\mathbf{p},S^{2}\right) .%
\mathbf{\hat{A}}^{+}\left( \mathbf{\alpha },\mathbf{p},S^{2}\right) \right)
\left\vert Vac\right\rangle  \notag
\end{eqnarray}%
\begin{equation*}
\mathbf{\hat{A}}^{\pm }\left( \mathbf{\alpha },\mathbf{p},S^{2}\right)
=O^{-1}\mathbf{A}^{\pm }\left( \mathbf{\alpha },\mathbf{p},S^{2}\right)
\end{equation*}%
The states $\left\vert \Delta \mathbf{T}_{p}^{\alpha },\mathbf{\alpha },%
\mathbf{p},S^{2}\right\rangle $ are thus combinations of product of states $%
\left\vert \mathbf{\alpha },\mathbf{p},S^{2}\right\rangle $.

\subsubsection{General form of collective states}

Then, we define the most general state of the structure given spatial
extension $S^{2}$ for state $\mathbf{\alpha },\mathbf{p}$:

\begin{equation}
\left\vert \underline{\gamma }\left( \boldsymbol{\alpha },\mathbf{p}%
,S^{2}\right) \right\rangle =\int \underline{\gamma }\left( \mathbf{T},%
\mathbf{\alpha },\mathbf{p},S^{2}\right) \left\vert \Delta \mathbf{T}%
_{p}^{\alpha },\mathbf{\alpha },\mathbf{p},S^{2}\right\rangle d\mathbf{T}
\label{FLS}
\end{equation}%
or:%
\begin{equation}
\left\vert \underline{\gamma }\left( \boldsymbol{\alpha },\mathbf{p}%
,S^{2}\right) \right\rangle =\int \underline{\gamma }\left( \mathbf{\bar{T}},%
\mathbf{\alpha },\mathbf{p},S^{2}\right) \left\vert \Delta \mathbf{\bar{T}}%
_{p}^{\alpha },\mathbf{\alpha },\mathbf{p},S^{2}\right\rangle d\mathbf{\bar{T%
}}  \label{FLD}
\end{equation}%
and summing over all possible states:%
\begin{eqnarray*}
\left\vert \underline{\gamma }\left( S^{2}\right) \right\rangle
&=&\sum_{\alpha ,p}\left\vert \underline{\gamma }\left( \boldsymbol{\alpha },%
\mathbf{p},S^{2}\right) \right\rangle =\sum_{\alpha ,p}\int \underline{%
\gamma }\left( \mathbf{T},\mathbf{\alpha },\mathbf{p},S^{2}\right)
\left\vert \Delta \mathbf{\bar{T}}_{p}^{\alpha },\mathbf{\alpha },\mathbf{p}%
,S^{2}\right\rangle d\mathbf{T} \\
\left\vert \underline{\gamma }\left( S^{2}\right) \right\rangle
&=&\sum_{\alpha ,p}\left\vert \underline{\gamma }\left( \boldsymbol{\alpha },%
\mathbf{p},S^{2}\right) \right\rangle =\sum_{\alpha ,p}\int \underline{%
\gamma }\left( \mathbf{\bar{T}},\mathbf{\alpha },\mathbf{p},S^{2}\right)
\left\vert \Delta \mathbf{\bar{T}}_{p}^{\alpha },\mathbf{\alpha },\mathbf{p}%
,S^{2}\right\rangle d\mathbf{\bar{T}}
\end{eqnarray*}%
Such states do not in general satisfy saddle point equation, but are
necessary as transitory states. In fact, the components $\underline{\gamma }%
\left( \mathbf{\bar{T}},\mathbf{\alpha },\mathbf{p},S^{2}\right) $ are the
possible realizations of a field $\underline{\Gamma }\left( \mathbf{\bar{T}},%
\mathbf{\alpha },\mathbf{p},S^{2}\right) $.

\subsubsection{Action functional for a general state}

Once the states $\left\vert \underline{\gamma }\left( \mathbf{T,}\boldsymbol{%
\alpha },\mathbf{p},S^{2}\right) \right\rangle $ are defined as combination
of the fundamental states, the action of the field becomes a functional $%
S\left( \underline{\gamma }\left( \mathbf{T},\mathbf{\alpha },\mathbf{p}%
,S^{2}\right) \right) $:%
\begin{eqnarray}
&&S\left( \underline{\gamma }\left( \mathbf{T},\mathbf{\alpha },\mathbf{p}%
,S^{2}\right) \right)  \label{FCW} \\
&=&\int \left\langle \left( \Delta \mathbf{\bar{T}}_{p}^{\alpha }\right)
^{\prime },\mathbf{\alpha },\mathbf{p},S^{2}\right\vert \underline{\gamma }%
^{\dag }\left( \mathbf{T}^{\prime },\mathbf{\alpha },\mathbf{p},S^{2}\right)
\notag \\
&&\times \left( -\frac{1}{2}\nabla _{\left( \mathbf{T}\right) _{S^{2}}}^{2}+%
\frac{1}{2}\left( \Delta \mathbf{T}_{p}^{\alpha }\right) _{S^{2}}^{t}\mathbf{%
A}_{S^{2}}^{\alpha }\left( \Delta \mathbf{T}_{p}^{\alpha }\right) \right) 
\underline{\gamma }\left( \mathbf{T},\mathbf{\alpha },\mathbf{p}%
,S^{2}\right) \left\vert \Delta \mathbf{\bar{T}}_{p}^{\alpha },\mathbf{%
\alpha },\mathbf{p},S^{2}\right\rangle d\mathbf{T}d\mathbf{T}^{\prime } 
\notag
\end{eqnarray}%
Given that $\left\vert \Delta \mathbf{\bar{T}}_{p}^{\alpha },\mathbf{\alpha }%
,\mathbf{p},S^{2}\right\rangle $ are eigenstates of $\Delta \mathbf{\bar{T}}%
_{p}^{\alpha }$ and using (\ref{DVT}), this reduces to:%
\begin{equation*}
S\left( \underline{\gamma }\left( \mathbf{T,\alpha },\mathbf{p},S^{2},\theta
\right) \right) =\int \underline{\gamma }^{\dag }\left( \mathbf{T},\mathbf{%
\alpha },\mathbf{p},S^{2}\right) \left( -\frac{1}{2}\nabla _{\left( \mathbf{T%
}\right) _{S^{2}}}^{2}+\frac{1}{2}\left( \Delta \mathbf{T}_{p}^{\alpha
}\right) _{S^{2}}^{t}\mathbf{A}_{S^{2}}^{\alpha }\left( \Delta \mathbf{T}%
_{p}^{\alpha }\right) \right) \underline{\gamma }\left( \mathbf{T},\mathbf{%
\alpha },\mathbf{p},S^{2}\right) d\mathbf{T}
\end{equation*}%
Working rather with the diagonalized form and using (\ref{DRV}), this allows
to write:%
\begin{equation*}
S\left( \underline{\gamma }\left( \mathbf{\bar{T},\alpha },\mathbf{p}%
,S^{2},\theta \right) \right) =\int \underline{\gamma }^{\dag }\left( 
\mathbf{T},\mathbf{\alpha },\mathbf{p},S^{2}\right) \left( -\frac{1}{2}%
\nabla _{\left( \mathbf{\bar{T}}\right) _{S^{2}}}^{2}+\frac{1}{2}\left(
\Delta \mathbf{\bar{T}}_{p}^{\alpha }\right) _{S^{2}}^{t}\mathbf{\bar{D}}%
_{S^{2}}^{\alpha }\left( \Delta \mathbf{\bar{T}}_{p}^{\alpha }\right)
_{S^{2}}\right) \underline{\gamma }\left( \mathbf{T},\mathbf{\alpha },%
\mathbf{p},S^{2}\right) d\mathbf{\bar{T}}
\end{equation*}

\subsubsection{Field action functional}

We now replace $\gamma \left( \mathbf{\bar{T},\alpha },\mathbf{p}%
,S^{2},\theta \right) $ by a field $\underline{\Gamma }\left( \mathbf{%
T,\alpha },\mathbf{p},S^{2},\theta \right) $ which is a field whose $%
\underline{\gamma }\left( \mathbf{\bar{T},\alpha },\mathbf{p},S^{2},\theta
\right) $ are the realizations. To describe the system as a whole we
consider that set of multi-component field $\left\{ \underline{\Gamma }%
\left( \mathbf{T,\alpha },\mathbf{p},S^{2},\theta \right) \right\} $.
Including the constants $\mathbf{C}$ defined in (\ref{LC})\ and the
potential $U$ leads: 
\begin{eqnarray}
&&S\left( \left\{ \underline{\Gamma }\left( \mathbf{T},\mathbf{\alpha },%
\mathbf{p},S^{2},\theta \right) \right\} \right)  \label{FCZ} \\
&=&\sum_{\left\{ \mathbf{\alpha },\mathbf{p},S^{2}\right\} }\underline{%
\Gamma }^{\dag }\left( \mathbf{T},\mathbf{\alpha },\mathbf{p},S^{2},\theta
\right) \left( -\nabla _{\Delta \mathbf{T}}^{2}+\frac{1}{2}\left( \Delta 
\mathbf{T}_{p}^{\alpha }\right) _{S^{2}}^{t}\mathbf{A}_{S^{2}}^{\alpha
}\left( \Delta \mathbf{T}_{p}^{\alpha }\right) +\mathbf{C}\right) \underline{%
\Gamma }\left( \Delta \mathbf{T},\mathbf{\alpha },\mathbf{p},S^{2},\theta
\right)  \notag \\
&&+U\left( \left\Vert \underline{\Gamma }\left( \left( \Delta \mathbf{T}%
_{p}^{\alpha }\right) _{S^{2}},\mathbf{\alpha },\mathbf{p},S^{2},\theta
\right) \right\Vert ^{2}\right)  \notag
\end{eqnarray}%
where:%
\begin{equation*}
\left( \Delta \mathbf{T}_{p}^{\alpha }\right) _{S^{2}}=\mathbf{T}%
_{S^{2}}-\left\langle \mathbf{T}_{p}^{\alpha }\right\rangle _{S^{2}}
\end{equation*}%
\begin{equation*}
\mathbf{C}=\sum_{\left( Z,Z^{\prime }\right) \in S\times S}C\left(
Z,Z^{\prime }\right)
\end{equation*}

Note that $S\left( \underline{\Gamma }\left( \mathbf{\bar{T},\alpha },%
\mathbf{p},S^{2},\theta \right) \right) $ is similar to $S\left( \underline{%
\gamma }\left( \mathbf{\bar{T},\alpha },\mathbf{p},S^{2},\theta \right)
\right) $ but with the difference that $\underline{\Gamma }\left( \mathbf{%
\bar{T},\alpha },\mathbf{p},S^{2},\theta \right) $ is a randomvrb with
realizations $\underline{\gamma }\left( \mathbf{\bar{T},\alpha },\mathbf{p}%
,S^{2},\theta \right) $. The dynamics of the system will thus be encompassed
in the partition function:%
\begin{equation}
\int \exp \left( -S\left( \underline{\Gamma }\left( \mathbf{\bar{T},\alpha },%
\mathbf{p},S^{2},\theta \right) \right) \right) D\left\{ \underline{\Gamma }%
\left( \mathbf{T},\mathbf{\alpha },\mathbf{p},S^{2},\theta \right) \right\}
\label{PF}
\end{equation}%
Moreover, since $S\left( \underline{\Gamma }\left( \mathbf{\bar{T},\alpha },%
\mathbf{p},S^{2},\theta \right) \right) $ describes the whole set of
possible stts, it depends on the collection $\left\{ \underline{\Gamma }%
\left( \mathbf{T},\mathbf{\alpha },\mathbf{p},S^{2},\theta \right) \right\} $%
. This will allow to include interactions between collective states in (\ref%
{PF}).

Note also that the action $S\left( \left\{ \underline{\Gamma }\left( \mathbf{%
T},\mathbf{\alpha },\mathbf{p},S^{2},\theta \right) \right\} \right) $ is
similar to $\hat{S}\left( \Delta \Gamma \left( T,\hat{T},\theta ,Z,Z^{\prime
}\right) \right) $ where the dynamics for $\hat{T}$ has been neglected but
with the replacement:%
\begin{equation*}
\Delta \Gamma \left( T,\hat{T},\theta ,Z,Z^{\prime }\right) \rightarrow 
\underline{\Gamma }\left( \mathbf{T},\mathbf{\alpha },\mathbf{p}%
,S^{2},\theta \right)
\end{equation*}%
where $S$ is the spatial extension of the collective state. This replacement
stresses that the fundamental object are now the states made of the set of
activated interacting connections and producing the activities $\Delta
\omega _{p}^{\alpha }\left( \theta ,\mathbf{Z}\right) $.

\subsubsection{Oprtrs descriptn}

Alternatively, the field action $S\left( \underline{\Gamma }\left( \mathbf{%
T,\alpha },\mathbf{p},S^{2},\theta \right) \right) $ can be considered as a
matrix element of an operator as in (\ref{FCW}). Actually, given (\ref{DRT}%
), the free part of the action (\ref{FCZ}) can be written:%
\begin{eqnarray}
&&S_{f}\left( \underline{\Gamma }\left( \mathbf{T,\alpha },\mathbf{p}%
,S^{2},\theta \right) \right) \\
&=&\int \left\langle \left( \Delta \mathbf{\bar{T}}_{p}^{\alpha }\right)
^{\prime },\mathbf{\alpha },\mathbf{p},S^{2}\right\vert \underline{\Gamma }%
^{\dag }\left( \mathbf{T}^{\prime },\mathbf{\alpha },\mathbf{p},S^{2}\right)
\notag \\
&&\times \left( -\frac{1}{2}\nabla _{\left( \mathbf{T}\right) _{S^{2}}}^{2}+%
\frac{1}{2}\left( \Delta \mathbf{T}_{p}^{\alpha }\right) _{S^{2}}^{t}\mathbf{%
A}_{S^{2}}^{\alpha }\left( \Delta \mathbf{T}_{p}^{\alpha }\right) +\mathbf{C}%
\right) \underline{\Gamma }\left( \mathbf{T},\mathbf{\alpha },\mathbf{p}%
,S^{2}\right) \left\vert \Delta \mathbf{\bar{T}}_{p}^{\alpha },\mathbf{%
\alpha },\mathbf{p},S^{2}\right\rangle d\mathbf{T}d\mathbf{T}^{\prime } 
\notag \\
&=&\left\langle \underline{\Gamma }\left( \mathbf{T,}\boldsymbol{\alpha },%
\mathbf{p},S^{2}\right) \right\vert \left( \mathbf{\bar{D}}_{S^{2}}^{\alpha
}\left( \mathbf{A}^{+}\left( \alpha ,p,S^{2}\right) \mathbf{A}^{-}\left(
\alpha ,p,S^{2}\right) +\frac{1}{2}+\mathbf{C}\right) \right) \left\vert 
\underline{\Gamma }\left( \mathbf{T,}\boldsymbol{\alpha },\mathbf{p}%
,S^{2}\right) \right\rangle  \notag
\end{eqnarray}%
Equivalently, using the initial variables:%
\begin{equation*}
S_{f}\left( \underline{\Gamma }\left( \mathbf{T,\alpha },\mathbf{p}%
,S^{2},\theta \right) \right) =\left\langle \underline{\Gamma }\left( 
\mathbf{T,}\boldsymbol{\alpha },\mathbf{p},S^{2}\right) \right\vert \left(
\left( \mathbf{\hat{A}}^{+}\left( \alpha ,p,S^{2}\right) \right) ^{t}\mathbf{%
A}_{S^{2}}^{\alpha }\mathbf{\hat{A}}^{-}\left( \alpha ,p,S^{2}\right) +\frac{%
1}{2}+\mathbf{C}\right) \left\vert \underline{\Gamma }\left( \mathbf{T,}%
\boldsymbol{\alpha },\mathbf{p},S^{2}\right) \right\rangle
\end{equation*}%
with:%
\begin{equation*}
\mathbf{\hat{A}}^{\pm -}\left( \alpha ,p,S^{2}\right) =O\mathbf{A}^{\pm
-}\left( \alpha ,p,S^{2}\right)
\end{equation*}%
Thus, integrating over the degrees of freedom for $\underline{\Gamma }\left(
\Delta \mathbf{T},\mathbf{\alpha },\mathbf{p},S^{2},\theta \right) $ is
equivalent to compute transition elements of operators.

Considering the potential $U\left( \left\Vert \underline{\Gamma }\left(
\left( \Delta \mathbf{T}_{p}^{\alpha }\right) _{S^{2}},\mathbf{\alpha },%
\mathbf{p},S^{2},\theta \right) \right\Vert ^{2}\right) $ in action (\ref%
{FCZ}),\ it is a matrix element between tensor products of states. Actually,
starting with a series expansion for $U$:%
\begin{eqnarray}
&&U\left( \left\Vert \underline{\Gamma }\left( \left( \Delta \mathbf{T}%
_{p}^{\alpha }\right) _{S^{2}},\mathbf{\alpha },\mathbf{p},S^{2},\theta
\right) \right\Vert ^{2}\right)  \label{PL} \\
&=&\sum_{k}\int \prod\limits_{l=1}^{k}\underline{\Gamma }^{\dag }\left(
\left( \Delta \mathbf{T}_{p}^{\alpha }\right) _{S^{2}}^{\left( l\right) },%
\mathbf{\alpha },\mathbf{p},S^{2},\theta \right) \hat{U}_{k}\left( \left(
\left( \Delta \mathbf{T}_{p}^{\alpha }\right) _{S^{2}}^{\left( l\right)
}\right) _{l=1...k}\right) \prod\limits_{l=1}^{k}\underline{\Gamma }\left(
\left( \Delta \mathbf{T}_{p}^{\alpha }\right) _{S^{2}}^{\left( l\right) },%
\mathbf{\alpha },\mathbf{p},S^{2},\theta \right)  \notag
\end{eqnarray}%
this writes;%
\begin{eqnarray*}
&&U\left( \left\Vert \underline{\Gamma }\left( \left( \Delta \mathbf{T}%
_{p}^{\alpha }\right) _{S^{2}},\mathbf{\alpha },\mathbf{p},S^{2},\theta
\right) \right\Vert ^{2}\right) \\
&=&\sum_{k}\int \prod\limits_{l=1}^{k}\left\langle \underline{\Gamma }\left(
\left( \Delta \mathbf{T}_{p}^{\alpha }\right) _{S^{2}}^{\left( l\right) }%
\mathbf{,}\boldsymbol{\alpha },\mathbf{p},S^{2}\right) \right\vert \hat{U}%
_{k}\left( \left( \left( \Delta \mathbf{T}_{p}^{\alpha }\right)
_{S^{2}}^{\left( l\right) }\right) _{l=1...k}\right)
\prod\limits_{l=1}^{k}\left\vert \underline{\Gamma }\left( \left( \Delta 
\mathbf{T}_{p}^{\alpha }\right) _{S^{2}}^{\left( l\right) }\mathbf{,}%
\boldsymbol{\alpha },\mathbf{p},S^{2}\right) \right\rangle
\end{eqnarray*}%
so that the potential is the matrix elements of the operator:%
\begin{equation*}
\hat{U}=\sum_{k}\int \prod\limits_{l=1}^{k}\left\vert \left( \Delta \mathbf{T%
}_{p}^{\alpha }\right) _{S^{2}}^{\left( l\right) },\mathbf{\alpha },\mathbf{p%
},S^{2}\right\rangle \hat{U}_{k}\left( \left( \left( \Delta \mathbf{T}%
_{p}^{\alpha }\right) _{S^{2}}^{\left( l\right) }\right) _{l=1...k}\right)
\prod\limits_{l=1}^{k}\left\langle \left( \Delta \mathbf{T}_{p}^{\alpha
}\right) _{S^{2}}^{\left( l\right) },\mathbf{\alpha },\mathbf{p}%
,S^{2}\right\vert
\end{equation*}%
between tensor product of states. This operator can be written in terms of
creation annihilation operators by a change of basis. Defining:%
\begin{eqnarray*}
&&\bar{U}_{mn}\left( \mathbf{\alpha },\mathbf{p},S^{2}\right)
=\sum_{k}\left( \left\langle \mathbf{\alpha },\mathbf{p},S^{2}\right\vert
\right) ^{\otimes m}\hat{U}_{k}\left( \left( \left( \Delta \mathbf{T}%
_{p}^{\alpha }\right) _{S^{2}}^{\left( l\right) }\right) _{l=1...k}\right)
\left( \left\vert \mathbf{\alpha },\mathbf{p},S^{2}\right\rangle \right)
^{\otimes n} \\
&=&\sum_{k}\int \hat{U}_{k}\left( \left( \left( \Delta \mathbf{T}%
_{p}^{\alpha }\right) _{S^{2}}^{\left( l\right) }\right) _{l=1...k}\right) \\
&&\times \left( \prod\limits_{l=1}^{k}\left\langle \left( \Delta \mathbf{T}%
_{p}^{\alpha }\right) _{S^{2}}^{\left( l\right) },\mathbf{\alpha },\mathbf{p}%
,S^{2}\right\vert \left( \left\vert \mathbf{\alpha },\mathbf{p}%
,S^{2}\right\rangle \right) ^{\otimes n}\right) \left(
\prod\limits_{l=1}^{k}\left\langle \left( \Delta \mathbf{T}_{p}^{\alpha
}\right) _{S^{2}}^{\left( l\right) },\mathbf{\alpha },\mathbf{p}%
,S^{2}\right\vert \left( \left\vert \mathbf{\alpha },\mathbf{p}%
,S^{2}\right\rangle \right) ^{\otimes m}\right) ^{\dag }
\end{eqnarray*}%
We can thus replace:%
\begin{equation*}
\hat{U}=\sum_{m,n}\left( \left\vert \mathbf{\alpha },\mathbf{p}%
,S^{2}\right\rangle \right) ^{\otimes n}\bar{U}_{mn}\left( \mathbf{\alpha },%
\mathbf{p},S^{2}\right) \left( \left\langle \mathbf{\alpha },\mathbf{p}%
,S^{2}\right\vert \right) ^{\otimes m}
\end{equation*}%
and the operator becomes:%
\begin{equation*}
\hat{U}=\sum_{m,n}\bar{U}_{mn}\left( \mathbf{\alpha },\mathbf{p}%
,S^{2}\right) \left( \mathbf{\hat{A}}^{+}\left( \mathbf{\alpha },\mathbf{p}%
,S^{2}\right) \right) ^{m}\left( \mathbf{\hat{A}}^{-}\left( \mathbf{\alpha },%
\mathbf{p},S^{2}\right) \right) ^{n}
\end{equation*}%
As a consequence, field action $S\left( \left\{ \underline{\gamma }\left( 
\mathbf{T},\mathbf{\alpha },\mathbf{p},S^{2},\theta \right) \right\} \right) 
$ has the same content as the operator:%
\begin{eqnarray}
\mathbf{S} &=&\left( \mathbf{\hat{A}}^{+}\left( \alpha ,p,S^{2}\right)
\right) ^{t}\mathbf{A}_{S^{2}}^{\alpha }\mathbf{\hat{A}}^{-}\left( \alpha
,p,S^{2}\right) +\frac{1}{2}+\mathbf{C}  \label{RT} \\
&&+\sum_{m,n}\bar{U}_{mn}\left( \mathbf{\alpha },\mathbf{p},S^{2}\right)
\left( \mathbf{\hat{A}}^{+}\left( \mathbf{\alpha },\mathbf{p},S^{2}\right)
\right) ^{m}\left( \mathbf{\hat{A}}^{-}\left( \mathbf{\alpha },\mathbf{p}%
,S^{2}\right) \right) ^{n}  \notag
\end{eqnarray}%
and this operator will compute the same transitions between states as the
integration over the field degrees of freedom of the field $\underline{%
\Gamma }\left( \Delta \mathbf{T},\mathbf{\alpha },\mathbf{p},S^{2},\theta
\right) $ in the partition function (\ref{PF}).

\subsection{Condition for collective state}

\subsubsection{General condition}

In the single state formalism, the condition for the existence of a
collective stat can be directly studied using the action (\ref{FC}):%
\begin{equation*}
S\left( \Delta \Gamma \left( T,\hat{T},\theta ,Z,Z^{\prime }\right) \right)
\end{equation*}%
and study the condition that a solution for the saddle point equations:%
\begin{equation*}
\frac{\delta S\left( \Delta \Gamma \left( T,\hat{T},\theta ,Z,Z^{\prime
}\right) \right) }{\Delta \Gamma \left( T,\hat{T},\theta ,Z,Z^{\prime
}\right) }=\frac{\delta S\left( \Delta \Gamma \left( T,\hat{T},\theta
,Z,Z^{\prime }\right) \right) }{\Delta \Gamma ^{\dag }\left( T,\hat{T}%
,\theta ,Z,Z^{\prime }\right) }=0
\end{equation*}%
with:%
\begin{equation*}
\Delta \Gamma \left( T,\hat{T},\theta ,Z,Z^{\prime }\right) \neq 0
\end{equation*}%
for a finite number of points $\left( Z,Z^{\prime }\right) $ satisfy:%
\begin{equation*}
S\left( \Delta \Gamma \left( T,\hat{T},\theta ,Z,Z^{\prime }\right) \right)
<0
\end{equation*}%
Equivalently, this can be done more in the field formalism directly by
minimizing:%
\begin{eqnarray}
&&S\left( \left\{ \underline{\Gamma }\left( \Delta \mathbf{T},\mathbf{\alpha 
},\mathbf{p},S^{2},\theta \right) \right\} \right)  \label{FCT} \\
&=&\underline{\Gamma }^{\dag }\left( \Delta \mathbf{T},\mathbf{\alpha },%
\mathbf{p},S^{2},\theta \right) \left( -\nabla _{\Delta \mathbf{T}}^{2}+%
\frac{1}{2}\left( \Delta \mathbf{T}_{p}^{\alpha }\right) _{S^{2}}^{t}\mathbf{%
A}_{S^{2}}^{\alpha }\left( \Delta \mathbf{T}_{p}^{\alpha }\right) +\mathbf{C}%
\right) \underline{\Gamma }\left( \Delta \mathbf{T},\mathbf{\alpha },\mathbf{%
p},S^{2},\theta \right)  \notag \\
&&+U\left( \left\Vert \underline{\Gamma }\left( \Delta \mathbf{T},\mathbf{%
\alpha },\mathbf{p},S^{2},\theta \right) \right\Vert ^{2}\right)  \notag
\end{eqnarray}%
with equation:%
\begin{equation*}
\left( -\nabla _{\Delta \mathbf{T}}^{2}+\frac{1}{2}\left( \Delta \mathbf{T}%
_{p}^{\alpha }\right) _{S^{2}}^{t}\mathbf{A}_{S^{2}}^{\alpha }\left( \Delta 
\mathbf{T}_{p}^{\alpha }\right) +\mathbf{C}\right) \underline{\Gamma }\left(
\Delta \mathbf{T},\mathbf{\alpha },\mathbf{p},S^{2},\theta \right)
+U^{\prime }\left( \left\Vert \underline{\Gamma }\left( \Delta \mathbf{T},%
\mathbf{\alpha },\mathbf{p},S^{2},\theta \right) \right\Vert ^{2}\right) 
\underline{\Gamma }\left( \Delta \mathbf{T},\mathbf{\alpha },\mathbf{p}%
,S^{2},\theta \right) =0
\end{equation*}%
which is understood as:%
\begin{equation}
\left( \left( -\sum_{\left( Z,Z^{\prime }\right) }\nabla _{\Delta \mathbf{T}%
\left( Z,Z^{\prime }\right) }^{2}+\frac{1}{2}\left( \Delta \mathbf{T}%
_{p}^{\alpha }\right) _{S^{2}}^{t}\mathbf{A}_{S^{2}}^{\alpha }\left( \Delta 
\mathbf{T}_{p}^{\alpha }\right) +\mathbf{C}\right) +U^{\prime }\left(
\left\Vert \underline{\Gamma }\left( \Delta \mathbf{T},\mathbf{\alpha },%
\mathbf{p},S^{2},\theta \right) \right\Vert ^{2}\right) \right) \underline{%
\Gamma }\left( \Delta \mathbf{T},\mathbf{\alpha },\mathbf{p},S^{2},\theta
\right) =0  \label{SDL}
\end{equation}%
\bigskip As in (\cite{GLt}) we have to compare $C$, considered as some
threshold, to the potential

If we diagonalize:%
\begin{equation*}
\mathbf{A}_{S^{2}}^{\alpha }=U\mathbf{D}U^{-1}
\end{equation*}%
using that $Tr\mathbf{D}=Tr\mathbf{A}_{S^{2}}^{\alpha }$, the lowest
eigenvalue of operator:%
\begin{equation*}
-\sum_{\left( Z,Z^{\prime }\right) }\nabla _{\Delta \mathbf{T}\left(
Z,Z^{\prime }\right) }^{2}+\frac{1}{2}\left( \Delta \mathbf{T}_{p}^{\alpha
}\right) _{S^{2}}^{t}\mathbf{A}_{S^{2}}^{\alpha }\left( \Delta \mathbf{T}%
_{p}^{\alpha }\right) +\mathbf{C}
\end{equation*}%
is:%
\begin{equation*}
Tr\mathbf{A}_{S^{2}}^{\alpha }+\mathbf{C}
\end{equation*}%
We then rewrite (\ref{SDL}) for the state with lowest eigenvalue:%
\begin{equation}
Tr\mathbf{A}_{S^{2}}^{\alpha }+\mathbf{C}+U^{\prime }\left( \left\Vert 
\underline{\Gamma }\left( \Delta \mathbf{T},\mathbf{\alpha },\mathbf{p}%
,S^{2},\theta \right) \right\Vert ^{2}\right) =0  \label{SDP}
\end{equation}%
and the existence of such states is the condition for a collective state.

Equation (\ref{SDP}) yields the norm of this state:%
\begin{equation*}
\left\Vert \underline{\Gamma }\left( \Delta \mathbf{T},\mathbf{\alpha },%
\mathbf{p},S^{2},\theta \right) \right\Vert ^{2}=\left( U^{\prime }\right)
^{-1}\left( -\left( Tr\mathbf{A}_{S^{2}}^{\alpha }+\mathbf{C}\right) \right)
\end{equation*}%
and the corresponding action writes:%
\begin{eqnarray*}
&&S\left( \left\Vert \Gamma \left( \left( \mathbf{T},\mathbf{\hat{T}},%
\mathbf{Z}\right) _{S_{\alpha }^{2}},\theta \right) \right\Vert ^{2}\right)
\\
&=&\underline{\Gamma }^{\dag }\left( \Delta \mathbf{T},\mathbf{\alpha },%
\mathbf{p},S^{2},\theta \right) \left( -\nabla _{\Delta \mathbf{T}}^{2}+%
\frac{1}{2}\left( \Delta \mathbf{T}_{p}^{\alpha }\right) _{S^{2}}^{t}\mathbf{%
A}_{S^{2}}^{\alpha }\left( \Delta \mathbf{T}_{p}^{\alpha }\right) +\mathbf{C}%
\right) \underline{\Gamma }\left( \Delta \mathbf{T},\mathbf{\alpha },\mathbf{%
p},S^{2},\theta \right) \\
&&+U\left( \left\Vert \underline{\Gamma }\left( \Delta \mathbf{T},\mathbf{%
\alpha },\mathbf{p},S^{2},\theta \right) \right\Vert ^{2}\right) \\
&=&\underline{\Gamma }^{\dag }\left( \Delta \mathbf{\bar{T}},\mathbf{\alpha }%
,\mathbf{p},S^{2},\theta \right) \left( Tr\mathbf{A}_{S^{2}}^{\alpha }+%
\mathbf{C}\right) \underline{\Gamma }\left( \Delta \mathbf{\bar{T}},\mathbf{%
\alpha },\mathbf{p},S^{2},\theta \right) +U\left( \left\Vert \underline{%
\Gamma }\left( \Delta \mathbf{\bar{T}},\mathbf{\alpha },\mathbf{p}%
,S^{2},\theta \right) \right\Vert ^{2}\right)
\end{eqnarray*}%
with:%
\begin{equation*}
\Delta \mathbf{\bar{T}=}U^{-1}\Delta \mathbf{T}
\end{equation*}%
Given (\ref{SDP}) this simplifies as:%
\begin{eqnarray}
S\left( \left\Vert \Gamma \left( \left( \mathbf{T},\mathbf{\hat{T}},\mathbf{Z%
}\right) _{S_{\alpha }^{2}},\theta \right) \right\Vert ^{2}\right)
&=&U\left( \left\Vert \underline{\Gamma }\left( \Delta \mathbf{T},\mathbf{%
\alpha },\mathbf{p},S^{2},\theta \right) \right\Vert ^{2}\right)  \label{CND}
\\
&&-U^{\prime }\left( \left\Vert \underline{\Gamma }\left( \Delta \mathbf{T},%
\mathbf{\alpha },\mathbf{p},S^{2},\theta \right) \right\Vert ^{2}\right)
\left\Vert \underline{\Gamma }\left( \Delta \mathbf{T},\mathbf{\alpha },%
\mathbf{p},S^{2},\theta \right) \right\Vert ^{2}  \notag
\end{eqnarray}%
This has to be inferior to $0$, for a collective state to exist.

\subsubsection{Particular form of the potential:}

We assume a potential of the form:%
\begin{equation*}
\left( \left\Vert \underline{\Gamma }\left( \Delta \mathbf{T},\mathbf{\alpha 
},\mathbf{p},S^{2},\theta \right) \right\Vert ^{2}\right) =\frac{b}{2}\left(
\left\Vert \underline{\Gamma }\left( \Delta \mathbf{T},\mathbf{\alpha },%
\mathbf{p},S^{2},\theta \right) \right\Vert ^{2}\right) ^{2}-a\left\Vert 
\underline{\Gamma }\left( \Delta \mathbf{T},\mathbf{\alpha },\mathbf{p}%
,S^{2},\theta \right) \right\Vert ^{2}
\end{equation*}%
with $b>0$ to ensure a potential bounded from below. We assume a potential
whose form increases with background activity:%
\begin{equation*}
a=a\left( \left\vert \Psi _{0}\left( \mathbf{Z}\right) \right\vert
^{2}\omega _{0}\left( \mathbf{Z}\right) \right) \text{ with }a^{\prime }>0
\end{equation*}%
The equation (\ref{SDP}) writes:%
\begin{equation}
0=A+b\left\Vert \underline{\Gamma }\left( \Delta \mathbf{T},\mathbf{\alpha },%
\mathbf{p},S^{2},\theta \right) \right\Vert ^{2}-a  \label{PRL}
\end{equation}%
with:%
\begin{equation*}
A=Tr\mathbf{A}_{S^{2}}^{\alpha }+\mathbf{C}
\end{equation*}%
Given (\ref{CND}), we have:%
\begin{eqnarray*}
S\left( \left\Vert \underline{\Gamma }\left( \Delta \mathbf{T},\mathbf{%
\alpha },\mathbf{p},S^{2},\theta \right) \right\Vert ^{2}\right) &=&U\left(
\left\Vert \underline{\Gamma }\left( \Delta \mathbf{T},\mathbf{\alpha },%
\mathbf{p},S^{2},\theta \right) \right\Vert ^{2}\right) \\
&&-U^{\prime }\left( \left\Vert \underline{\Gamma }\left( \Delta \mathbf{T},%
\mathbf{\alpha },\mathbf{p},S^{2},\theta \right) \right\Vert ^{2}\right) 
\frac{a-A}{b}
\end{eqnarray*}%
tht simplifies as:%
\begin{equation*}
S\left( \left\Vert \underline{\Gamma }\left( \Delta \mathbf{T},\mathbf{%
\alpha },\mathbf{p},S^{2},\theta \right) \right\Vert ^{2}\right) =-\frac{%
\left( A-a\right) ^{2}}{2b}<0
\end{equation*}%
Thus a collective state exists, only if (\ref{PRL}) has a solution. Since:%
\begin{equation*}
\left\Vert \underline{\Gamma }\left( \Delta \mathbf{T},\mathbf{\alpha },%
\mathbf{p},S^{2},\theta \right) \right\Vert ^{2}>0
\end{equation*}%
the condition for a collective state becomes: 
\begin{equation*}
A=Tr\mathbf{A}_{S^{2}}^{\alpha }+\mathbf{C}<a
\end{equation*}%
Remark that $A$ is an increasing function of $\left\vert S^{2}\right\vert $,
the number of connections involved in the stt. The larger the state, the
more unlikely its emergence. Moreover, given our assumption that $a$ is
dependent on the background activity, the emergence of a collective states
depends on the level of background activity.

\section{Interactions between collective states}

So far, we have examined fields describing independent collective states. In
this section, we introduce interaction terms and explore their implications
for transitions.

\subsection{Principle}

Previous mechanism translates in term of fields by considering $n$
multicmpnnts fields correspdng to the strctrs:%
\begin{eqnarray}
&&S\left( \left\{ \underline{\Gamma }\left( \Delta \mathbf{T},\mathbf{\alpha 
},\mathbf{p},S_{i}^{2},\theta \right) \right\} \right)  \notag \\
&=&-\underline{\Gamma }^{\dag }\left( \Delta \mathbf{T},\mathbf{\alpha },%
\mathbf{p},S_{i}^{2},\theta \right) \left( -\nabla _{\mathbf{T}}^{2}+\frac{1%
}{2}\left( \Delta \mathbf{T}_{p}^{\alpha }\right) _{S_{i}^{2}}^{t}\mathbf{A}%
_{S_{i}^{2}}^{\alpha }\left( \Delta \mathbf{T}_{p}^{\alpha }\right)
_{S_{i}^{2}}+\mathbf{C}\right) \underline{\Gamma }\left( \Delta \mathbf{T},%
\mathbf{\alpha },\mathbf{p},S_{i}^{2},\theta \right) \\
&&+U\left( \left\Vert \underline{\Gamma }\left( \Delta \mathbf{T},\mathbf{%
\alpha },\mathbf{p},S_{i}^{2},\theta \right) \right\Vert ^{2}\right)  \notag
\end{eqnarray}%
The set $S_{i}^{2}$ characterizes the structure localization along with its
possible states. The multi-components labelled by $\mathbf{\alpha },\mathbf{p%
}$ transcribes the possible averages and frequencies. The structure emerging
from interactions is described by the action functionl:%
\begin{eqnarray}
&&S\left( \left\{ \underline{\Gamma }\left( \Delta \mathbf{T},\mathbf{\alpha 
},\mathbf{p},\left( \cup S_{i}\right) \times \left( \cup S_{i}\right)
,\theta \right) \right\} \right) \\
&=&\underline{\Gamma }^{\dag }\left( \Delta \mathbf{T},\mathbf{\alpha },%
\mathbf{p},\left( \cup S_{i}\right) \times \left( \cup S_{i}\right) ,\theta
\right) \left( -\nabla _{\mathbf{T}}^{2}+\frac{1}{2}\left( \Delta \mathbf{T}%
_{p}^{\alpha }\right) _{\left( \cup S_{i}\right) ^{2}}^{t}\mathbf{A}_{\left(
\cup S_{i}\right) ^{2}}^{\alpha }\left( \Delta \mathbf{T}_{p}^{\alpha
}\right) _{\left( \cup S_{i}\right) ^{2}}+\mathbf{C}\right)  \notag \\
&&\underline{\Gamma }\left( \Delta \mathbf{T},\mathbf{\alpha },\mathbf{p}%
,\left( \cup S_{i}\right) \times \left( \cup S_{i}\right) ,\theta \right) 
\notag \\
&&+U\left( \left\Vert \underline{\Gamma }\left( \Delta \mathbf{T},\mathbf{%
\alpha },\mathbf{p},\left( \cup S_{i}\right) \times \left( \cup S_{i}\right)
,\theta \right) \right\Vert ^{2}\right)  \notag
\end{eqnarray}%
Implicitly, this structure has relatively low average connectivities for
elements of:%
\begin{equation*}
S_{i}\times S_{j}
\end{equation*}%
and $j\neq i$. We will relax this assumption in the next section. However,
the mechanisms described above allow to understand the dynamical aspects of
interactions between structures.

The full action for the system described above should be a sum of individual
actions:%
\begin{equation*}
\sum_{i}S\left( \left\{ \underline{\Gamma }\left( \Delta \mathbf{T},\mathbf{%
\alpha },\mathbf{p},S_{i}^{2},\theta \right) \right\} \right) +S\left(
\left\{ \underline{\Gamma }\left( \Delta \mathbf{T},\mathbf{\alpha },\mathbf{%
p},\left( \cup S_{i}\right) \times \left( \cup S_{i}\right) ,\theta \right)
\right\} \right)
\end{equation*}%
with additional interaction terms. These terms allow transition from states
over $S_{i}^{2}$ to states over $\left( \cup S_{i}\right) \times \left( \cup
S_{i}\right) $. This is possible when considering interaction terms of the
type:%
\begin{equation}
V\left( \left( \Delta \mathbf{T},\mathbf{\alpha },\mathbf{p},\left( \cup
S_{i}\right) \times \left( \cup S_{i}\right) \right) ,\left\{ \Delta \mathbf{%
T},\mathbf{\alpha },\mathbf{p},S_{i}^{2}\right\} \right) \underline{\Gamma }%
^{\dag }\left( \Delta \mathbf{T},\mathbf{\alpha },\mathbf{p},\left( \cup
S_{i}\right) \times \left( \cup S_{i}\right) ,\theta \right) \prod 
\underline{\Gamma }\left( \Delta \mathbf{T},\mathbf{\alpha },\mathbf{p}%
,S_{i}^{2},\theta \right)  \label{DSP}
\end{equation}%
where $V$ is a potential. We will detail the formalism in the next section,
but given our previous description, we may expect the potential:%
\begin{equation*}
V\left( \left( \Delta \mathbf{T},\mathbf{\alpha },\mathbf{p},\left( \cup
S_{i}\right) \times \left( \cup S_{i}\right) \right) ,\left\{ \Delta \mathbf{%
T},\mathbf{\alpha },\mathbf{p},S_{i}^{2}\right\} \right)
\end{equation*}%
to depend on the states frequencies, to allow to switch from some
equilibrium over the $S_{i}^{2}$ to equilibrium over the $\left( \cup
S_{i}\right) \times \left( \cup S_{i}\right) $. If the $S_{i}^{2}$ have
frqncs $\Upsilon _{i,l_{i}}$ we may expect $\left( \cup S_{i}\right) ^{2}$
to be characterized by the set of frequencies $\Upsilon _{\left(
i,l_{i}\right) }$. Similarly, $\Delta \mathbf{T}\left( \left( \cup
S_{i}\right) \times \left( \cup S_{i}\right) \right) $ on the diagonal
should be close to the $\Delta \mathbf{T}\left( \left( \cup S_{i}\right)
\times \left( \cup S_{i}\right) \right) $.

\subsection{Interactions}

\subsubsection{Different structures}

We described previously the collective state resulting by "merging"
different types of structures. To describe dynamically this transition in
terms of fields, we add to the action a term of the form:%
\begin{equation*}
\sum_{nn^{\prime }}\sum_{\substack{ k=1...n  \\ l=1,...,n^{\prime }}}%
\sum_{\left\{ S_{k},S_{l}\right\} _{\substack{ l=1,...,n^{\prime }  \\ %
k=1...n }}}\prod_{l}\underline{\Gamma }^{\dag }\left( \mathbf{T}_{l}^{\prime
},\mathbf{\alpha }_{l}^{\prime },\mathbf{p}_{l}^{\prime },S_{l}^{\prime
2}\right) V\left( \left\{ \left\vert \mathbf{T}_{l}^{\prime },\mathbf{\alpha 
}_{l}^{\prime },\mathbf{p}_{l}^{\prime },S_{l}^{\prime 2}\right\rangle
\right\} ,\left\{ \left\vert \mathbf{T}_{l},\mathbf{\alpha }_{k},\mathbf{p}%
_{k},S_{k}^{2}\right\rangle \right\} \right) \prod_{k}\underline{\Gamma }%
\left( \mathbf{T}_{k},\mathbf{\alpha }_{k},\mathbf{p}_{k},S_{k}^{2}\right)
\end{equation*}%
where:%
\begin{equation*}
V\left( \left\{ \left\vert \mathbf{T}_{l}^{\prime },\mathbf{\alpha }%
_{l}^{\prime },\mathbf{p}_{l}^{\prime },S_{l}^{\prime 2}\right\rangle
\right\} ,\left\{ \left\vert \mathbf{T}_{k},\mathbf{\alpha }_{k},\mathbf{p}%
_{k},S_{k}^{2}\right\rangle \right\} \right) =V\left( \left\{ \mathbf{T}%
_{l}^{\prime },\mathbf{\alpha }_{l}^{\prime },\mathbf{p}_{l}^{\prime
},S_{l}^{\prime 2}\right\} ,\left\{ \mathbf{T}_{k},\mathbf{\alpha }_{k},%
\mathbf{p}_{k},S_{k}^{2}\right\} \right)
\end{equation*}%
and the action for interacting structures becomes:%
\begin{eqnarray}
S &=&\sum_{S}\underline{\Gamma }^{\dag }\left( \mathbf{T},\mathbf{\alpha },%
\mathbf{p},S^{2}\right) \left( -\frac{1}{2}\nabla _{\left( \mathbf{\hat{T}}%
\right) _{S^{2}}}^{2}+\frac{1}{2}\left( \Delta \mathbf{T}_{p}^{\alpha
}\right) _{S^{2}}^{t}\mathbf{A}_{S^{2}}^{\alpha }\left( \Delta \mathbf{T}%
_{p}^{\alpha }\right) +\mathbf{C}\right) \underline{\Gamma }\left( \mathbf{T}%
,\mathbf{\alpha },\mathbf{p},S^{2}\right)  \label{RCN} \\
&&+\sum_{nn^{\prime }}\sum_{\substack{ k=1...n  \\ l=1,...,n^{\prime }}}%
\sum_{\left\{ S_{k},S_{l}\right\} _{\substack{ l=1,...,n^{\prime }  \\ %
k=1...n }}}\prod_{l}\underline{\Gamma }^{\dag }\left( \mathbf{T}_{l}^{\prime
},\mathbf{\alpha }_{l}^{\prime },\mathbf{p}_{l}^{\prime },S_{l}^{\prime
2}\right)  \notag \\
&&\times V\left( \left\{ \left\vert \mathbf{T}^{\prime },\mathbf{\alpha }%
_{l}^{\prime },\mathbf{p}_{l}^{\prime },S_{l}^{\prime 2}\right\rangle
\right\} ,\left\{ \left\vert \mathbf{T},\mathbf{\alpha }_{k},\mathbf{p}%
_{k},S_{k}^{2}\right\rangle \right\} \right) \prod_{k}\underline{\Gamma }%
\left( \mathbf{T}_{k},\mathbf{\alpha }_{k},\mathbf{p}_{k},S_{k}^{2}\right) 
\notag
\end{eqnarray}%
allowing for transitions between sets of several collective states. The form
of $V$ is conditionned by frequencies of oscillation:%
\begin{equation*}
V\left( \left\{ \mathbf{T}_{l}^{\prime },\mathbf{\alpha }_{l}^{\prime },%
\mathbf{p}_{l}^{\prime },S_{l}^{\prime 2}\right\} ,\left\{ \mathbf{T}_{k},%
\mathbf{\alpha }_{k},\mathbf{p}_{k},S_{k}^{2}\right\} \right) =V\left(
\Upsilon _{l}^{\mathbf{p}_{l}^{\prime }}\left( \mathbf{T}_{l}^{\prime
}\right) ,\Upsilon _{k}^{\mathbf{p}_{k}}\left( \mathbf{T}_{k}\right) \right)
\end{equation*}%
and models the results of the first part of this article, transitions depend
both on initial states characteritics and that of the merged ones.

\subsubsection{Substructures}

We also consider the possibility of activation by a substructure. This is a
particular case of interaction where the activation of some substructure
induces the full structure activation. To describe this type of transition,
the term:%
\begin{equation*}
\underline{\Gamma }^{\dag }\left( \mathbf{T},\mathbf{\alpha },\mathbf{p}%
,S^{2}\right) \left( -\frac{1}{2}\nabla _{\left( \mathbf{\hat{T}}\right)
_{S^{2}}}^{2}+\frac{1}{2}\left( \Delta \mathbf{T}_{p}^{\alpha }\right)
_{S^{2}}^{t}\mathbf{A}_{S^{2}}^{\alpha }\left( \Delta \mathbf{T}_{p}^{\alpha
}\right) +\mathbf{C}\right) \underline{\Gamma }\left( \mathbf{T},\mathbf{%
\alpha },\mathbf{p},S^{2}\right)
\end{equation*}%
is generalized by including free action terms for each substructures plus
interaction terms between these substructures, including the full one:%
\begin{eqnarray*}
&&\sum_{S_{1}\subseteq S}\underline{\Gamma }^{\dag }\left( \mathbf{T},%
\mathbf{\alpha },\mathbf{p},S_{1}^{2}\right) \left( -\frac{1}{2}\nabla
_{\left( \mathbf{\hat{T}}\right) _{S_{1}^{2}}}^{2}+\frac{1}{2}\left( \Delta 
\mathbf{T}_{p}^{\alpha }\right) _{S_{1}^{2}}^{t}\mathbf{A}%
_{S_{1}^{2}}^{\alpha }\left( \Delta \mathbf{T}_{p}^{\alpha }\right) +\mathbf{%
C}\right) \underline{\Gamma }\left( \mathbf{T},\mathbf{\alpha },\mathbf{p}%
,S_{1}^{2}\right) \\
&&+\sum_{n}\sum_{S_{1},...,S_{n}\subseteq S}\sum_{\left( \alpha
_{1},p_{1}\right) ,...,\left( \alpha _{n},p_{n}\right) }\sum_{k}V_{k}\left(
\left( \left\{ \mathbf{T}_{i},\mathbf{\alpha }_{i},\mathbf{p}_{i}\right\}
,\left( S_{i}\right) ^{2}\right) \right) \prod\limits_{i\leqslant k}%
\underline{\Gamma }^{\dag }\left( \mathbf{T}_{i},\mathbf{\alpha }_{i},%
\mathbf{p}_{i}\right) \prod\limits_{k+1\leqslant i\leqslant n}\underline{%
\Gamma }\left( \mathbf{T}_{i},\mathbf{\alpha }_{i},\mathbf{p}%
_{i},S_{i}^{2}\right)
\end{eqnarray*}

The potential $V_{k}\left( \left( \left\{ \mathbf{T}_{i},\mathbf{\alpha }%
_{i},\mathbf{p}_{i}\right\} ,\left( S_{i}\right) ^{2}\right) \right) $
induces transition from some state with $k$ substructures towards a state
with $n-k$ subtructures. It may include the transition from one or several
subsets to the full activated structure. This situation is depicted by a
potential of the type:%
\begin{equation*}
V_{k}\left( \left( \left\{ \mathbf{T}_{i},\mathbf{\alpha }_{i},\mathbf{p}%
_{i},\left( S_{i}\right) ^{2}\right\} \right) \right) =V_{k}\left( \left(
\left\{ \mathbf{T}_{i},\mathbf{\alpha }_{i},\mathbf{p}_{i},\left(
S_{i}\right) ^{2}\right\} _{i\leqslant k},\left\{ \mathbf{T}_{i},\mathbf{%
\alpha }_{i},\mathbf{p}_{i},\left( S_{i}\right) ^{2}\right\} _{k+1\leqslant
i\leqslant n}\right) \right)
\end{equation*}%
the group $\mathbf{Z}$ of possible states is defined, at least partly, by
initial background.

\subsection{Operator formalism for interactions}

\subsubsection{General case}

We have seen in (\ref{RT}) the operator formulation for the dynamic of one
type of structure:%
\begin{eqnarray}
\mathbf{S} &=&\left( \mathbf{\hat{A}}^{+}\left( \alpha ,p,S^{2}\right)
\right) ^{t}\mathbf{A}_{S^{2}}^{\alpha }\mathbf{\hat{A}}^{-}\left( \alpha
,p,S^{2}\right) +\frac{1}{2}+\mathbf{C} \\
&&+\sum_{m,n}\bar{U}_{mn}\left( \mathbf{\alpha },\mathbf{p},S^{2}\right)
\left( \mathbf{\hat{A}}^{+}\left( \mathbf{\alpha },\mathbf{p},S^{2}\right)
\right) ^{m}\left( \mathbf{\hat{A}}^{-}\left( \mathbf{\alpha },\mathbf{p}%
,S^{2}\right) \right) ^{n}  \notag
\end{eqnarray}

As for the field version, we can consider interaction potential between
different structures. The operator counterpart of the potential term in (\ref%
{RCN}) is:%
\begin{equation*}
\sum_{n,n^{\prime }}\sum_{\left\{ S_{k},S_{l}\right\} _{\substack{ %
l=1,...,n^{\prime }  \\ k=1...n}}}\prod_{l=1}^{n^{\prime }}\underline{\Gamma 
}^{\dag }\left( \mathbf{T}_{l}^{\prime },\mathbf{\alpha }_{l}^{\prime },%
\mathbf{p}_{l}^{\prime },S_{l}^{\prime 2}\right) V_{n,n^{\prime }}\left(
\left\{ \mathbf{T}_{l}^{\prime },\mathbf{\alpha }_{l}^{\prime },\mathbf{p}%
_{l}^{\prime },S_{l}^{\prime 2}\right\} _{l\leqslant n^{\prime }},\left\{ 
\mathbf{T}_{k},\mathbf{\alpha }_{k},\mathbf{p}_{k},S_{k}^{2}\right\}
_{l\leqslant n}\right) \prod_{k=1}^{n}\underline{\Gamma }\left( \mathbf{T}%
_{k},\mathbf{\alpha }_{k},\mathbf{p}_{k},S_{k}^{2}\right)
\end{equation*}%
is found by applying the same technique as for individual potential $U$ in (%
\ref{PL}). We change the basis by defining:%
\begin{eqnarray}
&&V_{n,n^{\prime }}\left( \left\{ \mathbf{\alpha }_{l}^{\prime },\mathbf{p}%
_{l}^{\prime },S_{l}^{\prime 2},m_{l}^{\prime }\right\} _{l\leqslant
n^{\prime }},\left\{ \mathbf{\alpha }_{k},\mathbf{p}_{k},S_{k}^{2},m_{k}%
\right\} _{l\leqslant n}\right)  \label{VN} \\
&=&\int d\left( \Delta \mathbf{T}_{p_{l}}^{\prime \alpha _{l}}\right)
_{S_{l}^{\prime 2}}d\left( \Delta \mathbf{T}_{p_{k}}^{\alpha _{k}}\right)
_{S_{k}^{2}}V_{n,n^{\prime }}\left( \left\{ \left( \Delta \mathbf{T}%
_{p_{l}}^{\prime \alpha _{l}}\right) _{S_{l}^{\prime 2}},\mathbf{\alpha }%
_{l}^{\prime },\mathbf{p}_{l}^{\prime },S_{l}^{\prime 2}\right\}
_{l\leqslant n^{\prime }},\left\{ \left( \Delta \mathbf{T}_{p_{k}}^{\alpha
_{k}}\right) _{S_{k}^{2}},\mathbf{\alpha }_{k},\mathbf{p}_{k},S_{k}^{2}%
\right\} _{l\leqslant n}\right)  \notag \\
&&\times \prod_{k=1}^{n}\left\langle \left( \Delta \mathbf{T}%
_{p_{k}}^{\alpha _{k}}\right) _{S_{k}^{2}},\mathbf{\alpha }_{k},\mathbf{p}%
_{k},S_{k}^{2}\right\vert \left( \left\vert \mathbf{\alpha }_{k},\mathbf{p}%
_{k},S_{k}^{2}\right\rangle \right) ^{\otimes m_{k}}\left(
\prod_{l=1}^{n^{\prime }}\left\langle \left( \Delta \mathbf{T}%
_{p_{l}}^{\prime \alpha _{l}}\right) _{S_{l}^{\prime 2}},\mathbf{\alpha }%
_{l}^{\prime },\mathbf{p}_{l}^{\prime },S_{l}^{\prime 2}\right\vert \left(
\left\vert \mathbf{\alpha }_{k},\mathbf{p}_{k},S_{k}^{2}\right\rangle
\right) ^{\otimes m_{l}^{\prime }}\right) ^{\dag }  \notag
\end{eqnarray}%
and the interaction operator writes in terms of annihilation and creation
operators:%
\begin{eqnarray}
&&\hat{V}=\sum_{n,n^{\prime }}\sum_{\left\{ S_{k},S_{l}\right\} _{\substack{ %
l=1,...,n^{\prime }  \\ k=1...n}}}\sum_{\left\{ m_{l}^{\prime
},m_{k}\right\} }\prod_{l=1}^{n^{\prime }}\left( \mathbf{\hat{A}}^{+}\left( 
\mathbf{\alpha }_{l}^{\prime },\mathbf{p}_{l}^{\prime },S_{l}^{\prime
2}\right) \right) ^{m_{l}^{\prime }}  \label{NT} \\
&&\times V_{n,n^{\prime }}\left( \left\{ \mathbf{\alpha }_{l}^{\prime },%
\mathbf{p}_{l}^{\prime },S_{l}^{\prime 2},m_{l}^{\prime }\right\}
_{l\leqslant n^{\prime }},\left\{ \mathbf{\alpha }_{k},\mathbf{p}%
_{k},S_{k}^{2},m_{k}\right\} _{l\leqslant n}\right) \prod_{k=1}^{n}\left( 
\mathbf{\hat{A}}^{-}\left( \mathbf{\alpha }_{k},\mathbf{p}%
_{k},S_{k}^{2}\right) \right) ^{m_{k}}  \notag
\end{eqnarray}%
Thus the corresponding operator is to $S\left( \left\{ \underline{\Gamma }%
\left( \mathbf{T},\mathbf{\alpha },\mathbf{p},S^{2},\theta \right) \right\}
\right) $ is:%
\begin{eqnarray}
&&\mathbf{S}=\sum_{S\times S}\mathbf{\bar{D}}_{S^{2}}^{\alpha }\left( 
\mathbf{A}^{+}\left( \alpha ,p,S^{2}\right) \mathbf{A}^{-}\left( \alpha
,p,S^{2}\right) +\frac{1}{2}\right)  \label{SP} \\
&&+\sum_{m,n}\bar{U}_{mn}\left( \mathbf{\alpha },\mathbf{p},S^{2}\right)
\left( \mathbf{\hat{A}}^{+}\left( \mathbf{\alpha },\mathbf{p},S^{2}\right)
\right) ^{m}\left( \mathbf{\hat{A}}^{-}\left( \mathbf{\alpha },\mathbf{p}%
,S^{2}\right) \right) ^{n}+\hat{V}  \notag
\end{eqnarray}

The advantage of this formulation is to directly translate the dynamics in
terms of creation and destruction of structures, describing the transition
resulting from such operators. It also allows straightforward computations
at the lowest order of approximation, presenting a direct interpretation as
transitions of structures.

In the sequel we will simplify the notation:%
\begin{equation*}
V_{n,n^{\prime }}\left( \left\{ \mathbf{\alpha }_{l}^{\prime },\mathbf{p}%
_{l}^{\prime },S_{l}^{\prime 2},m_{l}^{\prime }\right\} _{l\leqslant
n^{\prime }},\left\{ \mathbf{\alpha }_{k},\mathbf{p}_{k},S_{k}^{2},m_{k}%
\right\} _{l\leqslant n}\right) \rightarrow V_{n,n^{\prime }}\left( \left\{ 
\mathbf{\alpha }_{l}^{\prime },\mathbf{p}_{l}^{\prime },S_{l}^{\prime
2},m_{l}^{\prime }\right\} ,\left\{ \mathbf{\alpha }_{k},\mathbf{p}%
_{k},S_{k}^{2},m_{k}\right\} \right)
\end{equation*}

\subsubsection{Internal perturbations for one structure}

We can consider the particular case of several activated states for one
structure. This corresponds to several processes arising within the same
collective state. This case is intermediate between a single type of
structure and multiple interacting structures. In this case the pntl in (\ref%
{SP}) is replaced by:%
\begin{eqnarray}
\hat{V} &=&\sum_{n,n^{\prime }}\sum_{\left\{ m_{l}^{\prime },m_{k}\right\}
}\prod_{l=1}^{n^{\prime }}\left( \mathbf{\hat{A}}^{+}\left( \mathbf{\alpha }%
_{l}^{\prime },\mathbf{p}_{l}^{\prime },S^{2}\right) \right) ^{m_{l}^{\prime
}}  \notag \\
&&\times V_{n,n^{\prime }}\left( \left\{ \mathbf{\alpha }_{l}^{\prime },%
\mathbf{p}_{l}^{\prime },m_{l}^{\prime }\right\} _{l\leqslant n^{\prime
}},\left\{ \mathbf{\alpha }_{k},\mathbf{p}_{k},m_{k}\right\} _{l\leqslant
n},S^{2}\right) \prod_{k=1}^{n}\left( \mathbf{\hat{A}}^{-}\left( \mathbf{%
\alpha }_{k},\mathbf{p}_{k},S^{2}\right) \right) ^{m_{k}}  \notag
\end{eqnarray}%
where:%
\begin{equation*}
V_{n,n^{\prime }}\left( \left\{ \mathbf{\alpha }_{l}^{\prime },\mathbf{p}%
_{l}^{\prime },m_{l}^{\prime }\right\} _{l\leqslant n^{\prime }},\left\{ 
\mathbf{\alpha }_{k},\mathbf{p}_{k},m_{k}\right\} _{l\leqslant
n},S^{2}\right)
\end{equation*}%
is given by (\ref{VN}) with $S_{k}=S_{l}^{\prime }=S$ for all $l$ and $k$.

\subsection{External perturbation}

In (\cite{GLr}) and (\cite{GLs}), we have studied the effect of external
sources on the connectivity functions. External signals induce modified
activities and as a consequence, modifications in equilibrium states of
structures (see appendix 2 for a detailed account). In terms of collective
state formalism, this situation can be described through a modification of
the effective action or by an operator description.

\subsubsection{Effective action}

In the present context, external perturbations can be modelled by adding
extra terms in the action inducing switches in background states, and
dynamical transitions between stts. We show in appendix 0 that we can
consider a modified action:%
\begin{eqnarray}
&&S\left( \left\{ \underline{\Gamma }\left( \mathbf{T},\mathbf{\alpha },%
\mathbf{p},S^{2},\theta \right) \right\} \right)  \label{FCWZ} \\
&=&\sum_{\left\{ \mathbf{\alpha },\mathbf{p},S^{2}\right\} }\underline{%
\Gamma }^{\dag }\left( \mathbf{T},\mathbf{\alpha },\mathbf{p},S^{2},\theta
\right) \left( -\nabla _{\Delta \mathbf{T}}^{2}+\frac{1}{2}\left( \Delta 
\mathbf{T}_{p}^{\alpha }\right) _{S^{2}}^{t}\mathbf{A}_{S^{2}}^{\alpha
}\left( \Delta \mathbf{T}_{p}^{\alpha }\right) +\mathbf{C}\right) \underline{%
\Gamma }\left( \Delta \mathbf{T},\mathbf{\alpha },\mathbf{p},S^{2},\theta
\right)  \notag \\
&&+U\left( \left\Vert \underline{\Gamma }\left( \left( \Delta \mathbf{T}%
_{p}^{\alpha }\right) _{S^{2}},\mathbf{\alpha },\mathbf{p},S^{2},\theta
\right) \right\Vert ^{2}\right) +\underline{\Gamma }^{\dag }\left( \Delta 
\mathbf{T},\mathbf{\alpha },\mathbf{p},S^{2},\theta \right) J\left( \Delta 
\mathbf{T},S^{2},\theta \right) \underline{\Gamma }\left( \Delta \mathbf{T},%
\mathbf{\alpha },\mathbf{p},S^{2},\theta \right)  \notag
\end{eqnarray}%
The source term $J\left( S^{2},\theta \right) $ can be considered as a sum
of contributions acting at each points of $S$:%
\begin{equation*}
J\left( S^{2},\theta \right) =\sum_{Z\in S}J\left( Z,\theta \right)
\end{equation*}%
The introduction of $J\left( S^{2},\theta \right) $ modifies the saddle
points equations of the system, and may induce some structures to be
switched off or on. They may also induce some different structures to
combine through effective interaction.

\paragraph{Terms inducing types transition}

The presence of:%
\begin{equation}
\underline{\Gamma }^{\dag }\left( \Delta \mathbf{T},\mathbf{\alpha },\mathbf{%
p},S^{2},\theta \right) J\left( S^{2},\theta \right) \underline{\Gamma }%
\left( \Delta \mathbf{T},\mathbf{\alpha },\mathbf{p},S^{2},\theta \right)
\label{DC}
\end{equation}%
in the action accounts for the possibility to turn on or off the structure.
More generally, an external source may induce transition between two states
of the same strcture, and this is described by:%
\begin{equation}
\underline{\Gamma }^{\dag }\left( \Delta \mathbf{T}^{\prime },\mathbf{\alpha 
}^{\prime },\mathbf{p}^{\prime },S^{2},\theta \right) J\left( S^{2},\theta
\right) \underline{\Gamma }\left( \Delta \mathbf{T},\mathbf{\alpha },\mathbf{%
p},S^{2},\theta \right)  \label{TR2}
\end{equation}%
switching the connectivity states from $\left( \Delta \mathbf{T},\mathbf{%
\alpha },\mathbf{p}\right) $ to $\left( \Delta \mathbf{T}^{\prime },\mathbf{%
\alpha }^{\prime },\mathbf{p}^{\prime }\right) $.

In an effective formalism, if the signals modify several structures we may
assume that switches between states may be modelled directly by current
induced interactions. In this case, the interactions (\ref{DC}) or (\ref{TR2}%
) can be replaced in (\ref{FCWZ}) by:%
\begin{equation}
\sum_{\left\{ \mathbf{\alpha },\mathbf{p},S^{2}\right\} ,\left\{ \mathbf{%
\alpha }^{\prime },\mathbf{p}^{\prime },S^{\prime 2}\right\} }\underline{%
\Gamma }^{\dag }\left( \Delta \mathbf{T}^{\prime },\mathbf{\alpha }^{\prime
},\mathbf{p}^{\prime },S^{\prime 2},\theta \right) J\left( S^{\prime
2},\theta ^{\prime }\right) J\left( S^{2},\theta \right) \underline{\Gamma }%
\left( \Delta \mathbf{T},\mathbf{\alpha },\mathbf{p},S^{2},\theta \right)
\label{NC}
\end{equation}%
and this terms ensures dynamics switching between different states. This
corresponds to the integration of some "faster" structure, connecting
different states through some perturbations.

\paragraph{Activation or deactivation terms}

This possiility represents activation or deactivation of structures due to
the source terms. It is modelled by field linear terms of the form:%
\begin{equation}
J\left( S^{2},\theta \right) \left( \underline{\Gamma }\left( \Delta \mathbf{%
T},\mathbf{\alpha },\mathbf{p},S^{2},\theta \right) +\underline{\Gamma }%
^{\dagger }\left( \Delta \mathbf{T},\mathbf{\alpha },\mathbf{p},S^{2},\theta
\right) \right)  \label{CD}
\end{equation}

\subsubsection{operator formalism}

In term of operators the additional terms (\ref{DC}) and (\ref{NC}) are
modeled using the operators technique described above.

\paragraph{Transition type trms}

The operator representing (\ref{DC}) is:%
\begin{equation*}
\sum_{m,m^{\prime }}\left( \mathbf{\hat{A}}^{+}\left( \mathbf{\alpha },%
\mathbf{p},S^{2}\right) \right) ^{m^{\prime }}J_{m,m^{\prime }}\left( 
\mathbf{\alpha },\mathbf{p},S^{2},\theta \right) \left( \mathbf{\hat{A}}%
^{-}\left( \mathbf{\alpha },\mathbf{p},S^{2}\right) \right) ^{m}
\end{equation*}%
with:%
\begin{equation*}
J_{m,m^{\prime }}\left( \mathbf{\alpha },\mathbf{p},S^{2},\theta \right)
=\left( \left\langle \mathbf{\alpha },\mathbf{p},S^{2}\right\vert \right)
^{\otimes m^{\prime }}J\left( S^{2},\theta \right) \left( \left\vert \mathbf{%
\alpha },\mathbf{p},S^{2}\right\rangle \right) ^{\otimes m}
\end{equation*}%
and (\ref{NC}) is translated to:%
\begin{eqnarray*}
&&\sum_{m,m^{\prime }}\left( \mathbf{\hat{A}}^{+}\left( \mathbf{\alpha }%
_{l}^{\prime },\mathbf{p}_{l}^{\prime },S_{l}^{\prime 2}\right) \right)
^{m^{\prime }}V_{m,m^{\prime }}\left( \mathbf{\alpha }^{\prime },\mathbf{p}%
^{\prime },S^{\prime 2},\theta ^{\prime },\mathbf{\alpha },\mathbf{p}%
,S^{2},\theta \right) \left( \mathbf{\hat{A}}^{-}\left( \mathbf{\alpha }_{k},%
\mathbf{p}_{k},S_{k}^{2}\right) \right) ^{m} \\
&&+\left( \mathbf{\hat{A}}^{+}\left( \mathbf{\alpha }_{k},\mathbf{p}%
_{k},S_{k}^{2}\right) \right) ^{m}V_{m^{\prime },m}\left( \mathbf{\alpha },%
\mathbf{p},S^{2},\theta ,\mathbf{\alpha }^{\prime },\mathbf{p}^{\prime
},S^{\prime 2},\theta ^{\prime }\right) \left( \mathbf{\hat{A}}^{-}\left( 
\mathbf{\alpha }_{l}^{\prime },\mathbf{p}_{l}^{\prime },S_{l}^{\prime
2}\right) \right) ^{m^{\prime }}
\end{eqnarray*}%
with:%
\begin{equation*}
V_{m,m^{\prime }}\left( \mathbf{\alpha }^{\prime },\mathbf{p}^{\prime
},S^{\prime 2},\theta ^{\prime },\mathbf{\alpha },\mathbf{p},S^{2},\theta
\right) =\sum_{k}J_{m,k}\left( S^{\prime 2},\theta \right) J_{k,m^{\prime
}}\left( S^{2},\theta \right)
\end{equation*}

\paragraph{Activation or deactivation terms}

The operator equivalent of (\ref{CD}) is:%
\begin{equation*}
J\left( S^{2},\theta \right) \left( \mathbf{\hat{A}}^{+}\left( \mathbf{%
\alpha },\mathbf{p},S^{2}\right) +\mathbf{\hat{A}}^{-}\left( \mathbf{\alpha }%
,\mathbf{p},S^{2}\right) \right)
\end{equation*}

\part*{III Approaches to transitions, examples and extensions}

The formalism presented above enables the computation of transition
mechanisms between various states. Nevertheless, exact computation of the
path integral is not feasible. We introduce three approaches for these
computations. Two of them are perturbative in nature (perturbation expansion
and operator formalism), while the third one, effective field theory, is
not. Each of these methods has its own advantages. We present some
illustrative examples of these methods

\section{Mechanisms of transition}

We present three different complementary approaches to the interactions and
transitions of states. The perturbation expansion is the most
straightforward method for studying transitions. By considering the "free"
action obtained by neglecting the interaction terms, we can compute the
Green functions of individual structures. These Green functions describe the
dynamic fluctuations of a structure in the absence of interactions.
Subsequently, the interaction terms are incorporated to compute the
transitions of such free states to others. These transitions may involve the
activation of some structures and the deactivation of others. The advantage
of this approach lies in its clarity, as it directly calculates the
probabilities of the considered transitions. However, it falls short in
addressing non-perturbative effects, which encompass global effects where
some structures act as a background for others. These effects differ from
perturbative ones since they correspond to the impact of a permanent
landscape in which structures evolve. Nevertheless, since it mainly operates
at the background field level, it may not fully capture the precise
mechanisms by which some structures constitute effective interactions
between others. The third approach, operator formalism, addresses this issue
to some extent by integrating certain interactions to establish indirect
ones. Consequently, this approach combines elements of the other two methods.

\subsection{Perturbation expansion}

The first approach to transitions of states is the most direct and the most
readable due to its direct computation of transition functions based on
series expansion.

As explained in the introduction of this section, it begins by calculating
the Green functions, which are the transition functions for the free
structures. Then, perturbations that enable transitions are introduced in an
ordered manner to compute the transitions induced by interactions.

\subsubsection{Green Functions}

Grn fnctns cmptd wth free part in (\ref{RCN}):%
\begin{equation*}
\underline{\Gamma }^{\dag }\left( \Delta \mathbf{T},\mathbf{\alpha },\mathbf{%
p},S^{2},\theta \right) \left( -\nabla _{\Delta \mathbf{T}}^{2}+\frac{1}{2}%
\left( \Delta \mathbf{T}_{p}^{\alpha }\right) _{S^{2}}^{t}\mathbf{A}%
_{S^{2}}^{\alpha }\left( \Delta \mathbf{T}_{p}^{\alpha }\right) +\mathbf{C}%
\right) \underline{\Gamma }\left( \Delta \mathbf{T},\mathbf{\alpha },\mathbf{%
p},S^{2},\theta \right)
\end{equation*}%
and the trnstn prtr s gvn b:%
\begin{equation*}
\left( -\nabla _{\Delta \mathbf{T}}^{2}+\frac{1}{2}\left( \Delta \mathbf{T}%
_{p}^{\alpha }\right) _{S^{2}}^{t}\mathbf{A}_{S^{2}}^{\alpha }\left( \Delta 
\mathbf{T}_{p}^{\alpha }\right) +\mathbf{C}\right) ^{-1}
\end{equation*}%
It is diagonalized as:%
\begin{equation*}
U\left( -\frac{1}{2}\nabla _{\left( \mathbf{\bar{T}}\right) _{S^{2}}}^{2}+%
\frac{1}{2}\left( \mathbf{\bar{D}}_{S^{2}}\mathbf{\bar{T}}_{p}^{\alpha
}\right) ^{2}+\mathbf{C}\right) ^{-1}U^{-1}
\end{equation*}%
For lrg nmbr f vrbls, lowest order krnl wth grnd stts:%
\begin{equation}
G\left( \mathbf{\bar{T}}_{p}^{\prime \alpha },\mathbf{\bar{T}}_{p}^{\alpha
}\right) =\frac{\exp \left( -\left( \mathbf{\bar{T}}_{p}^{\prime \alpha
}\right) ^{t}\mathbf{\bar{D}}_{S^{2}}\mathbf{\bar{T}}_{p}^{\prime \alpha
}-\left( \mathbf{\bar{T}}_{p}^{\alpha }\right) ^{t}\mathbf{\bar{D}}_{S^{2}}%
\mathbf{\bar{T}}_{p}^{\alpha }\right) }{\sqrt{\det \left( \mathbf{\bar{D}}%
_{S^{2}}\right) }}  \label{GRF}
\end{equation}%
and coming back to the original varianles:%
\begin{equation}
G\left( \mathbf{T}_{p}^{\prime \alpha },\mathbf{T}_{p}^{\alpha }\right) =%
\frac{\exp \left( -\left( \mathbf{T}_{p}^{\prime \alpha }\right) ^{t}\mathbf{%
A}_{S^{2}}^{\alpha }\mathbf{T}_{p}^{\prime \alpha }-\left( \mathbf{T}%
_{p}^{\alpha }\right) ^{t}\mathbf{A}_{S^{2}}^{\alpha }\mathbf{T}_{p}^{\alpha
}\right) }{\sqrt{\det \left( \mathbf{A}_{S^{2}}^{\alpha }\right) }}
\label{GRK}
\end{equation}%
This computes the probability of a transition for a specific structure
defined by its data $\mathbf{\bar{T}}_{p}^{\alpha }$ , including thus
average connectivities and internal frequencies of activities, to an other
state $\mathbf{\bar{T}}_{p}^{\prime \alpha }$. This transition occurs in an
average timespan normalized here to $1$

When there are no interactions or external signals, the structure remains
unchanged. The characteristics represented by the parameters $\alpha $ and $%
p $ remain unchanged. The structure only undergoes fluctuations around its
average values.

However, these fluctuations can, in the presence of interactions, lead to a
switch from one state to another, cause the structure to turn off, or enable
it to combine with another. In practice, a deviation induced by the
interaction may bring the values $\mathbf{\bar{T}}_{p}^{\alpha }$ closer to
a new average determined by some parameters $\alpha ^{\prime }$ and $%
p^{\prime }$, resulting in a state transition. As a consequence, the
quantity (\ref{GRF}) (or (\ref{GRK})) will be the main component in all
calculations of transition functions. Mathematically, it represents the
lowest-order expansion for transitions, and interactions will modify this
formula, leading to transition effects.

\subsubsection{Perturbative contributions}

Rewrite the potential in terms of diagonalized variables in (\ref{RCN}):%
\begin{equation*}
V\left( \left\{ \mathbf{T}_{l}^{\prime },\mathbf{\alpha }_{l}^{\prime },%
\mathbf{p}_{l}^{\prime },S_{l}^{\prime 2}\right\} ,\left\{ \mathbf{T}_{k},%
\mathbf{\alpha }_{k},\mathbf{p}_{k},S_{k}^{2}\right\} \right) =\bar{V}\left(
\left\{ \mathbf{\bar{T}}_{l}^{\prime },\mathbf{\alpha }_{l}^{\prime },%
\mathbf{p}_{l}^{\prime },S_{l}^{\prime 2}\right\} ,\left\{ \mathbf{\bar{T}}%
_{k},\mathbf{\alpha }_{k},\mathbf{p}_{k},S_{k}^{2}\right\} \right)
\end{equation*}

This potential induces corrections to the free Green functions, through the
perturbative expansion of the partition function. Actually:%
\begin{eqnarray*}
&&\exp \left( -S\right) \\
&=&\exp \left( -\sum_{S}\underline{\Gamma }^{\dag }\left( \mathbf{T},\mathbf{%
\alpha },\mathbf{p},S^{2}\right) \left( -\frac{1}{2}\nabla _{\left( \mathbf{%
\hat{T}}\right) _{S^{2}}}^{2}+\frac{1}{2}\left( \Delta \mathbf{T}%
_{p}^{\alpha }\right) _{S^{2}}^{t}\mathbf{A}_{S^{2}}^{\alpha }\left( \Delta 
\mathbf{T}_{p}^{\alpha }\right) +\mathbf{C}\right) \underline{\Gamma }\left( 
\mathbf{T},\mathbf{\alpha },\mathbf{p},S^{2}\right) \right) \\
&&\times \sum \frac{1}{r!}\left( \sum_{nn^{\prime }}\sum_{\substack{ k=1...n 
\\ l=1,...,n^{\prime }}}\sum_{\left\{ S_{k},S_{l}\right\} _{\substack{ %
l\leqslant n^{\prime }  \\ k\leqslant n}}}\prod_{l}\underline{\Gamma }^{\dag
}\left( \mathbf{T}_{l}^{\prime },\mathbf{\alpha }_{l}^{\prime },\mathbf{p}%
_{l}^{\prime },S_{l}^{\prime 2}\right) \bar{V}\left( \left\{ \mathbf{\bar{T}}%
_{l}^{\prime },\mathbf{\alpha }_{l}^{\prime },\mathbf{p}_{l}^{\prime
},S_{l}^{\prime 2}\right\} ,\left\{ \mathbf{\bar{T}}_{k},\mathbf{\alpha }%
_{k},\mathbf{p}_{k},S_{k}^{2}\right\} \right) \prod_{k}\underline{\Gamma }%
\left( \mathbf{T}_{k},\mathbf{\alpha }_{k},\mathbf{p}_{k},S_{k}^{2}\right)
\right) ^{r}
\end{eqnarray*}%
and the contributions to transition are computed with graphs. The vertices
are given by the potential and the legs correspond to the free Green
functions (\ref{GRF}):%
\begin{eqnarray}
&&G\left( \left\{ \mathbf{\bar{T}}_{l},\mathbf{\alpha }_{l},\mathbf{p}%
_{l},S_{l}^{2}\right\} ,\left\{ \mathbf{\bar{T}}_{k},\mathbf{\alpha }_{k},%
\mathbf{p}_{k},S_{k}^{2}\right\} \right)  \label{GRD} \\
&=&\int \prod\limits_{l}\frac{\exp \left( -\left( \mathbf{\bar{T}}%
_{l}^{\prime }\right) ^{t}\mathbf{\bar{D}}_{S_{l}^{2}}\mathbf{\bar{T}}%
_{l}^{\prime }-\left( \mathbf{\bar{T}}_{l}\right) ^{t}\mathbf{\bar{D}}%
_{S_{l}^{2}}\mathbf{\bar{T}}_{l}\right) }{\sqrt{\det \left( \mathbf{\bar{D}}%
_{S^{2}}\right) }}  \notag \\
&&\times \bar{V}\left( \left\{ \mathbf{\bar{T}}_{l}^{\prime },\mathbf{\alpha 
}_{l},\mathbf{p}_{l},S_{l}^{\prime 2}\right\} ,\left\{ \mathbf{\bar{T}}%
_{k}^{\prime },\mathbf{\alpha }_{k},\mathbf{p}_{k},S_{k}^{2}\right\} \right)
\prod\limits_{k}\frac{\exp \left( -\left( \mathbf{\bar{T}}_{k}^{\prime
}\right) ^{t}\mathbf{\bar{D}}_{S_{k}^{2}}\mathbf{\bar{T}}_{k}^{\prime
}-\left( \mathbf{\bar{T}}_{k}\right) ^{t}\mathbf{\bar{D}}_{S_{k}^{2}}\mathbf{%
\bar{T}}_{k}\right) }{\sqrt{\det \left( \mathbf{\bar{D}}_{S^{2}}\right) }}%
d\Delta \mathbf{\bar{T}}_{l}^{\prime }d\Delta \mathbf{\bar{T}}_{k}^{\prime }
\notag
\end{eqnarray}

Assuming the following form for the potentials:%
\begin{equation*}
\bar{V}\left( \left\{ \mathbf{\bar{T}}_{l}^{\prime },\mathbf{\alpha }_{l},%
\mathbf{p}_{l},S_{l}^{\prime 2}\right\} ,\left\{ \mathbf{\bar{T}}%
_{k}^{\prime },\mathbf{\alpha }_{k},\mathbf{p}_{k},S_{k}^{2}\right\} \right)
=V\left( \left( \left\Vert \mathbf{\bar{T}}_{l}^{\prime }\mathbf{-\bar{T}}%
_{k}^{\prime }\right\Vert ^{2}\right) \right)
\end{equation*}%
the integrals in (\ref{GRD}) can be computed.

Compared to the Green functions (\ref{GRF}), the terms (\ref{GRD}) introduce
a probability of transition between the two structures $S_{k}^{2}$ and $%
S_{l}^{\prime 2}$. If $S_{k}^{2}=S_{l}^{\prime 2}$, it represents a state
transition for structure $k$. The same cells are involved and interact, but
there is a change in parameters from $\left( \mathbf{\alpha }_{k},\mathbf{p}%
_{k}\right) $ to $\left( \mathbf{\alpha }_{l},\mathbf{p}_{l}\right) $. This
change affects the frequencies of activity.

If the two structures are different, potential $V$\ inducesa probability of
transitioning from one structure to another. The first one can be considered
switched off, while the second one switches on. It's worth noting that among
other possibilities, this may model spatial transitions of the same
structures along the thread. Some information is retained but not
necessarily at a fixed location.

\subsubsection{Exemple of perturbative transition}

Here, we study a transition due to sources, and consider (\ref{TR2}):%
\begin{equation}
\underline{\Gamma }^{\dag }\left( \Delta \mathbf{T}^{\prime },\mathbf{\alpha 
}^{\prime },\mathbf{p}^{\prime },S^{2},\theta \right) J\left( S^{2},\theta
\right) \underline{\Gamma }\left( \Delta \mathbf{T},\mathbf{\alpha },\mathbf{%
p},S^{2},\theta \right)
\end{equation}%
along with the following interaction term between two structures, each of
them in a given state $\left\{ \mathbf{T}_{l}^{\prime },\mathbf{\alpha }_{l},%
\mathbf{p}_{l},S_{l}^{2}\right\} $ and $\left\{ \mathbf{T}_{k},\mathbf{%
\alpha }_{k},\mathbf{p}_{k},S_{k}^{2}\right\} $ respectively:%
\begin{equation*}
V\left( \left\{ \mathbf{T}_{l}^{\prime },\mathbf{\alpha }_{l},\mathbf{p}%
_{l},S_{l}^{2}\right\} ,\left\{ \mathbf{T}_{k},\mathbf{\alpha }_{k},\mathbf{p%
}_{k},S_{k}^{2}\right\} \right) =\bar{V}\left( \left\{ \mathbf{\bar{T}}%
_{l}^{\prime },\mathbf{\alpha }_{l},\mathbf{p}_{l},S_{l}^{\prime 2}\right\}
,\left\{ \mathbf{\bar{T}}_{k},\mathbf{\alpha }_{k},\mathbf{p}%
_{k},S_{k}^{2}\right\} \right)
\end{equation*}%
The combination of these two contibutions in the effective action induces
the possibility of a transition:%
\begin{equation*}
\left\vert \mathbf{\bar{T}}_{k},\mathbf{\alpha }_{k},\mathbf{p}%
_{k},S_{k}^{2}\right\rangle \overset{J}{\rightarrow }\left\vert \mathbf{\bar{%
T}}_{k}^{\prime },\mathbf{\alpha }_{k}^{\prime },\mathbf{p}_{k}^{\prime
},S_{k}^{2}\right\rangle \overset{\bar{V}}{\rightarrow }\left\vert \mathbf{%
\bar{T}}_{l},\mathbf{\alpha }_{l},\mathbf{p}_{l},S_{l}^{2}\right\rangle
\end{equation*}%
The first arrow corresponds to the transition due to the external source,
and the second arrow describes the transition due to the structure-structure
interaction.

In terms of amplitude, this corresponds to the perturbative expansion of the
Green function:%
\begin{eqnarray}
&&G\left( \left\{ \mathbf{\bar{T}}_{l},\mathbf{\alpha }_{l},\mathbf{p}%
_{l},S_{l}^{2}\right\} ,\left\{ \mathbf{\bar{T}}_{k},\mathbf{\alpha }_{k},%
\mathbf{p}_{k},S_{k}^{2}\right\} \right)  \label{GRN} \\
&=&\int \frac{\exp \left( -\left( \mathbf{\bar{T}}_{l}^{\prime }\right) ^{t}%
\mathbf{\bar{D}}_{S_{l}^{2}}\mathbf{\bar{T}}_{l}^{\prime }-\left( \mathbf{%
\bar{T}}_{l}\right) ^{t}\mathbf{\bar{D}}_{S_{l}^{2}}\mathbf{\bar{T}}%
_{l}\right) }{\sqrt{\det \left( \mathbf{\bar{D}}_{S^{2}}\right) }}  \notag \\
&&\times \bar{V}\left( \left\{ \mathbf{\bar{T}}_{l}^{\prime },\mathbf{\alpha 
}_{l},\mathbf{p}_{l},S_{l}^{2}\right\} ,\left\{ \mathbf{\bar{T}}_{k}^{\prime
},\mathbf{\alpha }_{k}^{\prime },\mathbf{p}_{k}^{\prime },S_{k}^{2}\right\}
\right) \frac{\exp \left( -\left( \mathbf{\bar{T}}_{k}^{\prime }\right) ^{t}%
\mathbf{\bar{D}}_{S^{2}}\mathbf{\bar{T}}_{k}^{\prime }-\left( \mathbf{\bar{T}%
}_{k}^{\prime \prime }\right) ^{t}\mathbf{\bar{D}}_{S^{2}}\mathbf{\bar{T}}%
_{k}^{\prime \prime }\right) }{\sqrt{\det \left( \mathbf{\bar{D}}%
_{S^{2}}\right) }}  \notag \\
&&\times J\left( S_{k}^{2},\theta \right) \frac{\exp \left( -\left( \mathbf{%
\bar{T}}_{k}\right) ^{t}\mathbf{\bar{D}}_{S^{2}}\mathbf{\bar{T}}_{k}-\left( 
\mathbf{\bar{T}}_{k}^{\prime }\right) ^{t}\mathbf{\bar{D}}_{S^{2}}\mathbf{%
\bar{T}}_{k^{\prime }}^{\prime }\right) }{\sqrt{\det \left( \mathbf{\bar{D}}%
_{S^{2}}\right) }}d\Delta \mathbf{\bar{T}}_{l}^{\prime }d\Delta \mathbf{\bar{%
T}}_{k}^{\prime }d\Delta \mathbf{\bar{T}}_{k}^{\prime \prime }  \notag
\end{eqnarray}%
Assuming the potential $\bar{V}$ to be proportionl to a Dirac delta, so that
the transition between structures occur for states in which the structures'
frequencies in activity are similar:%
\begin{equation*}
\delta \left( \left\{ \mathbf{\bar{T}}_{l}^{\prime },\mathbf{\alpha }_{l},%
\mathbf{p}_{l},S_{l}^{\prime 2}\right\} -\left\{ \mathbf{\bar{T}}%
_{k}^{\prime },\mathbf{\alpha }_{k}^{\prime },\mathbf{p}_{k}^{\prime
},S_{k}^{2}\right\} \right)
\end{equation*}%
the integration over $\Delta \mathbf{\bar{T}}_{k}^{\prime }$ and $\Delta 
\mathbf{\bar{T}}_{k}^{\prime \prime }$ reduces the amplitude (\ref{GRN}) to:%
\begin{equation*}
G\left( \left\{ \mathbf{\bar{T}}_{l},\mathbf{\alpha }_{l},\mathbf{p}%
_{l},S_{l}^{2}\right\} ,\left\{ \mathbf{\bar{T}}_{k},\mathbf{\alpha }_{k},%
\mathbf{p}_{k},S_{k}^{2}\right\} \right) =AJ\left( S_{k}^{2},\theta \right) 
\frac{\exp \left( -\left( \mathbf{\bar{T}}_{l}\right) ^{t}\mathbf{\bar{D}}%
_{S_{l}^{2}}\mathbf{\bar{T}}_{l}-\left( \mathbf{\bar{T}}_{k}\right) ^{t}%
\mathbf{\bar{D}}_{S^{2}}\mathbf{\bar{T}}_{k}\right) }{\sqrt{\det \left( 
\mathbf{\bar{D}}_{S^{2}}\right) }}
\end{equation*}%
with $A$ an integration constant. As a consequence, an apparent transition
from one structure to a different one has arisen from the source signal.
However, the transition is feasible only through a decay of the initial
structure towards a state synchronized with the new emerging state.

It is noteworthy that the same mechanism arises if the extrrnal source
activates a substructure and that substructure, in turn, activates the full
structure:

\begin{equation*}
\left\vert \mathbf{vac}\right\rangle \overset{J}{\rightarrow }\left\vert 
\mathbf{\bar{T}}_{l},\mathbf{\alpha }_{l},\mathbf{p}_{l},S_{l}^{2}\subset
S_{k}^{2}\right\rangle \overset{\bar{V}}{\rightarrow }\left\vert \mathbf{%
\bar{T}}_{k},\mathbf{\alpha }_{k},\mathbf{p}_{k},S_{k}^{2}\right\rangle
\end{equation*}%
The first arrow represents the activation of the substructure through $J$,
while $\bar{V}$ induces the transition to the structure $\left\vert \mathbf{%
\bar{T}}_{k},\mathbf{\alpha }_{k},\mathbf{p}_{k},S_{k}^{2}\right\rangle $.
The associated amplitude is:%
\begin{eqnarray}
&&G\left( \left\{ \mathbf{vac}\right\} ,\left\{ \mathbf{\bar{T}}_{k},\mathbf{%
\alpha }_{k},\mathbf{p}_{k},S_{k}^{2}\right\} \right) \\
&=&\int \frac{\exp \left( -\left( \mathbf{\bar{T}}_{k}\right) ^{t}\mathbf{%
\bar{D}}_{S^{2}}\mathbf{\bar{T}}_{k}-\left( \mathbf{\bar{T}}_{k}^{\prime
}\right) ^{t}\mathbf{\bar{D}}_{S^{2}}\mathbf{\bar{T}}_{k^{\prime }}^{\prime
}\right) }{\sqrt{\det \left( \mathbf{\bar{D}}_{S^{2}}\right) }}\bar{V}\left(
\left\{ \mathbf{\bar{T}}_{l}^{\prime },\mathbf{\alpha }_{l},\mathbf{p}%
_{l},S_{l}^{2}\right\} ,\left\{ \mathbf{\bar{T}}_{k}^{\prime },\mathbf{%
\alpha }_{k}^{\prime },\mathbf{p}_{k}^{\prime },S_{k}^{2}\right\} \right) 
\notag \\
&&\times J\left( S_{k}^{2},\theta \right) \frac{\exp \left( -\left( \mathbf{%
\bar{T}}_{l}^{\prime }\right) ^{t}\mathbf{\bar{D}}_{S_{l}^{2}}\mathbf{\bar{T}%
}_{l}^{\prime }-\left( \mathbf{\bar{T}}_{l}\right) ^{t}\mathbf{\bar{D}}%
_{S_{l}^{2}}\mathbf{\bar{T}}_{l}\right) }{\sqrt{\det \left( \mathbf{\bar{D}}%
_{S^{2}}\right) }}d\Delta \mathbf{\bar{T}}_{l}^{\prime }d\Delta \mathbf{\bar{%
T}}_{k}^{\prime }  \notag
\end{eqnarray}

\subsection{Effective field approach}

Transition from two structures of different types to a third combined one
may be described in several ways. The perturbation expansion enables an
understanding of the transition mechanism but may not capture
non-perturbative or long-lasting aspects. The effective field approach
overcomes this limitation by concentrating on the initial and final
background states. This approach allows for the consideration of permanent
effects of some background structures on others. By determining the
background state for some structure, we can rewrite the effective action for
the remaining components.

\subsubsection{Principle}

We start with the partition function describing the full potential system
including the entire set of structures, independent and composed n:%
\begin{equation}
\int \exp \left( \sum_{i}S\left( \underline{\Gamma _{i}}\right) +S\left( 
\underline{\Gamma _{\left[ 1,n\right] }}\right) +U\left( \left( \underline{%
\Gamma _{i}}\right) ,\underline{\Gamma _{\left[ 1,n\right] }}\right) \right)
\prod\limits_{i}\mathcal{D}\Gamma _{i}\mathcal{D}\Gamma _{\left[ 1,n\right] }
\label{PRT}
\end{equation}

In this context, we have divided the fields into two sets. Our objective is
to perform an integration over the degrees of freedom of $\Gamma _{i}$ to
derive the effective action for the fields $\Gamma _{\left[ 1,n\right] }$.
Consequently, the interaction between $\underline{\Gamma _{i}}$ and $%
\underline{\Gamma _{\left[ 1,n\right] }}$ is integrated out, which in turn
modifies the action of the $\underline{\Gamma _{\left[ 1,n\right] }}$. This
formulation is based on the assumption that certain potential structures are
present, some of which are activated (represented by the $\underline{\Gamma
_{i}}$), while others are not. The interactions among the $\underline{\Gamma
_{i}}$, whose overall impact is encompassed in the solutions of the saddle
point equations, create a landscape that facilitates the emergence of the
combined structure described by an action $S_{e}\left( \underline{\Gamma _{%
\left[ 1,n\right] }}\right) $, so that after integration over $\Gamma _{i}$,
the effective action $S_{e}\left( \underline{\Gamma _{\left[ 1,n\right] }}%
\right) $ has an associated partition function:%
\begin{equation*}
\int \exp \left( S_{e}\left( \underline{\Gamma _{\left[ 1,n\right] }}\right)
\right) \mathcal{D}\Gamma _{\left[ 1,n\right] }
\end{equation*}

Practically, at the lowest order of approximation, we solve the saddle point
equations for the $\underline{\Gamma _{i}}$:%
\begin{equation*}
\frac{\delta S\left( \underline{\Gamma _{i}}\right) }{\delta \underline{%
\Gamma _{i}}\left( \mathbf{T}_{i},\mathbf{\alpha }_{i},\mathbf{p}%
_{i},S_{i}^{2},\theta \right) }+\frac{\delta U\left( \left( \underline{%
\Gamma _{i}}\right) ,\underline{\Gamma _{\left[ 1,n\right] }}\right) }{%
\delta \underline{\Gamma _{i}}\left( \mathbf{T}_{i},\mathbf{\alpha }_{i},%
\mathbf{p}_{i},S_{i}^{2},\theta \right) }=0
\end{equation*}%
This set of equation allows to express $\Gamma _{i}$ as a functional of $%
\Gamma _{\left[ 1,n\right] }$ and (\ref{PRT}) becoms:%
\begin{equation*}
\int \exp \left( S\left( \underline{\Gamma _{\left[ 1,n\right] }}\right)
+\sum_{i}S\left( \underline{\Gamma _{i}}\left( \Gamma _{\left[ 1,n\right]
}\right) \right) +U\left( \left( \underline{\Gamma _{i}}\left( \Gamma _{%
\left[ 1,n\right] }\right) \right) ,\underline{\Gamma _{\left[ 1,n\right] }}%
\right) \right) \mathcal{D}\Gamma _{\left[ 1,n\right] }
\end{equation*}

The interaction $U\left( \left( \underline{\Gamma _{i}}\right) ,\underline{%
\Gamma _{\left[ 1,n\right] }}\right) $ between structures $\underline{\Gamma
_{i}}$ and $\Gamma _{\left[ 1,n\right] }$ is replaced by an effective
potential:%
\begin{equation*}
U_{e}\left( \underline{\Gamma _{\left[ 1,n\right] }}\right) =U\left( \left( 
\underline{\Gamma _{i}}\left( \Gamma _{\left[ 1,n\right] }\right) \right) ,%
\underline{\Gamma _{\left[ 1,n\right] }}\right)
\end{equation*}%
This modified potential may induce a non-trivial stable background for $%
\Gamma _{\left[ 1,n\right] }$ that is, a set of activated structures.

For a potential such that the interaction depends on some compatibility
conditions between the $\left( \underline{\Gamma _{i}}\right) $ and the $%
\left( \underline{\Gamma _{\left[ 1,n\right] }}\right) $, we may assume that:%
\begin{equation*}
U\left( \left( \underline{\Gamma _{i}}\right) ,\underline{\Gamma _{\left[ 1,n%
\right] }}\right)
\end{equation*}%
is proportional to:%
\begin{equation}
\delta \left( f\left( \left( \mathbf{\alpha }_{i},\mathbf{p}%
_{i},S_{i}^{2}\right) ,\mathbf{\alpha }_{\left[ 1,n\right] },\mathbf{p}_{%
\left[ 1,n\right] }\right) \right)  \label{CS}
\end{equation}%
the Dirac function $\delta $ implementing the condition between the
structures characteristics to interaction.

We can deduce the condition of activation for the $\Gamma _{\left[ 1,n\right]
}$. Actually, given the condition (\ref{CS}), the $\left( \underline{\Gamma
_{i}}\right) $ remains neutral, that is unactivated, if the following
condition:%
\begin{equation}
\frac{\delta U\left( \left( \underline{\Gamma _{i}}\right) ,\underline{%
\Gamma _{\left[ 1,n\right] }}\right) }{\delta \underline{\Gamma _{i}}\left( 
\mathbf{T}_{i},\mathbf{\alpha }_{i},\mathbf{p}_{i},S_{i}^{2},\theta \right) }%
=0
\end{equation}%
is not satisfied.

If on the contrry the $\left( \underline{\Gamma _{i}}\right) $ are
activated, their effective action at their minima is negative:%
\begin{equation}
\sum_{i}S\left( \underline{\Gamma _{i}}\left( \Gamma _{\left[ 1,n\right]
}\right) \right) +U\left( \left( \underline{\Gamma _{i}}\left( \Gamma _{%
\left[ 1,n\right] }\right) \right) ,\underline{\Gamma _{\left[ 1,n\right] }}%
\right) <0  \label{TV}
\end{equation}%
The full effective action for the $\underline{\Gamma _{\left[ 1,n\right] }}$
is 
\begin{equation*}
S_{e}\left( \underline{\Gamma _{\left[ 1,n\right] }}\right) =S\left( 
\underline{\Gamma _{\left[ 1,n\right] }}\right) +\sum_{i}S\left( \underline{%
\Gamma _{i}}\left( \Gamma _{\left[ 1,n\right] }\right) \right) +U\left(
\left( \underline{\Gamma _{i}}\left( \Gamma _{\left[ 1,n\right] }\right)
\right) ,\underline{\Gamma _{\left[ 1,n\right] }}\right)
\end{equation*}%
and as a consequence of (\ref{TV}):%
\begin{equation*}
S_{e}\left( \underline{\Gamma _{\left[ 1,n\right] }}\right) <S\left( 
\underline{\Gamma _{\left[ 1,n\right] }}\right)
\end{equation*}%
This lower value of the effective action for $\Gamma _{\left[ 1,n\right] }$
leads to a possibility of activation, even if the $\Gamma _{\left[ 1,n\right]
}$ were initially neutral.

\subsubsection{First example}

Assume two initial structures, i.e., $n=2$ and the following action
functional for the system:%
\begin{eqnarray}
&&\sum_{i=1,2}\underline{\Gamma _{i}}^{\dag }\left( \mathbf{T}_{i},\mathbf{%
\alpha }_{i},\mathbf{p}_{i},S_{i}^{2}\right) \left( -\frac{1}{2}\nabla
_{\left( \mathbf{\hat{T}}\right) _{S_{i}^{2}}}^{2}+\frac{1}{2}\left( \left( 
\mathbf{D}_{S_{i}^{2}}+\left( \mathbf{M}_{p_{i}}^{\mathbf{\alpha }%
_{i}}\right) _{S_{i}^{2}}\right) \Delta \mathbf{T}_{p_{i}}^{\mathbf{\alpha }%
_{i}}\right) ^{2}\right) \underline{\Gamma _{i}}\left( \mathbf{T}_{i},%
\mathbf{\alpha }_{i},\mathbf{p}_{i},S_{i}^{2}\right)  \label{TC} \\
&&+V_{i}\left( \left\vert \underline{\Gamma _{i}}\left( \mathbf{T}_{i},%
\mathbf{\alpha }_{i},\mathbf{p}_{i},S_{i}^{2}\right) \right\vert ^{2}\right)
\notag \\
&&+\underline{\Gamma _{12}}^{\dag }\left( \mathbf{T}_{1,2},\mathbf{\alpha }%
_{1,2},\mathbf{p}_{1,2},S_{1,2}^{2}\right) \left( -\frac{1}{2}\nabla
_{\left( \Delta \mathbf{T}_{p}^{\alpha \beta }\right) }^{2}+\frac{1}{2}%
\left( \left( \mathbf{D}+\left( \mathbf{M}_{p}^{\alpha \beta }\right)
\right) \left( \Delta \mathbf{T}_{p}^{\alpha \beta }-\left\langle \Delta 
\mathbf{T}\right\rangle _{p}^{\alpha \beta }\right) \right) ^{2}\right) 
\underline{\Gamma _{12}}\left( \mathbf{T}_{1,2},\mathbf{\alpha }_{1,2},%
\mathbf{p}_{1,2},S_{1,2}^{2}\right)  \notag \\
&&+V_{12}\left( \left\vert \underline{\Gamma _{12}}\left( \mathbf{T}_{1,2},%
\mathbf{\alpha }_{1,2},\mathbf{p}_{1,2},S_{1,2}^{2}\right) \right\vert
^{2}\right)  \notag \\
&&+U\left( \prod\limits_{i=1,2}\left\vert \underline{\Gamma _{i}}\left( 
\mathbf{T}_{i},\mathbf{\alpha }_{i},\mathbf{p}_{i},S_{i}^{2}\right)
\right\vert ^{2}f\left( \left( \mathbf{T}_{1,2},\mathbf{\alpha }_{1,2},%
\mathbf{p}_{1,2},S_{1,2}^{2}\right) ,\left\{ \left( \mathbf{T}_{i},\mathbf{%
\alpha }_{i},\mathbf{p}_{i},S_{i}^{2}\right) \right\} _{1,2}\right)
,\left\vert \underline{\Gamma _{12}}\left( \mathbf{T}_{1,2},\mathbf{\alpha }%
_{1,2},\mathbf{p}_{1,2},S_{1,2}^{2}\right) \right\vert ^{2}\right)  \notag
\end{eqnarray}%
It describes two independent structures $\underline{\Gamma _{i}}$ along with
a potential composite structure $\underline{\Gamma _{12}}$. The potential $U$
transcribes the possible transitions between $\underline{\Gamma _{i}}$ and $%
\Gamma _{12}$.

Assume that $V_{i}$, $V_{12}$ are such that $V_{12}\left( 0\right) =0$ and
the ground states $\underline{\Gamma _{i}}$ are activated, but $\underline{%
\Gamma _{12}}$ is not. Thus, the stable state for the composite structure is 
$\underline{\Gamma _{12}}=0$.

\paragraph{Ground state without interaction}

In absence of interaction, integrating the $\Gamma _{i}$' degrees of freedom
corresponds in first approximation to replace the dields $\underline{\Gamma
_{i}}$ by their saddle point solution $\underline{\Gamma _{i}}^{\left(
0\right) }$ where these saddle points equations are independent:%
\begin{equation*}
\left( -\frac{1}{2}\nabla _{\left( \mathbf{T}\right) _{S_{i}^{2}}}^{2}+\frac{%
1}{2}\left( \left( \mathbf{D}_{S_{i}^{2}}+\left( \mathbf{M}_{p_{i}}^{\mathbf{%
\alpha }_{i}}\right) _{S_{i}^{2}}\right) \Delta \mathbf{T}_{p_{i}}^{\mathbf{%
\alpha }_{i}}\right) ^{2}\right) \underline{\Gamma _{i}}^{\left( 0\right) }+%
\frac{\delta }{\delta \left\vert \underline{\Gamma _{i}}\right\vert ^{2}}%
V_{i}\left( \left\vert \underline{\Gamma _{i}^{\left( 0\right) }}\left( 
\mathbf{T}_{i},\mathbf{\alpha }_{i},\mathbf{p}_{i},S_{i}^{2}\right)
\right\vert ^{2}\right) \underline{\Gamma _{i}}^{\left( 0\right) }=0
\end{equation*}%
as explained above, in this state, the $\underline{\Gamma _{i}}$ are
activated. They do not interact with the field $\Gamma _{12}$, and the
combined structure $\underline{\Gamma _{12}}$ is not activated. Actually,
since $V_{12}\left( 0\right) =0$, the ground state for this structure is $%
\underline{\Gamma _{12}}=0$.

\paragraph{Interactions modified background\protect\bigskip}

To describe the interactions we first define:%
\begin{equation*}
f\left( \left\{ i\right\} ,\left( 1,2\right) \right) =f\left( \left( \mathbf{%
T}_{1,2},\mathbf{\alpha }_{1,2},\mathbf{p}_{1,2},S_{1,2}^{2}\right) ,\left\{
\left( \mathbf{T}_{i},\mathbf{\alpha }_{i},\mathbf{p}_{i},S_{i}^{2}\right)
\right\} _{1,2}\right)
\end{equation*}%
Then, including the interactions in the action, the saddle point equation
rewrites:%
\begin{eqnarray*}
0 &=&\left( -\frac{1}{2}\nabla _{\left( \mathbf{T}\right) _{S_{i}^{2}}}^{2}+%
\frac{1}{2}\left( \left( \mathbf{D}_{S_{i}^{2}}+\left( \mathbf{M}_{p_{i}}^{%
\mathbf{\alpha }_{i}}\right) _{S_{i}^{2}}\right) \Delta \mathbf{T}_{p_{i}}^{%
\mathbf{\alpha }_{i}}\right) ^{2}\right) \underline{\Gamma _{i}}+\frac{%
\delta }{\delta \left\vert \underline{\Gamma _{i}}\right\vert ^{2}}%
V_{i}\left( \left\vert \underline{\Gamma _{i}}\left( \mathbf{T}_{i},\mathbf{%
\alpha }_{i},\mathbf{p}_{i},S_{i}^{2}\right) \right\vert ^{2}\right) 
\underline{\Gamma _{i}} \\
&&+\frac{\delta }{\delta \left\vert \underline{\Gamma _{i}}\right\vert ^{2}}%
U\left( \prod\limits_{i=1,2}\left\vert \underline{\Gamma _{i}}\left( \mathbf{%
T}_{i},\mathbf{\alpha }_{i},\mathbf{p}_{i},S_{i}^{2}\right) \right\vert
^{2}f\left( \left\{ i\right\} ,\left( 1,2\right) \right) ,\left\vert 
\underline{\Gamma _{12}}\left( \mathbf{T}_{1,2},\mathbf{\alpha }_{1,2},%
\mathbf{p}_{1,2},S_{1,2}^{2}\right) \right\vert ^{2}\right) \underline{%
\Gamma _{i}}
\end{eqnarray*}%
This is solved in first approximation, by decomposing the field $\underline{%
\Gamma _{i}}$ as the background plus a fluctuation: 
\begin{equation*}
\underline{\Gamma _{i}}\simeq \underline{\Gamma _{i}}^{\left( 0\right)
}+\Delta \underline{\Gamma _{i}}
\end{equation*}%
and we are led to:%
\begin{eqnarray*}
0 &\simeq &\left( \left( -\frac{1}{2}\nabla _{\left( \mathbf{T}\right)
_{S_{i}^{2}}}^{2}+\frac{1}{2}\left( \left( \mathbf{D}_{S_{i}^{2}}+\left( 
\mathbf{M}_{p_{i}}^{\mathbf{\alpha }_{i}}\right) _{S_{i}^{2}}\right) \Delta 
\mathbf{T}_{p_{i}}^{\mathbf{\alpha }_{i}}\right) ^{2}\right) +\frac{\delta
V_{i}\left( \left\vert \underline{\Gamma _{i}^{\left( 0\right) }}\left( 
\mathbf{T}_{i},\mathbf{\alpha }_{i},\mathbf{p}_{i},S_{i}^{2}\right)
\right\vert ^{2}\right) }{\delta \left\vert \underline{\Gamma _{i}^{\left(
0\right) }}\right\vert ^{2}}\right) \Delta \underline{\Gamma _{i}} \\
&&+\frac{\delta ^{2}V_{i}\left( \left\vert \underline{\Gamma _{i}^{\left(
0\right) }}\left( \mathbf{T}_{i},\mathbf{\alpha }_{i},\mathbf{p}%
_{i},S_{i}^{2}\right) \right\vert ^{2}\right) }{\delta ^{2}\left( \left\vert 
\underline{\Gamma _{i}^{\left( 0\right) }}\right\vert ^{2}\right) ^{2}}%
\left\vert \underline{\Gamma _{i}^{\left( 0\right) }}\left( \mathbf{T}_{i},%
\mathbf{\alpha }_{i},\mathbf{p}_{i},S_{i}^{2}\right) \right\vert ^{2}\Delta 
\underline{\Gamma _{i}} \\
&&+\frac{\delta }{\delta \left\vert \underline{\Gamma _{i}^{\left( 0\right) }%
}\right\vert ^{2}}U\left( \prod\limits_{i=1,2}\left\vert \underline{\Gamma
_{i}^{\left( 0\right) }}\left( \mathbf{T}_{i},\mathbf{\alpha }_{i},\mathbf{p}%
_{i},S_{i}^{2}\right) \right\vert ^{2}f\left( \left\{ i\right\} ,\left(
1,2\right) \right) ,\left\vert \underline{\Gamma _{12}}\left( \mathbf{T}%
_{1,2},\mathbf{\alpha }_{1,2},\mathbf{p}_{1,2},S_{1,2}^{2}\right)
\right\vert ^{2}\right) \Gamma _{i}^{\left( 0\right) }
\end{eqnarray*}%
with solution:%
\begin{equation}
\underline{\Gamma _{i}}\simeq \underline{\Gamma _{i}}^{\left( 0\right)
}+O_{i}^{-1}\frac{\delta }{\delta \Gamma _{i}^{\dag \left( 0\right) }}%
U\left( \prod\limits_{i=1,2}\left\vert \underline{\Gamma _{i}^{\left(
0\right) }}^{\dag }\left( \mathbf{T}_{i},\mathbf{\alpha }_{i},\mathbf{p}%
_{i},S_{i}^{2}\right) \right\vert ^{2}f\left( \left\{ i\right\} ,\left(
1,2\right) \right) ,\left\vert \underline{\Gamma _{12}}\left( \mathbf{T}%
_{1,2},\mathbf{\alpha }_{1,2},\mathbf{p}_{1,2},S_{1,2}^{2}\right)
\right\vert ^{2}\right)  \label{SM}
\end{equation}%
where:%
\begin{equation*}
O_{\alpha }^{-1}=\left( \left( -\frac{1}{2}\nabla _{\left( \mathbf{\hat{T}}%
\right) _{S_{i}^{2}}}^{2}+\frac{1}{2}\left( \left( \mathbf{D}%
_{S_{i}^{2}}+\left( \mathbf{M}_{p_{i}}^{\mathbf{\alpha }_{i}}\right)
_{S_{i}^{2}}\right) \Delta \mathbf{T}_{p_{i}}^{\mathbf{\alpha }_{i}}\right)
^{2}\right) +\frac{\delta V_{i}\left( \left\vert \underline{\Gamma _{i}}%
^{\left( 0\right) }\right\vert ^{2}\right) }{\delta \left\vert \underline{%
\Gamma _{i}}^{\left( 0\right) }\right\vert ^{2}}+\frac{\delta
^{2}V_{i}\left( \left\vert \underline{\Gamma _{i}^{\left( 0\right) }}%
\right\vert ^{2}\right) }{\delta \left( \left\vert \underline{\Gamma
_{i}^{\left( 0\right) }}\right\vert ^{2}\right) ^{2}}\left\vert \underline{%
\Gamma _{i}^{\left( 0\right) }}\right\vert ^{2}\right) ^{-1}
\end{equation*}

\paragraph{Effective action for $\Gamma _{12}$}

Once the saddle point solution is known, we can rewrite the effective action
for $\Gamma _{i}$ as a function of $\Gamma _{12}$, thereby integrating the
degrees of freedom associated with $\Gamma _{i}$ to obtain an effective
potential for $\Gamma _{12}$:%
\begin{eqnarray}
&&\underline{\Gamma _{i}}^{\dag }\left( \mathbf{T}_{i},\mathbf{\alpha }_{i},%
\mathbf{p}_{i},S_{i}^{2}\right) \left( -\frac{1}{2}\nabla _{\left( \mathbf{%
\hat{T}}\right) _{S_{i}^{2}}}^{2}+\frac{1}{2}\left( \left( \mathbf{D}%
_{S_{i}^{2}}+\left( \mathbf{M}_{p_{i}}^{\mathbf{\alpha }_{i}}\right)
_{S_{i}^{2}}\right) \Delta \mathbf{T}_{p_{i}}^{\mathbf{\alpha }_{i}}\right)
^{2}\right) \underline{\Gamma _{i}}\left( \mathbf{T}_{i},\mathbf{\alpha }%
_{i},\mathbf{p}_{i},S_{i}^{2}\right)  \label{FL} \\
&&+V_{i}\left( \left\vert \underline{\Gamma _{i}}\left( \mathbf{T}_{i},%
\mathbf{\alpha }_{i},\mathbf{p}_{i},S_{i}^{2}\right) \right\vert ^{2}\right)
\notag \\
&=&V_{i}\left( \left\vert \underline{\Gamma _{i}}\left( \mathbf{T}_{i},%
\mathbf{\alpha }_{i},\mathbf{p}_{i},S_{i}^{2}\right) \right\vert ^{2}\right)
\notag \\
&&-\underline{\Gamma _{i}}^{\dag }\left( \mathbf{T}_{i},\mathbf{\alpha }_{i},%
\mathbf{p}_{i},S_{i}^{2}\right) \left( \frac{\delta V_{i}\left( \left\vert 
\underline{\Gamma _{i}}\right\vert ^{2}\right) }{\delta \left\vert 
\underline{\Gamma _{i}}\right\vert ^{2}}+\frac{\delta }{\delta \left\vert 
\underline{\Gamma _{i}}\right\vert ^{2}}U\left(
\prod\limits_{i=1,2}\left\vert \underline{\Gamma _{i}}\right\vert
^{2}f\left( \left\{ i\right\} ,\left( 1,2\right) \right) ,\left\vert 
\underline{\Gamma _{12}}\right\vert ^{2}\right) \right) \underline{\Gamma
_{i}}\left( \mathbf{T}_{i},\mathbf{\alpha }_{i},\mathbf{p}%
_{i},S_{i}^{2}\right)  \notag \\
&\simeq &V_{i}\left( \left\vert \underline{\Gamma _{i}^{\left( 0\right) }}%
\left( \mathbf{T}_{i},\mathbf{\alpha }_{i},\mathbf{p}_{i},S_{i}^{2}\right)
\right\vert ^{2}\right) +\frac{1}{2}\frac{\delta ^{2}V_{i}\left( \left\vert 
\underline{\Gamma _{i}^{\left( 0\right) }}\right\vert ^{2}\right) }{\delta
\left( \left\vert \underline{\Gamma _{i}^{\left( 0\right) }}\right\vert
^{2}\right) ^{2}}\left\vert \Delta \underline{\Gamma _{i}}\right\vert ^{2} 
\notag \\
&&-\underline{\Gamma _{i}^{\left( 0\right) }}^{\dag }\left( \mathbf{T}_{i},%
\mathbf{\alpha }_{i},\mathbf{p}_{i},S_{i}^{2}\right) \frac{\delta }{\delta
\left\vert \underline{\Gamma _{i}^{0}}\right\vert ^{2}}U\left(
\prod\limits_{i=1,2}\left\vert \underline{\Gamma _{i}}\right\vert
^{2}f\left( \left\{ i\right\} ,\left( 1,2\right) \right) ,\left\vert 
\underline{\Gamma _{12}}\right\vert ^{2}\right) \underline{\Gamma
_{i}^{\left( 0\right) }}\left( \mathbf{T}_{i},\mathbf{\alpha }_{i},\mathbf{p}%
_{i},S_{i}^{2}\right)  \notag
\end{eqnarray}%
The effective action for $\underline{\Gamma _{12}}$ combines the $\Gamma
_{12}$ part of (\ref{TC}) plus (\ref{FL}), so that it writes:%
\begin{eqnarray*}
&&\underline{\Gamma _{12}}^{\dag }\left( \mathbf{T}_{1,2},\mathbf{\alpha }%
_{1,2},\mathbf{p}_{1,2},S_{1,2}^{2}\right) \left( -\frac{1}{2}\nabla
_{\left( \Delta \mathbf{T}_{p}^{\alpha \beta }\right) }^{2}+\frac{1}{2}%
\left( \left( \mathbf{D}+\left( \mathbf{M}_{p}^{\alpha \beta }\right)
\right) \left( \Delta \mathbf{T}_{p}^{\alpha \beta }-\left\langle \Delta 
\mathbf{T}\right\rangle _{p}^{\alpha \beta }\right) \right) ^{2}\right) 
\underline{\Gamma _{12}}\left( \mathbf{T}_{1,2},\mathbf{\alpha }_{1,2},%
\mathbf{p}_{1,2},S_{1,2}^{2}\right) \\
&&+V_{12}\left( \left\vert \underline{\Gamma _{12}}\left( \mathbf{T}_{1,2},%
\mathbf{\alpha }_{1,2},\mathbf{p}_{1,2},S_{1,2}^{2}\right) \right\vert
^{2}\right) +U\left( \prod\limits_{i=1,2}\left\vert \underline{\Gamma _{i}}%
^{\left( 0\right) }\right\vert ^{2}f\left( \left\{ i\right\} ,\left(
1,2\right) \right) ,\left\vert \underline{\Gamma _{12}}\right\vert
^{2}\right) \\
&&+\frac{1}{2}\frac{\delta ^{2}V_{i}\left( \left\vert \underline{\Gamma
_{i}^{\left( 0\right) }}\right\vert ^{2}\right) }{\delta \left( \left\vert 
\underline{\Gamma _{i}^{\left( 0\right) }}\right\vert ^{2}\right) ^{2}}%
\left\vert \Delta \underline{\Gamma _{i}}\right\vert ^{2}-\underline{\Gamma
_{i}^{\left( 0\right) }}^{\dag }\left( \mathbf{T}_{i},\mathbf{\alpha }_{i},%
\mathbf{p}_{i},S_{i}^{2}\right) \frac{\delta U\left(
\prod\limits_{i=1,2}\left\vert \underline{\Gamma _{i}}\right\vert
^{2}f\left( \left\{ i\right\} ,\left( 1,2\right) \right) ,\left\vert 
\underline{\Gamma _{12}}\right\vert ^{2}\right) }{\delta \left\vert 
\underline{\Gamma _{i}^{0}}\right\vert ^{2}}\underline{\Gamma _{i}^{\left(
0\right) }}\left( \mathbf{T}_{i},\mathbf{\alpha }_{i},\mathbf{p}%
_{i},S_{i}^{2}\right)
\end{eqnarray*}

Assume that $U$ is proportional to:%
\begin{equation*}
\prod\limits_{i=1,2}\left\vert \underline{\Gamma _{i}}^{\left( 0\right) \dag
}\right\vert ^{2}
\end{equation*}%
Thus,the effective action for $\underline{\Gamma _{12}}\left( \mathbf{T}%
_{1,2},\mathbf{\alpha }_{1,2},\mathbf{p}_{1,2},S_{1,2}^{2}\right) $ becomes:%
\begin{eqnarray*}
&&\underline{\Gamma _{12}}^{\dag }\left( \mathbf{T}_{1,2},\mathbf{\alpha }%
_{1,2},\mathbf{p}_{1,2},S_{1,2}^{2}\right) \left( -\frac{1}{2}\nabla
_{\left( \Delta \mathbf{T}_{p}^{\alpha \beta }\right) }^{2}+\frac{1}{2}%
\left( \left( \mathbf{D}+\left( \mathbf{M}_{p}^{\alpha \beta }\right)
\right) \left( \Delta \mathbf{T}_{p}^{\alpha \beta }-\left\langle \Delta 
\mathbf{T}\right\rangle _{p}^{\alpha \beta }\right) \right) ^{2}\right) 
\underline{\Gamma _{12}}\left( \mathbf{T}_{1,2},\mathbf{\alpha }_{1,2},%
\mathbf{p}_{1,2},S_{1,2}^{2}\right) \\
&&+V_{12}^{\left( e\right) }\left( \left\vert \underline{\Gamma _{12}}\left( 
\mathbf{T}_{1,2},\mathbf{\alpha }_{1,2},\mathbf{p}_{1,2},S_{1,2}^{2}\right)
\right\vert ^{2}\right)
\end{eqnarray*}%
with the effective potential defined by:%
\begin{eqnarray}
&&V_{12}^{\left( e\right) }\left( \left\vert \underline{\Gamma _{12}}\left( 
\mathbf{T}_{1,2},\mathbf{\alpha }_{1,2},\mathbf{p}_{1,2},S_{1,2}^{2}\right)
\right\vert ^{2}\right)  \label{FP} \\
&=&V_{12}\left( \left\vert \underline{\Gamma _{12}}\left( \mathbf{T}_{1,2},%
\mathbf{\alpha }_{1,2},\mathbf{p}_{1,2},S_{1,2}^{2}\right) \right\vert
^{2}\right) +\frac{1}{2}\frac{\delta ^{2}V_{i}\left( \left\vert \underline{%
\Gamma _{i}^{\left( 0\right) }}\right\vert ^{2}\right) }{\delta \left(
\left\vert \underline{\Gamma _{i}^{\left( 0\right) }}\right\vert ^{2}\right)
^{2}}\left\vert \Delta \underline{\Gamma _{i}}\right\vert ^{2}-U\left(
\prod\limits_{i=1,2}\left\vert \underline{\Gamma _{i}}^{\left( 0\right)
}\right\vert ^{2}f\left( \left\{ i\right\} ,\left( 1,2\right) \right)
,\left\vert \underline{\Gamma _{12}}\right\vert ^{2}\right)  \notag
\end{eqnarray}%
The first term in (\ref{FP}) is the potential for structure $12$. It is not
sufficient to allow the emergenc of the structure. The second term: 
\begin{equation*}
\frac{1}{2}\frac{\delta ^{2}V_{i}\left( \left\vert \underline{\Gamma
_{i}^{\left( 0\right) }}\right\vert ^{2}\right) }{\delta \left( \left\vert 
\underline{\Gamma _{i}^{\left( 0\right) }}\right\vert ^{2}\right) ^{2}}%
\left\vert \Delta \underline{\Gamma _{i}}\right\vert ^{2}
\end{equation*}%
stabilizes the structures $i$ in its stable equilibrium and prevent the
transition of states.

However, the third term:%
\begin{equation*}
-U\left( \prod\limits_{i=1,2}\left\vert \underline{\Gamma _{i}}^{\left(
0\right) }\right\vert ^{2}f\left( \left\{ i\right\} ,\left( 1,2\right)
\right) ,\left\vert \underline{\Gamma _{12}}\right\vert ^{2}\right)
\end{equation*}%
represents a gain of transitioning from $1$ and $2$ to $12$.

As a consequence, if:%
\begin{equation*}
\frac{1}{2}\frac{\delta ^{2}V_{i}\left( \left\vert \underline{\Gamma
_{i}^{\left( 0\right) }}\right\vert ^{2}\right) }{\delta \left( \left\vert 
\underline{\Gamma _{i}^{\left( 0\right) }}\right\vert ^{2}\right) ^{2}}%
\left\vert \Delta \underline{\Gamma _{i}}\right\vert ^{2}-U\left(
\prod\limits_{i=1,2}\left\vert \underline{\Gamma _{i}}^{\left( 0\right)
}\right\vert ^{2}f\left( \left\{ i\right\} ,\left( 1,2\right) \right)
,\left\vert \underline{\Gamma _{12}}\right\vert ^{2}\right) <0
\end{equation*}%
then $V_{12}^{\left( e\right) }<V_{12}$ and there is a possibility for a
composed stable state. One condition is that $f\left( \left\{ i\right\}
,\left( 1,2\right) \right) >>1$.

\subsubsection{Second example, structures activation by substructures:}

We study an example in which the activation of a substructure may activate
the full structure. Consider the transitions:%
\begin{equation}
\left\vert \mathbf{T},\mathbf{\alpha },\mathbf{p},\left( S^{\prime }\subset
S\right) ^{2}\right\rangle \rightarrow \left\vert \mathbf{T}^{\prime },%
\mathbf{\alpha }^{\prime },\mathbf{p}^{\prime },S^{2}\right\rangle
\label{TRSN}
\end{equation}%
The effective action for $S_{1}^{2}\subset S^{2}$ writes:%
\begin{eqnarray*}
&&\underline{\Gamma }^{\dag }\left( \mathbf{T},\mathbf{\alpha },\mathbf{p}%
,S^{2}\right) \left( -\frac{1}{2}\nabla _{\left( \mathbf{\hat{T}}\right)
_{S^{2}}}^{2}+\frac{1}{2}\left( \left( \mathbf{D}_{S^{2}}+\left( \mathbf{M}%
_{p}^{\alpha }\right) _{S^{2}}\right) \Delta \mathbf{T}_{p}^{\alpha }\right)
^{2}\right) \underline{\Gamma }\left( \mathbf{T},\mathbf{\alpha },\mathbf{p}%
,S^{2}\right) +V\left( \left\vert \underline{\Gamma }\left( \mathbf{T},%
\mathbf{\alpha },\mathbf{p},S^{2}\right) \right\vert ^{2}\right) \\
&&+\underline{\Gamma }^{\dag }\left( \mathbf{T}_{1},\mathbf{\alpha }_{1},%
\mathbf{p}_{1},S_{1}^{2}\right) \left( -\frac{1}{2}\nabla _{\left( \mathbf{%
\hat{T}}\right) _{S_{1}^{2}}}^{2}+\frac{1}{2}\left( \left( \mathbf{D}%
_{S_{1}^{2}}+\left( \mathbf{M}_{p}^{\alpha }\right) _{S_{1}^{2}}\right)
\Delta \mathbf{T}_{p}^{\alpha }\right) ^{2}\right) \underline{\Gamma }\left( 
\mathbf{T}_{1},\mathbf{\alpha }_{1},\mathbf{p}_{1},S_{1}^{2}\right) \\
&&+V_{1}\left( \left\vert \underline{\Gamma }\left( \mathbf{T}_{1},\mathbf{%
\alpha }_{1},\mathbf{p}_{1},S_{1}^{2}\right) \right\vert ^{2}\right) \\
&&+I\left( \left( \mathbf{\alpha },\mathbf{p}\right) ,\left( \mathbf{\alpha }%
_{i}^{\prime },\mathbf{p}_{i}^{\prime }\right) ,\left( \mathbf{\alpha },%
\mathbf{p}\right) ,S^{2},S_{1}^{2}\right) \underline{\Gamma _{i}}^{\dag
}\left( \mathbf{T},\mathbf{\alpha },\mathbf{p},S_{i}^{2}\right) \underline{%
\Gamma }_{1}\left( \mathbf{T},\mathbf{\alpha }^{\prime },\mathbf{p}^{\prime
},S_{i}^{2}\right) \underline{\Gamma }\left( \mathbf{T}_{1},\mathbf{\alpha }%
_{1},\mathbf{p}_{1}S_{1}^{2}\right)
\end{eqnarray*}

To model the activation (\ref{TRSN}), we assume that in absence of
interaction term, the state describing the substructure is more stable than
the state describing the entire structure. Thus, we assume that for $I=0$:%
\begin{equation*}
S\left( \left\vert \underline{\Gamma }\left( \mathbf{T}_{1},\mathbf{\alpha }%
_{1},\mathbf{p}_{1},S_{1}^{2}\right) \right\vert ^{2},0\right) <S\left(
\left\vert \underline{\Gamma }\left( \mathbf{T}_{1},\mathbf{\alpha }_{1},%
\mathbf{p}_{1},S_{1}^{2}\right) \right\vert ^{2},\left\vert \underline{%
\Gamma }\left( \mathbf{T},\mathbf{\alpha },\mathbf{p},S^{2}\right)
\right\vert ^{2}\right)
\end{equation*}%
so that the background states can be assumed to satisfy:%
\begin{equation*}
\left\vert \underline{\Gamma }\left( \mathbf{T}_{1},\mathbf{\alpha }_{1},%
\mathbf{p}_{1},S_{1}^{2}\right) \right\vert ^{2}\neq 0
\end{equation*}%
and:%
\begin{equation*}
\left\vert \underline{\Gamma }\left( \mathbf{T},\mathbf{\alpha },\mathbf{p}%
,S^{2}\right) \right\vert ^{2}=0
\end{equation*}

However, if interaction $I$ induces activation of $\underline{\Gamma }\left( 
\mathbf{T},\mathbf{\alpha },\mathbf{p},S^{2}\right) $ then the saddle point
equation for both structures:%
\begin{eqnarray}
\frac{\delta S\left( \left\vert \underline{\Gamma }_{1}\right\vert
^{2},\left\vert \underline{\Gamma }\right\vert ^{2}\right) }{\delta 
\underline{\Gamma }_{1}} &=&\left( -\frac{1}{2}\nabla _{\left( \mathbf{\hat{T%
}}\right) _{S_{1}^{2}}}^{2}+\frac{1}{2}\left( \left( \mathbf{D}%
_{S_{1}^{2}}+\left( \mathbf{M}_{p}^{\alpha }\right) _{S_{1}^{2}}\right)
\Delta \mathbf{T}_{p}^{\alpha }\right) ^{2}\right) \underline{\Gamma }\left( 
\mathbf{T}_{1},\mathbf{\alpha }_{1},\mathbf{p}_{1},S_{1}^{2}\right)
\label{SPN} \\
&&+\frac{\delta V_{1}\left( \left\vert \underline{\Gamma }\left( \mathbf{T}%
_{1},\mathbf{\alpha }_{1},\mathbf{p}_{1},S_{1}^{2}\right) \right\vert
^{2}\right) }{\delta \underline{\Gamma }_{1}^{\dag }}  \notag \\
&&+I\left( \left( \mathbf{\alpha },\mathbf{p}\right) ,\left( \mathbf{\alpha }%
_{i}^{\prime },\mathbf{p}_{i}^{\prime }\right) ,\left( \mathbf{\alpha },%
\mathbf{p}\right) ,S^{2},S_{1}^{2}\right) \underline{\Gamma }_{1}\left( 
\mathbf{T},\mathbf{\alpha }^{\prime },\mathbf{p}^{\prime },S_{i}^{2}\right) 
\underline{\Gamma }\left( \mathbf{T}_{1},\mathbf{\alpha }_{1},\mathbf{p}%
_{1}S_{1}^{2}\right)  \notag
\end{eqnarray}%
and:%
\begin{eqnarray}
\frac{\delta S\left( \left\vert \underline{\Gamma }_{1}\right\vert
^{2},\left\vert \underline{\Gamma }\right\vert ^{2}\right) }{\delta 
\underline{\Gamma }} &=&\left( -\frac{1}{2}\nabla _{\left( \mathbf{\hat{T}}%
\right) _{S^{2}}}^{2}+\frac{1}{2}\left( \left( \mathbf{D}_{S^{2}}+\left( 
\mathbf{M}_{p}^{\alpha }\right) _{S^{2}}\right) \Delta \mathbf{T}%
_{p}^{\alpha }\right) ^{2}\right) \underline{\Gamma }\left( \mathbf{T},%
\mathbf{\alpha },\mathbf{p},S^{2}\right)  \label{SPD} \\
&&+\frac{\delta V\left( \left\vert \underline{\Gamma }\left( \mathbf{T},%
\mathbf{\alpha },\mathbf{p},S^{2}\right) \right\vert ^{2}\right) }{\delta 
\underline{\Gamma }^{\dag }}  \notag \\
&&+I\left( \left( \mathbf{\alpha },\mathbf{p}\right) ,\left( \mathbf{\alpha }%
_{i}^{\prime },\mathbf{p}_{i}^{\prime }\right) ,\left( \mathbf{\alpha },%
\mathbf{p}\right) ,S^{2},S_{1}^{2}\right) \left\vert \underline{\Gamma }%
_{1}\left( \mathbf{T},\mathbf{\alpha }^{\prime },\mathbf{p}^{\prime
},S_{1}^{2}\right) \right\vert ^{2}  \notag
\end{eqnarray}%
present a stable minimum with:%
\begin{eqnarray*}
\left\vert \underline{\Gamma }\left( \mathbf{T}_{1},\mathbf{\alpha }_{1},%
\mathbf{p}_{1},S_{1}^{2}\right) \right\vert ^{2} &\neq &0 \\
\left\vert \underline{\Gamma }\left( \mathbf{T},\mathbf{\alpha },\mathbf{p}%
,S^{2}\right) \right\vert ^{2} &\neq &0
\end{eqnarray*}%
To find the condition fo such stable state We first look at (\ref{SPN}). Due
to interaction, the background substructure:%
\begin{equation*}
\underline{\Gamma }_{0}\left( \mathbf{T}_{1},\mathbf{\alpha }_{1},\mathbf{p}%
_{1},S_{1}^{2}\right)
\end{equation*}%
is shifted by the interaction to:%
\begin{equation*}
\underline{\Gamma }_{0}\left( \mathbf{T}_{1},\mathbf{\alpha }_{1},\mathbf{p}%
_{1},S_{1}^{2}\right) +\delta \underline{\Gamma }_{0}\left( \mathbf{T}_{1},%
\mathbf{\alpha }_{1},\mathbf{p}_{1},S_{1}^{2}\right)
\end{equation*}%
and the saddle point equation (\ref{SPD}) for the structure $\Gamma $
becomes in turn: 
\begin{eqnarray*}
\frac{\delta S\left( \left\vert \underline{\Gamma }_{1}\right\vert
^{2},\left\vert \underline{\Gamma }\right\vert ^{2}\right) }{\delta 
\underline{\Gamma }} &=&\left( -\frac{1}{2}\nabla _{\left( \mathbf{\hat{T}}%
\right) _{S^{2}}}^{2}+\frac{1}{2}\left( \left( \mathbf{D}_{S^{2}}+\left( 
\mathbf{M}_{p}^{\alpha }\right) _{S^{2}}\right) \Delta \mathbf{T}%
_{p}^{\alpha }\right) ^{2}\right) \underline{\Gamma }\left( \mathbf{T},%
\mathbf{\alpha },\mathbf{p},S^{2}\right) \\
&&+\frac{\delta V\left( \left\vert \underline{\Gamma }\left( \mathbf{T},%
\mathbf{\alpha },\mathbf{p},S^{2}\right) \right\vert ^{2}\right) }{\delta 
\underline{\Gamma }^{\dag }} \\
&&+I\left( \left( \mathbf{\alpha },\mathbf{p}\right) ,\left( \mathbf{\alpha }%
_{i}^{\prime },\mathbf{p}_{i}^{\prime }\right) ,\left( \mathbf{\alpha },%
\mathbf{p}\right) ,S^{2},S_{1}^{2}\right) \left\vert \underline{\Gamma }%
_{10}\left( \mathbf{T},\mathbf{\alpha }^{\prime },\mathbf{p}^{\prime
},S_{1}^{2}\right) \right\vert ^{2}
\end{eqnarray*}%
Given this equation, we can conclude that if:%
\begin{equation*}
\frac{\delta V\left( \left\vert \underline{\Gamma }\left( \mathbf{T},\mathbf{%
\alpha },\mathbf{p},S^{2}\right) \right\vert ^{2}\right) }{\delta \underline{%
\Gamma }^{\dag }}+I\left( \left( \mathbf{\alpha },\mathbf{p}\right) ,\left( 
\mathbf{\alpha }_{i}^{\prime },\mathbf{p}_{i}^{\prime }\right) ,\left( 
\mathbf{\alpha },\mathbf{p}\right) ,S^{2},S_{1}^{2}\right) \left\vert 
\underline{\Gamma }_{10}\left( \mathbf{T},\mathbf{\alpha }^{\prime },\mathbf{%
p}^{\prime },S_{1}^{2}\right) \right\vert ^{2}<0
\end{equation*}%
then:%
\begin{equation*}
\underline{\Gamma }\left( \mathbf{T},\mathbf{\alpha },\mathbf{p}%
,S^{2}\right) \neq 0
\end{equation*}%
and the structure may be activated.

\subsection{Operators formalism and transformations from independent
structures to composite ones}

By translating the dynamics in terms of creation and destruction of
structures, the advantge of operator formalism, is to read directly which
terms induce transitions between structures and thus to understand the
dynamical mechanisms of transitions. We found in (\ref{SP}):

\begin{eqnarray}
&&S^{\left( O\right) }\left( \left( \mathbf{\alpha },\mathbf{p},S^{2}\right)
\right)  \label{SPR} \\
&=&\sum_{S\times S}\mathbf{\bar{D}}_{S^{2}}^{\alpha }\left( \mathbf{A}%
^{+}\left( \alpha ,p,S^{2}\right) \mathbf{A}^{-}\left( \alpha
,p,S^{2}\right) +\frac{1}{2}\right)  \notag \\
&&+\sum_{m,n}\bar{U}_{mn}\left( \mathbf{\alpha },\mathbf{p},S^{2}\right)
\left( \mathbf{\hat{A}}^{+}\left( \mathbf{\alpha },\mathbf{p},S^{2}\right)
\right) ^{m}\left( \mathbf{\hat{A}}^{-}\left( \mathbf{\alpha },\mathbf{p}%
,S^{2}\right) \right) ^{n}+\hat{V}  \notag
\end{eqnarray}%
This formulation shows the instablity of a state since the form of
interaction always allows apriori for transitions. However, some change of
basis makes possible to integrate out the overall results of these
interactions and to reveal the appearance of resulting stable dressed
structures, having included the action of some structures considered as
auxiliary in this perspective. This change of basis is similar to the
effective action formalism but is more precise in the present approach.

\subsubsection{Transformation of $S^{\left( O\right) }$\ and emergence of
composed structures}

Starting with operators describing transitions between structures, the idea
is to perform a transformation that modifies $S^{\left( O\right) }\left(
\left( \mathbf{\alpha },\mathbf{p},S^{2}\right) \right) $. The
transformation is performed through an operator $\exp \left( -F\right) $,
with $F$ to be determined in order, at least in first approximation, to
diagonalize partially (\ref{SPR}) and cancel the interaction terms between
two types of structures. This terms will be replaced by an effective
interaction terms between a subset of remaining bound structures.

\paragraph{Interaction terms}

Technically, we divide the structures into two sets. The first one labelled
by indices $k$ and $l$ describes the strctrs for which we aim at finding an
effective description. The second set labelled by indices $c$ and $d$
corresponds to structures that will be integrated out to produce effective
interactions in the remaining subset. The interaction between the subsets
takes the form: 
\begin{eqnarray*}
&&\sum_{n,n^{\prime }}\sum_{\left\{ S_{k/c},S_{l/d}\right\} _{\substack{ %
l/d=1,...,n^{\prime }  \\ k/c=1...n}}}\sum_{\left\{ m_{l/d}^{\prime
},m_{k/c}\right\} }\prod_{l/d=1}^{n^{\prime }}\left( \mathbf{\hat{A}}%
^{+}\left( \mathbf{\alpha }_{l/d}^{\prime },\mathbf{p}_{l/d}^{\prime
},S_{l/d}^{\prime 2}\right) \right) ^{m_{l/d}^{\prime }} \\
&&V_{n,n^{\prime }}\left( \left\{ \mathbf{\alpha }_{l/d}^{\prime },\mathbf{p}%
_{l/d}^{\prime },S_{l/d}^{\prime 2},m_{l/d}^{\prime }\right\} ,\left\{ 
\mathbf{\alpha }_{k/c},\mathbf{p}_{k/c},S_{k/c}^{2},m_{k/c}\right\} \right)
\prod_{k/c=1}^{n}\left( \mathbf{\hat{A}}^{-}\left( \mathbf{\alpha }_{k/c},%
\mathbf{p}_{k/c},S_{k/c}^{2}\right) \right) ^{m_{k}}
\end{eqnarray*}%
where indices $l/d$ or $k/c$ indicate that the structures can be of either
type. Or goal is to integrate the crossed interactions:%
\begin{equation}
\prod\limits_{l=1}^{n^{\prime }}\left( \mathbf{\hat{A}}^{+}\left( \mathbf{%
\alpha }_{l}^{\prime },\mathbf{p}_{l}^{\prime },S_{l}^{\prime 2}\right)
\right) ^{m_{l}^{\prime }}V_{n,n^{\prime }}\left( \left\{ \mathbf{\alpha }%
_{l}^{\prime },\mathbf{p}_{l}^{\prime },S_{l}^{\prime 2},m_{l}^{\prime
}\right\} ,\left\{ \mathbf{\alpha }_{c},\mathbf{p}_{c},S_{c}^{2},m_{c}\right%
\} \right) \prod_{k/c=1}^{n}\left( \mathbf{\hat{A}}^{-}\left( \mathbf{\alpha 
}_{c},\mathbf{p}_{c},S_{c}^{2}\right) \right) ^{m_{c}}+\left( \left(
l,c\right) \leftrightarrow \left( d,k\right) \right)  \label{SNR}
\end{equation}%
to obtain an effective action for structures $S_{k}^{2}$, $S_{l}^{\prime 2}$.

\paragraph{Transformation operator}

To integrate the crossed interactions and obtain the required effective
action, we consider the following transformation: 
\begin{equation}
\left( S^{\left( O\right) }\left( \left( \mathbf{\alpha },\mathbf{p}%
,S^{2}\right) \right) \right) ^{\prime }=\exp \left( -F\right) S^{\left(
O\right) }\left( \left( \mathbf{\alpha },\mathbf{p},S^{2}\right) \right)
\exp \left( F\right)  \label{TRS}
\end{equation}%
where $F$ will be found to cancel the interaction term (\ref{SNR}) after
transformation. Doing so modifies $S^{\left( O\right) }\left( \left( \mathbf{%
\alpha },\mathbf{p},S^{2}\right) \right) $ to a descrption of composed,
stable structures.

Disregarding the potential $\bar{U}_{mm^{\prime }}\left( \mathbf{\alpha },%
\mathbf{p},S^{2}\right) $ which can be considered as slowly varying and
expanding (\ref{TRS}) at the lowest order in interactions, while canceling
the interaction term (\ref{SNR}) leads to the relation:%
\begin{equation}
\left[ F,S_{0}\right] +I=0  \label{CNT}
\end{equation}%
with:%
\begin{equation*}
S_{0}=\mathbf{\bar{D}}_{S^{2}}^{\alpha }\left( \mathbf{A}^{+}\left( \alpha
,p,S^{2}\right) \mathbf{A}^{-}\left( \alpha ,p,S^{2}\right) +\frac{1}{2}%
\right)
\end{equation*}%
so that $S_{0}+I$ is transformed into: 
\begin{eqnarray*}
\left( S^{\left( O\right) }\right) ^{\prime } &=&S_{0}+I+\left[ F,S_{0}%
\right] +\left[ F,I\right] +\frac{1}{2}\left[ F,\left[ F,S_{0}+I\right] %
\right] \\
&=&S_{0}+I+\left[ F,S_{0}\right] +\left[ F,I\right] +\frac{1}{2}\left[ F,%
\left[ F,S_{0}\right] \right]
\end{eqnarray*}%
Then, using (\ref{CNT}), we find at lowest order:%
\begin{equation}
\left( S^{\left( O\right) }\right) ^{\prime }=S_{0}+\frac{1}{2}\left[ I,F%
\right]  \label{TM}
\end{equation}%
To find this effective action, we solve (\ref{CNT}) by postulating that $F$
has the same form as $I$:%
\begin{eqnarray*}
F &=&\sum_{nn^{\prime }}\sum_{\substack{ k=1...n  \\ l=1,...,n^{\prime }}}%
\sum_{\left\{ S_{k},S_{l}\right\} _{\substack{ l=1,...,n^{\prime }  \\ %
k=1...n }}}\prod_{l=1}^{n^{\prime }}\prod\limits_{s=1}^{m_{l}^{\prime }}%
\mathbf{A}^{+}\left( \mathbf{\alpha }_{l}^{\prime },\mathbf{p}_{l}^{\prime
},S_{l}^{\prime 2}\right) \\
&&\times F\left( \left\{ \mathbf{\alpha }_{l}^{\prime },\mathbf{p}%
_{l}^{\prime },S_{l}^{\prime 2},m_{l}^{\prime }\right\} ,\left\{ \mathbf{%
\alpha }_{k},\mathbf{p}_{k},S_{k}^{2},m_{k}\right\} \right)
\prod_{k=1}^{n}\prod\limits_{s=1}^{m_{k}}\mathbf{A}^{-}\left( \mathbf{\alpha 
}_{k},\mathbf{p}_{k},S_{k}^{2}\right)
\end{eqnarray*}%
and in appendix 4 the solution of (\ref{CNT}) writes:%
\begin{equation}
F\left( \left\{ \mathbf{\alpha }_{l}^{\prime },\mathbf{p}_{l}^{\prime
},S_{l}^{\prime 2},m_{l}^{\prime }\right\} ,\left\{ \mathbf{\alpha }_{k},%
\mathbf{p}_{k},S_{k}^{2},m_{k}\right\} \right) =-\frac{V_{n,n^{\prime
}}\left( \left\{ \mathbf{\alpha }_{l}^{\prime },\mathbf{p}_{l}^{\prime
},S_{l}^{\prime 2},m_{l}^{\prime }\right\} ,\left\{ \mathbf{\alpha }_{k},%
\mathbf{p}_{k},S_{k}^{2},m_{k}\right\} \right) }{\sum_{l=1}^{n^{\prime
}}m_{l}^{\prime }\mathbf{\bar{D}}_{S_{l}^{\prime 2}}^{\mathbf{\alpha }%
_{l}^{\prime }}-\sum_{k=1}^{n}m_{k}\mathbf{\bar{D}}_{S_{k}^{2}}^{\mathbf{%
\alpha }_{k}}}  \label{CMT}
\end{equation}

\paragraph{Correction (\protect\ref{TM}) to the action}

We also obtain the matrix elements of $\left[ I,F\right] $ that modify (\ref%
{TM}). Defining:%
\begin{eqnarray*}
\mathbf{\Lambda }_{k} &=&\left( \mathbf{\alpha }_{k},\mathbf{p}%
_{k},S_{k}^{2}\right) \\
\mathbf{\Lambda }_{l}^{\prime } &=&\left( \mathbf{\alpha }_{l}^{\prime },%
\mathbf{p}_{l}^{\prime },S_{l}^{\prime 2}\right)
\end{eqnarray*}%
and:%
\begin{eqnarray*}
\mathbf{\bar{\Lambda}}_{\bar{k}} &=&\left( \mathbf{\bar{\alpha}}_{\bar{k}},%
\mathbf{\bar{p}}_{\bar{k}},\bar{S}_{\bar{k}}^{2}\right) \\
\mathbf{\bar{\Lambda}}_{\bar{l}}^{\prime } &=&\left( \mathbf{\bar{\alpha}}_{%
\bar{l}}^{\prime },\mathbf{\bar{p}}_{\bar{l}}^{\prime },\bar{S}_{\bar{l}%
}^{\prime 2}\right)
\end{eqnarray*}%
we find:

\begin{eqnarray}
&&\left[ I,F\right] \left( \left\{ \mathbf{\Lambda }_{L}^{\prime
},M_{L}^{\prime }\right\} ,\left\{ \mathbf{\Lambda }_{K},M_{K}\right\}
\right)  \label{CMR} \\
&=&-\sum_{P_{K},P_{L}}\sum_{\left\{ \epsilon _{d}^{\prime }\right\} ,\left\{
\epsilon _{c}\right\} }\sum_{\left\{ \delta _{k}\right\} ,\left\{ \delta
_{l}^{\prime }\right\} }\prod \left( \epsilon _{d}^{\prime }!\epsilon
_{c}!\right) ^{2}\prod\limits_{\bar{k},\bar{l},k,l}\left( -1\right) ^{\delta
_{l}^{\prime }}C_{m_{l}^{\prime }+\delta _{l}^{\prime }}^{\delta
_{l}^{\prime }}C_{m_{k}+\delta _{k}}^{\delta _{k}}  \notag \\
&&\times \mathbf{\bar{D}}_{S_{k}^{2}}^{\mathbf{\alpha }_{k}}\mathbf{\bar{D}}%
_{S_{k}^{2}}^{\mathbf{\alpha }_{l}^{\prime }}\mathbf{\bar{D}}_{S_{c}^{2}}^{%
\mathbf{\alpha }_{c}}\mathbf{\bar{D}}_{S_{d}^{2}}^{\mathbf{\alpha }%
_{d}^{\prime }}\delta \left( \mathbf{\Lambda }_{k}-\mathbf{\bar{\Lambda}}_{%
\bar{l}}^{\prime }\right) \delta \left( \mathbf{\Lambda }_{l}^{\prime }-%
\mathbf{\bar{\Lambda}}_{\bar{k}}\right) \delta \left( \mathbf{\Lambda }_{c}-%
\mathbf{\bar{\Lambda}}_{\bar{d}}^{\prime }\right) \delta \left( \mathbf{%
\Lambda }_{d}^{\prime }-\mathbf{\bar{\Lambda}}_{\bar{c}}\right)  \notag \\
&&\times \frac{V^{\left( 2\right) }\left( \left\{ \left( \mathbf{\Lambda }%
_{l\cup d}^{\prime },m_{l}^{\prime }+\delta _{l}^{\prime },\epsilon
_{d}^{\prime }\right) ,\left( \mathbf{\Lambda }_{k\cup c},m_{k}+\delta
_{k},\epsilon _{c}\right) ,\left( \mathbf{\bar{\Lambda}}_{\bar{l}\cup \bar{d}%
}^{\prime },\bar{m}_{^{\bar{l}}}^{\prime }+\delta _{k},\epsilon _{d}\right)
,\left( \mathbf{\bar{\Lambda}}_{\bar{k}\cup \bar{c}},\bar{m}_{^{\bar{k}%
}}+\delta _{l}^{\prime },\epsilon _{c}^{\prime }\right) \right\} \right) }{%
\sum_{l=1}^{n^{\prime }}\left( m_{l}^{\prime }+\delta _{l^{\prime }}^{\prime
}\right) \mathbf{\bar{D}}_{S_{l}^{\prime 2}}^{\mathbf{\alpha }_{l}^{\prime
}}-\sum_{k=1}^{n}\left( m_{k}+\delta _{k}\right) \mathbf{\bar{D}}%
_{S_{k}^{2}}^{\mathbf{\alpha }_{k}}+\sum_{d=1}^{p^{\prime }}\epsilon
_{d}^{\prime }\mathbf{\bar{D}}_{S_{d}^{\prime 2}}^{\mathbf{\alpha }%
_{d}^{\prime }}-\sum_{c=1}^{p}\epsilon _{c}\mathbf{\bar{D}}_{S_{c}^{2}}^{%
\mathbf{\alpha }_{c}}}  \notag
\end{eqnarray}%
where:%
\begin{eqnarray*}
&&V^{\left( 2\right) }\left( \left\{ \left( \mathbf{\Lambda }_{l\cup
d}^{\prime },m_{l}^{\prime }+\delta _{l}^{\prime },\epsilon _{d}^{\prime
}\right) ,\left( \mathbf{\Lambda }_{k\cup c},m_{k}+\delta _{k},\epsilon
_{c}\right) ,\left( \mathbf{\bar{\Lambda}}_{\bar{l}\cup \bar{d}}^{\prime },%
\bar{m}_{^{\bar{l}}}^{\prime }+\delta _{k},\epsilon _{d}\right) ,\left( 
\mathbf{\bar{\Lambda}}_{\bar{k}\cup \bar{c}},\bar{m}_{^{\bar{k}}}+\delta
_{l}^{\prime },\epsilon _{c}^{\prime }\right) \right\} \right) \\
&=&V\left( \left\{ \mathbf{\Lambda }_{l}^{\prime },m_{l}^{\prime }+\delta
_{l}^{\prime }\right\} \cup \left\{ \mathbf{\Lambda }_{d}^{\prime },\epsilon
_{d}^{\prime }\right\} ,\left\{ \mathbf{\Lambda }_{k},m_{k}+\delta
_{k}\right\} \cup \left\{ \mathbf{\Lambda }_{c},\epsilon _{c}\right\} \right)
\\
&&\times V\left( \left\{ \mathbf{\bar{\Lambda}}_{\bar{l}}^{\prime },\bar{m}%
_{^{\bar{l}}}^{\prime }+\delta _{k}\right\} \cup \left\{ \mathbf{\bar{\Lambda%
}}_{\bar{d}}^{\prime },\epsilon _{c}\right\} ,\left\{ \mathbf{\bar{\Lambda}}%
_{\bar{k}},\bar{m}_{^{\bar{k}}}+\delta _{l}^{\prime }\right\} \cup \left\{ 
\mathbf{\bar{\Lambda}}_{\bar{c}},\epsilon _{d}^{\prime }\right\} \right)
\end{eqnarray*}%
with $P_{K},P_{L}$ are partitions of $\left\{ \mathbf{\alpha }_{K},\mathbf{p}%
_{K},S_{K}^{2},M_{K}\right\} $, $\left\{ \mathbf{\alpha }_{L}^{\prime },%
\mathbf{p}_{L}^{\prime },S_{L}^{\prime 2},M_{L}^{\prime }\right\} $:%
\begin{eqnarray*}
\left\{ \mathbf{\alpha }_{L}^{\prime },\mathbf{p}_{L}^{\prime
},S_{L}^{\prime 2},M_{L}^{\prime }\right\} &=&\left\{ \mathbf{\alpha }%
_{l}^{\prime },\mathbf{p}_{l}^{\prime },S_{l}^{\prime 2},m_{l^{\prime
}}^{\prime }\right\} \cup \left\{ \mathbf{\bar{\alpha}}_{\bar{l}}^{\prime },%
\mathbf{\bar{p}}_{\bar{l}}^{\prime },\bar{S}_{\bar{l}}^{\prime 2},\bar{m}_{%
\bar{l}^{\prime }}^{\prime }\right\} \\
\left\{ \mathbf{\alpha }_{K},\mathbf{p}_{K},S_{K}^{2},M_{K}\right\}
&=&\left\{ \mathbf{\alpha }_{k},\mathbf{p}_{k},S_{k}^{2},m_{k}\right\} \cup
\left\{ \mathbf{\bar{\alpha}}_{\bar{k}},\mathbf{\bar{p}}_{\bar{k}},\bar{S}_{%
\bar{k}}^{2},\bar{m}_{\bar{k}}\right\}
\end{eqnarray*}%
In the commutator (\ref{CMR}), we sum over all of these possible partitions.

\subsubsection{Effective structures}

After transformation, the operator version of the action writes:%
\begin{eqnarray}
&&\left( S^{\left( O\right) }\right) ^{\prime }=\sum_{S\times S}\mathbf{\bar{%
D}}_{S^{2}}^{\alpha }\left( \mathbf{A}^{+}\left( \alpha ,p,S^{2}\right) 
\mathbf{A}^{-}\left( \alpha ,p,S^{2}\right) +\frac{1}{2}\right)  \label{FST}
\\
&&+\sum_{n,n^{\prime }}\sum_{\left\{ S_{k},S_{l}\right\} _{\substack{ %
l=1,...,n^{\prime }  \\ k=1...n}}}\sum_{\left\{ m_{l}^{\prime
},m_{k}\right\} }\prod_{l=1}^{n^{\prime }}\left( \mathbf{\hat{A}}^{+}\left( 
\mathbf{\alpha }_{l}^{\prime },\mathbf{p}_{l}^{\prime },S_{l}^{\prime
2}\right) \right) ^{m_{l}^{\prime }}  \notag \\
&&\times V_{n,n^{\prime }}\left( \left\{ \mathbf{\alpha }_{l}^{\prime },%
\mathbf{p}_{l}^{\prime },S_{l}^{\prime 2},m_{l}^{\prime }\right\} ,\left\{ 
\mathbf{\alpha }_{k},\mathbf{p}_{k},S_{k}^{2},m_{k}\right\} \right)
\prod_{k=1}^{n}\left( \mathbf{\hat{A}}^{-}\left( \mathbf{\alpha }_{k},%
\mathbf{p}_{k},S_{k}^{2}\right) \right) ^{m_{k}}  \notag \\
&&+\frac{1}{2}\prod_{L=1}^{n^{\prime }}\prod\limits_{s^{\prime
}=1}^{M_{L}^{\prime }}\mathbf{A}^{+}\left( \mathbf{\alpha }_{L}^{\prime },%
\mathbf{p}_{L}^{\prime },S_{L}^{\prime 2}\right) \left[ I,F\right] \left(
\left\{ \mathbf{\alpha }_{L}^{\prime },\mathbf{p}_{L}^{\prime
},S_{L}^{\prime 2},M_{L}^{\prime }\right\} ,\left\{ \mathbf{\alpha }_{K},%
\mathbf{p}_{K},S_{K}^{2},M_{K}\right\} \right)
\prod\limits_{K=1}^{n}\prod\limits_{s=1}^{M_{k}^{\prime }}\mathbf{A}%
^{-}\left( \mathbf{\alpha }_{K},\mathbf{p}_{K},S_{K}^{2}\right)  \notag
\end{eqnarray}%
In effective action:%
\begin{equation*}
\left[ I,F\right] \left( \left\{ \mathbf{\alpha }_{L}^{\prime },\mathbf{p}%
_{L}^{\prime },S_{L}^{\prime 2},M_{L}^{\prime }\right\} ,\left\{ \mathbf{%
\alpha }_{K},\mathbf{p}_{K},S_{K}^{2},M_{K}\right\} \right)
\end{equation*}%
the structures $\left\{ \mathbf{\alpha }_{c},\mathbf{p}_{c},S_{c}^{2}\right%
\} $ $\left\{ \mathbf{\bar{\alpha}}_{\bar{d}}^{\prime },\mathbf{\bar{p}}_{%
\bar{d}}^{\prime },\bar{S}_{\bar{d}}^{\prime 2}\right\} $ have been
integrated and do not appear anymore in the interaction. They have glued
structures $\left\{ \mathbf{\alpha }_{l}^{\prime },\mathbf{p}_{l}^{\prime
},S_{l}^{\prime 2}\right\} $ and $\left\{ \mathbf{\alpha }_{k},\mathbf{p}%
_{k},S_{k}^{2}\right\} $ even if this ones were not interacting initially
that is, even if:%
\begin{equation}
V_{n,n^{\prime }}\left( \left\{ \mathbf{\alpha }_{l}^{\prime },\mathbf{p}%
_{l}^{\prime },S_{l}^{\prime 2},m_{l}^{\prime }\right\} _{l\leqslant
n^{\prime }},\left\{ \mathbf{\alpha }_{k},\mathbf{p}_{k},S_{k}^{2},m_{k}%
\right\} _{l\leqslant n}\right) =0  \label{VZ}
\end{equation}%
Depending on the form of the resulting interaction (\ref{FST}), some new
combined structures may appear.

\subsubsection{Bound states}

Assuming that condition (\ref{VZ}) is satisfied, we can describe the
combined structures by computing the

the eigenstates of (\ref{FST}) with lowest eigenvalues. It is written as a
series involving the $n$ types of sructures:%
\begin{equation*}
\left\vert \prod\limits_{K}\left( \left( \mathbf{\alpha }_{K},\mathbf{p}%
_{K},S_{K}^{2},M_{K}\right) \right) \right\rangle =\sum_{\left( M_{K}\right)
}A\left( \left( \mathbf{\alpha }_{K},\mathbf{p}_{K},S_{K}^{2},M_{K}\right)
\right) \prod\limits_{K=1}^{n}\prod\limits_{s=1}^{M_{K}}\mathbf{A}^{+}\left( 
\mathbf{\alpha }_{K},\mathbf{p}_{K},S_{K}^{2}\right)
\prod\limits_{K}\left\vert Vac\right\rangle _{K}
\end{equation*}

\paragraph{Stt wth lwst gnvls}

At the lowest order of the series expansion, the eigenstate writes:%
\begin{equation*}
\left\vert \prod\limits_{K}\left( \mathbf{\alpha }_{K},\mathbf{p}%
_{K},S_{K}^{2}\right) \right\rangle =\sum_{K}A\left( \left( \mathbf{\alpha }%
_{K},\mathbf{p}_{K},S_{K}^{2}\right) \right) \mathbf{A}^{+}\left( \mathbf{%
\alpha }_{K},\mathbf{p}_{K},S_{K}^{2}\right) \prod\limits_{K}\left\vert
Vac\right\rangle _{K}
\end{equation*}%
The coefficients $A\left( \left( \mathbf{\alpha }_{K},\mathbf{p}%
_{K},S_{K}^{2}\right) \right) $ are obtained by writing the eigenvalues
equation for $\left\vert \prod\limits_{K}\left( \mathbf{\alpha }_{K},\mathbf{%
p}_{K},S_{K}^{2}\right) \right\rangle $. The action of $\left( S^{\left(
O\right) }\right) $ on $\left\vert \prod\limits_{K}\left( \mathbf{\alpha }%
_{K},\mathbf{p}_{K},S_{K}^{2}\right) \right\rangle $ yields: 
\begin{eqnarray*}
&&\left( S^{\left( O\right) }\right) \left\vert \prod\limits_{K}\left( 
\mathbf{\alpha }_{K},\mathbf{p}_{K},S_{K}^{2}\right) \right\rangle \\
&=&\sum_{K}A\left( \left( \mathbf{\alpha }_{K},\mathbf{p}_{K},S_{K}^{2}%
\right) \right) \left( \sum_{K}\mathbf{\bar{D}}_{S_{K^{\prime }}^{2}}^{%
\mathbf{\alpha }_{K^{\prime }}}+2\mathbf{\bar{D}}_{S_{K}^{2}}^{\mathbf{%
\alpha }_{K}}\right) \mathbf{A}^{+}\left( \mathbf{\alpha }_{K},\mathbf{p}%
_{K},S_{K}^{2}\right) \prod\limits_{K}\left\vert Vac\right\rangle _{K} \\
&&+\sum_{K,L}A\left( \left( \mathbf{\alpha }_{K},\mathbf{p}%
_{K},S_{K}^{2}\right) \right) \left[ I,F\right] \left( \left\{ \mathbf{%
\alpha }_{L}^{\prime },\mathbf{p}_{L}^{\prime },S_{L}^{\prime 2},1\right\}
,\left\{ \mathbf{\alpha }_{K},\mathbf{p}_{K},S_{K}^{2},1\right\} ,\left( 
\mathbf{\alpha }_{P},\mathbf{p}_{P},S_{P}^{2},0\right) \right) \mathbf{A}%
^{+}\left( \mathbf{\alpha }_{L}^{\prime },\mathbf{p}_{L}^{\prime
},S_{L}^{\prime 2}\right) \prod\limits_{K}\left\vert Vac\right\rangle _{K}
\end{eqnarray*}%
writing the eigenvalues:%
\begin{equation*}
\eta \left( \mathbf{\alpha }_{K},\mathbf{p}_{K},S_{K}^{2}\right)
=\sum_{K^{\prime }}\mathbf{\bar{D}}_{S_{K^{\prime }}^{2}}^{\mathbf{\alpha }%
_{K^{\prime }}}+2\mathbf{\bar{D}}_{S_{K}^{2}}^{\mathbf{\alpha }%
_{K}}-\varepsilon \left( \mathbf{\alpha }_{K},\mathbf{p}_{K},S_{K}^{2}\right)
\end{equation*}%
We show in appendix 4 that for weak interactions, i.e.: 
\begin{equation*}
\left[ I,F\right] \left( \left( \mathbf{\alpha }_{K},\mathbf{p}%
_{K},S_{K}^{2},1\right) ,\left( \mathbf{\alpha }_{K^{\prime }},\mathbf{p}%
_{K^{\prime }},S_{K^{\prime }}^{2},1\right) \right) <<1
\end{equation*}
the eigenvalues are:

\begin{equation*}
\varepsilon \left( \mathbf{\alpha }_{K},\mathbf{p}_{K},S_{K}^{2}\right)
=\sum_{K^{\prime }}\frac{\left[ I,F\right] \left( \left( \mathbf{\alpha }%
_{K},\mathbf{p}_{K},S_{K}^{2},1\right) ,\left( \mathbf{\alpha }_{K^{\prime
}},\mathbf{p}_{K^{\prime }},S_{K^{\prime }}^{2},1\right) \right) \left[ I,F%
\right] \left( \left( \mathbf{\alpha }_{K^{\prime }},\mathbf{p}_{K^{\prime
}},S_{K^{\prime }}^{2},1\right) ,\left( \mathbf{\alpha }_{K},\mathbf{p}%
_{K},S_{K}^{2},1\right) \right) }{\left( \mathbf{\bar{D}}_{S_{K^{\prime
}}^{2}}^{\mathbf{\alpha }_{K^{\prime }}}+2\mathbf{\bar{D}}_{S_{K^{\prime
}}^{2}}^{\mathbf{\alpha }_{K^{\prime }}}-\eta \right) }
\end{equation*}%
and the corresponding eigenstates are defined by the following relation on
coefficients:%
\begin{equation*}
A\left( \mathbf{\alpha }_{L},\mathbf{p}_{L},S_{L}^{2}\right) \simeq -\frac{%
\left[ I,F\right] \left( \left( \mathbf{\alpha }_{L},\mathbf{p}%
_{L},S_{L}^{2},1\right) ,\left( \mathbf{\alpha }_{K^{\prime }},\mathbf{p}%
_{K^{\prime }},S_{K^{\prime }}^{2},1\right) \right) }{\mathbf{\bar{D}}%
_{S_{L}^{2}}^{\mathbf{\alpha }_{L}}+2\mathbf{\bar{D}}_{S_{L}^{2}}^{\mathbf{%
\alpha }_{L}}}A\left( \mathbf{\alpha }_{K},\mathbf{p}_{K},S_{K}^{2}\right)
\end{equation*}%
Given one value $A\left( \mathbf{\alpha }_{K},\mathbf{p}_{K},S_{K}^{2}%
\right) $, all other coefficients of the series are known. In first
approximation;%
\begin{equation*}
\eta _{K}=\sum_{K^{\prime }}\mathbf{\bar{D}}_{S_{K^{\prime }}^{2}}^{\mathbf{%
\alpha }_{K^{\prime }}}+2\mathbf{\bar{D}}_{S_{K}^{2}}^{\mathbf{\alpha }%
_{K}}+\varepsilon _{K}
\end{equation*}%
where:%
\begin{equation*}
\varepsilon _{K}A\left( \left( \mathbf{\alpha }_{K},\mathbf{p}%
_{K},S_{K}^{2}\right) \right) =\sum_{L}A\left( \left( \mathbf{\alpha }%
_{L}^{\prime },\mathbf{p}_{L}^{\prime },S_{L}^{\prime 2}\right) \right) %
\left[ I,F\right] \left( \left\{ \mathbf{\alpha }_{K},\mathbf{p}%
_{K},S_{K}^{2},1\right\} ,\left\{ \mathbf{\alpha }_{L}^{\prime },\mathbf{p}%
_{L}^{\prime },S_{L}^{\prime 2},1\right\} ,\left( \mathbf{\alpha }_{P},%
\mathbf{p}_{P},S_{P}^{2},0\right) \right)
\end{equation*}%
and the eigenstates writes:%
\begin{eqnarray}
&&\left\vert \prod\limits_{K}\left( \mathbf{\alpha }_{K},\mathbf{p}%
_{K},S_{K}^{2}\right) \right\rangle  \label{ST} \\
&=&\mathcal{N}\left\{ \sum_{K}A\left( \left( \mathbf{\alpha }_{K},\mathbf{p}%
_{K},S_{K}^{2}\right) \right) H_{1}\left( \frac{\sigma _{T}}{2\sqrt{2}}%
\left( \mathbf{\Delta T}_{\mathbf{p}_{K}}^{\mathbf{\alpha }_{K}}\right)
^{t}\left( \mathbf{A}_{\mathbf{p}_{K}}^{\mathbf{\alpha }_{K}}\right)
_{S_{K}^{2}}\left( \mathbf{\Delta T}_{\mathbf{p}_{K}}^{\mathbf{\alpha }%
_{K}}\right) _{S_{K}^{2}}\right) \right\}  \notag \\
&&\times \exp \left( -\frac{1}{2}\sum_{K}\left( \mathbf{\Delta T}_{\mathbf{p}%
_{K}}^{\mathbf{\alpha }_{K}}\right) ^{t}\left( \mathbf{A}_{\mathbf{p}_{K}}^{%
\mathbf{\alpha }_{K}}\right) _{S_{K}^{2}}\left( \mathbf{\Delta T}_{\mathbf{p}%
_{K}}^{\mathbf{\alpha }_{K}}\right) _{S_{K}^{2}}\right)  \notag
\end{eqnarray}%
where $H_{1}$ is the Hermite polynomial:%
\begin{equation*}
H_{1}\left( \frac{\sigma _{T}}{2\sqrt{2}}\left( \mathbf{\Delta T}_{\mathbf{p}%
_{K}}^{\mathbf{\alpha }_{K}}\right) ^{t}\left( \mathbf{A}_{\mathbf{p}_{K}}^{%
\mathbf{\alpha }_{K}}\right) _{S_{K}^{2}}\left( \mathbf{\Delta T}_{\mathbf{p}%
_{K}}^{\mathbf{\alpha }_{K}}\right) _{S_{K}^{2}}\right) =\frac{\sigma _{T}}{2%
\sqrt{2}}\left( \mathbf{\Delta T}_{\mathbf{p}_{K}}^{\mathbf{\alpha }%
_{K}}\right) ^{t}\left( \mathbf{A}_{\mathbf{p}_{K}}^{\mathbf{\alpha }%
_{K}}\right) _{S_{K}^{2}}\left( \mathbf{\Delta T}_{\mathbf{p}_{K}}^{\mathbf{%
\alpha }_{K}}\right) _{S_{K}^{2}}
\end{equation*}%
and $\mathcal{N}$ normalization factor. The form of this state is similar to
the states obtained while describing independent structures. However, in the
present approach, the structures obtained combine initial independent
structures. We thus obtain, through the interaction with a third part, an
integrated structure encompassing the characteristics of some "primary"
collective states.

\paragraph{Full series expansion}

More generally, the state is determined by a series:%
\begin{eqnarray*}
&&\left\vert \prod\limits_{K}\left( \mathbf{\alpha }_{K},\mathbf{p}%
_{K},S_{K}^{2},M_{K}\right) \right\rangle \\
&=&\sum_{\left( M_{K}\right) }A\left( \left( \mathbf{\alpha }_{K},\mathbf{p}%
_{K},S_{K}^{2},M_{K}\right) \right)
\prod\limits_{K=1}^{n}\prod\limits_{s=1}^{M_{K}}\mathbf{A}^{+}\left( \mathbf{%
\alpha }_{K},\mathbf{p}_{K},S_{K}^{2}\right) \prod\limits_{K}\left\vert
Vac\right\rangle _{K}
\end{eqnarray*}%
Given the definition of the creation operators, this writes:%
\begin{eqnarray*}
\left\vert \prod\limits_{K}\left( \mathbf{\alpha }_{K},\mathbf{p}%
_{K},S_{K}^{2},M_{K}\right) \right\rangle &=&\mathcal{N}\sum_{\left(
M_{K}\right) }A\left( \left( \mathbf{\alpha }_{K},\mathbf{p}%
_{K},S_{K}^{2},M_{k}\right) \right) \prod\limits_{K=1}^{n}H_{n}\left( \frac{%
\sigma _{T}}{2\sqrt{2}}\left( \mathbf{\Delta T}_{\mathbf{p}_{K}}^{\mathbf{%
\alpha }_{K}}\right) ^{t}\left( \mathbf{A}_{\mathbf{p}_{K}}^{\mathbf{\alpha }%
_{K}}\right) _{S_{K}^{2}}\left( \mathbf{\Delta T}_{\mathbf{p}_{K}}^{\mathbf{%
\alpha }_{K}}\right) _{S_{K}^{2}}\right) \\
&&\times \exp \left( -\frac{1}{2}\sum_{K}\left( \mathbf{\Delta T}_{\mathbf{p}%
_{K}}^{\mathbf{\alpha }_{K}}\right) ^{t}\left( \mathbf{A}_{\mathbf{p}_{K}}^{%
\mathbf{\alpha }_{K}}\right) _{S_{K}^{2}}\left( \mathbf{\Delta T}_{\mathbf{p}%
_{K}}^{\mathbf{\alpha }_{K}}\right) _{S_{K}^{2}}\right)
\end{eqnarray*}%
where $H_{n}$ is the n-th Hermite polynomial. The coefficients are computed
in appendix 6. These states are similar to (\ref{ST}) but include higher
level of activty.

\section{Exemple with 3 structures}

We present both the effective formalism and the operator formalism approach
to study the binding of two independent structures through the
intermediation of a third one. Binding structures through a third part is
described by the diagram:

\begin{equation*}
\left\vert \mathbf{T}_{k},\mathbf{\alpha }_{k},\mathbf{p}_{k},S_{k}^{2}%
\right\rangle \rightleftarrows \left\vert \mathbf{T}_{l},\mathbf{\alpha }%
_{l},\mathbf{p}_{l},S_{l}^{2}\right\rangle \rightleftarrows \left\vert 
\mathbf{T}_{k^{\prime }},\mathbf{\alpha }_{k^{\prime }},\mathbf{p}%
_{k^{\prime }},S_{k^{\prime }}^{2}\right\rangle \overset{\text{effective}}{%
\Longrightarrow }\left\vert \mathbf{T}_{k},\mathbf{\alpha }_{k},\mathbf{p}%
_{k},S_{k}^{2}\right\rangle \rightleftarrows \left\vert \mathbf{T}%
_{k^{\prime }},\mathbf{\alpha }_{k^{\prime }},\mathbf{p}_{k^{\prime
}},S_{k^{\prime }}^{2}\right\rangle
\end{equation*}%
Initially, structures $\left\vert \mathbf{T}_{l},\mathbf{\alpha }_{l},%
\mathbf{p}_{l},S_{l}^{2}\right\rangle $ interacts both with $\left\vert 
\mathbf{T}_{k},\mathbf{\alpha }_{k},\mathbf{p}_{k},S_{k}^{2}\right\rangle $
and $\left\vert \mathbf{T}_{k^{\prime }},\mathbf{\alpha }_{k^{\prime }},%
\mathbf{p}_{k^{\prime }},S_{k^{\prime }}^{2}\right\rangle $ that are a
priori not related. If we consider that the time scale of $\left\vert 
\mathbf{T}_{l},\mathbf{\alpha }_{l},\mathbf{p}_{l},S_{l}^{2}\right\rangle $
is longer than that of the others, it can be integrated out, which yields
the effective interaction between $\left\vert \mathbf{T}_{k},\mathbf{\alpha }%
_{k},\mathbf{p}_{k},S_{k}^{2}\right\rangle $ and $\left\vert \mathbf{T}%
_{k^{\prime }},\mathbf{\alpha }_{k^{\prime }},\mathbf{p}_{k^{\prime
}},S_{k^{\prime }}^{2}\right\rangle $.

\subsection{Effective formalism}

Within the effective field formalism the situation is described as the
follwing. Consider three structures. First, we study independently the three
structures:%
\begin{eqnarray*}
&&S_{0}=\sum_{i=1,2}\underline{\Gamma _{i}}^{\dag }\left( \mathbf{T}_{i},%
\mathbf{\alpha }_{i},\mathbf{p}_{i},S_{i}^{2}\right) \left( -\frac{1}{2}%
\nabla _{\left( \mathbf{\hat{T}}\right) _{S_{i}^{2}}}^{2}+\frac{1}{2}\left(
\left( \mathbf{D}_{S_{i}^{2}}+\left( \mathbf{M}_{p}^{\alpha }\right)
_{S_{i}^{2}}\right) \Delta \mathbf{T}_{p}^{\alpha }\right) ^{2}\right) 
\underline{\Gamma _{i}}\left( \mathbf{T}_{i},\mathbf{\alpha }_{i},\mathbf{p}%
_{i},S_{i}^{2}\right) \\
&&+V_{i}\left( \left\vert \underline{\Gamma _{i}}\left( \mathbf{T}_{i},%
\mathbf{\alpha }_{i},\mathbf{p}_{i},S_{i}^{2}\right) \right\vert ^{2}\right)
\\
&&+\underline{\Gamma _{0}}^{\dag }\left( \mathbf{T}_{0},\mathbf{\alpha }_{0},%
\mathbf{p}_{0},S_{0}^{2}\right) \left( -\frac{1}{2}\nabla _{\left( \mathbf{%
\hat{T}}\right) _{S_{0}^{2}}}^{2}+\frac{1}{2}\left( \left( \mathbf{D}%
_{S_{0}^{2}}+\left( \mathbf{M}_{p}^{\alpha }\right) _{S_{0}^{2}}\right)
\Delta \mathbf{T}_{p}^{\alpha }\right) ^{2}\right) \underline{\Gamma _{0}}%
\left( \mathbf{T}_{0},\mathbf{\alpha }_{0},\mathbf{p}_{0},S_{0}^{2}\right) \\
&&+V_{0}\left( \left\vert \underline{\Gamma _{0}}\left( \mathbf{T}_{0},%
\mathbf{\alpha }_{0},\mathbf{p}_{0},S_{0}^{2}\right) \right\vert ^{2}\right)
\end{eqnarray*}%
The structure denoted $0$ has a dynamic with lower frequencies. We assume
that $\underline{\Gamma _{i}}$ do not interact with each other. Interactions
arise through structure $0$. We consider the interaction term: 
\begin{eqnarray*}
&&I=\sum_{i=0,1,2}\underline{\Gamma _{i}}^{\dag }\left( \mathbf{T}_{i},%
\mathbf{\alpha }_{i},\mathbf{p}_{i},S_{i}^{2}\right) \left( -\frac{1}{2}%
\nabla _{\left( \mathbf{\hat{T}}\right) _{S_{i}^{2}}}^{2}+\frac{1}{2}\left(
\left( \mathbf{D}_{S_{i}^{2}}+\left( \mathbf{M}_{p_{i}}^{\mathbf{\alpha }%
_{i}}\right) _{S_{i}^{2}}\right) \Delta \mathbf{T}_{p_{i}}^{\mathbf{\alpha }%
_{i}}\right) ^{2}\right) \underline{\Gamma _{i}}\left( \mathbf{T}_{i},%
\mathbf{\alpha }_{i},\mathbf{p}_{i},S_{i}^{2}\right) \\
&&+V_{i}\left( \left\vert \underline{\Gamma _{i}}\left( \mathbf{T}_{i},%
\mathbf{\alpha }_{i},\mathbf{p}_{i},S_{i}^{2}\right) \right\vert ^{2}\right)
\\
&&+\sum_{i=1,2}I\left( \left( \mathbf{\alpha }_{i},\mathbf{p}_{i}\right)
,\left( \mathbf{\alpha }_{i}^{\prime },\mathbf{p}_{i}^{\prime }\right)
,\left( \mathbf{\alpha }_{0},\mathbf{p}_{0}\right)
,S_{i}^{2},S_{0}^{2}\right) \underline{\Gamma _{i}}^{\dag }\left( \mathbf{T}%
_{i},\mathbf{\alpha }_{i},\mathbf{p}_{i},S_{i}^{2}\right) \underline{\Gamma
_{i}}\left( \mathbf{T}_{i},\mathbf{\alpha }_{i}^{\prime },\mathbf{p}%
_{i}^{\prime },S_{i}^{2}\right) \underline{\Gamma _{0}}\left( \mathbf{T}_{0},%
\mathbf{\alpha }_{0},\mathbf{p}_{0},S_{0}^{2}\right)
\end{eqnarray*}

We compute the vaccum state for $\underline{\Gamma _{0}}\left( \mathbf{T}%
_{0},\mathbf{\alpha }_{0},\mathbf{p}_{0},S_{0}^{2}\right) $ as a function of
the $\underline{\Gamma _{i}}$. These two structures act as source terms. The
saddle point equations for $S_{0}+I$ with respect to $\underline{\Gamma _{0}}
$ yield:%
\begin{eqnarray}
\underline{\Gamma _{0}}\left( \mathbf{T}_{0},\mathbf{\alpha }_{0},\mathbf{p}%
_{0},S_{0}^{2}\right) &=&-\sum_{i=1,2}\int G\left( \left( \mathbf{\alpha }%
_{0},\mathbf{p}_{0}\right) ,\left( \mathbf{\alpha }_{0}^{\prime },\mathbf{p}%
_{0}^{\prime }\right) \right) d\left( \mathbf{\alpha }_{0}^{\prime },\mathbf{%
p}_{0}^{\prime }\right)  \label{GM} \\
&&I\left( \left( \mathbf{\alpha }_{i},\mathbf{p}_{i}\right) ,\left( \mathbf{%
\alpha }_{i}^{\prime },\mathbf{p}_{i}^{\prime }\right) ,\left( \mathbf{%
\alpha }_{0}^{\prime },\mathbf{p}_{0}^{\prime }\right)
,S_{i}^{2},S_{0}^{2}\right) \underline{\Gamma _{i}}^{\dag }\left( \mathbf{T}%
_{i},\mathbf{\alpha }_{i},\mathbf{p}_{i},S_{i}^{2}\right) \underline{\Gamma
_{i}}\left( \mathbf{T}_{i},\mathbf{\alpha }_{i}^{\prime },\mathbf{p}%
_{i}^{\prime },S_{i}^{2}\right)  \notag
\end{eqnarray}%
with the Green function $G\left( \left( \mathbf{\alpha }_{0},\mathbf{p}%
_{0}\right) ,\left( \mathbf{\alpha }_{0}^{\prime },\mathbf{p}_{0}^{\prime
}\right) \right) $ is the inverse of the following operator:%
\begin{equation*}
G\left( \left( \mathbf{\alpha }_{0},\mathbf{p}_{0}\right) ,\left( \mathbf{%
\alpha }_{0}^{\prime },\mathbf{p}_{0}^{\prime }\right) \right) =\left( -%
\frac{1}{2}\nabla _{\left( \mathbf{\hat{T}}\right) _{S_{0}^{2}}}^{2}+\frac{1%
}{2}\left( \left( \mathbf{D}_{S_{0}^{2}}+\left( \mathbf{M}_{p}^{\alpha
}\right) _{S_{0}^{2}}\right) \Delta \mathbf{T}_{p}^{\alpha }\right)
^{2}\right) ^{-1}
\end{equation*}%
The effective action for $\underline{\Gamma _{i}}\left( \mathbf{T}_{i},%
\mathbf{\alpha }_{i},\mathbf{p}_{i},S_{i}^{2}\right) $, $i=1,2$ is obtained
by replacing (\ref{GM}) in $S_{0}+I$:%
\begin{eqnarray*}
&&S_{f}=\sum_{i=0,1,2}\underline{\Gamma _{i}}^{\dag }\left( \mathbf{T}_{i},%
\mathbf{\alpha }_{i},\mathbf{p}_{i},S_{i}^{2}\right) \left( -\frac{1}{2}%
\nabla _{\left( \mathbf{\hat{T}}\right) _{S_{i}^{2}}}^{2}+\frac{1}{2}\left(
\left( \mathbf{D}_{S_{i}^{2}}+\left( \mathbf{M}_{p_{i}}^{\mathbf{\alpha }%
_{i}}\right) _{S_{i}^{2}}\right) \Delta \mathbf{T}_{p_{i}}^{\mathbf{\alpha }%
_{i}}\right) ^{2}\right) \underline{\Gamma _{i}}\left( \mathbf{T}_{i},%
\mathbf{\alpha }_{i},\mathbf{p}_{i},S_{i}^{2}\right) \\
&&+V_{i}\left( \left\vert \underline{\Gamma _{i}}\left( \mathbf{T}_{i},%
\mathbf{\alpha }_{i},\mathbf{p}_{i},S_{i}^{2}\right) \right\vert ^{2}\right)
\\
&&+\sum_{\left( i,j\right) \in \left\{ 1,2\right\} ^{2}}I_{i,j}\left( \left( 
\mathbf{\alpha }_{i},\mathbf{p}_{i}\right) ,\left( \mathbf{\alpha }%
_{i}^{\prime },\mathbf{p}_{i}^{\prime }\right) ,\left( \mathbf{\beta }_{j},%
\mathbf{t}_{j}\right) ,\left( \mathbf{\beta }_{j}^{\prime },\mathbf{t}%
_{j}^{\prime }\right) ,S_{i}^{2},S_{j}^{2}\right) \\
&&\times \underline{\Gamma _{i}}^{\dag }\left( \mathbf{T}_{i},\mathbf{\alpha 
}_{i},\mathbf{p}_{i},S_{i}^{2}\right) \underline{\Gamma _{i}}\left( \mathbf{T%
}_{i},\mathbf{\alpha }_{i}^{\prime },\mathbf{p}_{i}^{\prime
},S_{i}^{2}\right) \underline{\Gamma _{i}}^{\dag }\left( \mathbf{T}_{j},%
\mathbf{\beta }_{j},\mathbf{t}_{j},S_{j}^{2}\right) \underline{\Gamma _{i}}%
\left( \mathbf{T}_{j}^{\prime },\mathbf{\beta }_{j}^{\prime },\mathbf{t}%
_{j}^{\prime },S_{j}^{2}\right)
\end{eqnarray*}%
where:%
\begin{eqnarray*}
&&I_{i,j}\left( \left( \mathbf{\alpha }_{i},\mathbf{p}_{i}\right) ,\left( 
\mathbf{\alpha }_{i}^{\prime },\mathbf{p}_{i}^{\prime }\right) ,\left( 
\mathbf{\beta }_{j},\mathbf{t}_{j}\right) ,\left( \mathbf{\beta }%
_{j}^{\prime },\mathbf{t}_{j}^{\prime }\right) ,S_{i}^{2},S_{j}^{2}\right) \\
&=&\int I\left( \left( \mathbf{\alpha }_{i},\mathbf{p}_{i}\right) ,\left( 
\mathbf{\alpha }_{i}^{\prime },\mathbf{p}_{i}^{\prime }\right) ,\left( 
\mathbf{\alpha }_{0},\mathbf{p}_{0}\right) ,S_{i}^{2},S_{0}^{2}\right) \\
&&\times G\left( \left( \mathbf{\alpha }_{0},\mathbf{p}_{0}\right) ,\left( 
\mathbf{\alpha }_{0}^{\prime },\mathbf{p}_{0}^{\prime }\right) \right)
I\left( \left( \mathbf{\beta }_{j},\mathbf{t}_{j}\right) ,\left( \mathbf{%
\beta }_{j}^{\prime },\mathbf{t}_{j}^{\prime }\right) ,\left( \mathbf{\alpha 
}_{0}^{\prime },\mathbf{p}_{0}^{\prime }\right) ,S_{j}^{2},S_{0}^{2}\right)
d\left( \mathbf{\alpha }_{0},\mathbf{p}_{0}\right) d\left( \mathbf{\alpha }%
_{0}^{\prime },\mathbf{p}_{0}^{\prime }\right)
\end{eqnarray*}%
so that the resulting interaction binds the structures.

\subsection{Operator formalism}

To be more precise, we write the systm in terms of operators:

\begin{eqnarray*}
S &=&\sum_{i=0,1,2}\sqrt{\left( \mathbf{\bar{D}}_{S_{i}^{2}}\right) ^{2}}%
\left( \mathbf{A}^{+}\left( \mathbf{\alpha }_{i},\mathbf{p}%
_{i},S_{i}^{2}\right) \mathbf{A}^{-}\left( \mathbf{\alpha }_{i},\mathbf{p}%
_{i},S_{j}^{2}\right) +\frac{1}{2}\right) \\
&&+\sum_{i=1,2}I\left( \left( \mathbf{\alpha }_{i},\mathbf{p}_{i}\right)
,\left( \mathbf{\alpha }_{i}^{\prime },\mathbf{p}_{i}^{\prime }\right)
,\left( \mathbf{\alpha }_{0},\mathbf{p}_{0}\right)
,S_{i}^{2},S_{0}^{2}\right) \mathbf{A}^{+}\left( \mathbf{\alpha }_{i},%
\mathbf{p}_{i},S_{i}^{2}\right) \mathbf{A}^{-}\left( \mathbf{\alpha }%
_{i}^{\prime },\mathbf{p}_{i}^{\prime },S_{i}^{2}\right) \\
&&\times \left( \mathbf{A}^{-}\left( \mathbf{\alpha }_{0},\mathbf{p}%
_{0},S_{0}^{2}\right) +\mathbf{A}^{+}\left( \mathbf{\alpha }_{0},\mathbf{p}%
_{0},S_{0}^{2}\right) \right)
\end{eqnarray*}%
We will integrate over degrees of freedom of $\underline{\Gamma _{0}}\left( 
\mathbf{T}_{0},\mathbf{\alpha }_{0},\mathbf{p}_{0},S_{0}^{2}\right) $
through a transformation canceling these degrees of freedom.

As explained before, we perform the transformation:%
\begin{equation*}
\exp \left( -F\right) S\exp \left( F\right)
\end{equation*}%
such that:%
\begin{equation*}
I+\left[ F,S_{0}\right] =0
\end{equation*}%
We show in appendix 4 that using (\ref{CMT}),the operator $F$ is given by:%
\begin{eqnarray}
F=- &&\frac{I\left( \left( \mathbf{\alpha }_{i},\mathbf{p}_{i}\right)
,\left( \mathbf{\alpha }_{i}^{\prime },\mathbf{p}_{i}^{\prime }\right)
,\left( \mathbf{\alpha }_{0},\mathbf{p}_{0}\right)
,S_{i}^{2},S_{0}^{2}\right) }{\left( \sqrt{\left( \mathbf{\bar{D}}_{\mathbf{%
\alpha }_{i},\mathbf{p}_{i},S_{i}^{2}}\right) ^{2}}-\sqrt{\left( \mathbf{%
\bar{D}}_{\mathbf{\alpha }_{i}^{\prime },\mathbf{p}_{i}^{\prime
},S_{i}^{2}}\right) ^{2}}\right) ^{2}-\left( \mathbf{\bar{D}}_{\mathbf{%
\alpha }_{0},\mathbf{p}_{0},S_{0}^{2}}\right) ^{2}}  \label{RSL} \\
&&\times \left( \left( \sqrt{\left( \mathbf{\bar{D}}_{\mathbf{\alpha }_{i},%
\mathbf{p}_{i},S_{i}^{2}}\right) ^{2}}-\sqrt{\left( \mathbf{\bar{D}}_{%
\mathbf{\alpha }_{i}^{\prime },\mathbf{p}_{i}^{\prime },S_{i}^{2}}\right)
^{2}}+\sqrt{\left( \mathbf{\bar{D}}_{\mathbf{\alpha }_{0},\mathbf{p}%
_{0},S_{0}^{2}}\right) ^{2}}\right) \mathbf{A}^{-}\left( \mathbf{\alpha }%
_{0},\mathbf{p}_{0},S_{0}^{2}\right) \right.  \notag \\
&&\left. +\left( \sqrt{\left( \mathbf{\bar{D}}_{\mathbf{\alpha }_{i},\mathbf{%
p}_{i},S_{i}^{2}}\right) ^{2}}-\sqrt{\left( \mathbf{\bar{D}}_{\mathbf{\alpha 
}_{i}^{\prime },\mathbf{p}_{i}^{\prime },S_{i}^{2}}\right) ^{2}}-\sqrt{%
\left( \mathbf{\bar{D}}_{\mathbf{\alpha }_{0},\mathbf{p}_{0},S_{0}^{2}}%
\right) ^{2}}\right) \mathbf{A}^{+-}\left( \mathbf{\alpha }_{0},\mathbf{p}%
_{0},S_{0}^{2}\right) \right)  \notag
\end{eqnarray}%
Ultimately, we obtain:%
\begin{equation*}
S^{\prime }=S_{0}+\frac{1}{2}\left[ I,F\right]
\end{equation*}%
with:%
\begin{eqnarray*}
\frac{1}{2}\left[ I,F\right] &=&-\frac{1}{2}\sum_{\left( i,j\right) \in
\left\{ 1,2\right\} ^{2}}\frac{1}{2}\Delta \left( \left( \mathbf{\alpha }%
_{i},\mathbf{p}_{i}\right) ,\left( \mathbf{\alpha }_{i}^{\prime },\mathbf{p}%
_{i}^{\prime }\right) ,\left( \mathbf{\alpha }_{0},\mathbf{p}_{0}\right)
,\left( \mathbf{\beta }_{j},\mathbf{t}_{j}\right) ,\left( \mathbf{\beta }%
_{j}^{\prime },\mathbf{t}_{j}^{\prime }\right)
,S_{i}^{2},S_{i}^{2},S_{0}^{2}\right) \\
&&\times I\left( \left( \mathbf{\alpha }_{i},\mathbf{p}_{i}\right) ,\left( 
\mathbf{\alpha }_{i}^{\prime },\mathbf{p}_{i}^{\prime }\right) ,\left( 
\mathbf{\alpha }_{0},\mathbf{p}_{0}\right) ,S_{i}^{2},S_{0}^{2}\right)
I\left( \left( \mathbf{\beta }_{j},\mathbf{t}_{j}\right) ,\left( \mathbf{%
\beta }_{j}^{\prime },\mathbf{t}_{j}^{\prime }\right) ,\left( \mathbf{\alpha 
}_{0},\mathbf{p}_{0}\right) ,S_{j}^{2},S_{0}^{2}\right) \\
&&\times \mathbf{A}^{+}\left( \mathbf{\alpha }_{i},\mathbf{p}%
_{i},S_{i}^{2}\right) \mathbf{A}^{-}\left( \mathbf{\alpha }_{i}^{\prime },%
\mathbf{p}_{i}^{\prime },S_{i}^{2}\right) \mathbf{A}^{+}\left( \mathbf{\beta 
}_{j},\mathbf{t}_{j},S_{j}^{2}\right) \mathbf{A}^{-}\left( \mathbf{\beta }%
_{j}^{\prime },\mathbf{t}_{j}^{\prime },S_{j}^{2}\right) \\
&=&\hat{I}\left( \left( \mathbf{\alpha }_{i},\mathbf{p}_{i}\right) ,\left( 
\mathbf{\alpha }_{i}^{\prime },\mathbf{p}_{i}^{\prime }\right) ,\left( 
\mathbf{\alpha }_{0},\mathbf{p}_{0}\right) ,S_{i}^{2},S_{0}^{2}\right)
I\left( \left( \mathbf{\beta }_{j},\mathbf{t}_{j}\right) ,\left( \mathbf{%
\beta }_{j}^{\prime },\mathbf{t}_{j}^{\prime }\right) ,\left( \mathbf{\alpha 
}_{0},\mathbf{p}_{0}\right) ,S_{j}^{2},S_{0}^{2}\right) \\
&&\times \mathbf{A}^{+}\left( \mathbf{\alpha }_{i},\mathbf{p}%
_{i},S_{i}^{2}\right) \mathbf{A}^{-}\left( \mathbf{\alpha }_{i}^{\prime },%
\mathbf{p}_{i}^{\prime },S_{i}^{2}\right) \mathbf{A}^{+}\left( \mathbf{\beta 
}_{j},\mathbf{t}_{j},S_{j}^{2}\right) \mathbf{A}^{-}\left( \mathbf{\beta }%
_{j}^{\prime },\mathbf{t}_{j}^{\prime },S_{j}^{2}\right)
\end{eqnarray*}%
and where:%
\begin{eqnarray*}
&&\Delta \left( \left( \mathbf{\alpha }_{i},\mathbf{p}_{i}\right) ,\left( 
\mathbf{\alpha }_{i}^{\prime },\mathbf{p}_{i}^{\prime }\right) ,\left( 
\mathbf{\alpha }_{0},\mathbf{p}_{0}\right) ,\left( \mathbf{\beta }_{j},%
\mathbf{t}_{j}\right) ,\left( \mathbf{\beta }_{j}^{\prime },\mathbf{t}%
_{j}^{\prime }\right) ,S_{i}^{2},S_{i}^{2},S_{0}^{2}\right) \\
&=&\frac{\sqrt{\left( \mathbf{\bar{D}}_{\mathbf{\alpha }_{0},\mathbf{p}%
_{0},S_{0}^{2}}\right) ^{2}}}{\left( \sqrt{\left( \mathbf{\bar{D}}_{\mathbf{%
\alpha }_{i},\mathbf{p}_{i},S_{i}^{2}}\right) ^{2}}-\sqrt{\left( \mathbf{%
\bar{D}}_{\mathbf{\alpha }_{i}^{\prime },\mathbf{p}_{i}^{\prime
},S_{i}^{2}}\right) ^{2}}\right) ^{2}-\left( \mathbf{\bar{D}}_{\mathbf{%
\alpha }_{0},\mathbf{p}_{0},S_{0}^{2}}\right) ^{2}} \\
&&+\frac{\sqrt{\left( \mathbf{\bar{D}}_{\mathbf{\alpha }_{0},\mathbf{p}%
_{0},S_{0}^{2}}\right) ^{2}}}{\left( \sqrt{\left( \mathbf{\bar{D}}_{\mathbf{%
\beta }_{j},\mathbf{t}_{j},S_{j}^{2}}\right) ^{2}}-\sqrt{\left( \mathbf{\bar{%
D}}_{\mathbf{\beta }_{j}^{\prime },\mathbf{t}_{j}^{\prime
},S_{j}^{2}}\right) ^{2}}\right) ^{2}-\left( \mathbf{\bar{D}}_{\mathbf{%
\alpha }_{0},\mathbf{p}_{0},S_{0}^{2}}\right) ^{2}}
\end{eqnarray*}

As a consequence, two states $\left( \mathbf{\alpha }_{i},\mathbf{p}%
_{i},S_{i}^{2}\right) $ and $\left( \mathbf{\beta }_{j},\mathbf{t}%
_{j},S_{j}^{2}\right) $ that where independent become bound through $\left( 
\mathbf{\alpha }_{0},\mathbf{p}_{0},S_{0}^{2}\right) $.

Ultimately, the effective operator writes:%
\begin{eqnarray*}
S_{f}^{O} &=&\sum_{i=1,2}\sqrt{\left( \mathbf{\bar{D}}_{S_{i}^{2}}\right)
^{2}}\left( \mathbf{A}^{+}\left( \mathbf{\alpha }_{i},\mathbf{p}%
_{i},S_{i}^{2}\right) \mathbf{A}^{-}\left( \mathbf{\alpha }_{i},\mathbf{p}%
_{i},S_{j}^{2}\right) +\frac{1}{2}\right) \\
&&-\frac{1}{2}\sum_{\left( i,j\right) \in \left\{ 1,2\right\} ^{2}}\frac{1}{2%
}\Delta \left( \left( \mathbf{\alpha }_{i},\mathbf{p}_{i}\right) ,\left( 
\mathbf{\alpha }_{i}^{\prime },\mathbf{p}_{i}^{\prime }\right) ,\left( 
\mathbf{\alpha }_{0},\mathbf{p}_{0}\right) ,\left( \mathbf{\beta }_{j},%
\mathbf{t}_{j}\right) ,\left( \mathbf{\beta }_{j}^{\prime },\mathbf{t}%
_{j}^{\prime }\right) ,S_{i}^{2},S_{i}^{2},S_{0}^{2}\right) \\
&&\times I\left( \left( \mathbf{\alpha }_{i},\mathbf{p}_{i}\right) ,\left( 
\mathbf{\alpha }_{i}^{\prime },\mathbf{p}_{i}^{\prime }\right) ,\left( 
\mathbf{\alpha }_{0},\mathbf{p}_{0}\right) ,S_{i}^{2},S_{0}^{2}\right)
I\left( \left( \mathbf{\beta }_{j},\mathbf{t}_{j}\right) ,\left( \mathbf{%
\beta }_{j}^{\prime },\mathbf{t}_{j}^{\prime }\right) ,\left( \mathbf{\alpha 
}_{0},\mathbf{p}_{0}\right) ,S_{j}^{2},S_{0}^{2}\right) \\
&&\times \mathbf{A}^{+}\left( \mathbf{\alpha }_{i},\mathbf{p}%
_{i},S_{i}^{2}\right) \mathbf{A}^{-}\left( \mathbf{\alpha }_{i}^{\prime },%
\mathbf{p}_{i}^{\prime },S_{i}^{2}\right) \mathbf{A}^{+}\left( \mathbf{\beta 
}_{j},\mathbf{t}_{j},S_{j}^{2}\right) \mathbf{A}^{-}\left( \mathbf{\beta }%
_{j}^{\prime },\mathbf{t}_{j}^{\prime },S_{j}^{2}\right)
\end{eqnarray*}

\section{Extension: non localized structures}

\subsection{Describing structures with variable spatial extension}

We have considered the activities and activities' oscillations of the states
as endogeneous, given that they depend solely on the connectivities of the
states. Similarly, the spatial extension of the structures is considered as
fixed in first approximation. However, we may assume that depending on
background, some connections may break and others may be created while
maintaining the same properties of the state. This corresponds to describe
states experiencing some displacement $S^{2}\rightarrow S^{2}+\delta S^{2}$.

This possibility can be taken into account in the initial formultion (\ref%
{CST}), where collective states have the form:%
\begin{equation}
\prod\limits_{Z,Z^{\prime }}\left\vert \Delta T\left( Z,Z^{\prime }\right)
,\Delta \hat{T}\left( Z,Z^{\prime }\right) ,\alpha \left( Z,Z^{\prime
}\right) ,p\left( Z,Z^{\prime }\right) \right\rangle \equiv \left\vert 
\mathbf{\alpha },\mathbf{p},S^{2}\right\rangle
\end{equation}%
which implies that switching from $S^{2}\rightarrow S^{2}+\delta S^{2}$,
amounts to replace formally:%
\begin{eqnarray*}
&&\prod\limits_{\left( Z,Z^{\prime }\right) \in S^{2}}\left\vert \Delta
T\left( Z,Z^{\prime }\right) ,\Delta \hat{T}\left( Z,Z^{\prime }\right)
,\alpha \left( Z,Z^{\prime }\right) ,p\left( Z,Z^{\prime }\right)
\right\rangle \\
&\rightarrow &\prod\limits_{\left( Z,Z^{\prime }\right) \in S^{2}}\left\vert
\Delta T\left( Z,Z^{\prime }\right) ,\Delta \hat{T}\left( Z,Z^{\prime
}\right) ,\alpha \left( Z,Z^{\prime }\right) ,p\left( Z,Z^{\prime }\right)
\right\rangle \\
&&\times \prod\limits_{\left( Z,Z^{\prime }\right) \in S^{2}+\delta
S^{2}/S^{2}}\left\vert \Delta T\left( Z,Z^{\prime }\right) ,\Delta \hat{T}%
\left( Z,Z^{\prime }\right) ,\alpha \left( Z,Z^{\prime }\right) ,p\left(
Z,Z^{\prime }\right) \right\rangle \\
&&\times \prod\limits_{\left( Z,Z^{\prime }\right) \in S^{2}/S^{2}+\delta
S^{2}}\left\vert \Delta T\left( Z,Z^{\prime }\right) ,\Delta \hat{T}\left(
Z,Z^{\prime }\right) ,\alpha \left( Z,Z^{\prime }\right) ,p\left(
Z,Z^{\prime }\right) \right\rangle ^{-1}
\end{eqnarray*}%
Practically, the power $-1$, amounts to divide $\left\vert \mathbf{\alpha },%
\mathbf{p},S^{2}\right\rangle $ by a factor (\ref{SF}) for each point $%
\left( Z,Z^{\prime }\right) \in S^{2}/S^{2}+\delta S^{2}$. In terms of
operator formalism, this translates by considering two different structure
with spatial extension $S^{2}$ and $S^{2}+\delta S^{2}$, the switching $%
S^{2}\rightarrow S^{2}+\delta S^{2}$ being described by an interaction term
between the different structures whose form is:%
\begin{equation*}
\int I\left( \mathbf{\alpha }^{\prime },\mathbf{p}^{\prime },S^{2},\mathbf{%
\alpha }^{\prime },\mathbf{p}^{\prime },S^{2}+\delta S^{2}\right) \mathbf{A}%
^{+}\left( \mathbf{\alpha }^{\prime },\mathbf{p}^{\prime },S^{2}+\delta
S^{2}\right) \mathbf{A}^{-}\left( \mathbf{\alpha },\mathbf{p},S^{2}\right)
\end{equation*}%
However, to model the possibility of some permanence in the structures
characteristics, independently of any localization, we rather consider the
dynamic for the same structure described by the states $\left\vert \mathbf{%
\alpha },\mathbf{p},S^{2}\right\rangle $ with $S^{2}$ varying. To do so we
will consider the fields $\underline{\Gamma }\left( \mathbf{T},\mathbf{%
\alpha },\mathbf{p},S^{2}\right) $ with $S^{2}$ seen as a dynamic variable.

\subsection{Fild action with variable spatial extension}

We write the following action by considering the frequencies and position as
a dynamic variable:%
\begin{eqnarray*}
S &=&\sum_{S}\underline{\Gamma }^{\dag }\left( \mathbf{T},\mathbf{\alpha },%
\mathbf{p},S^{2}\right) \left( -\frac{1}{2}\nabla _{\Upsilon _{S^{2}}}^{2}-%
\frac{1}{2}\frac{\delta ^{2}}{\delta \left( S^{2}\right) ^{2}}+V\left(
S^{2}\right) +U\left( \nabla S^{2}\right) \right) \underline{\Gamma }\left( 
\mathbf{T},\mathbf{\alpha },\mathbf{p},S^{2}\right) \\
&&+\sum_{S}\sum_{\delta S^{2}}\underline{\Gamma }^{\dag }\left( \mathbf{T},%
\mathbf{\alpha }+\delta \mathbf{\alpha },\mathbf{p+}\delta \mathbf{p}%
,S^{2}+\delta S^{2}\right) U\left( S^{2},S^{2}+\delta S^{2}\right) 
\underline{\Gamma }\left( \mathbf{T},\mathbf{\alpha },\mathbf{p},S^{2}\right)
\\
&&+\sum_{S}\underline{\Gamma }^{\dag }\left( \mathbf{T},\mathbf{\alpha },%
\mathbf{p},S^{2}\right) \left( -\frac{1}{2}\nabla _{\left( \mathbf{T}\right)
_{S^{2}}}^{2}+\frac{1}{2}\left( \Delta \mathbf{T}_{p}^{\alpha }\right)
^{t}\left( \mathbf{A}_{p}^{\alpha }\right) _{S^{2}}\Delta \mathbf{T}%
_{p}^{\alpha }+\mathbf{C}\right) \underline{\Gamma }\left( \mathbf{T},%
\mathbf{\alpha },\mathbf{p},S^{2}\right) \\
&&+\sum_{S_{k},S_{l}}\prod \underline{\Gamma }^{\dag }\left( \mathbf{T}%
_{l}^{\prime },\mathbf{\alpha }_{l}^{\prime },\mathbf{p}_{l}^{\prime
},S_{l}^{\prime 2}\right) V\left( \left\{ \left\vert \mathbf{T}^{\prime },%
\mathbf{\alpha }_{l}^{\prime },\mathbf{p}_{l}^{\prime },S_{l}^{\prime
2}\right\rangle \right\} ,\left\{ \left\vert \mathbf{T},\mathbf{\alpha }_{k},%
\mathbf{p}_{k},S_{k}^{2}\right\rangle \right\} \right) \prod \underline{%
\Gamma }\left( \mathbf{T}_{k},\mathbf{\alpha }_{k},\mathbf{p}%
_{k},S_{k}^{2}\right)
\end{eqnarray*}%
with:%
\begin{equation*}
\frac{\delta }{\delta S^{2}}=\int_{S^{2}}d\left( Z,Z^{\prime }\right) \nabla
_{\left( Z,Z^{\prime }\right) }
\end{equation*}%
and:%
\begin{equation*}
\frac{\delta ^{2}}{\delta \left( S^{2}\right) ^{2}}=\int_{S^{2}}d\left(
Z,Z^{\prime }\right) \nabla _{\left( Z,Z^{\prime }\right) }^{2}
\end{equation*}%
The potential $V\left( S^{2}\right) $ should depend on the background on
which the structure emerged. It conditions the possible displacement of the
structures. \ The potential:%
\begin{equation}
I=\sum_{\delta S^{2}}\underline{\Gamma }^{\dag }\left( \mathbf{T},\mathbf{%
\alpha }+\delta \mathbf{\alpha },\mathbf{p+}\delta \mathbf{p},S^{2}+\delta
S^{2}\right) U\left( S^{2},S^{2}+\delta S^{2}\right) \underline{\Gamma }%
\left( \mathbf{T},\mathbf{\alpha },\mathbf{p},S^{2}\right)  \label{PTL}
\end{equation}%
should encompass the structural change arising while displacing the content
from one spatial zone to anothr.

Appendix 5 shows that in first approximation we can write the potential:%
\begin{eqnarray*}
I &=&\int_{S^{2}}\bar{U}\left( S^{2},Z,Z^{\prime }\right) \nabla _{\left(
Z,Z^{\prime }\right) }\underline{\Gamma }\left( \mathbf{T},\mathbf{\alpha },%
\mathbf{p},S^{2}\right) \\
&\equiv &-\bar{U}\left( S^{2}\right) \frac{\delta }{\delta S^{2}}\underline{%
\Gamma }\left( \mathbf{T},\mathbf{\alpha },\mathbf{p},S^{2}\right)
\end{eqnarray*}%
The expression for $\bar{U}\left( S^{2},Z,Z^{\prime }\right) $ is given in
this appendix.

The Green function of:%
\begin{equation}
-\frac{1}{2}\nabla _{\Upsilon _{S^{2}}}^{2}-\frac{1}{2}\frac{\delta ^{2}}{%
\delta \left( S^{2}\right) ^{2}}-\bar{U}\left( S^{2}\right) \frac{\delta }{%
\delta S^{2}}+V\left( S^{2}\right) +U_{\Upsilon }  \label{DPRT}
\end{equation}%
should describe the motion of the structure along the whole thread, without
breaking the main characteristics of this structure. Since this has to be
realized at each points of the trajectory, the green function should itself
be written as some path integral.

For the displacement to happen without modifying the structures content,
some type of\ topological content should arise.

Note that such motion along gradient of potential should reproduce
continuously what would be obtained by a sequence of deactivations and
activations of structures (see discussion after (\ref{GRK})).

We will neglect the activities variations and only consider fluctuations in
connectivities and coordinates. The "free" dynamics for single structures is
given by the operator (\ref{DPRT}). Moreover, to study the intrinsic
displacement we neglect $V\left( S^{2}\right) $ and the structures
interaction and we should only consider the systm described by:%
\begin{equation}
-\frac{1}{2}\frac{\delta ^{2}}{\delta \left( S^{2}\right) ^{2}}-\bar{U}%
\left( S^{2}\right) \frac{\delta }{\delta S^{2}}-\frac{1}{2}\nabla _{\left( 
\mathbf{T}\right) _{S^{2}}}^{2}+\frac{1}{2}\left( \Delta \mathbf{T}%
_{p}^{\alpha }\right) ^{t}\left( \mathbf{A}_{p}^{\alpha }\right)
_{S^{2}}\Delta \mathbf{T}_{p}^{\alpha }+\mathbf{C}  \label{CVS}
\end{equation}%
The global term:%
\begin{equation*}
-\frac{1}{2}\frac{\delta ^{2}}{\delta \left( S^{2}\right) ^{2}}+\bar{U}%
\left( S^{2}\right)
\end{equation*}%
should favour displacement from $S^{2}$ to $S^{2}+\delta S^{2}$ such that
the average connections $\mathbf{\bar{T}}_{p}^{\alpha }\left( Z,Z^{\prime
}\right) $ are overall decreasing along the displacement. This corresponds
to a loss of content. Moreover this modification of $\mathbf{\bar{T}}%
_{p}^{\alpha }\left( Z,Z^{\prime }\right) $ may induce a transition of a
state $\left\vert \mathbf{T},\mathbf{\alpha },\mathbf{p},S^{2}\right\rangle $
towards $\left\vert \mathbf{T},\mathbf{\alpha }^{\prime },\mathbf{p}^{\prime
},S^{2}\right\rangle $ wth $\mathbf{\bar{T}}_{p^{\prime }}^{\alpha ^{\prime
}}\left( Z,Z^{\prime }\right) $ closer to the modfied $\mathbf{\bar{T}}%
_{p}^{\alpha }\left( Z,Z^{\prime }\right) $.

However, the potential:%
\begin{equation*}
\frac{1}{2}\left( \Delta \mathbf{T}_{p}^{\alpha }\right) ^{t}\left( \mathbf{A%
}_{p}^{\alpha }\right) _{S^{2}}\Delta \mathbf{T}_{p}^{\alpha }+\mathbf{C}
\end{equation*}%
may avoid this tendency. The first term $\frac{1}{2}\left( \Delta \mathbf{T}%
_{p}^{\alpha }\right) ^{t}\left( \mathbf{A}_{p}^{\alpha }\right)
_{S^{2}}\Delta \mathbf{T}_{p}^{\alpha }$ is quadratic in $\Delta \mathbf{T}%
_{p}^{\alpha }$. It increases if $\mathbf{\bar{T}}_{p}^{\alpha }\left(
Z,Z^{\prime }\right) $ decreases. At least it slows down the decrease in $%
\mathbf{\bar{T}}_{p}^{\alpha }\left( Z,Z^{\prime }\right) $. The second term
is a function of $S^{2}$:%
\begin{equation*}
\mathbf{C=}\int_{S^{2}}C\left( Z,Z^{\prime }\right) d\left( Z,Z^{\prime
}\right)
\end{equation*}%
so that motion should be favoured in a direction of a decreasing $\mathbf{C}$%
. We write $\mathbf{C}\left( S^{2}\right) $ for $\mathbf{C}$ to account for
this dependency.

The value of $\mathbf{C}\left( S^{2}\right) $ is given by (\ref{LC}):%
\begin{equation}
\mathbf{C}\left( S^{2}\right) \mathbf{=}\int_{S^{2}}\left( \frac{\tau \omega
_{0}\left( Z\right) }{2}+\frac{\rho \left( C\left( \theta \right) \left\vert
\Psi _{0}\left( Z\right) \right\vert ^{2}\omega _{0}\left( Z\right) +D\left(
\theta \right) \left\vert \Psi _{0}\left( Z^{\prime }\right) \right\vert
^{2}\omega _{0}\left( Z^{\prime }\right) \right) }{2\omega _{0}\left(
Z\right) }\right) d\left( Z,Z^{\prime }\right)
\end{equation}%
so that switching to region with lower average activity is favoured.

\subsection{Green functions}

Action (\ref{CVS}) modifies the Green functions (\ref{GRK}). The computation
is presented in appndx 5. We obtain in first approximation:%
\begin{eqnarray}
G\left( S^{\prime 2},\mathbf{T}_{p}^{\prime \alpha },S^{2},\mathbf{T}%
_{p}^{\alpha }\right) &\simeq &\exp \left( \frac{\delta }{\delta S^{2}}\bar{U%
}\left( S^{2}\right) -\frac{\delta }{\delta S^{\prime 2}}\bar{U}\left(
S^{\prime 2}\right) \mathbf{-}\left( \mathbf{C}\left( S^{\prime 2}\right) -%
\mathbf{C}\left( S^{2}\right) \right) \right) \\
&&\frac{\exp \left( -\left( \left( \mathbf{Z,Z}^{\prime }\right) ^{\prime
}\right) ^{t}\left\langle \mathbf{A}\right\rangle _{S^{\prime 2}}\left( 
\mathbf{Z,Z}^{\prime }\right) ^{\prime }-\left( \mathbf{Z,Z}^{\prime
}\right) ^{t}\left\langle \mathbf{A}\right\rangle _{S^{2}}\left( \mathbf{Z,Z}%
^{\prime }\right) ^{\prime }\right) }{\left( \left\langle \mathbf{A}%
\right\rangle _{S^{\prime 2}}\left\langle \mathbf{A}\right\rangle
_{S^{2}}\right) ^{\frac{1}{4}}}  \notag \\
&&\times \frac{\exp \left( -\left( \mathbf{T}_{p}^{\prime \alpha }\right)
^{t}\mathbf{A}_{S^{\prime 2}}^{\alpha }\mathbf{T}_{p}^{\prime \alpha
}-\left( \mathbf{T}_{p}^{\alpha }\right) ^{t}\mathbf{A}_{S^{2}}^{\alpha }%
\mathbf{T}_{p}^{\alpha }\right) }{\sqrt{\det \left( \mathbf{A}%
_{S^{2}}^{\alpha }\right) }}  \notag
\end{eqnarray}%
where $\left( \mathbf{Z,Z}^{\prime }\right) ^{\prime }=S^{\prime 2}$ and $%
\left( \mathbf{Z,Z}^{\prime }\right) =S^{2}$ and $\left\langle \mathbf{A}%
\right\rangle _{S^{2}}$ is the average of $A\left( Z,Z^{\prime }\right) $ in 
$S^{2}$ and $\left\langle \mathbf{A}\right\rangle _{S^{\prime 2}}$ is the
avrage of $A\left( Z,Z^{\prime }\right) $ in $S^{\prime 2}$. The first
exponential factor is a trend term that confirms the arguments presented in
the previous paragraph: the potential:%
\begin{equation*}
\frac{\delta }{\delta S^{2}}\bar{U}\left( S^{2}\right) -\frac{\delta }{%
\delta S^{\prime 2}}\bar{U}\left( S^{\prime 2}\right)
\end{equation*}%
favours displacement towards lower connections, while the second term:%
\begin{equation*}
\mathbf{-}\left( \mathbf{C}\left( S^{\prime 2}\right) -\mathbf{C}\left(
S^{2}\right) \right)
\end{equation*}%
implies a higher probability of displacement towards lower activity region.

\subsection{Displacement induced transitions}

Assume localzid interactions of the form:%
\begin{equation*}
\underline{\Gamma }^{\dag }\left( \mathbf{T}_{l}^{\prime },\mathbf{\alpha }%
_{l}^{\prime },\mathbf{p}_{l}^{\prime },S_{l}^{\prime 2}\right) V\left(
\left\{ \left\vert \mathbf{T}^{\prime },\mathbf{\alpha }_{l}^{\prime },%
\mathbf{p}_{l}^{\prime },S_{l}^{\prime 2}\right\rangle \right\} ,\left\{
\left\vert \mathbf{T},\mathbf{\alpha }_{k},\mathbf{p}_{k},S_{k}^{2}\right%
\rangle \right\} \right) \prod \underline{\Gamma }\left( \mathbf{T}_{k},%
\mathbf{\alpha }_{k},\mathbf{p}_{k},S_{k}^{2}\right)
\end{equation*}%
where $V$ includes factors $\delta \left( S_{l}^{\prime 2}-S_{k}^{2}\right) $%
. Structures experiencing a displacement:%
\begin{equation*}
\left\vert \mathbf{T},\mathbf{\alpha },\mathbf{p},S^{2}\right\rangle
\rightarrow \left\vert \mathbf{T},\mathbf{\alpha },\mathbf{p},S_{l}^{\prime
2}\right\rangle
\end{equation*}%
may be modified to:%
\begin{equation*}
\left\vert \mathbf{T}_{l}^{\prime },\mathbf{\alpha }_{l}^{\prime },\mathbf{p}%
_{l}^{\prime },S_{l}^{\prime 2}\right\rangle
\end{equation*}%
with amplitude:%
\begin{equation*}
G\left( S_{l}^{\prime 2},\mathbf{T}_{p_{l}}^{\prime \alpha
_{l}},S_{l}^{\prime 2},\mathbf{T}_{p}^{\prime \alpha }\right) V\left(
\left\{ \left\vert \mathbf{T}_{p_{l}}^{\prime \alpha _{l}},S_{l}^{\prime
2}\right\rangle \right\} ,\left\{ \left\vert \mathbf{T}_{p}^{\prime \alpha
},S_{l}^{\prime 2}\right\rangle \right\} \right) G\left( S_{l}^{\prime 2},%
\mathbf{T}_{p}^{\prime \alpha },S^{2},\mathbf{T}_{p}^{\alpha }\right)
\end{equation*}

This amplitude includes a term:%
\begin{eqnarray*}
&&\frac{\exp \left( -\left( \mathbf{T}_{p_{l}}^{\prime \alpha _{l}}\right)
^{t}\mathbf{A}_{S_{l}^{\prime 2}}^{\alpha }\mathbf{T}_{p_{l}}^{\prime \alpha
_{l}}-\left( \mathbf{T}_{p}^{\prime \alpha }\right) ^{t}\mathbf{A}%
_{S_{l}^{\prime 2}}^{\alpha }\mathbf{T}_{p}^{\prime \alpha }\right) }{\sqrt{%
\det \left( \mathbf{A}_{S^{2}}^{\alpha }\right) }} \\
&&\times \exp \left( \frac{\delta }{\delta S^{2}}\bar{U}\left( S^{2}\right) -%
\frac{\delta }{\delta S^{\prime 2}}\bar{U}\left( S_{l}^{\prime 2}\right) 
\mathbf{-}\left( \mathbf{C}\left( S_{l}^{\prime 2}\right) -\mathbf{C}\left(
S^{2}\right) \right) \right) \frac{\exp \left( -\left( \mathbf{T}%
_{p}^{\prime \alpha }\right) ^{t}\mathbf{A}_{S_{l}^{\prime 2}}^{\alpha }%
\mathbf{T}_{p}^{\prime \alpha }-\left( \mathbf{T}_{p}^{\alpha }\right) ^{t}%
\mathbf{A}_{S^{2}}^{\alpha }\mathbf{T}_{p}^{\alpha }\right) }{\sqrt{\det
\left( \mathbf{A}_{S^{2}}^{\alpha }\right) }}
\end{eqnarray*}%
which, after convolution becomes:%
\begin{eqnarray*}
&&\frac{\exp \left( -\left( \mathbf{T}_{p_{l}}^{\prime \alpha _{l}}\right)
^{t}\mathbf{A}_{S_{l}^{\prime 2}}^{\alpha }\mathbf{T}_{p_{l}}^{\prime \alpha
_{l}}\right) }{\sqrt{\det \left( \mathbf{A}_{S^{2}}^{\alpha }\right) }}\hat{V%
}\left( \left\{ \left\vert \mathbf{T}_{p_{l}}^{\prime \alpha
_{l}},S_{l}^{\prime 2}\right\rangle \right\} \right) \\
&&\times \exp \left( \frac{\delta }{\delta S^{2}}\bar{U}\left( S^{2}\right) -%
\frac{\delta }{\delta S^{\prime 2}}\bar{U}\left( S_{l}^{\prime 2}\right) 
\mathbf{-}\left( \mathbf{C}\left( S_{l}^{\prime 2}\right) -\mathbf{C}\left(
S^{2}\right) \right) \right) \frac{\exp \left( -\left( \mathbf{T}%
_{p}^{\alpha }\right) ^{t}\mathbf{A}_{S^{2}}^{\alpha }\mathbf{T}_{p}^{\alpha
}\right) }{\sqrt{\det \left( \mathbf{A}_{S^{2}}^{\alpha }\right) }}
\end{eqnarray*}%
with:%
\begin{equation*}
\hat{V}\left( \left\{ \left\vert \mathbf{T}_{p_{l}}^{\prime \alpha
_{l}},S_{l}^{\prime 2}\right\rangle \right\} \right) =\int V\left( \left\{
\left\vert \mathbf{T}_{p_{l}}^{\prime \alpha _{l}},S_{l}^{\prime
2}\right\rangle \right\} ,\left\{ \left\vert \mathbf{T}_{p}^{\prime \alpha
},S_{l}^{\prime 2}\right\rangle \right\} \right) \exp \left( -2\left( 
\mathbf{T}_{p}^{\prime \alpha }\right) ^{t}\mathbf{A}_{S_{l}^{\prime
2}}^{\alpha }\mathbf{T}_{p}^{\prime \alpha }\right) d\mathbf{T}%
_{p_{l}}^{\prime \alpha _{l}}
\end{equation*}%
If $V$ is decreasing function of $\left\vert \mathbf{\bar{T}}%
_{p_{l}}^{\prime \alpha _{l}}-\mathbf{\bar{T}}_{p}^{\prime \alpha
}\right\vert =\left\vert \mathbf{\bar{T}}_{p_{l}}^{\prime \alpha _{l}}-%
\mathbf{T}_{p}^{\prime \alpha }\right\vert $, transition is possible under
two conditions:%
\begin{equation*}
\frac{\delta }{\delta S^{2}}\bar{U}\left( S^{2}\right) -\frac{\delta }{%
\delta S^{\prime 2}}\bar{U}\left( S_{l}^{\prime 2}\right) \mathbf{-}\left( 
\mathbf{C}\left( S_{l}^{\prime 2}\right) -\mathbf{C}\left( S^{2}\right)
\right) >0
\end{equation*}%
and:%
\begin{equation*}
\left\vert \mathbf{\bar{T}}_{p_{l}}^{\prime \alpha _{l}}-\mathbf{T}%
_{p}^{\prime \alpha }\right\vert \simeq 0
\end{equation*}%
That is, transition occurs for motion towards relatively low actvity region
with similar level of connectivity.

\section{Conclusion}

The final section of the present work has extended our formalism to
encompass a field-based representation for non-localized interacting
structures. Activated states can undergo shifts from one region to another,
the driving forces behind such displacements being the interactions. The
system's background state, conditions these displacements and functions as a
landscape. In our formalism, we have primarily focused on activity and
frequencies as the key elements. Consequently, the spatial extension of a
state should be determined by these variables, and the spatial extent of a
state remains either undefined or, at the very least, variable. Exploring
these possibilities is a topic for future research.

\section*{Appendix 1}

\subsection*{1.1 Full system action}

In addition to the neurons activity field action, we add the functionals
that describe the dynamics for connectivity dynamics (see (\cite{GLt})). 
\begin{equation}
S_{\Gamma }^{\left( 1\right) }=\int \Gamma ^{\dag }\left( T,\hat{T},\theta
,Z,Z^{\prime },C,D\right) \nabla _{T}\left( \frac{\sigma _{T}^{2}}{2}\nabla
_{T}+O_{T}\right) \Gamma \left( T,\hat{T},\theta ,Z,Z^{\prime },C,D\right)
\label{wGT}
\end{equation}%
\begin{equation}
S_{\Gamma }^{\left( 2\right) }=\int \Gamma ^{\dag }\left( T,\hat{T},\theta
,Z,Z^{\prime },C,D\right) \nabla _{\hat{T}}\left( \frac{\sigma _{\hat{T}}^{2}%
}{2}\nabla _{\hat{T}}+O_{\hat{T}}\right) \Gamma \left( T,\hat{T},\theta
,Z,Z^{\prime },C,D\right)  \label{WGT}
\end{equation}%
\begin{equation*}
S_{\Gamma }^{\left( 3\right) }=\Gamma ^{\dag }\left( T,\hat{T},\theta
,Z,Z^{\prime },C,D\right) \nabla _{C}\left( \frac{\sigma _{C}^{2}}{2}\nabla
_{C}+O_{C}\right) \Gamma \left( T,\hat{T},\theta ,Z,Z^{\prime },C,D\right)
\end{equation*}%
\begin{equation}
S_{\Gamma }^{\left( 4\right) }=\Gamma ^{\dag }\left( T,\hat{T},\theta
,Z,Z^{\prime },C,D\right) \nabla _{D}\left( \frac{\sigma _{D}^{2}}{2}\nabla
_{D}+O_{D}\right) \Gamma \left( T,\hat{T},\theta ,Z,Z^{\prime },C,D\right)
\label{wGC}
\end{equation}%
where:%
\begin{eqnarray}
O_{C} &=&\frac{C}{\tau _{C}\omega \left( J,\theta ,Z,\left\vert \Psi
\right\vert ^{2}\right) }-\frac{\alpha _{C}\left( 1-C\right) \omega \left(
J,\theta -\frac{\left\vert Z-Z^{\prime }\right\vert }{c},Z^{\prime
},\left\vert \Psi \right\vert ^{2}\right) \left\vert \Psi \left( \theta -%
\frac{\left\vert Z-Z^{\prime }\right\vert }{c},Z^{\prime }\right)
\right\vert ^{2}}{\omega \left( J,\theta ,Z,\left\vert \Psi \right\vert
^{2}\right) }  \label{DP} \\
O_{D} &=&\frac{D}{\tau _{D}\omega \left( J,\theta ,Z,\left\vert \Psi
\right\vert ^{2}\right) }-\alpha _{D}\left( 1-D\right) \left\vert \Psi
\left( \theta ,Z\right) \right\vert ^{2}  \notag \\
O_{\hat{T}} &=&-\frac{\rho }{\omega \left( J,\theta ,Z,\left\vert \Psi
\right\vert ^{2}\right) }\left( \left( h\left( Z,Z^{\prime }\right) -\hat{T}%
\right) C\left\vert \Psi \left( \theta ,Z\right) \right\vert ^{2}h_{C}\left(
\omega \left( J,\theta ,Z,\left\vert \Psi \right\vert ^{2}\right) \right)
\right.  \notag \\
&&\left. -D\hat{T}\left\vert \Psi \left( \theta -\frac{\left\vert
Z-Z^{\prime }\right\vert }{c},Z^{\prime }\right) \right\vert ^{2}h_{D}\left(
\omega \left( J,\theta -\frac{\left\vert Z-Z^{\prime }\right\vert }{c}%
,Z^{\prime },\left\vert \Psi \right\vert ^{2}\right) \right) \right)  \notag
\\
O_{T} &=&-\left( -\frac{1}{\tau \omega \left( J,\theta ,Z,\left\vert \Psi
\right\vert ^{2}\right) }T+\frac{\lambda }{\omega \left( J,\theta
,Z,\left\vert \Psi \right\vert ^{2}\right) }\hat{T}\right)  \notag
\end{eqnarray}

The functional $S_{\Gamma }^{\left( 1\right) }$ describes the
connectivities, while $S_{\Gamma }^{\left( i\right) }$ describe the
accumulation of incoming and outcoming current that influence the
connectivity dynamics.

\subsection*{1.2 Replacing activity field}

To find the background state for connectivity, we first derived in (\cite%
{GLr}) that activty can be in average replaced as functional of the
connectivities. The inverse activities satisfy:%
\begin{eqnarray}
&&\omega ^{-1}\left( J,\theta ,Z,\left\vert \Psi \right\vert ^{2}\right)  \\
&=&G\left( J\left( \theta ,Z\right) +\int \frac{\kappa }{N}\frac{\omega
\left( J,\theta -\frac{\left\vert Z-Z_{1}\right\vert }{c},Z_{1},\Psi \right)
T\left\vert \Gamma \left( T,\hat{T},\theta ,Z,Z_{1}\right) \right\vert ^{2}}{%
\omega \left( J,\theta ,Z,\left\vert \Psi \right\vert ^{2}\right) }%
\left\vert \Psi \left( \theta -\frac{\left\vert Z-Z_{1}\right\vert }{c}%
,Z_{1}\right) \right\vert ^{2}dZ_{1}\right)   \notag
\end{eqnarray}%
where we have assumed that the field $\left\vert \Psi \left( \theta
,Z\right) \right\vert ^{2}$ is constrained by a potential limiting the
activity around some average $\left\vert \Psi _{0}\left( Z\right)
\right\vert ^{2}$. We choose:%
\begin{equation*}
V=\frac{1}{2}\left( \left\vert \Psi \left( Z\right) \right\vert ^{2}-\int
T\left( Z^{\prime },Z_{1}\right) \left\vert \Psi _{0}\left( Z\right)
\right\vert ^{2}dZ_{1}\right) ^{2}
\end{equation*}%
We showed in (\cite{GLr})\ that in a static first approximation,
corresponding to an averaging over individual neurons signls time-scale:

\begin{equation}
\left\vert \Psi \left( Z\right) \right\vert ^{2}=\frac{2T\left( Z\right)
\left\langle \left\vert \Psi _{0}\left( Z^{\prime }\right) \right\vert
^{2}\right\rangle _{Z}}{\left( 1+\sqrt{1+4\left( \frac{\lambda \tau \nu
c-T\left( Z\right) }{\left( \frac{1}{\tau _{D}\alpha _{D}}+\frac{1}{\tau
_{C}\alpha _{C}}+\Omega \right) T\left( Z\right) -\frac{1}{\tau _{D}\alpha
_{D}}\lambda \tau \nu c}\right) ^{2}T\left( Z\right) \left\langle \left\vert
\Psi _{0}\left( Z^{\prime }\right) \right\vert ^{2}\right\rangle _{Z}}%
\right) }  \label{Ps}
\end{equation}%
which allows to derive the connectivities background states.

\subsection*{1.3 Full system background state}

Replacing $\omega ^{-1}\left( J,\theta ,Z,\left\vert \Psi \right\vert
^{2}\right) $ and $\left\vert \Psi \left( Z\right) \right\vert ^{2}$ allowed
to minimize $\sum_{i}S_{\Gamma }^{\left( 1\right) }$. The background state
for connectivity functions may have two forms, for activated connections or
unactivated ones (those with $\left\langle T\right\rangle =0$). In
quasi-static approximation, we find:

\begin{eqnarray}
&&\left\vert \Gamma \right\vert _{a}^{2}\left( T,\hat{T},\theta ,C,D\right) 
\label{gmv} \\
&\simeq &\left\{ \mathcal{N}\exp \left( -\frac{a_{C}\left( Z,Z^{\prime
}\right) }{2}\left( C-C\left( \theta \right) \right) ^{2}\right) \exp \left(
-\frac{a_{D}\left( Z,Z^{\prime }\right) }{2}\left( D-D\left( \theta \right)
\right) ^{2}\right) \right.   \notag \\
&&\times \exp \left( -\frac{\rho \left\vert \bar{\Psi}\left( \theta
,Z,Z^{\prime }\right) \right\vert ^{2}}{2}\left( \hat{T}-\left\langle \hat{T}%
\right\rangle \right) ^{2}\right)   \notag \\
&&\times \left. \left\Vert \Gamma _{0}\left( \theta ,Z,Z^{\prime }\right)
\right\Vert \exp \left( -\frac{\left\vert \Psi \left( \theta ,Z\right)
\right\vert ^{2}}{2\tau \omega }\left( T-\left\langle T\right\rangle \right)
^{2}\right) \right\} _{\left( Z,Z^{\prime }\right) ,\left\langle T\left(
Z,Z^{\prime }\right) \right\rangle \neq 0}  \notag
\end{eqnarray}%
\begin{eqnarray*}
&&\left\vert \Gamma \right\vert _{u}^{2}\left( T,\hat{T},\theta ,C,D\right) 
\\
&\simeq &\left\{ \mathcal{N}\exp \left( -\frac{a_{C}\left( Z,Z^{\prime
}\right) }{2}\left( C-C\left( \theta \right) \right) ^{2}\right) \exp \left(
-\frac{a_{D}\left( Z,Z^{\prime }\right) }{2}\left( D-\left\langle
D\right\rangle \right) ^{2}\right) \right.  \\
&&\left. \times \exp \left( -\frac{\rho \left\vert \bar{\Psi}\left( \theta
,Z,Z^{\prime }\right) \right\vert ^{2}}{2}\left( \hat{T}-\left\langle \hat{T}%
\right\rangle \right) ^{2}\right) \times \delta \left( T\right) \right\}
_{\left( Z,Z^{\prime }\right) ,\left\langle T\left( Z,Z^{\prime }\right)
\right\rangle \neq 0}
\end{eqnarray*}%
where $\mathcal{N}$ is a normalization factor ensuring that the constraint
over the number of connections is satisfied and where:%
\begin{eqnarray*}
\left\vert \bar{\Psi}\left( \theta ,Z,Z^{\prime }\right) \right\vert ^{2}
&=&C\left( \theta \right) \left\vert \Psi \left( \theta ,Z\right)
\right\vert ^{2}h_{C}+D\left( \theta \right) \left\vert \Psi \left( \theta -%
\frac{\left\vert Z-Z^{\prime }\right\vert }{c},Z^{\prime }\right)
\right\vert ^{2}h_{D} \\
a_{C}\left( Z,Z^{\prime }\right)  &=&\frac{1}{\tau _{C}\omega }+\alpha _{C}%
\frac{\omega ^{\prime }\left\vert \Psi \left( \theta -\frac{\left\vert
Z-Z^{\prime }\right\vert }{c},Z^{\prime }\right) \right\vert ^{2}}{\omega }
\\
a_{D}\left( Z,Z^{\prime }\right)  &=&\frac{1}{\tau _{D}\omega }+\alpha
_{D}\left\vert \Psi \left( \theta ,Z\right) \right\vert ^{2}
\end{eqnarray*}%
The averages for $C$ and $D$ are: 
\begin{eqnarray}
C &\rightarrow &\left\langle C\left( \theta \right) \right\rangle =\frac{%
\alpha _{C}\frac{\omega ^{\prime }\left\vert \Psi \left( \theta -\frac{%
\left\vert Z-Z^{\prime }\right\vert }{c},Z^{\prime }\right) \right\vert ^{2}%
}{\omega }}{\frac{1}{\tau _{C}\omega }+\alpha _{C}\frac{\omega ^{\prime
}\left\vert \Psi \left( \theta -\frac{\left\vert Z-Z^{\prime }\right\vert }{c%
},Z^{\prime }\right) \right\vert ^{2}}{\omega }}=\frac{\alpha _{C}\omega
^{\prime }\left\vert \Psi \left( \theta -\frac{\left\vert Z-Z^{\prime
}\right\vert }{c},Z^{\prime }\right) \right\vert ^{2}}{\frac{1}{\tau _{C}}%
+\alpha _{C}\omega ^{\prime }\left\vert \Psi \left( \theta -\frac{\left\vert
Z-Z^{\prime }\right\vert }{c},Z^{\prime }\right) \right\vert ^{2}}\equiv
C\left( \theta \right)   \label{vrG} \\
D &\rightarrow &\left\langle D\left( \theta \right) \right\rangle =\frac{%
\alpha _{D}\omega \left\vert \Psi \left( \theta ,Z\right) \right\vert ^{2}}{%
\frac{1}{\tau _{D}}+\alpha _{D}\omega \left\vert \Psi \left( \theta
,Z\right) \right\vert ^{2}}\equiv D\left( \theta \right)   \label{vRG}
\end{eqnarray}%
We showed that under some approximations, the average values in this
background states present several possble patterns:%
\begin{eqnarray}
T\left( Z_{-},Z_{+}^{\prime }\right)  &=&\frac{\lambda \tau \exp \left( -%
\frac{\left\vert Z-Z^{\prime }\right\vert }{\nu c}\right) \left( \frac{1}{%
\tau _{D}\alpha _{D}}+\frac{1}{b\bar{T}^{2}\left\langle \left\vert \Psi
_{0}\left( Z^{\prime }\right) \right\vert ^{2}\right\rangle _{Z}}\right) }{%
\frac{1}{\tau _{D}\alpha _{D}}+\frac{1}{\alpha _{C}\tau _{C}}+\frac{1}{b\bar{%
T}^{2}\left\langle \left\vert \Psi _{0}\left( Z^{\prime }\right) \right\vert
^{2}\right\rangle _{Z}}+b\bar{T}\left( \bar{T}\left\langle \left\vert \Psi
_{0}\left( Z^{\prime }\right) \right\vert ^{2}\right\rangle _{Z^{\prime
}}^{2}\right) ^{2}}\simeq 0  \label{LV} \\
T\left( Z_{+},Z_{+}^{\prime }\right)  &=&\frac{\lambda \tau \exp \left( -%
\frac{\left\vert Z-Z^{\prime }\right\vert }{\nu c}\right) \left( \frac{1}{%
\tau _{D}\alpha _{D}}+b\bar{T}\left( \bar{T}\left\langle \left\vert \Psi
_{0}\left( Z^{\prime }\right) \right\vert ^{2}\right\rangle _{Z}^{2}\right)
^{2}\right) }{\frac{1}{\tau _{D}\alpha _{D}}+\frac{1}{\alpha _{C}\tau _{C}}+b%
\bar{T}\left( \bar{T}\left\langle \left\vert \Psi _{0}\left( Z^{\prime
}\right) \right\vert ^{2}\right\rangle _{Z}^{2}\right) ^{2}+b\bar{T}\left( 
\bar{T}\left\langle \left\vert \Psi _{0}\left( Z^{\prime }\right)
\right\vert ^{2}\right\rangle _{Z^{\prime }}^{2}\right) ^{2}}\simeq \frac{%
\lambda \tau \exp \left( -\frac{\left\vert Z-Z^{\prime }\right\vert }{\nu c}%
\right) }{2}  \notag \\
T\left( Z_{+},Z_{-}^{\prime }\right)  &=&\frac{\lambda \tau \exp \left( -%
\frac{\left\vert Z-Z^{\prime }\right\vert }{\nu c}\right) \left( \frac{1}{%
\tau _{D}\alpha _{D}}+b\bar{T}\left( \bar{T}\left\langle \left\vert \Psi
_{0}\left( Z^{\prime }\right) \right\vert ^{2}\right\rangle _{Z}^{2}\right)
^{2}\right) }{\frac{1}{\tau _{D}\alpha _{D}}+\frac{1}{\alpha _{C}\tau _{C}}+b%
\bar{T}\left( \bar{T}\left\langle \left\vert \Psi _{0}\left( Z^{\prime
}\right) \right\vert ^{2}\right\rangle _{Z}^{2}\right) ^{2}+\frac{1}{b\bar{T}%
^{2}\left\langle \left\vert \Psi _{0}\left( Z^{\prime }\right) \right\vert
^{2}\right\rangle _{Z^{\prime }}}}\simeq \lambda \tau \exp \left( -\frac{%
\left\vert Z-Z^{\prime }\right\vert }{\nu c}\right)   \notag \\
T\left( Z_{-},Z_{-}^{\prime }\right)  &\simeq &\frac{\lambda \tau \exp
\left( -\frac{\left\vert Z-Z^{\prime }\right\vert }{\nu c}\right) +\frac{1}{b%
\bar{T}^{2}\left\langle \left\vert \Psi _{0}\left( Z^{\prime }\right)
\right\vert ^{2}\right\rangle _{Z}}}{1+\frac{\tau _{D}\alpha _{D}}{\alpha
_{C}\tau _{C}}+\frac{1}{b\bar{T}^{2}\left\langle \left\vert \Psi _{0}\left(
Z^{\prime }\right) \right\vert ^{2}\right\rangle _{Z}}+\frac{1}{b\bar{T}%
^{2}\left\langle \left\vert \Psi _{0}\left( Z^{\prime }\right) \right\vert
^{2}\right\rangle _{Z^{\prime }}}}\simeq \frac{\lambda \tau \exp \left( -%
\frac{\left\vert Z-Z^{\prime }\right\vert }{\nu c}\right) }{2}  \notag
\end{eqnarray}%
with $\bar{T}=\frac{\lambda \tau \nu cb}{2}$, $b$ a coefficient
characterizing the function $G$ in the linear approximation\footnote{$%
b\simeq G^{\prime }\left( 0\right) $} and $\left\langle \left\vert \Psi
_{0}\left( Z^{\prime }\right) \right\vert ^{2}\right\rangle _{Z}^{2}$, \ $%
\left\langle \left\vert \Psi _{0}\left( Z^{\prime }\right) \right\vert
^{2}\right\rangle _{Z}^{2}$ are some averaged background fields in regions
surrounding $Z$ and $Z^{\prime }$ respctvl, given by a potential describing
some average activity depending on the points. These results are under the
hypothesis the field $\Psi _{0}\left( Z\right) $ is static.

\bigskip 

\subsection*{1.4 Derivation of the effective action}

We present the main points of the derivation of \ the effective action.
Details can be found in \cite{GLt}.

\subsubsection*{1.4.1 Replacing auxiliary variables by averages and
simplified action for $\Gamma $}

To write the effective action for $\Gamma \left( T,\hat{T},C,D\right) $, we
start with a simplification by replacing $C$ and $D$ by averages: 
\begin{eqnarray}
C &\rightarrow &\left\langle C\left( \theta \right) \right\rangle =\frac{%
\alpha _{C}\omega ^{\prime }\left\langle \left\vert \Psi \left( \theta -%
\frac{\left\vert Z-Z^{\prime }\right\vert }{c},Z^{\prime }\right)
\right\vert ^{2}\right\rangle }{\frac{1}{\tau _{C}}+\alpha _{C}\omega
^{\prime }\left\langle \left\vert \Psi \left( \theta -\frac{\left\vert
Z-Z^{\prime }\right\vert }{c},Z^{\prime }\right) \right\vert
^{2}\right\rangle }\equiv C\left( \theta \right)  \\
D &\rightarrow &\left\langle D\left( \theta \right) \right\rangle =\frac{%
\alpha _{D}\omega \left\langle \left\vert \Psi \left( \theta ,Z\right)
\right\vert ^{2}\right\rangle }{\frac{1}{\tau _{D}}+\alpha _{D}\omega
\left\langle \left\vert \Psi \left( \theta ,Z\right) \right\vert
^{2}\right\rangle }\equiv D\left( \theta \right) 
\end{eqnarray}%
(see \cite{GLs}). We also choose:%
\begin{eqnarray*}
h_{C}\left( \omega \left( \theta ,Z,\left\vert \Psi \right\vert ^{2}\right)
\right)  &=&\omega \left( \theta ,Z,\left\vert \Psi \right\vert ^{2}\right) 
\\
h_{D}\left( \omega \left( \theta -\frac{\left\vert Z-Z^{\prime }\right\vert 
}{c},Z^{\prime },\left\vert \Psi \right\vert ^{2}\right) \right)  &=&\omega
\left( \theta -\frac{\left\vert Z-Z^{\prime }\right\vert }{c},Z^{\prime
},\left\vert \Psi \right\vert ^{2}\right) 
\end{eqnarray*}%
and the action for the connectivity field is:%
\begin{eqnarray}
&&\Gamma ^{\dag }\left( T,\hat{T},\theta ,Z,Z^{\prime }\right) \left( \nabla
_{T}\left( \nabla _{T}-\left( -\frac{1}{\tau \omega }T+\frac{\lambda }{%
\omega }\hat{T}\right) \left\vert \Psi \left( \theta ,Z\right) \right\vert
^{2}\right) \right) \Gamma \left( T,\hat{T},\theta ,Z,Z^{\prime }\right) 
\label{tCD} \\
&&+\Gamma ^{\dag }\left( T,\hat{T},\theta ,Z,Z^{\prime }\right) \left(
\nabla _{\hat{T}}\left( \nabla _{\hat{T}}-\frac{\rho }{\omega \left(
J,\theta ,Z,\left\vert \Psi \right\vert ^{2}\right) }\left( \left( h\left(
Z,Z^{\prime }\right) -\hat{T}\right) C\left( \theta \right) \left\vert \Psi
\left( \theta ,Z\right) \right\vert ^{2}h_{C}\left( \omega \left( \theta
,Z,\left\vert \Psi \right\vert ^{2}\right) \right) \right. \right. \right.  
\notag \\
&&\times \left. \left. \left. -D\left( \theta \right) \hat{T}\left\vert \Psi
\left( \theta -\frac{\left\vert Z-Z^{\prime }\right\vert }{c},Z^{\prime
}\right) \right\vert ^{2}h_{D}\left( \omega \left( \theta -\frac{\left\vert
Z-Z^{\prime }\right\vert }{c},Z^{\prime },\left\vert \Psi \right\vert
^{2}\right) \right) \right) \right) \right) \Gamma \left( T,\hat{T},\theta
,Z,Z^{\prime }\right)   \notag
\end{eqnarray}

\subsubsection*{1.4.2 Decomposition of $\Gamma \left( T,\hat{T},\protect%
\theta ,Z,Z^{\prime }\right) $ as background state plus fluctuation}

Starting with (\ref{tCD}), our aim is to obtain an effective action for
fluctuations of $\Gamma $ around some quasi-static background fld. For this
purpose, we decompose the field into the background field and the
fluctuations:%
\begin{eqnarray}
\Gamma \left( T,\hat{T},\theta ,Z,Z^{\prime }\right)  &=&\Gamma _{0}\left( T,%
\hat{T},\theta ,Z,Z^{\prime }\right) +\Delta \Gamma \left( T,\hat{T},\theta
,Z,Z^{\prime }\right)   \label{dcp} \\
\Gamma ^{\dag }\left( T,\hat{T},\theta ,Z,Z^{\prime }\right)  &=&\Gamma
_{0}^{\dag }\left( T,\hat{T},\theta ,Z,Z^{\prime }\right) +\Delta \Gamma
^{\dag }\left( T,\hat{T},\theta ,Z,Z^{\prime }\right)   \notag
\end{eqnarray}

\subsubsection*{1.4.3 Activities as functions of connectivities and activity
induced by fluctuations $\Delta \Gamma \left( T,\hat{T},\protect\theta %
,Z,Z^{\prime }\right) $}

To compute the effective action, we have to take into account the dependency
of $\omega \left( \theta ,Z,\left\vert \Psi \right\vert ^{2}\right) $ in the
connectivity field. Actually, in (\ref{tCD}) activity $\omega \left(
J,\theta ,Z,\left\vert \Psi \right\vert ^{2}\right) $ is itself a functional
of $\Gamma \left( T,\hat{T},\theta ,Z,Z^{\prime }\right) $ whose derivation
was given in (\cite{GLs}). We introduce the variation in activity $\delta
\omega \left( \theta ,Z,\left\vert \Psi \right\vert ^{2}\right) $, depending
on $\Delta \Gamma \left( T,\hat{T},\theta ,Z,Z^{\prime }\right) $ and $%
\Delta \Gamma ^{\dag }\left( T,\hat{T},\theta ,Z,Z^{\prime }\right) $, that
can be expressed as a series expansion:%
\begin{equation*}
\delta \omega \left( \theta ,Z,\left\vert \Psi \right\vert ^{2}\right) =\int
\left( \frac{\delta \left( \delta \omega \left( \theta ,Z,\left\vert \Psi
\right\vert ^{2}\right) \right) }{\delta \left\vert \Delta \Gamma \left( T,%
\hat{T},\theta ,Z_{1},Z_{1}^{\prime }\right) \right\vert ^{2}}\right)
_{\Delta \Gamma \left( T,\hat{T},\theta ,Z,Z^{\prime }\right) =0}\left\vert
\Delta \Gamma \left( T,\hat{T},\theta ,Z_{1},Z_{1}^{\prime }\right)
\right\vert ^{2}+...
\end{equation*}%
and this quantity arises while computing the effective action. More details
are given in (\cite{GLt}). 

\subsubsection*{1.4.4 Effective action for $\Delta \Gamma \left( T,\hat{T},%
\protect\theta ,Z,Z^{\prime }\right) $}

In this paragraph, we expand the action (\ref{tCD}) using the averages
values of $T$ and $\hat{T}$ in the background state. We show in (\cite{GLt})
that the fluctuation part of action (\ref{tCD}) around the background state:%
\begin{equation*}
\Gamma _{0}\left( T,\hat{T},\theta ,Z,Z^{\prime }\right) ,\Gamma _{0}^{\dag
}\left( T,\hat{T},\theta ,Z,Z^{\prime }\right) 
\end{equation*}%
is decomposed into three contributions:%
\begin{equation*}
\Gamma _{0}^{\dag }\left( T,\hat{T},\theta ,Z,Z^{\prime }\right) \Xi \Gamma
_{0}\left( T,\hat{T},\theta ,Z,Z^{\prime }\right) +S_{f}+V\left( \Delta
\Gamma ,\Delta \Gamma ^{\dag }\right) 
\end{equation*}%
with:%
\begin{eqnarray}
&&S_{f}=\Delta \Gamma ^{\dag }\left( T,\hat{T},\theta ,Z,Z^{\prime }\right)
\left( \nabla _{T}\left( \nabla _{T}-\left( -\frac{1}{\tau \omega _{0}\left(
Z\right) }T+\frac{\lambda }{\omega _{0}}\hat{T}\right) \right) \right)
\Delta \Gamma \left( T,\hat{T},\theta ,Z,Z^{\prime }\right)   \label{trm} \\
&&+\Delta \Gamma ^{\dag }\left( T,\hat{T},\theta ,Z,Z^{\prime }\right)
\left( \nabla _{\hat{T}}\left( \nabla _{\hat{T}}-\Delta \right) \right)
\Delta \Gamma \left( T,\hat{T},\theta ,Z,Z^{\prime }\right)   \notag
\end{eqnarray}%
\begin{equation}
V\left( \Delta \Gamma ,\Delta \Gamma ^{\dag }\right) =\Delta \Gamma ^{\dag
}\left( T,\hat{T},\theta ,Z,Z^{\prime }\right) \left( \nabla _{T}\left( \Phi
\right) -\nabla _{\hat{T}}\left( \Xi -\Upsilon \right) \right) \Delta \Gamma
\left( T,\hat{T},\theta ,Z,Z^{\prime }\right)   \label{trl}
\end{equation}%
and where we defined:%
\begin{eqnarray*}
\Xi  &=&\frac{\rho }{\omega _{0}^{2}\left( Z\right) }\left( D\left( \theta
\right) \left\langle \hat{T}\right\rangle \left\vert \Psi _{0}\left(
Z^{\prime }\right) \right\vert ^{2}\left( \omega _{0}\left( Z\right) \delta
\omega \left( \theta -\frac{\left\vert Z-Z^{\prime }\right\vert }{c}%
,Z^{\prime },\left\vert \Psi \right\vert ^{2}\right) -\omega _{0}\left(
Z^{\prime }\right) \delta \omega \left( \theta ,Z,\left\vert \Psi
\right\vert ^{2}\right) \right) \right)  \\
\Delta  &=&\frac{\rho }{\omega _{0}\left( Z\right) }\left( \left( h\left(
Z,Z^{\prime }\right) -\hat{T}\right) C\left( \theta \right) \left\vert \Psi
_{0}\left( Z\right) \right\vert ^{2}\omega _{0}\left( Z\right) -D\left(
\theta \right) \hat{T}\left\vert \Psi _{0}\left( Z^{\prime }\right)
\right\vert ^{2}\omega _{0}\left( Z^{\prime }\right) \right)  \\
\Upsilon  &=&\frac{\rho }{\omega _{0}\left( Z\right) }\left( \left( C\left(
\theta \right) \delta \omega \left( \theta ,Z,\left\vert \Psi \right\vert
^{2}\right) +D\left( \theta \right) \delta \omega \left( \theta -\frac{%
\left\vert Z-Z^{\prime }\right\vert }{c},Z^{\prime },\left\vert \Psi
\right\vert ^{2}\right) \right) \left\vert \Psi _{0}\left( Z\right)
\right\vert ^{2}\left( \hat{T}-\left\langle \hat{T}\right\rangle \right)
\left\vert \Psi _{0}\left( Z^{\prime }\right) \right\vert ^{2}\right)  \\
\Phi  &=&\left( -\frac{\delta \omega \left( \theta ,Z,\left\vert \Psi
\right\vert ^{2}\right) }{\tau \omega _{0}^{2}\left( Z\right) }\left(
T-\left\langle T\right\rangle \right) +\frac{\lambda \delta \omega \left(
\theta ,Z,\left\vert \Psi \right\vert ^{2}\right) }{\omega _{0}^{2}}\left( 
\hat{T}-\left\langle \hat{T}\right\rangle \right) \right) 
\end{eqnarray*}%
The first contribution:%
\begin{equation*}
\Gamma _{0}^{\dag }\left( T,\hat{T},\theta ,Z,Z^{\prime }\right) \Xi \Gamma
_{0}\left( T,\hat{T},\theta ,Z,Z^{\prime }\right) 
\end{equation*}%
translates the modifications in the background state dynamics due to
fluctuations in the connectivities. This can be neglected in first
approximation. The second contribution $S_{f}$ computes the free transition
functions in the backgroundstate, i.e. the transitions due to internal
fluctuations in absence of interactions. The third contribution $V\left(
\Delta \Gamma ,\Delta \Gamma ^{\dag }\right) $\ is an interaction term.

The term (\ref{trm}) is the free part of the effective action, while (\ref%
{trl}) includes the interaction terms describing the self interaction of the
connectivities system, through the fluctuations in activities.

We demonstrate in (\cite{GLt})\ that the first term in (\ref{trl}) may be
neglected and the second term can be approximated using estimates of $\delta
\omega \left( \theta -\frac{\left\vert Z-Z^{\prime }\right\vert }{c}%
,Z^{\prime },\left\vert \Psi \right\vert ^{2}\right) $ provided in appendix
1 and 2 of (\cite{GLs}). Therefore, we can alternatively use for
interactions:%
\begin{equation}
V\left( \Delta \Gamma ,\Delta \Gamma ^{\dag }\right) =-\Delta \Gamma ^{\dag
}\left( T,\hat{T},\theta ,Z,Z^{\prime }\right) \left( \nabla _{\hat{T}}\Xi
\right) \Delta \Gamma \left( T,\hat{T},\theta ,Z,Z^{\prime }\right) 
\end{equation}%
or its continuous approximation:%
\begin{equation}
V\left( \Delta \Gamma ,\Delta \Gamma ^{\dag }\right) =\Delta \Gamma ^{\dag
}\left( T,\hat{T},\theta ,Z,Z^{\prime }\right) \left( \nabla _{\hat{T}}\hat{%
\Xi}\right) \Delta \Gamma \left( T,\hat{T},\theta ,Z,Z^{\prime }\right) 
\label{Trm}
\end{equation}%
where:%
\begin{equation*}
\hat{\Xi}=\frac{\rho }{\omega _{0}\left( Z\right) }\left( D\left( \theta
\right) \left\langle \hat{T}\right\rangle \left\vert \Psi _{0}\left(
Z^{\prime }\right) \right\vert ^{2}\left( \left( \left( Z-Z^{\prime }\right)
\left( \nabla _{Z}+\nabla _{Z}\omega _{0}\left( Z\right) \right) +\frac{%
\left\vert Z-Z^{\prime }\right\vert }{c}\right) \delta \omega \left( \theta
,Z,\left\vert \Psi \right\vert ^{2}\right) \right) \right) 
\end{equation*}

Ultimately, we show in (\cite{GLt}) that the free action can be rewritten
using the background state equations, so that the effective actin for $%
\Gamma \left( T,\hat{T},\theta ,Z,Z^{\prime }\right) $ becomes:%
\begin{eqnarray}
&&S\left( \Delta \Gamma \left( T,\hat{T},\theta ,Z,Z^{\prime }\right)
\right)   \label{fcgg} \\
&=&\Delta \Gamma ^{\dag }\left( T,\hat{T},\theta ,Z,Z^{\prime }\right)
\left( \nabla _{T}\left( \nabla _{T}+\frac{\left( T-\left\langle
T\right\rangle \right) }{\tau \omega _{0}\left( Z\right) }\left\vert \Psi
\left( \theta ,Z\right) \right\vert ^{2}\right) \right) \Delta \Gamma \left(
T,\hat{T},\theta ,Z,Z^{\prime }\right)   \notag \\
&&+\Delta \Gamma ^{\dag }\left( T,\hat{T},\theta ,Z,Z^{\prime }\right)
\nabla _{\hat{T}}\left( \nabla _{\hat{T}}+\rho \left\vert \bar{\Psi}%
_{0}\left( Z,Z^{\prime }\right) \right\vert ^{2}\left( \hat{T}-\left\langle 
\hat{T}\right\rangle \right) \right) \Delta \Gamma \left( T,\hat{T},\theta
,Z,Z^{\prime }\right) +V\left( \Delta \Gamma ,\Delta \Gamma ^{\dag }\right) 
\notag
\end{eqnarray}%
where $\left\vert \bar{\Psi}_{0}\left( Z,Z^{\prime }\right) \right\vert ^{2}$
is a weighted sum of the field over the two connected points: 
\begin{equation}
\left\vert \bar{\Psi}_{0}\left( Z,Z^{\prime }\right) \right\vert ^{2}=\frac{%
C\left( \theta \right) \left\vert \Psi _{0}\left( Z\right) \right\vert
^{2}\omega _{0}\left( Z\right) +D\left( \theta \right) \left\vert \Psi
_{0}\left( Z^{\prime }\right) \right\vert ^{2}\omega _{0}\left( Z^{\prime
}\right) }{\omega _{0}\left( Z\right) }  \label{PS}
\end{equation}

\subsection*{1.5 Eigenstates of the effective action}

\subsubsection*{1.5.1 Rewriting saddle point equation}

After change of variable:%
\begin{eqnarray}
\Delta \Gamma \left( T,\hat{T},\theta ,Z,Z^{\prime }\right) &\rightarrow
&\exp \left( -\frac{\rho \left\vert \bar{\Psi}\left( Z,Z^{\prime }\right)
\right\vert ^{2}\left( \hat{T}-\left\langle \hat{T}\right\rangle \right) ^{2}%
}{4\sigma _{\hat{T}}^{2}}-\frac{V_{0}\left( \hat{T}-\left\langle \hat{T}%
\right\rangle \right) }{2\sigma _{\hat{T}}^{2}\omega \left( \theta
,Z,\left\vert \Psi \right\vert ^{2}\right) }\right)  \label{Cgv} \\
&&\times \exp \left( -\frac{\left( \left( T-\left\langle T\right\rangle
\right) ^{2}-2\lambda \tau \left( \hat{T}-\left\langle \hat{T}\right\rangle
\right) \left( T-\left\langle T\right\rangle \right) \right) }{4\sigma
_{T}^{2}\tau \omega }\right) \Delta \Gamma \left( T,\hat{T},\theta
,Z,Z^{\prime }\right)  \notag
\end{eqnarray}%
and:%
\begin{eqnarray}
\Delta \Gamma ^{\dag }\left( T,\hat{T},\theta ,Z,Z^{\prime }\right)
&\rightarrow &\exp \left( \frac{\rho \left\vert \bar{\Psi}\left( Z,Z^{\prime
}\right) \right\vert ^{2}\left( \hat{T}-\left\langle \hat{T}\right\rangle
\right) ^{2}}{4\sigma _{\hat{T}}^{2}}+\frac{V_{0}\left( \hat{T}-\left\langle 
\hat{T}\right\rangle \right) }{2\sigma _{\hat{T}}^{2}\omega \left( \theta
,Z,\left\vert \Psi \right\vert ^{2}\right) }\right)  \label{Cgr} \\
&&\times \exp \left( \frac{\left( \frac{\left( T-\left\langle T\right\rangle
\right) ^{2}}{\tau }-2\lambda \left( \hat{T}-\left\langle \hat{T}%
\right\rangle \right) \left( T-\left\langle T\right\rangle \right) \right) }{%
4\sigma _{T}^{2}\tau \omega }\right) \Delta \Gamma ^{\dag }\left( T,\hat{T}%
,\theta ,Z,Z^{\prime }\right)  \notag
\end{eqnarray}%
with:%
\begin{equation*}
V_{0}=\left( \frac{\rho D\left( \theta \right) \left\langle \hat{T}%
\right\rangle \left\vert \Psi _{0}\left( Z^{\prime }\right) \right\vert ^{2}%
}{\omega _{0}\left( Z\right) }\hat{T}\left( 1-\left( 1+\left\langle
\left\vert \Psi _{\Gamma }\right\vert ^{2}\right\rangle \right) \hat{T}%
\right) ^{-1}\left[ O\frac{\Delta T\left\vert \Delta \Gamma \left( \theta
_{1},Z_{1},Z_{1}^{\prime }\right) \right\vert ^{2}}{T}\right] \right)
\end{equation*}%
the saddle point equation of the action (\ref{fcp}) for the connectivity
fields $\Delta \Gamma $ and\ $\Delta \Gamma ^{\dag }$ becomes:

\begin{eqnarray}
&&0=\left( -\sigma _{\hat{T}}^{2}\nabla _{\hat{T}}^{2}+\frac{1}{4\sigma _{%
\hat{T}}^{2}}\left( \left\vert \bar{\Psi}\left( Z,Z^{\prime }\right)
\right\vert ^{2}\Delta \hat{T}+\frac{\rho \left( V_{0}-\frac{\sigma _{\hat{T}%
}^{2}}{\sigma _{T}^{2}}\lambda \Delta T\left\vert \Psi \left( Z\right)
\right\vert ^{2}\right) }{\omega _{0}\left( Z\right) }\right) ^{2}\right)
\Delta \Gamma \left( T,\hat{T},\theta ,Z,Z^{\prime }\right)  \label{fcn} \\
&&+\left( -\sigma _{T}^{2}\nabla _{T}^{2}+\frac{1}{4\sigma _{T}^{2}}\left( 
\frac{\Delta T-\lambda \tau \Delta \hat{T}}{\tau \omega _{0}\left( Z\right) }%
\right) ^{2}\right.  \notag \\
&&\left. -\left( \frac{\left\vert \bar{\Psi}\left( Z,Z^{\prime }\right)
\right\vert ^{2}}{2}+\frac{\left\vert \Psi \left( Z\right) \right\vert ^{2}}{%
2\tau \omega _{0}\left( Z\right) }+V\left( \theta ,Z,Z^{\prime },\Delta
\Gamma \right) \Delta T-\alpha \right) \right) \Delta \Gamma \left( T,\hat{T}%
,\theta ,Z,Z^{\prime }\right)  \notag
\end{eqnarray}%
with:%
\begin{equation}
V_{0}\left( Z,Z^{\prime }\right) =\left( \frac{\rho D\left( \theta \right)
\left\langle \hat{T}\right\rangle \left\vert \Psi _{0}\left( Z^{\prime
}\right) \right\vert ^{2}}{\omega _{0}\left( Z\right) }\hat{T}\left(
1-\left( 1+\left\langle \left\vert \Psi _{\Gamma }\right\vert
^{2}\right\rangle \right) \hat{T}\right) ^{-1}\left[ O\frac{\Delta
T\left\vert \Delta \Gamma \left( \theta _{1},Z_{1},Z_{1}^{\prime }\right)
\right\vert ^{2}}{T}\right] \right)  \label{cfh}
\end{equation}%
\begin{equation*}
V\left( \theta ,Z,Z^{\prime },\Delta \Gamma \right) =V_{1}\left( \theta
,Z,Z^{\prime },\Delta \Gamma \right) \left( 1+V_{2}\left( \theta
,Z,Z^{\prime },\Delta \Gamma \right) \right)
\end{equation*}%
with $V_{1}$\ and $V_{2}$\ given by:%
\begin{eqnarray}
&&V_{1}\left( \theta ,Z,Z^{\prime },\Delta \Gamma \right)  \label{Vfn} \\
&=&\int \Delta \Gamma ^{\dag }\left( T_{2},\hat{T}_{2},\theta
_{2},Z_{2},Z_{2}^{\prime }\right) \nabla _{\hat{T}_{2}}\left( \frac{\rho
D\left( \theta \right) \left\langle \hat{T}_{2}\right\rangle \left\vert \Psi
_{0}\left( Z_{2}^{\prime }\right) \right\vert ^{2}}{\omega _{0}\left(
Z_{2}\right) }\left[ \check{T}\left( 1-\left( 1+\left\langle \left\vert \Psi
_{\Gamma }\right\vert ^{2}\right\rangle \right) \check{T}\right) ^{-1}O%
\right] _{\left( T,\hat{T},\theta ,Z,Z^{\prime }\right) }^{\left( T_{2},\hat{%
T}_{2},\theta _{2},Z_{2},Z_{2}^{\prime }\right) }\right)  \notag \\
&&\times \Delta \Gamma \left( T_{2},\hat{T}_{2},\theta
_{2},Z_{2},Z_{2}^{\prime }\right) d\left( T_{2},\hat{T}_{2},\theta
_{2},Z_{2},Z_{2}^{\prime }\right)  \notag
\end{eqnarray}%
and:

\begin{equation*}
V_{2}\left( \theta ,Z,Z^{\prime },\Delta \Gamma \right) =\int \left[ \check{T%
}\left( 1-\left( 1+\left\langle \left\vert \Psi _{\Gamma }\right\vert
^{2}\right\rangle \right) \check{T}\right) ^{-1}\right] _{\left( T_{1},\hat{T%
}_{1},\theta _{1},Z_{1},Z_{1}^{\prime }\right) }^{\left( T,\hat{T},\theta
,Z,Z^{\prime }\right) }\left[ \frac{\Delta T\left\vert \Delta \Gamma \left(
\theta _{1},Z_{1},Z_{1}^{\prime }\right) \right\vert ^{2}}{T}\right] d\left(
T_{1},\hat{T}_{1},\theta _{1},Z_{1},Z_{1}^{\prime }\right)
\end{equation*}%
respectively.

Equation (\ref{fcn}) encapsulates the main characteristics of connectivity
states:%
\begin{equation*}
\Delta \Gamma \left( T,\hat{T},\theta ,Z,Z^{\prime }\right)
\end{equation*}%
The two first terms describes the individual connectivities. The term $%
V\left( \theta ,Z,Z^{\prime },\Delta \Gamma \right) $ measures the
interactions between connectivities

Operator $O$ is defined by (\ref{fms}). Recall that $\alpha $ implements the
constraint $\left\Vert \Delta \Gamma \right\Vert =\overline{\left\Vert
\Delta \Gamma \right\Vert }$. As in the previous paragraph $\alpha $ stands
for:%
\begin{equation}
\alpha _{0}+U^{\prime }\left( \left\vert \Delta \Gamma \left( Z,Z^{\prime
}\right) \right\vert ^{2}\right)  \label{lpg}
\end{equation}%
where $\alpha _{0}$ is the Lagrange multiplier for the overall constraint,
and $U\left( \Delta \Gamma \left( Z,Z^{\prime }\right) \right) $ the
potential.

After diagonalization of the potential by a matrix $P=\left( 
\begin{array}{cc}
w_{1} & w_{2} \\ 
w_{1}^{\prime } & w_{2}^{\prime }%
\end{array}%
\right) $, whose components are:%
\begin{eqnarray*}
w_{1} &=&\sqrt{\frac{1}{2}\left( 1+\sqrt{\frac{\left( \frac{\left(
v-u\right) }{2s}\right) ^{2}}{1+\left( \frac{\left( v-u\right) }{2s}\right)
^{2}}}\right) }\text{, }w_{2}=\sqrt{\frac{1}{2}\left( 1-\sqrt{\frac{\left( 
\frac{\left( v-u\right) }{2s}\right) ^{2}}{1+\left( \frac{\left( v-u\right) 
}{2s}\right) ^{2}}}\right) } \\
w_{1}^{\prime } &=&-\sqrt{\frac{1}{2}\left( 1-\sqrt{\frac{\left( \frac{%
\left( v-u\right) }{2s}\right) ^{2}}{1+\left( \frac{\left( v-u\right) }{2s}%
\right) ^{2}}}\right) }\text{, }w_{2}^{\prime }=\sqrt{\frac{1}{2}\left( 1+%
\sqrt{\frac{\left( \frac{\left( v-u\right) }{2s}\right) ^{2}}{1+\left( \frac{%
\left( v-u\right) }{2s}\right) ^{2}}}\right) }
\end{eqnarray*}%
we show that this background state equation becomes: 
\begin{eqnarray}
0 &=&\left( -\sigma _{\hat{T}}^{2}\nabla _{\hat{T}^{\prime }}^{2}+\frac{%
\lambda _{+}^{2}}{4\sigma _{\hat{T}}^{2}}\left( \Delta \hat{T}^{\prime
}-\Delta \hat{T}_{0}^{\prime }-\frac{w_{2}}{\lambda _{+}}V\right)
^{2}\right) \Delta \Gamma \left( T,\hat{T},\theta ,Z,Z^{\prime }\right)
\label{sdl} \\
&&+\left( -\sigma _{T}^{2}\nabla _{T^{\prime }}^{2}+\frac{\lambda _{-}^{2}}{%
\sigma _{T}^{2}}\left( \Delta T^{\prime }-\Delta T_{0}^{\prime }-\frac{w_{1}%
}{\lambda _{-}}V\right) ^{2}\right) \Delta \Gamma \left( T,\hat{T},\theta
,Z,Z^{\prime }\right)  \notag \\
&&-\left( u+v+\left( \frac{w_{1}^{2}}{\lambda _{+}}V^{2}+\frac{w_{2}^{2}}{%
\lambda _{-}}V^{2}\right) -\alpha \right) \Delta \Gamma \left( T,\hat{T}%
,\theta ,Z,Z^{\prime }\right)  \notag
\end{eqnarray}%
where:%
\begin{equation*}
\lambda _{\pm }=\sqrt{\frac{1}{2}\left( u^{2}+v^{2}\right) +s^{2}\pm \frac{%
\left( u+v\right) }{2}\sqrt{\left( u-v\right) ^{2}+4s^{2}}}
\end{equation*}%
\begin{eqnarray*}
u &=&\frac{\left\vert \Psi _{0}\left( Z\right) \right\vert ^{2}}{\tau \omega
_{0}\left( Z\right) } \\
v &=&\rho \left\vert \bar{\Psi}_{0}\left( Z,Z^{\prime }\right) \right\vert
^{2} \\
s &=&-\frac{\lambda \left\vert \Psi _{0}\left( Z\right) \right\vert ^{2}}{%
\omega _{0}\left( Z\right) }\frac{\sigma _{\hat{T}}}{\sigma _{T}}
\end{eqnarray*}%
\begin{equation}
\left( \Delta T_{0},\Delta \hat{T}_{0}\right) \simeq \left( -\frac{\lambda
\tau V_{0}}{\sigma _{T}\omega _{0}\left( Z\right) \left\vert \bar{\Psi}%
_{0}\left( Z,Z^{\prime }\right) \right\vert ^{2}},\frac{\Delta T_{0}}{%
\lambda \tau }\frac{\sigma _{T}}{\sigma _{\hat{T}}}\right)  \label{Sfh}
\end{equation}%
and $\left( X^{\prime },\hat{X}^{\prime }\right) $ are the coordinates of
any vector in the diagonal basis of the potential:%
\begin{equation*}
\left( X^{\prime },\hat{X}^{\prime }\right) ^{t}=P^{-1}\left( X,\hat{X}%
\right)
\end{equation*}%
Note that:%
\begin{equation*}
\lambda _{+}+\lambda _{-}=u+v
\end{equation*}

The eigenstates and averages are obtained in (\cite{GLt}). Under several
approximations we obtain that:

\subsubsection*{1.5.2 Eigenstates}

The solutions of (\ref{fcn}) become:%
\begin{eqnarray}
&&\Delta \Gamma _{\delta }\left( T,\hat{T},\theta ,Z,Z^{\prime }\right)
\label{Sdn} \\
&=&\exp \left( -\frac{1}{2}\left( \mathbf{\Delta T-}\underline{\left\langle 
\mathbf{\Delta T}\right\rangle }\right) ^{t}\hat{U}\left( \mathbf{\Delta T-}%
\underline{\left\langle \mathbf{\Delta T}\right\rangle }\right) \right) 
\notag \\
&&\times H_{p}\left( \left( \mathbf{\Delta T}^{\prime }\mathbf{-}\underline{%
\left\langle \mathbf{\Delta T}\right\rangle }^{\prime }\right) _{2}\frac{%
\sigma _{T}\lambda _{+}}{2\sqrt{2}}\left( \mathbf{\Delta T}^{\prime }\mathbf{%
-}\underline{\left\langle \mathbf{\Delta T}\right\rangle }^{\prime }\right)
_{2}\right) H_{p-\delta }\left( \left( \mathbf{\Delta T}^{\prime }\mathbf{-}%
\underline{\left\langle \mathbf{\Delta T}\right\rangle }^{\prime }\right)
_{2}\frac{\sigma _{\hat{T}}\lambda _{-}}{2\sqrt{2}}\left( \mathbf{\Delta T}%
^{\prime }\mathbf{-}\underline{\left\langle \mathbf{\Delta T}\right\rangle }%
^{\prime }\right) _{2}\right)  \notag
\end{eqnarray}%
and their conjugate:%
\begin{eqnarray}
&&\Delta \Gamma _{\delta }^{\dagger }\left( T,\hat{T},\theta ,Z,Z^{\prime
}\right)  \label{Sdd} \\
&=&H_{p}\left( \left( \mathbf{\Delta T}^{\prime }\mathbf{-}\underline{%
\left\langle \mathbf{\Delta T}\right\rangle }^{\prime }\right) _{2}\frac{%
\sigma _{T}\lambda _{+}}{2\sqrt{2}}\left( \mathbf{\Delta T}^{\prime }\mathbf{%
-}\underline{\left\langle \mathbf{\Delta T}\right\rangle }^{\prime }\right)
_{2}\right) H_{p-\delta }\left( \left( \mathbf{\Delta T}^{\prime }\mathbf{-}%
\underline{\left\langle \mathbf{\Delta T}\right\rangle }^{\prime }\right)
_{2}\frac{\sigma _{\hat{T}}\lambda _{-}}{2\sqrt{2}}\left( \mathbf{\Delta T}%
^{\prime }\mathbf{-}\underline{\left\langle \mathbf{\Delta T}\right\rangle }%
^{\prime }\right) _{2}\right)  \notag
\end{eqnarray}%
where $H_{p}$ and $H_{p-\delta }$ are Hermite polynomials and the variables
are:%
\begin{eqnarray}
\mathbf{\Delta T-}\underline{\left\langle \mathbf{\Delta T}\right\rangle }
&=&\left( 
\begin{array}{c}
\Delta T-\left\langle \Delta T\right\rangle \\ 
\Delta \hat{T}-\left\langle \Delta \hat{T}\right\rangle%
\end{array}%
\right)  \label{shn} \\
\mathbf{\Delta T}^{\prime }\mathbf{-}\underline{\left\langle \mathbf{\Delta T%
}\right\rangle }^{\prime } &=&P^{-1}\left( \mathbf{\Delta T-}\underline{%
\left\langle \mathbf{\Delta T}\right\rangle }\right)  \notag
\end{eqnarray}%
with parameters:%
\begin{eqnarray*}
\Delta T_{0} &\simeq &-\frac{\lambda \tau V_{0}}{\omega _{0}\left( Z\right)
\left\vert \Psi _{0}\left( Z\right) \right\vert ^{2}} \\
\Delta \hat{T}_{0} &\simeq &\frac{\Delta T_{0}}{\lambda \tau }
\end{eqnarray*}%
\begin{equation*}
\left( 
\begin{array}{c}
\Delta T_{1} \\ 
\Delta \hat{T}_{1}%
\end{array}%
\right) =PD^{-1}P^{-1}\left( 
\begin{array}{c}
V \\ 
0%
\end{array}%
\right) =U^{-1}\left( 
\begin{array}{c}
V \\ 
0%
\end{array}%
\right)
\end{equation*}%
and the matrix $\hat{U}$ given by:%
\begin{equation*}
\hat{U}=\left( 
\begin{array}{cc}
\frac{1}{\sigma _{T}} & 0 \\ 
0 & \frac{1}{\sigma _{\hat{T}}}%
\end{array}%
\right) U\left( 
\begin{array}{cc}
\frac{1}{\sigma _{T}} & 0 \\ 
0 & \frac{1}{\sigma _{\hat{T}}}%
\end{array}%
\right) =\left( 
\begin{array}{cc}
\frac{s^{2}+u^{2}}{\sigma _{T}^{2}} & -\frac{s\left( u+v\right) }{\sigma
_{T}\sigma _{\hat{T}}} \\ 
-\frac{s\left( u+v\right) }{\sigma _{T}\sigma _{\hat{T}}} & \frac{s^{2}+v^{2}%
}{\sigma _{\hat{T}}^{2}}%
\end{array}%
\right)
\end{equation*}%
The potential background field of the sytem are thus defined by considering
the set:%
\begin{equation}
W=\left\{ \left( Z,Z^{\prime }\right) ,p_{1}\left( Z,Z^{\prime }\right)
\lambda _{+}+p_{2}\left( Z,Z^{\prime }\right) \lambda _{-}=\frac{u+v}{2}%
+\left( \frac{w_{1}^{2}}{\lambda _{+}}+\frac{w_{2}^{2}}{\lambda _{-}}\right)
V^{2}-\alpha >0\right\}  \label{WSt}
\end{equation}%
and associating to each function $\delta \left( Z,Z^{\prime }\right)
:W\rightarrow $ $\left[ 0,p\left( Z,Z^{\prime }\right) \right] $, the
potential background state:%
\begin{equation*}
\prod\limits_{W}\Delta \Gamma _{\delta \left( Z,Z^{\prime }\right) }\left( T,%
\hat{T},\theta ,Z,Z^{\prime }\right)
\end{equation*}%
and:%
\begin{equation*}
\prod\limits_{W}\Delta \Gamma _{\delta \left( Z,Z^{\prime }\right)
}^{\dagger }\left( T,\hat{T},\theta ,Z,Z^{\prime }\right)
\end{equation*}

We conclude by considering an example of solutions of (\ref{sdl}), and
consider $\left( Z,Z^{\prime }\right) $ such that:%
\begin{equation*}
\frac{u+v}{2}+\left( \frac{w_{1}^{2}}{\lambda _{+}}+\frac{w_{2}^{2}}{\lambda
_{-}}\right) V^{2}-\alpha =0
\end{equation*}%
Taking into account (\ref{Cgv}) and (\ref{Cgr}), the background state for (%
\ref{sdl}) is:%
\begin{equation*}
\Delta \Gamma \left( T,\hat{T},Z,Z^{\prime }\right) =\exp \left( -\frac{1}{2}%
\left( 
\begin{array}{c}
\Delta T^{\prime }-\Delta T_{0}^{\prime }-\frac{w_{2}}{\lambda _{+}}V \\ 
\Delta \hat{T}^{\prime }-\Delta \hat{T}_{0}^{\prime }-\frac{w_{1}}{\lambda
_{-}}V%
\end{array}%
\right) ^{t}D\left( 
\begin{array}{c}
\Delta T^{\prime }-\Delta T_{0}^{\prime }-\frac{w_{2}}{\lambda _{+}}V \\ 
\Delta \hat{T}^{\prime }-\Delta \hat{T}_{0}^{\prime }-\frac{w_{1}}{\lambda
_{-}}V%
\end{array}%
\right) \right)
\end{equation*}%
Coming back to the initial variables and reintroducing $\sigma _{T}$ and $%
\sigma _{\hat{T}}$, it yields:%
\begin{equation}
\Delta \Gamma \left( T,\hat{T},Z,Z^{\prime }\right) =\exp \left( -\frac{1}{2}%
\left( 
\begin{array}{c}
\Delta T-\Delta T_{0}-\Delta T_{1} \\ 
\Delta \hat{T}-\Delta \hat{T}_{0}-\Delta \hat{T}_{1}%
\end{array}%
\right) ^{t}\hat{U}\left( 
\begin{array}{c}
\Delta T-\Delta T_{0}-\Delta T_{1} \\ 
\Delta \hat{T}-\Delta \hat{T}_{0}-\Delta \hat{T}_{1}%
\end{array}%
\right) \right)  \label{bcs}
\end{equation}%
and for $\Delta \Gamma ^{\dagger }\left( T,\hat{T},Z,Z^{\prime }\right) $: 
\begin{subequations}
\begin{equation}
\Delta \Gamma ^{\dagger }\left( T,\hat{T},Z,Z^{\prime }\right) =1
\label{bcj}
\end{equation}

\subsubsection*{1.5.3 Estimation of $\Delta T\left( Z,Z^{\prime }\right) $
and $\Delta \hat{T}\left( Z,Z^{\prime }\right) $}

Average connectivities are given by: 
\end{subequations}
\begin{equation}
\left\langle \Delta T\right\rangle \simeq \frac{\omega _{0}\left( Z\right)
\left\langle T\right\rangle }{\rho D\left( \theta \right) \left\langle \hat{T%
}\right\rangle \left\vert \Psi _{0}\left( Z^{\prime }\right) \right\vert
^{2}k\underline{A_{1}}\left\Vert \Delta \Gamma \right\Vert ^{6}}\left\langle
\rho \frac{\left\vert \Psi _{0}\left( Z\right) \right\vert ^{2}}{A}%
\right\rangle ^{2}\left\langle T\right\rangle  \label{VRP}
\end{equation}%
with:%
\begin{equation*}
A_{1}\left( Z,Z^{\prime }\right) =\frac{\rho D\left( \theta \right)
\left\langle \hat{T}\right\rangle \left\vert \Psi _{0}\left( Z^{\prime
}\right) \right\vert ^{2}}{\omega _{0}\left( Z\right) }\underline{A_{1}}%
\left( Z,Z^{\prime }\right)
\end{equation*}%
and:%
\begin{equation*}
\underline{A_{1}}\left( Z,Z^{\prime }\right) =\left\langle \left[ F\left(
Z_{2},Z_{2}^{\prime }\right) \left[ \check{T}\left( 1-\left\langle
\left\vert \Psi _{0}\right\vert ^{2}\right\rangle \frac{\left\langle \Delta
T\right\rangle }{T}\left\Vert \Delta \Gamma \right\Vert ^{2}\right) ^{-1}O%
\right] \right] _{\left( T,\hat{T},\theta ,Z,Z^{\prime }\right)
}\right\rangle
\end{equation*}%
\begin{equation}
\left\langle \Delta \hat{T}\right\rangle =\hat{A}\left\langle \Delta
T\right\rangle  \label{dlt}
\end{equation}%
\begin{equation}
\hat{A}\simeq -\frac{1}{v}A\frac{\left\Vert \Delta \Gamma \right\Vert ^{2}}{%
\left\langle T\right\rangle }  \label{fht}
\end{equation}

With our approximations and using (\ref{fht}) this becomes:%
\begin{eqnarray*}
\Delta T\left( Z,Z^{\prime }\right) &=&\frac{A_{1}\left( Z,Z^{\prime
}\right) A\left( Z,Z^{\prime }\right) \left\langle v^{2}\right\rangle }{%
\left\langle A_{1}\left( Z,Z^{\prime }\right) \right\rangle \left\langle
A_{0}\left( Z,Z^{\prime }\right) \right\rangle v^{2}}\left\langle \Delta
T\right\rangle \\
\Delta \hat{T}\left( Z,Z^{\prime }\right) &=&\frac{A_{0}\left( Z,Z^{\prime
}\right) }{\left\langle A_{0}\left( Z,Z^{\prime }\right) \right\rangle }%
\left\langle \Delta \hat{T}\right\rangle
\end{eqnarray*}

\section*{Appendix 2. Dynamics in activities and emerging collective state}

This appendix incorporates several elements of (\cite{GL}), (\cite{GLr}) and
(\cite{GLs}). We revisit how activity may manifest stable oscillation
patterns, and that these oscillations, induced by external signals, may bind
individual elements to produce an emerging activated state. The link between
the connectivities of the state and its average activities is provided.
Ultimately, synthetizing the results, we justify the reverse point of view
adopted in the text, namely, that a collective state can be characterized by
some possible stable oscillating activities.

\subsection*{2.1 Dynamic wave equation for activities}

This section seeks dynamic solutions for (\ref{qf}). We use the relation (%
\ref{psf}) provided below to substitute the non-static part of the field $%
\Psi $ as a function of the activities, and subsequently derive a wave
equation for these activities. The outcomes of this section will be
instrumental in deducing the internal activity of a collective state and
justifying the possibility of sustained oscillating internal activity in
such states.

\subsubsection*{2.1.1 Differential equation for activities in the local
approximation}

A local approximation of (\ref{qf}) around a position-independent static
equilibrium can be derived for non-static activities. Assuming a static
background field $\Psi _{0}$, we showed in (\cite{GL}) (see (\cite{GLr}) for
an account) the following relation between the fluctuations $\delta \Psi
\left( \theta ,Z\right) $ around this background and the time-dependent part
of the activities $\omega \left( J,Z,\left\vert \Psi \right\vert ^{2}\right) 
$ is given by:%
\begin{equation}
\delta \Psi \left( \theta ,Z\right) \simeq \frac{\nabla _{\theta }\omega
\left( J,Z,\left\vert \Psi \right\vert ^{2}\right) }{V^{\prime \prime
}\left( \Psi _{0}\left( Z\right) \right) \omega _{0}^{2}\left(
J,Z,\left\vert \Psi \right\vert ^{2}\right) }\Psi _{0}  \label{psf}
\end{equation}%
where $\omega \left( J,Z,\left\vert \Psi \right\vert ^{2}\right) $ is the
time-dependent activity.

We can find a local approximation of (\ref{qf}) if we expand $\omega \left(
J\left( \theta \right) ,\theta ,Z,\mathcal{G}_{0}+\left\vert \Psi
\right\vert ^{2}\right) $ to the second-order in $Z-Z_{1}$, and consider the
other terms in the right-hand side of (\ref{qf}) as corrections. Neglecting
the perturbative corrections in the effective action for $\Psi \left( \theta
,Z\right) $, the cells'system is described by the "classical" action:%
\begin{equation}
\hat{S}=-\frac{1}{2}\Psi ^{\dagger }\left( \theta ,Z,\omega \right) \nabla
\left( \frac{\sigma _{\theta }^{2}}{2}\nabla -\omega ^{-1}\left( J,\theta
,Z,\left\vert \Psi \right\vert ^{2}\right) \right) \Psi \left( \theta
,Z\right) +V\left( \Psi \right)  \label{SCP}
\end{equation}%
and the equation for $\omega \left( J\left( \theta \right) ,\theta ,Z,%
\mathcal{G}_{0}+\left\vert \Psi \right\vert ^{2}\right) $ is:%
\begin{eqnarray}
F^{-1}\left( \omega \left( J\left( \theta \right) ,\theta \right) \right)
&=&J\left( \theta ,Z\right) +\int \frac{\kappa }{N}\frac{\omega \left(
J,\theta -\frac{\left\vert Z-Z_{1}\right\vert }{c},Z_{1},\Psi \right)
T\left( Z,\theta ,Z_{1},\theta -\frac{\left\vert Z-Z_{1}\right\vert }{c}%
\right) }{\omega \left( J,\theta ,Z,\left\vert \Psi \right\vert ^{2}\right) }
\label{vnq} \\
&&\times \left( \left\vert \Psi _{0}+\delta \Psi \left( \theta -\frac{%
\left\vert Z-Z_{1}\right\vert }{c},Z_{1}\right) \right\vert ^{2}\right)
dZ_{1}  \notag
\end{eqnarray}

We then expand $\omega \left( \theta -\frac{\left\vert Z-Z_{1}\right\vert }{c%
},Z_{1}\right) $ around $\omega \left( \theta ,Z\right) $ to the
second-order in $Z-Z_{1}$\ and compute the integrals, which yields for the
right-hand side of (\ref{vnq}):%
\begin{eqnarray*}
&&J\left( \theta \right) +\int \frac{\kappa }{N}\frac{\omega \left( \theta -%
\frac{\left\vert Z-Z_{1}\right\vert }{c},Z_{1}\right) }{\omega \left( \theta
,Z\right) }T\left( Z,\theta ,Z_{1},\theta -\frac{\left\vert
Z-Z_{1}\right\vert }{c}\right) \\
&&\times \left\vert \Psi _{0}\left( Z_{1}\right) +\delta \Psi \left( \theta -%
\frac{\left\vert Z-Z_{1}\right\vert }{c},Z_{1}\right) \right\vert ^{2}dZ_{1}
\\
&\simeq &J\left( \theta \right) +\frac{TW\left( 1\right) }{\bar{\Lambda}}+%
\frac{\hat{f}_{1}\nabla _{\theta }\omega \left( \theta ,Z\right) }{\omega
\left( \theta ,Z\right) }+\frac{\hat{f}_{3}\nabla _{\theta }^{2}\omega
\left( \theta ,Z\right) }{\omega \left( \theta ,Z\right) }+c^{2}\frac{\hat{f}%
_{3}\nabla _{Z}^{2}\omega \left( \theta ,Z\right) }{\omega \left( \theta
,Z\right) }+T\Psi _{0}\delta \Psi \left( \theta ,Z\right)
\end{eqnarray*}%
where we defined:%
\begin{eqnarray}
\hat{f}_{1} &=&-\frac{\Gamma _{1}}{c}\text{, }\hat{f}_{3}=\frac{\Gamma _{2}}{%
c^{2}}  \label{ctf} \\
\Gamma _{1} &=&\frac{\kappa }{NX_{r}}\int \left\vert Z-Z_{1}\right\vert
T\left( Z,Z_{1}\right) \left\vert \Psi _{0}\left( Z_{1}\right) \right\vert
^{2}dZ_{1}  \notag \\
\Gamma _{2} &=&\frac{\kappa }{2NX_{r}}\int \left( Z-Z_{1}\right) ^{2}T\left(
Z,Z_{1}\right) \left\vert \Psi _{0}\left( Z_{1}\right) \right\vert ^{2}dZ_{1}
\notag
\end{eqnarray}
and:%
\begin{equation*}
T\Psi _{0}\delta \Psi \left( \theta ,Z\right) =\int \frac{\kappa T\left(
Z,Z_{1}\right) }{N}\Psi _{0}\left( Z_{1}\right) \delta \Psi \left( \theta -%
\frac{\left\vert Z-Z_{1}\right\vert }{c},Z_{1}\right) dZ_{1}
\end{equation*}%
\begin{equation*}
\delta T\Psi _{0}=\int \frac{\kappa \delta T\left( Z,Z_{1},\theta \right) }{N%
}\Psi _{0}\left( Z_{1}\right) dZ_{1}
\end{equation*}

Substracting $F^{-1}\left( \omega _{0}\right) $, where $\omega _{0}$ is the
static solution for activity, equation (\ref{vnq}) becomes:%
\begin{equation}
F^{-1}\left( \omega \left( J\left( \theta \right) ,\theta \right) \right)
-F^{-1}\left( \omega _{0}\right) =J\left( \theta ,Z\right) +\frac{\hat{f}%
_{1}\nabla _{\theta }\omega \left( \theta ,Z\right) }{\omega \left( \theta
,Z\right) }+\frac{\hat{f}_{3}\nabla _{\theta }^{2}\omega \left( \theta
,Z\right) }{\omega \left( \theta ,Z\right) }+c^{2}\hat{f}_{3}\frac{\nabla
_{Z}^{2}\omega \left( \theta ,Z\right) }{\omega \left( \theta ,Z\right) }%
+T\Psi _{0}\delta \Psi \left( \theta ,Z\right)  \label{fsr}
\end{equation}%
Using also (\ref{psf}):%
\begin{equation}
\delta \Psi \left( \theta ,Z\right) \simeq \frac{\nabla _{\theta }\omega
\left( J,Z,\left\vert \Psi \right\vert ^{2}\right) }{V^{\prime \prime
}\left( \Psi _{0}\left( Z\right) \right) \omega _{0}^{2}\left(
J,Z,\left\vert \Psi _{0}\right\vert ^{2}\right) }\Psi _{0}  \label{bgd}
\end{equation}%
leads to rewrite the last term in (\ref{fsr}):%
\begin{eqnarray*}
T\delta \Psi \left( \theta ,Z\right) &\simeq &\delta \Psi \left( \theta
,Z\right) -\Gamma _{1}\nabla _{\theta }\delta \Psi \left( \theta ,Z\right) \\
&\simeq &N_{1}\nabla _{\theta }\omega _{0}\left( J,Z,\left\vert \Psi
_{0}\right\vert ^{2}\right) -N_{2}\nabla _{\theta }\omega _{0}\left(
J,Z,\left\vert \Psi _{0}\right\vert ^{2}\right)
\end{eqnarray*}%
with:%
\begin{eqnarray*}
N_{1} &=&\frac{\Psi _{0}\left( Z\right) }{U^{\prime \prime }\left(
X_{0}\right) \omega ^{2}\left( J,Z,\left\vert \Psi _{0}\right\vert
^{2}\right) } \\
N_{2} &=&\frac{\Gamma _{1}\Psi _{0}\left( Z\right) }{U^{\prime \prime
}\left( X_{0}\right) \omega ^{2}\left( J,Z,\left\vert \Psi _{0}\right\vert
^{2}\right) }
\end{eqnarray*}%
We assume that $F^{-1}$ is slowly varying, so that:%
\begin{equation*}
F^{-1}\left( \omega \left( J\left( \theta \right) ,\theta \right) \right)
-F^{-1}\left( \omega _{0}\right) \simeq \Gamma _{0}\left( \omega \left(
J\left( \theta \right) ,\theta \right) -\omega _{0}\right)
\end{equation*}%
with\footnote{%
Given our assumption that $F$ is an increasing function, $f>0$.
\par
{}}:%
\begin{equation*}
f=\left( F^{-1}\right) ^{\prime }\left( \frac{\kappa }{N}\int T\left(
Z,Z_{1}\right) W\left( 1\right) dZ_{1}\mathcal{\bar{G}}_{0}\left(
0,Z_{1}\right) \right)
\end{equation*}%
and define:%
\begin{equation*}
\Omega \left( \theta ,Z\right) =\omega \left( \theta ,Z\right) -\omega _{0}
\end{equation*}%
As a result, the expansion of (\ref{fsr}) for a non-static current is then:%
\begin{equation}
f\Omega \left( \theta ,Z\right) =J\left( \theta ,Z\right) +\left( \frac{\hat{%
f}_{1}}{\omega \left( \theta ,Z\right) }+N_{1}\right) \nabla _{\theta
}\Omega \left( \theta ,Z\right) +\left( \frac{\hat{f}_{3}}{\omega \left(
\theta ,Z\right) }-N_{2}\right) \nabla _{\theta }^{2}\Omega \left( \theta
,Z\right) +\frac{c^{2}\hat{f}_{3}}{\omega \left( \theta ,Z\right) }\nabla
_{Z}^{2}\Omega \left( \theta ,Z\right)  \label{pw}
\end{equation}

A careful study of this equation is performed in (\cite{GL}). We show that
this equation has non sinusoidal stable traveling wave solutions and that in
first approximation it can be replaced by an usual wave equation:%
\begin{equation}
f\Omega \left( \theta ,Z\right) -\left( \frac{\hat{f}_{3}}{\omega _{0}}%
-N_{2}\right) \nabla _{\theta }^{2}\Omega \left( \theta ,Z\right) -\frac{%
c^{2}\hat{f}_{3}}{\omega _{0}}\nabla _{Z}^{2}\Omega \left( \theta ,Z\right)
=J\left( \theta ,Z\right)  \label{wvn}
\end{equation}%
where $\omega _{0}$ is the average of the static activity.

\subsubsection*{2.1.2 Perturbative corrections to the local frequency
equation}

The perturbative expansion of the path integral for the field action (\ref%
{flt}) modifies the activities equation. We computed this effective action,
written $\Gamma \left( \Psi ^{\dagger },\Psi \right) $, in (\cite{GL}). It
is not equal to $\hat{S}\left( \Psi ^{\dagger },\Psi \right) $ defined in (%
\ref{flt}) since the dependency of $\omega ^{-1}\left( J,\theta
,Z,\left\vert \Psi \right\vert ^{2}\right) $ in $\left\vert \Psi \right\vert
^{2}$ introduces self interaction trms.

In the local approximation, this effective action corrects (\ref{flt}) by a
series expansion in field:

\begin{equation}
\Gamma \left( \Psi ^{\dagger },\Psi \right) \simeq \int \Psi ^{\dagger
}\left( \theta ,Z\right) \left( -\nabla _{\theta }\left( \frac{\sigma
_{\theta }^{2}}{2}\nabla _{\theta }-\omega ^{-1}\left( J\left( \theta
\right) ,\theta ,Z,\mathcal{G}_{0}+\left\vert \Psi \right\vert ^{2}\right)
\right) \delta \left( \theta _{f}-\theta _{i}\right) +\Omega \left( \theta
,Z\right) \right) \Psi \left( \theta ,Z\right)  \label{gmf}
\end{equation}%
where $\Omega \left( \theta ,Z\right) $ is a corrective term depending on
the successive\ derivatives of the field (the constants $a_{j}$ are derived
in (\cite{GL})):%
\begin{eqnarray}
\Omega \left( \theta ,Z\right) &=&\int \sum_{\substack{ j\geqslant 1  \\ %
m\geqslant 1}}\sum_{\substack{ \left( p_{l}^{i}\right) _{m\times j}  \\ %
p_{l}+\sum_{i}p_{l}^{i}\geqslant 2}}\frac{a_{j}}{j!}\frac{\delta
^{\sum_{l}p_{l}}\left[ -\frac{1}{2}\left( \nabla _{\theta }\left( \frac{%
\sigma _{\theta }^{2}}{2}\nabla _{\theta }-\omega ^{-1}\left( \left\vert
\Psi \left( \theta ,Z\right) \right\vert ^{2}\right) \right) \right) \right] 
}{\prod\limits_{l=1}^{j}\prod\limits_{k_{l}^{i}=1}^{p_{l}}\delta \left\vert
\Psi \left( \theta ^{\left( l\right) },Z_{_{l}}\right) \right\vert ^{2}}\Psi
\left( \theta ,Z\right)  \label{mG} \\
&&\times \left( \prod\limits_{l=1}^{j}\Psi ^{\dagger }\left( \theta
_{f}^{\left( l\right) },Z_{l}\right) \right) \prod\limits_{i=1}^{m}\left[ 
\frac{\delta ^{\sum_{l}p_{l}^{i}}\left[ \hat{S}_{cl,\theta }\left( \Psi
^{\dagger },\Psi \right) \right] }{\prod\limits_{l=1}^{j}\prod%
\limits_{k_{l}^{i}=1}^{p_{l}^{i}}\delta \left\vert \Psi \left( \theta
^{\left( l\right) },Z_{_{l}}\right) \right\vert ^{2}}\right] \left(
\prod\limits_{l=1}^{j}\Psi \left( \theta _{i}^{\left( l\right)
},Z_{l}\right) \right)  \notag
\end{eqnarray}%
The term $\mathcal{G}_{0}$ is a function of $Z$ and represents a two points
free Green function (see (\cite{GL})).

The previous equation (\ref{gmf}) defines an effective activity that can be
identified as:%
\begin{equation}
\omega _{e}^{-1}\left( J\left( \theta \right) ,\theta ,Z,\mathcal{G}%
_{0}+\left\vert \Psi \right\vert ^{2}\right) =\omega ^{-1}\left( J\left(
\theta \right) ,\theta ,Z,\mathcal{G}_{0}+\left\vert \Psi \right\vert
^{2}\right) +\int^{\theta }\Omega \left( \theta ,Z\right)  \label{fft}
\end{equation}%
where $\omega \left( J\left( \theta \right) ,\theta ,Z,\mathcal{\bar{G}}%
_{0}+\left\vert \Psi \right\vert ^{2}\right) $ is the solution of:%
\begin{eqnarray*}
\omega ^{-1}\left( J,\theta ,Z,\left\vert \Psi \right\vert ^{2}\right)
&=&G\left( J\left( \theta ,Z\right) +\int \frac{\kappa }{N}\frac{\omega
\left( J,\theta -\frac{\left\vert Z-Z_{1}\right\vert }{c},Z_{1},\Psi \right)
T\left( Z,\theta ,Z_{1},\theta -\frac{\left\vert Z-Z_{1}\right\vert }{c}%
\right) }{\omega \left( J,\theta ,Z,\left\vert \Psi \right\vert ^{2}\right) }%
\right. \\
&&\times \left. \left( \mathcal{\bar{G}}_{0}\left( 0,Z_{1}\right)
+\left\vert \Psi \left( \theta -\frac{\left\vert Z-Z_{1}\right\vert }{c}%
,Z_{1}\right) \right\vert ^{2}\right) dZ_{1}\right)
\end{eqnarray*}%
Which is the classical activity equation, up to the inclusion of the Green
function $\mathcal{\bar{G}}_{0}\left( 0,Z_{1}\right) $.

The second term $\int^{\theta }\Omega \left( \theta ,Z\right) $ in (\ref{fft}%
) represents corrections due to the interactions. Using (\ref{mG}), we can
find its expression as a series expansion in terms of activities and field.
The computations of these corrections to the classical equation are
presented in (\cite{GL}) and confirm the possibility of traveling wave
solutions. At the lowest order, we find:%
\begin{equation*}
\int^{\theta }\Omega \left( \theta ,Z\right) =\frac{1}{4}\int \int^{\theta }%
\frac{\delta \left( \nabla _{\theta }\omega ^{-1}\left( J\left( \theta
\right) ,\theta ,Z,\mathcal{G}_{0}+\left\vert \Psi \right\vert ^{2}\right)
\right) }{\delta \left\vert \Psi \left( \theta ^{\left( l\right)
},Z_{_{l}}\right) \right\vert ^{2}}\times \frac{\delta \left( \nabla
_{\theta }\omega ^{-1}\left( J\left( \theta \right) ,\theta ,Z,\mathcal{G}%
_{0}+\left\vert \Psi \right\vert ^{2}\right) \right) }{\delta \left\vert
\Psi \left( \theta ^{\left( l\right) },Z_{_{l}}\right) \right\vert ^{2}}%
\left\vert \Psi \left( \theta ^{\left( l\right) },Z_{_{l}}\right)
\right\vert ^{2}
\end{equation*}%
and we show that this term counters the variations of $\omega ^{-1}\left(
J\left( \theta \right) ,\theta ,Z,\mathcal{G}_{0}+\left\vert \Psi
\right\vert ^{2}\right) $ and thus stabilizes the oscillations.

To sum up, the perturbative corrections account for interactions between the
classical solutions and the entire thread. Moreover these interactions
stabilize the traveling waves.

\subsection*{2.3 External signals and connectivity switching}

We have given in (\ref{LV}) a first approximation of equilibrium values for
connectivity functions. Based on wave dynamics in the previous paragraph,
this section presents the effect of external signals on these values. The
complete derivation is given in part I and II. The result obtained will also
apply to the modified connectivities (\ref{VRP}) as these ones are
considered in this work as independent activated states.

\subsubsection*{2.3.1 Sources induced activities}

In the perspective of this work, we are looking at the solutions of (\ref%
{wvn}) induced by some ponctual sources. Assume several signals arising at
some points $\left( Z_{1},\theta _{1}\right) ,...,\left( Z_{N},\theta
_{N}\right) $.

The solution to (\ref{wvn}) are then:%
\begin{equation}
\Omega \left( Z,\theta \right) =\sum_{i=1}^{N}\emph{G}\left( \left( Z,\theta
\right) ,\left( Z_{i},\theta _{i}\right) \right) J\left( Z_{i},\theta
_{i}\right)  \label{lcs}
\end{equation}%
Where $\emph{G}\left( \left( Z,\theta \right) ,\left( Z_{i},\theta
_{i}\right) \right) $ is the Green function of:%
\begin{equation*}
f-\left( \frac{\hat{f}_{3}}{\omega _{0}}-N_{2}\right) \nabla _{\theta }^{2}-%
\frac{c^{2}\hat{f}_{3}}{\omega _{0}}\nabla _{Z}^{2}
\end{equation*}%
The non local equation (\ref{vnq}) in presence of sources has solutions:%
\begin{equation}
\Omega \left( \theta ,Z\right) \simeq \sum_{i=1}^{N}\emph{G}_{T}\left(
\left( Z,\theta \right) ,\left( Z_{i},\theta _{i}\right) \right) \frac{%
J\left( Z_{i},\theta _{i}\right) }{1+\left\langle T\right\rangle _{\Psi _{0}}%
}  \label{nls}
\end{equation}%
where $\emph{G}_{T}\left( \left( Z,\theta \right) ,\left( Z_{i},\theta
_{i}\right) \right) $ is defined by:%
\begin{equation*}
\emph{G}_{T}\left( \left( Z,\theta \right) ,\left( Z_{i},\theta _{i}\right)
\right) =\left( \frac{1}{1-G_{T}}\right) \left( \left( Z,\theta \right)
,\left( Z_{i},\theta _{i}\right) \right)
\end{equation*}%
and $G_{T}$ is the operator with kernel $G_{T}\left( Z,\theta ,Z_{1},\theta -%
\frac{\left\vert Z-Z_{1}\right\vert }{c}\right) $. Equation (\ref{nls}) is
the local version of (\ref{lcs}). Both solutions present interference
phenomena. When the number of sources is large, we may expect that solutions
of (\ref{nls}) and (\ref{lcs}) locate mainly at some maxima depending both
on the connectivity field $\left\vert \Gamma \left( T,\hat{T},\theta
,Z,Z^{\prime }\right) \right\vert ^{2}$ and neuron field. In the sequel, we
will write:

\begin{equation*}
Z_{M}^{\left( \varepsilon \right) }\left( \left\vert \Gamma \right\vert
^{2},\left\vert \Psi _{0}\right\vert ^{2}\right)
\end{equation*}%
the location of these maxima, with $\varepsilon =1,...$ indexing these
maxima. We will also assume that at these maxima, the activities are all
equal to some value:%
\begin{equation*}
\omega \simeq \omega ^{\prime }\simeq \omega _{M}
\end{equation*}%
so that:%
\begin{eqnarray*}
h_{C}\left( \omega \right) &\simeq &h_{C}\left( \omega _{M}\right) \\
h_{D}\left( \omega \right) &\simeq &h_{D}\left( \omega _{M}\right)
\end{eqnarray*}%
The precise derivation of the interference phenomenom has been presented in
a field theoretic context in part II. It is sufficient for the rest of the
section to build on the qualititative argument we presented here.

\subsubsection*{2.3.2 Effective action and background state for given
sources states}

We have seen that for a given external state, interference arise, and
activities localize at some points 
\begin{equation*}
Z_{M}^{\left( \varepsilon \right) }\left( \left\vert \Gamma \right\vert
^{2},\left\vert \Psi _{0}\right\vert ^{2}\right)
\end{equation*}%
with $\varepsilon =1,...$ indexing these points. We will also assumed for
the sake of simplicity, that that at these maxima, the activities are all
equal to some value:%
\begin{equation*}
\omega \simeq \omega ^{\prime }\simeq \omega _{M}
\end{equation*}%
so that:%
\begin{eqnarray*}
h_{C}\left( \omega \right) &\simeq &h_{C}\left( \omega _{M}\right) \\
h_{D}\left( \omega \right) &\simeq &h_{D}\left( \omega _{M}\right)
\end{eqnarray*}%
Assuming that functions $h_{C}\left( \omega \right) $ and $h_{D}\left(
\omega \right) $ are proportional to some positive power of $\omega $
implies that outside the set of points $U_{M}=\left\{ Z_{M}^{\left(
\varepsilon \right) }\left( \left\vert \Gamma \right\vert ^{2},\left\vert
\Psi _{0}\right\vert ^{2}\right) \right\} $, $h_{C}\left( \omega \right) $
and $h_{D}\left( \omega \right) $ can be considered as nul. We will write $Z$
the generic points of the complementary set of $U_{M}$, written $CU_{M}$. In
the context of this work, the points $U_{M}$ correspond to an activated
collective state.

We compute average connectivity between points of $U_{M}$, between points of 
$CU_{M}$, and between points of $U_{M}$ and $CU_{M}$.

\subsubsection*{2.3.3 Connectivity between points of $U_{M}$}

The background state at points $\left( Z_{M}^{\left( \varepsilon _{1}\right)
},Z_{M}^{\left( \varepsilon _{2}\right) }\right) \subset U_{M}$ is similar
to (\ref{gmv}):%
\begin{eqnarray*}
&&\Gamma \left( T,\hat{T},\theta ,Z_{M}^{\left( \varepsilon _{1}\right)
},Z_{M}^{\left( \varepsilon _{2}\right) }\right) \\
&=&\exp \left( -\left( \left( -\frac{1}{\tau \omega _{M}}T+\frac{\lambda }{%
\omega _{M}}\left\langle \hat{T}\right\rangle \right) \left\vert \Psi \left(
\theta ,Z_{M}^{\left( \varepsilon _{1}\right) }\right) \right\vert
^{2}\right) ^{2}\right) \\
&&\times \exp \left( -\left( \frac{\rho }{\omega _{M}}\left( \left( h\left(
Z_{M}^{\left( \varepsilon _{1}\right) },Z_{M}^{\left( \varepsilon
_{2}\right) }\right) -\hat{T}\right) C\left( \theta \right) h_{C}-D\left(
\theta \right) \hat{T}h_{D}\right) \left\vert \left[ \Psi .\Psi \right]
\left( \theta ,Z_{M}^{\left( \varepsilon _{1}\right) },Z_{M}^{\left(
\varepsilon _{2}\right) }\right) \right\vert ^{2}\right) ^{2}\right)
\end{eqnarray*}%
wth:%
\begin{equation*}
\left\vert \left[ \Psi .\Psi \right] \left( \theta ,Z_{M}^{\left(
\varepsilon _{1}\right) },Z_{M}^{\left( \varepsilon _{2}\right) }\right)
\right\vert ^{2}=\left\vert \Psi \left( \theta ,Z_{M}^{\left( \varepsilon
_{1}\right) }\right) \right\vert ^{2}\left\vert \Psi \left( \theta -\frac{%
\left\vert Z_{M}^{\left( \varepsilon _{1}\right) }-Z_{M}^{\left( \varepsilon
_{2}\right) }\right\vert }{c},Z_{M}^{\left( \varepsilon _{2}\right) }\right)
\right\vert ^{2}
\end{equation*}%
and the average values in this background states satisfy:

\begin{eqnarray*}
C_{Z_{M}^{\left( \varepsilon _{1}\right) },Z_{M}^{\left( \varepsilon
_{2}\right) }} &=&\frac{\alpha _{C}\omega _{M}\left\vert \Psi \left( \theta -%
\frac{\left\vert Z_{M}^{\left( \varepsilon _{1}\right) }-Z_{M}^{\left(
\varepsilon _{2}\right) }\right\vert }{c},Z_{M}^{\left( \varepsilon
_{2}\right) }\right) \right\vert ^{2}}{\frac{1}{\tau _{C}}+\alpha _{C}\omega
_{M}\left\vert \Psi \left( \theta -\frac{\left\vert Z_{M}^{\left(
\varepsilon _{1}\right) }-Z_{M}^{\left( \varepsilon _{2}\right) }\right\vert 
}{c},Z_{M}^{\left( \varepsilon _{2}\right) }\right) \right\vert ^{2}} \\
D_{Z_{M}^{\left( \varepsilon _{1}\right) },Z_{M}^{\left( \varepsilon
_{2}\right) }} &=&\frac{\alpha _{D}\omega _{M}}{\frac{1}{\tau _{D}}+\alpha
_{D}\omega _{M}\left\vert \Psi \left( \theta ,Z_{M}^{\left( \varepsilon
_{1}\right) }\right) \right\vert ^{2}}
\end{eqnarray*}

\begin{equation*}
T\left( Z_{M}^{\left( \varepsilon _{1}\right) },Z_{M}^{\left( \varepsilon
_{2}\right) }\right) =\lambda \tau \hat{T}\left( Z_{M}^{\left( \varepsilon
_{1}\right) },Z_{M}^{\left( \varepsilon _{2}\right) }\right) =\lambda \tau 
\frac{h\left( Z_{M}^{\left( \varepsilon _{1}\right) },Z_{M}^{\left(
\varepsilon _{2}\right) }\right) C_{Z_{M}^{\left( \varepsilon _{1}\right)
},Z_{M}^{\left( \varepsilon _{2}\right) }}\left( \theta \right) h_{C}}{%
C_{Z_{M}^{\left( \varepsilon _{1}\right) },Z_{M}^{\left( \varepsilon
_{2}\right) }}\left( \theta \right) h_{C}+D_{Z_{M}^{\left( \varepsilon
_{1}\right) },Z_{M}^{\left( \varepsilon _{2}\right) }}\left( \theta \right)
h_{D}}
\end{equation*}%
For an exponential dependency of connectivities in the distance between the
connected points: 
\begin{equation*}
h\left( Z_{M}^{\left( \varepsilon _{1}\right) },Z_{M}^{\left( \varepsilon
_{2}\right) }\right) \simeq \exp \left( -\frac{\left\vert Z_{M}^{\left(
\varepsilon _{1}\right) }-Z_{M}^{\left( \varepsilon _{2}\right) }\right\vert 
}{\nu c}\right)
\end{equation*}%
we obtain the average connectivity at points impacted by signals:%
\begin{equation}
T\left( Z_{M}^{\left( \varepsilon _{1}\right) },Z_{M}^{\left( \varepsilon
_{2}\right) }\right) =\frac{\lambda \tau \exp \left( -\frac{\left\vert
Z_{M}^{\left( \varepsilon _{1}\right) }-Z_{M}^{\left( \varepsilon
_{2}\right) }\right\vert }{\nu c}\right) \left\vert \Psi \left( \theta -%
\frac{\left\vert Z_{M}^{\left( \varepsilon _{1}\right) }-Z_{M}^{\left(
\varepsilon _{2}\right) }\right\vert }{c},Z_{M}^{\left( \varepsilon
_{2}\right) }\right) \right\vert ^{2}h_{C}}{\left\vert \Psi \left( \theta -%
\frac{\left\vert Z_{M}^{\left( \varepsilon _{1}\right) }-Z_{M}^{\left(
\varepsilon _{2}\right) }\right\vert }{c},Z_{M}^{\left( \varepsilon
_{2}\right) }\right) \right\vert ^{2}h_{C}+\frac{\left( \frac{1}{\alpha
_{C}\tau _{C}}+\omega _{M}\left\vert \Psi \left( \theta -\frac{\left\vert
Z_{M}^{\left( \varepsilon _{1}\right) }-Z_{M}^{\left( \varepsilon
_{2}\right) }\right\vert }{c},Z_{M}^{\left( \varepsilon _{2}\right) }\right)
\right\vert ^{2}\right) \alpha _{D}h_{D}}{\frac{1}{\tau _{D}}+\alpha
_{D}\omega _{M}\left\vert \Psi \left( \theta ,Z_{M}^{\left( \varepsilon
_{1}\right) }\right) \right\vert ^{2}}}  \label{dfM}
\end{equation}%
In a lon- run static perspective, it becomes:%
\begin{equation*}
T\left( Z_{M}^{\left( \varepsilon _{1}\right) },Z_{M}^{\left( \varepsilon
_{2}\right) }\right) =\frac{\lambda \tau \exp \left( -\frac{\left\vert
Z_{M}^{\left( \varepsilon _{1}\right) }-Z_{M}^{\left( \varepsilon
_{2}\right) }\right\vert }{\nu c}\right) \left\vert \Psi _{0}\left(
Z_{M}^{\left( \varepsilon _{2}\right) }\right) \right\vert ^{2}h_{C}}{%
\left\vert \Psi _{0}\left( Z_{M}^{\left( \varepsilon _{2}\right) }\right)
\right\vert ^{2}h_{C}+\left( \frac{1}{\alpha _{C}\tau _{C}}+\omega
_{M}\left\vert \Psi _{0}\left( Z_{M}^{\left( \varepsilon _{2}\right)
}\right) \right\vert ^{2}\right) \frac{\alpha _{D}h_{D}}{\frac{1}{\tau _{D}}%
+\alpha _{D}\omega _{M}\left\vert \Psi _{0}\left( Z_{M}^{\left( \varepsilon
_{1}\right) }\right) \right\vert ^{2}}}
\end{equation*}

The system has to be supplemented with long-term determination of activities 
$\omega _{M}$:%
\begin{eqnarray*}
\omega _{M}^{-1}\left( Z_{M}^{\left( \varepsilon _{1}\right) },\left\vert
\Psi \right\vert ^{2}\right) &\simeq &G\left( \frac{\kappa }{N}%
\sum_{Z_{M}^{\left( \varepsilon _{2}\right) }}T\left( Z_{M}^{\left(
\varepsilon _{1}\right) },Z_{M}^{\left( \varepsilon _{2}\right) }\right)
\left\vert \Psi _{0}\left( Z_{M}^{\left( \varepsilon _{2}\right) }\right)
\right\vert ^{2}\right) \\
&\simeq &G\left( C\frac{\left\vert \Psi _{0}\left( Z_{M}\right) \right\vert
^{4}h_{C}}{\left\vert \Psi _{0}\left( Z_{M}\right) \right\vert
^{2}h_{C}+\left( \frac{1}{\alpha _{C}\tau _{C}}+\omega _{M}\left\vert \Psi
_{0}\left( Z_{M}\right) \right\vert ^{2}\right) \frac{\alpha _{D}h_{D}}{%
\frac{1}{\tau _{D}}+\alpha _{D}\omega _{M}\left\vert \Psi _{0}\left(
Z_{M}\right) \right\vert ^{2}}}\right)
\end{eqnarray*}%
where:%
\begin{equation*}
C=\frac{\kappa \lambda \tau }{N\left( \sharp \left\{ Z_{M}^{\left(
\varepsilon _{1}\right) }\right\} \right) }\sum_{Z_{M}^{\left( \varepsilon
_{1}\right) },Z_{M}^{\left( \varepsilon _{2}\right) }}\exp \left( -\frac{%
\left\vert Z_{M}^{\left( \varepsilon _{1}\right) }-Z_{M}^{\left( \varepsilon
_{2}\right) }\right\vert }{\nu c}\right)
\end{equation*}%
and $\left\vert \Psi _{0}\left( Z_{M}\right) \right\vert ^{2}$ is the
average of $\left\vert \Psi _{0}\left( Z_{M}^{\left( \varepsilon _{2}\right)
}\right) \right\vert ^{2}$ over $\left\{ Z_{M}^{\left( \varepsilon
_{2}\right) }\right\} $. The value of $\left\vert \Psi _{0}\left(
Z_{M}\right) \right\vert ^{2}$ can be approximated in the following way. We
have seen in part II in the field theoretic approach to the interferences
that the signals modify the potential for $\left\vert \Psi _{0}\left(
Z\right) \right\vert ^{2}$ but that in first approximation, this
modification can be neglected. Thus, the value of $\left\vert \Psi
_{0}\left( Z\right) \right\vert ^{2}$ after interferences may be replaced in
first approximation by the background field before interferences. This is
formula (\ref{Ps}):%
\begin{equation*}
\left\vert \Psi _{0}\left( Z\right) \right\vert ^{2}=\frac{2T\left( Z\right)
\left\langle \left\vert \Psi _{0}\left( Z^{\prime }\right) \right\vert
^{2}\right\rangle _{Z}}{\left( 1+\sqrt{1+4\left( \frac{\lambda \tau \nu
c-T\left( Z\right) }{\left( \frac{1}{\tau _{D}\alpha _{D}}+\frac{1}{\tau
_{C}\alpha _{C}}+\Omega \right) T\left( Z\right) -\frac{1}{\tau _{D}\alpha
_{D}}\lambda \tau \nu c}\right) ^{2}T\left( Z\right) \left\langle \left\vert
\Psi _{0}\left( Z^{\prime }\right) \right\vert ^{2}\right\rangle _{Z}}%
\right) }
\end{equation*}%
where all quantities are computed in the initial background state. The
system emrging from the interferences thus depends on the whole initial
structure.

\subsubsection*{2.3.4 Connectivity between points of $CU_{M}$}

The connectivity function for two points in $CU_{M}$ is obtained by setting $%
\omega <<1$ and $\omega ^{\prime }<<1$:

\begin{equation*}
T\left( Z,Z^{\prime }\right) \simeq \frac{\alpha _{C}\omega \lambda \tau
\exp \left( -\frac{\left\vert Z-Z^{\prime }\right\vert }{\nu c}\right)
\left\vert \Psi \left( \theta -\frac{\left\vert Z-Z^{\prime }\right\vert }{c}%
,Z^{\prime }\right) \right\vert ^{2}h_{C}}{\alpha _{C}\omega \left\vert \Psi
\left( \theta -\frac{\left\vert Z-Z^{\prime }\right\vert }{c},Z^{\prime
}\right) \right\vert ^{2}h_{C}+\left( \frac{\omega ^{\prime }}{\tau _{C}}%
\right) \frac{\alpha _{D}h_{D}}{\frac{\left\vert \Psi \left( \theta
,Z\right) \right\vert ^{2}}{\tau _{D}}}}
\end{equation*}%
and these values are identitical to those computed for the static background
state in the previous section (see (\ref{cnv})), up to some global
modifications of the system by the interfering signals. These modifications
are encompassed in the values of the constants $\Omega $, $\bar{\Omega}$...
in (\ref{cnv}). These modifications are negligible in general.

\subsubsection*{2.3.5 Connectivity between points of $U_{M}$ and points of $%
CU_{M}$}

Two cases arise. The connectivity function for two points in $CU_{M}$ are
obtained by setting $\omega =\omega _{M}$ and $\omega ^{\prime }<<1$ or $%
\omega <<1$ and $\omega ^{\prime }=\omega _{M}$.%
\begin{equation}
T\left( Z_{M}^{\left( \varepsilon \right) },Z^{\prime }\right) \simeq \frac{%
\lambda \tau \exp \left( -\frac{\left\vert Z-Z^{\prime }\right\vert }{\nu c}%
\right) \left\vert \Psi \left( \theta -\frac{\left\vert Z-Z^{\prime
}\right\vert }{c},Z^{\prime }\right) \right\vert ^{2}h_{C}}{\left\vert \Psi
\left( \theta -\frac{\left\vert Z-Z^{\prime }\right\vert }{c},Z^{\prime
}\right) \right\vert ^{2}h_{C}+\left\vert \Psi \left( \theta -\frac{%
\left\vert Z-Z^{\prime }\right\vert }{c},Z^{\prime }\right) \right\vert ^{2}%
\frac{\alpha _{D}\omega _{M}h_{D}}{\frac{\left\vert \Psi \left( \theta
,Z\right) \right\vert ^{2}}{\tau _{D}}+\alpha _{D}\omega _{M}}}  \label{cnn}
\end{equation}%
\begin{equation}
T\left( Z,Z_{M}^{\left( \varepsilon \right) }\right) \simeq \frac{\alpha
_{C}\omega \lambda \tau \exp \left( -\frac{\left\vert Z-Z^{\prime
}\right\vert }{\nu c}\right) \left\vert \Psi \left( \theta -\frac{\left\vert
Z-Z^{\prime }\right\vert }{c},Z^{\prime }\right) \right\vert ^{2}h_{C}}{%
\alpha _{C}\omega \left\vert \Psi \left( \theta -\frac{\left\vert
Z-Z^{\prime }\right\vert }{c},Z^{\prime }\right) \right\vert ^{2}h_{C}+\frac{%
\omega _{M}}{\tau _{C}}\frac{\alpha _{D}h_{D}\tau _{D}}{\left\vert \Psi
\left( \theta ,Z\right) \right\vert ^{2}}}<<1  \label{nT}
\end{equation}%
As a consequence, points of the set $U$ do not connect with elements of $CU$%
. On the contrary, elements of $CU$ send signals and connect to elements of $%
U$ but their firing rate being slow, they do not influence the whole set
that remains unaffected.

In the perspective of this article, this means that we can consider external
sources affecting a single collective state.

\subsection*{2.4 Remark: Field theoretic transcription of source-induced
modifications}

Note that the switches:%
\begin{equation*}
T\left( Z_{M},Z_{M}\right) \rightarrow T\left( Z_{M}^{\left( \varepsilon
_{1}\right) },Z_{M}^{\left( \varepsilon _{2}\right) }\right)
\end{equation*}%
due to the sources can be implemented in the following manner. At points $%
\left( Z_{M}^{\left( \varepsilon _{1}\right) },Z_{M}^{\left( \varepsilon
_{2}\right) }\right) $, the saddle point equation (\ref{fcn}):%
\begin{eqnarray}
&&0=\left( -\sigma _{\hat{T}}^{2}\nabla _{\hat{T}}^{2}+\frac{1}{4\sigma _{%
\hat{T}}^{2}}\left( \left\vert \bar{\Psi}\left( Z,Z^{\prime }\right)
\right\vert ^{2}\Delta \hat{T}+\frac{\rho \left( V_{0}-\frac{\sigma _{\hat{T}%
}^{2}}{\sigma _{T}^{2}}\lambda \Delta T\left\vert \Psi \left( Z\right)
\right\vert ^{2}\right) }{\omega _{0}\left( Z\right) }\right) ^{2}\right)
\Delta \Gamma \left( T,\hat{T},\theta ,Z,Z^{\prime }\right) \\
&&+\left( -\sigma _{T}^{2}\nabla _{T}^{2}+\frac{1}{4\sigma _{T}^{2}}\left( 
\frac{\Delta T-\lambda \tau \Delta \hat{T}}{\tau \omega _{0}\left( Z\right) }%
\right) ^{2}\right.  \notag \\
&&\left. -\left( \frac{\left\vert \bar{\Psi}\left( Z,Z^{\prime }\right)
\right\vert ^{2}}{2}+\frac{\left\vert \Psi \left( Z\right) \right\vert ^{2}}{%
2\tau \omega _{0}\left( Z\right) }+V\left( \theta ,Z,Z^{\prime },\Delta
\Gamma \right) \Delta T-\alpha \right) \right) \Delta \Gamma \left( T,\hat{T}%
,\theta ,Z,Z^{\prime }\right)  \notag
\end{eqnarray}%
is quadratic in connectivities. As a consequence adding to this equation a
term:%
\begin{equation*}
\frac{\left( \Delta T\left( Z_{M},Z_{M}\right) -T\left( Z_{M}^{\left(
\varepsilon _{1}\right) },Z_{M}^{\left( \varepsilon _{2}\right) }\right)
\right) \Delta T}{2\sigma _{T}^{2}\left( \tau \omega _{0}\left( Z\right)
\right) ^{2}}\Delta \Gamma \left( T,\hat{T},\theta ,Z,Z^{\prime }\right)
\end{equation*}%
with:%
\begin{equation*}
\Delta T\left( Z_{M},Z_{M}\right) =T\left( Z_{M}^{\left( \varepsilon
_{1}\right) },Z_{M}^{\left( \varepsilon _{2}\right) }\right) -T\left(
Z_{M},Z_{M}\right)
\end{equation*}%
will shift the average, in first approximation, by an amount:%
\begin{equation*}
\Delta T\rightarrow \Delta T+\Delta T\left( Z_{M},Z_{M}\right)
\end{equation*}%
In terms of action functional, this corresponds to add a term in (\ref{FCWZ}%
):%
\begin{equation*}
\Delta \Gamma \left( T,\hat{T},\theta ,Z,Z^{\prime }\right) J\left( \Delta
T,Z,Z^{\prime }\right) \Delta \Gamma \left( T,\hat{T},\theta ,Z,Z^{\prime
}\right)
\end{equation*}%
where:%
\begin{equation*}
J\left( \Delta T,Z,Z^{\prime }\right) =\frac{\left( \Delta T\left(
Z_{M},Z_{M}\right) -T\left( Z_{M}^{\left( \varepsilon _{1}\right)
},Z_{M}^{\left( \varepsilon _{2}\right) }\right) \right) \Delta T}{2\sigma
_{T}^{2}\left( \tau \omega _{0}\left( Z\right) \right) ^{2}}
\end{equation*}%
This modificaion of action generalizes easily to the collective field
describing the collective states. We introduce in action (\ref{FCWZ}) a term:%
\begin{equation*}
\underline{\Gamma }^{\dag }\left( \Delta \mathbf{T},\mathbf{\alpha },\mathbf{%
p},S^{2},\theta \right) J\left( \Delta \mathbf{T},S^{2},\theta \right) 
\underline{\Gamma }\left( \Delta \mathbf{T},\mathbf{\alpha },\mathbf{p}%
,S^{2},\theta \right)
\end{equation*}%
where $J\left( \Delta \mathbf{T},S^{2},\theta \right) $ encompass the
contributions at every point of: 
\begin{equation*}
J\left( \Delta \mathbf{T},S^{2},\theta \right) =\sum_{Z\in S}J\left( \Delta 
\mathbf{T},Z,\theta \right) J\left( \Delta \mathbf{T},S^{2},\theta \right)
\end{equation*}

\subsection*{2.5 Synthesis: activities of activated states}

We can gather the various results from this section. External signals induce
oscillations in activity that propagate along the thread. Some cells may
bind together at points of positive interference. The activation of a
collective state corresponds to an average activity, as given by the formula
above. Reversing the point of view, we can consider that collective states
of bound cells are characterized by an activity characteristic of the
collective state. The average of this activity can be computed, but we may
also consider that these activities themselves may present oscillating
characteristics. In the next section, we will compute both average levels
and frequencies in activities. An important characteristic will emerge in
this derivation. The solutions for average levels of activity amplitudes and
frequencies may be multiple. A given collective state will thus possibly
exist in multiple (possible infinite) states, labelled by integers. Such
states may undergo transitions, and the collective state may switch from
state to another.

\section*{Appendix 3. Activities for collective field and averages}

As explained in the text, we begin with a certain collective state, i.e. we
consider a large number of elements connected with some average values of
connectivity functions that will be determined later, based on consistency
conditions. Writing the equations for activities in this state leads us to
discover multiple oscillating solutions characterizing the collective state.

To find the activities $\omega \left( \theta ,Z,\left\vert \Psi \right\vert
^{2}\right) $, we start with the defining equation: 
\begin{eqnarray*}
&&\left( \omega _{0}+\Delta \omega \right) ^{-1}\left( J,\theta
,Z,\left\vert \Psi \right\vert ^{2}\right) \\
&=&G\left( J\left( \theta ,Z\right) +\int \frac{\kappa }{N}\frac{\omega
\left( J,\theta -\frac{\left\vert Z-Z_{1}\right\vert }{c},Z_{1},\Psi \right)
\left( \left\langle T\right\rangle +\left\langle \Delta T\right\rangle
\right) \left( Z,\theta ,Z_{1},\theta -\frac{\left\vert Z-Z_{1}\right\vert }{%
c}\right) }{\left( \omega _{0}+\Delta \omega \right) \left( J,\theta
,Z,\left\vert \Psi \right\vert ^{2}\right) }\left\vert \Psi \left( \theta -%
\frac{\left\vert Z-Z_{1}\right\vert }{c},Z_{1}\right) \right\vert
^{2}dZ_{1}\right)
\end{eqnarray*}

\subsection*{Static part}

As in (\cite{GL}), we consider a static part that satisfies: 
\begin{equation*}
\left( \omega _{S}\right) ^{-1}\left( Z,\left\vert \Psi \right\vert
^{2}\right) =G\left( \int \frac{\kappa }{N}\frac{\omega _{S}\left(
Z_{1},\Psi \right) \left( \left\langle T\right\rangle +\left\langle \Delta
T\right\rangle \right) \left( Z,Z_{1}\right) }{\omega _{S}\left(
Z,\left\vert \Psi \right\vert ^{2}\right) }\left\vert \Psi \left(
Z_{1}\right) \right\vert ^{2}dZ_{1}\right)
\end{equation*}%
Using that the modifications states is a group over a bounded domain and
involve some finite number of points we find the equation for the
modification at each point $Z_{i}$ of this group:%
\begin{equation*}
\left( \omega _{S}\right) _{i}^{-1}\left( Z,\left\vert \Psi \right\vert
^{2}\right) =G\left( \sum_{j}\frac{\kappa }{N}\frac{\left( \omega
_{S}\right) _{j}\left( Z_{j},\Psi \right) \left( \left\langle T\right\rangle
+\left\langle \Delta T\right\rangle \right) \left( Z_{i},Z_{j}\right) }{%
\left( \omega _{S}\right) _{i}\left( Z_{i},\left\vert \Psi \right\vert
^{2}\right) }\left\vert \Psi \left( Z_{j}\right) \right\vert ^{2}\right)
\end{equation*}%
with solution:%
\begin{equation*}
\overline{\Delta \omega }\left( \mathbf{Z},\mathbf{T},\left\vert \Psi
\right\vert ^{2}\right)
\end{equation*}%
where $\overline{\Delta \omega }\left( \mathbf{Z},\mathbf{T},\left\vert \Psi
\right\vert ^{2}\right) $ is the vector with coordinates $\overline{\Delta
\omega }\left( Z_{i},T_{i},\left\vert \Psi \right\vert ^{2}\right) $. We
will write:%
\begin{equation*}
\omega _{S}\left( \mathbf{Z}\right) =\omega _{0}\left( \mathbf{Z}\right) +%
\overline{\Delta \omega }\left( \mathbf{Z},\mathbf{T},\left\vert \Psi
\right\vert ^{2}\right)
\end{equation*}%
in the sequel.

\subsection*{Dynamic part}

The first order variation around the static background plus modfcation
solution becomes:

\begin{eqnarray}
&&\Delta \omega _{D}\left( J,\theta ,Z,\left\vert \Psi \right\vert
^{2}\right)   \label{DMG} \\
&=&G^{\prime }\left( G^{-1}\left( \omega _{S}^{-1}\left( J,Z,\left\vert \Psi
\right\vert ^{2}\right) \right) \right)   \notag \\
&&\times \Delta \left( \int \frac{\kappa }{N}\frac{\omega \left( J,\theta -%
\frac{\left\vert Z-Z_{1}\right\vert }{c},Z_{1},\Psi \right) \left(
\left\langle T\right\rangle +\left\langle \Delta T\right\rangle \right)
\left( Z,\theta ,Z_{1},\theta -\frac{\left\vert Z-Z_{1}\right\vert }{c}%
\right) }{\omega \left( J,\theta ,Z,\left\vert \Psi \right\vert ^{2}\right) }%
\left\vert \Psi \left( \theta -\frac{\left\vert Z-Z_{1}\right\vert }{c}%
,Z_{1}\right) \right\vert ^{2}dZ_{1}\right)   \notag
\end{eqnarray}%
where the variation $\Delta $ is given by:%
\begin{eqnarray*}
&&\Delta \left( \int \frac{\kappa }{N}\frac{\left( \omega _{0}+\Delta \omega
\right) \left( J,\theta -\frac{\left\vert Z-Z_{1}\right\vert }{c},Z_{1},\Psi
\right) T\left( Z,\theta ,Z_{1},\theta -\frac{\left\vert Z-Z_{1}\right\vert 
}{c}\right) }{\left( \omega _{0}+\Delta \omega \right) \left( J,\theta
,Z,\left\vert \Psi \right\vert ^{2}\right) }\left\vert \Psi \left( \theta -%
\frac{\left\vert Z-Z_{1}\right\vert }{c},Z_{1}\right) \right\vert
^{2}dZ_{1}\right)  \\
&\simeq &\int \frac{\kappa }{N}\frac{T\left( Z,Z_{1}\right) \left\vert \Psi
_{0}\left( Z_{1}\right) \right\vert ^{2}dZ_{1}}{\omega _{S}\left(
J,Z,\left\vert \Psi \right\vert ^{2}\right) }\Delta \omega _{D}\left(
J,\theta -\frac{\left\vert Z-Z_{1}\right\vert }{c},Z_{1},\Psi \right)  \\
&&-\int \frac{\kappa }{N}\frac{T\left( Z,Z_{1}\right) \omega _{S}\left(
J,Z_{1},\Psi \right) \left\vert \Psi _{0}\left( Z_{1}\right) \right\vert
^{2}dZ_{1}}{\omega _{S}\left( J,Z,\left\vert \Psi \right\vert ^{2}\right) }%
\frac{\Delta \omega _{D}\left( J,\theta ,Z,\left\vert \Psi \right\vert
^{2}\right) }{\omega _{S}\left( J,Z,\left\vert \Psi \right\vert ^{2}\right) }
\\
&&+\int \frac{\kappa }{N}\frac{T\left( Z,Z_{1}\right) \omega _{S}\left(
J,Z_{1},\Psi \right) dZ_{1}}{\omega _{S}\left( J,Z,\left\vert \Psi
\right\vert ^{2}\right) }\Delta \left\vert \Psi _{0}\left( Z_{1},\theta
\right) \right\vert ^{2}
\end{eqnarray*}%
The important point at this stage is the following. We showed in (\cite{GL})
(see an account in appendix 2) that wave dynamics for connectivities implies
that the contribution $\Delta \left\vert \Psi _{0}\left( Z_{1},\theta
\right) \right\vert ^{2}$ stabilizes the dynamics for $\Delta \omega
_{D}^{-1}\left( J,\theta ,Z,\left\vert \Psi \right\vert ^{2}\right) $%
\footnote{%
See also the discussion after(\ref{FRQQ}).}. As a consequence, we can
consider in first approximation that $\Delta \left\vert \Psi _{0}\left(
Z_{1},\theta \right) \right\vert ^{2}=0$ and assume that activities will be
stably oscillating. Moreover, at the level of activities oscillations,
connectivities $T\left( Z,Z_{1}\right) $ will also be considered as static
for average oscillations:

The oscillation part is obtained by:

\begin{eqnarray*}
&&\Delta \left( \int \frac{\kappa }{N}\frac{\omega \left( J,\theta -\frac{%
\left\vert Z-Z_{1}\right\vert }{c},Z_{1},\Psi \right) T\left( Z,\theta
,Z_{1},\theta -\frac{\left\vert Z-Z_{1}\right\vert }{c}\right) }{\omega
\left( J,\theta ,Z,\left\vert \Psi \right\vert ^{2}\right) }\left\vert \Psi
\left( \theta -\frac{\left\vert Z-Z_{1}\right\vert }{c},Z_{1}\right)
\right\vert ^{2}dZ_{1}\right) \\
&\simeq &\int \frac{\kappa }{N}\frac{T\left( Z,Z_{1}\right) \left\vert \Psi
_{0}\left( Z_{1}\right) \right\vert ^{2}dZ_{1}}{\omega _{S}\left(
J,Z,\left\vert \Psi \right\vert ^{2}\right) }\Delta \omega _{D}\left(
J,\theta -\frac{\left\vert Z-Z_{1}\right\vert }{c},Z_{1},\Psi \right) \\
&&\int \frac{\kappa }{N}\frac{T\left( Z,Z_{1}\right) \omega _{S}\left(
J,Z_{1},\Psi \right) \left\vert \Psi _{0}\left( Z_{1}\right) \right\vert
^{2}dZ_{1}}{\omega _{S}\left( J,Z,\left\vert \Psi \right\vert ^{2}\right) }%
\frac{\Delta \omega _{D}\left( J,\theta ,Z,\left\vert \Psi \right\vert
^{2}\right) }{\omega _{S}\left( J,Z,\left\vert \Psi \right\vert ^{2}\right) }
\end{eqnarray*}%
Eand equation (\ref{DMG}) rewrites:

\begin{eqnarray*}
&&\left( 1+\frac{G^{\prime }\left( G^{-1}\left( \omega _{S}^{-1}\left(
J,Z,\left\vert \Psi \right\vert ^{2}\right) \right) \right) \omega
_{S}\left( J,Z_{1},\Psi \right) }{\omega _{S}^{2}\left( J,Z,\left\vert \Psi
\right\vert ^{2}\right) }\right) \Delta \omega _{D}\left( J,\theta
,Z,\left\vert \Psi \right\vert ^{2}\right)  \\
&\simeq &G^{\prime }\left( G^{-1}\left( \omega _{S}^{-1}\left(
J,Z,\left\vert \Psi \right\vert ^{2}\right) \right) \right) \int \frac{%
\kappa }{N}\frac{T\left( Z,Z_{1}\right) \omega _{S}\left( J,Z_{1},\Psi
\right) \left\vert \Psi _{0}\left( Z_{1}\right) \right\vert ^{2}dZ_{1}}{%
\omega _{S}\left( J,Z,\left\vert \Psi \right\vert ^{2}\right) }\frac{\Delta
\omega _{D}\left( J,\theta ,Z,\left\vert \Psi \right\vert ^{2}\right) }{%
\omega _{S}\left( J,Z,\left\vert \Psi \right\vert ^{2}\right) }
\end{eqnarray*}%
with solution:%
\begin{eqnarray*}
&&\Delta \omega _{D}\left( J,\theta ,Z,\left\vert \Psi \right\vert
^{2}\right)  \\
&=&G^{\prime }\left( G^{-1}\left( \omega _{S}^{-1}\left( J,Z,\left\vert \Psi
\right\vert ^{2}\right) \right) \right)  \\
&&\times \int \frac{\kappa }{N}\frac{T\left( Z,Z_{1}\right) \omega \left(
J,Z_{1},\Psi \right) \left\vert \Psi _{0}\left( Z_{1}\right) \right\vert
^{2}dZ_{1}}{G^{\prime }\left( G^{-1}\left( \omega _{S}^{-1}\left(
J,Z,\left\vert \Psi \right\vert ^{2}\right) \right) \right) \omega
_{S}\left( J,Z_{1},\Psi \right) +\omega _{S}^{2}\left( J,Z,\left\vert \Psi
\right\vert ^{2}\right) }\Delta \omega _{D}\left( J,\theta -\frac{\left\vert
Z-Z_{1}\right\vert }{c},Z_{1},\Psi \right) 
\end{eqnarray*}

We use that the modified state form a group over a bounded domain and
involve some finite number of points. For such group, we replace the
integral by a sum and we have:%
\begin{equation}
\Delta \omega _{D}\left( \theta ,Z_{i},\left\vert \Psi \right\vert
^{2}\right) =\sum_{j}\check{T}\left( Z_{i},Z_{j}\right) \Delta \omega \left(
\theta -\frac{\left\vert Z_{i}-Z_{j}\right\vert }{c},Z_{j},\Psi \right)
\label{SCL}
\end{equation}%
where:%
\begin{equation*}
\check{T}\left( Z_{i},Z_{j}\right) =\frac{\kappa }{N}G^{\prime }\left(
G^{-1}\left( \omega _{S}^{-1}\left( J,Z,\left\vert \Psi \right\vert
^{2}\right) \right) \right) \frac{T\left( Z_{i},Z_{j}\right) \omega
_{S}\left( J,Z_{j},\Psi \right) \left\vert \Psi _{0}\left( Z_{j}\right)
\right\vert ^{2}}{G^{\prime }\left( G^{-1}\left( \omega _{S}^{-1}\left(
J,Z,\left\vert \Psi \right\vert ^{2}\right) \right) \right) \omega
_{S}\left( J,Z_{1},\Psi \right) +\omega _{S}^{2}\left( J,Z,\left\vert \Psi
\right\vert ^{2}\right) }
\end{equation*}

We look for oscillatory solutions:

\begin{equation*}
\Delta \omega _{D}\left( J,\theta ,Z,\left\vert \Psi \right\vert ^{2}\right)
=A\left( Z_{i}\right) \exp \left( i\Upsilon \theta \right)
\end{equation*}%
and for such functions, equation (\ref{SCL}) writes:%
\begin{equation}
A\left( Z_{i}\right) =\sum_{j}\check{T}\left( Z_{i},Z_{j}\right) \exp \left(
-i\Upsilon \frac{\left\vert Z_{i}-Z_{j}\right\vert }{c}\right) A\left(
Z_{j}\right)  \label{FRQQ}
\end{equation}%
Equation (\ref{FRQQ}) has non-nul solutions for frequencies $\Upsilon _{p}$
satisfying:%
\begin{equation*}
\det \left( 1-\check{T}\left( Z_{i},Z_{j}\right) \exp \left( -i\Upsilon _{p}%
\frac{\left\vert Z_{i}-Z_{j}\right\vert }{c}\right) \right) =0
\end{equation*}%
Generally, the $\Upsilon _{p}$ are complex, and oscillations are dampened,
However we showed in (\cite{GL}), that the perturbative corrections to the
effective action for $\left\vert \Psi _{0}\left( Z_{l}\right) \right\vert
^{2}$ shifts the background:

\begin{equation*}
\left\vert \Psi _{0}\left( Z_{l}\right) \right\vert ^{2}\rightarrow
\left\vert \Psi _{0}\left( Z_{l}\right) \right\vert ^{2}+\Delta \left\vert
\Psi _{0}\left( Z_{l}\right) \right\vert ^{2}
\end{equation*}%
and this shifts allows for stable oscillations. In first proximation, this
is equivalent to consider that for some parameters, the $\Upsilon _{p}$ can
be considered as real, and $\left\vert \Psi _{0}\left( Z_{1}\right)
\right\vert ^{2}$ as time independent.

The possible osillatory activities associated to the assembly is thus given
by the sets:%
\begin{equation*}
\left\{ \left\{ A\left( Z_{i}\right) \right\} _{i=1,...n},\Upsilon
_{p}\left( \left\{ \check{T}\left( Z_{i},Z_{j}\right) \right\} \right)
\right\} _{p}
\end{equation*}%
where $p$ refers to the frequencies $\Upsilon $, and the $A\left(
Z_{i}\right) $ satisfy: 
\begin{equation*}
A\left( Z_{i}\right) =\sum_{j\neq i}A\left( Z_{j}\right) \check{T}\left(
Z_{i},Z_{j}\right) \exp \left( -i\Upsilon _{p}\left( \left\{ \check{T}\left(
Z_{i},Z_{j}\right) \right\} \right) \frac{\left\vert Z_{i}-Z_{j}\right\vert 
}{c}\right)
\end{equation*}

In these equation, one amplitude is a free parameter. We can choose $A\left(
Z_{1}\right) $ and set: 
\begin{equation*}
A_{1}\left( Z_{i}\right) =\frac{A\left( Z_{i}\right) }{A\left( Z_{1}\right) }
\end{equation*}%
Thus, the $A_{1}\left( Z_{i}\right) $ are solutions of $n-1$ systems of
equation with $i\geqslant 2$:

\begin{equation}
A_{1}\left( Z_{i}\right) =\sum_{j\neq i}A_{1}\left( Z_{j}\right) \check{T}%
\left( Z_{i},Z_{j}\right) \exp \left( -i\Upsilon _{p}\frac{\left\vert
Z_{i}-Z_{j}\right\vert }{c}\right)  \label{sm}
\end{equation}%
For $i\neq 1$, $j\neq 1$, the solutions of(\ref{sm}) are: 
\begin{equation}
\left( \delta _{ij}-\check{T}\left( Z_{i},Z_{j}\right) \exp \left(
-i\Upsilon _{p}\frac{\left\vert Z_{i}-Z_{j}\right\vert }{c}\right) \right)
A_{1}\left( Z_{j}\right) =\check{T}\left( Z_{i},Z_{1}\right) \exp \left(
-i\Upsilon _{p}\frac{\left\vert Z_{i}-Z_{1}\right\vert }{c}\right)
\label{DFT}
\end{equation}%
We rewrite these equations (\ref{DFT}) matricially by setting:%
\begin{equation*}
\left( 1-\mathbf{\check{T}}\exp \left( -i\Upsilon _{p}\frac{\left\vert 
\mathbf{\Delta Z}\right\vert }{c}\right) \right) _{ij}=\left( \delta _{ij}-%
\check{T}\left( Z_{i},Z_{j}\right) \exp \left( -i\Upsilon _{p}\frac{%
\left\vert Z_{i}-Z_{j}\right\vert }{c}\right) \right)
\end{equation*}%
and%
\begin{equation*}
\left( \check{T}_{1}\left( \mathbf{Z}\right) \exp \left( -i\Upsilon _{p}%
\frac{\left\vert \mathbf{\Delta Z}_{1}\right\vert }{c}\right) \right) _{i}=%
\check{T}\left( Z_{i},Z_{1}\right) \exp \left( -i\Upsilon _{p}\frac{%
\left\vert Z_{i}-Z_{1}\right\vert }{c}\right)
\end{equation*}%
As a consequence, equation (\ref{DFT}) has the form:%
\begin{equation*}
A_{1}\left( \mathbf{Z}\right) =\left( 1-\mathbf{\check{T}}\exp \left(
-i\Upsilon _{p}\frac{\left\vert \mathbf{\Delta Z}\right\vert }{c}\right)
\right) ^{-1}\check{T}_{1}\left( \mathbf{Z}\right) \exp \left( -i\Upsilon
_{p}\frac{\left\vert \mathbf{\Delta Z}_{1}\right\vert }{c}\right)
\end{equation*}%
and the modified dynamic of activities writes;%
\begin{equation*}
\Delta \omega _{D}\left( \theta ,\mathbf{Z},\left\vert \Psi \right\vert
^{2}\right) =A\left( Z_{1}\right) \left( 1,\left( 1-\mathbf{\check{T}}\exp
\left( -i\Upsilon _{p}\frac{\left\vert \mathbf{\Delta Z}\right\vert }{c}%
\right) \right) ^{-1}\check{T}_{1}\left( \mathbf{Z}\right) \exp \left(
-i\Upsilon _{p}\frac{\left\vert \mathbf{\Delta Z}_{1}\right\vert }{c}\right)
\right) ^{t}\exp \left( i\Upsilon _{p}\left( \mathbf{\check{T}}\right)
\theta \right)
\end{equation*}%
where:%
\begin{equation*}
\Upsilon _{p}\left( \mathbf{\check{T}}\right) =\Upsilon _{p}\left( \left\{ 
\check{T}\left( Z_{i},Z_{j}\right) \right\} \right)
\end{equation*}%
Ultimately, adding also the average static solution with respect to $\mathbf{%
\check{T}}$ yields the result for the additnal activity:%
\begin{eqnarray*}
&&\overline{\Delta \omega }\left( Z,\mathbf{\check{T}}\right) \\
&&+A\left( Z_{1}\right) \left( 1,\left( 1-\mathbf{\check{T}}\exp \left(
-i\Upsilon _{p}\left( \mathbf{\check{T}}\right) \frac{\left\vert \mathbf{%
\Delta Z}\right\vert }{c}\right) \right) _{1}^{-1}\check{T}_{1}\left( 
\mathbf{Z}\right) \exp \left( -i\Upsilon _{p}\left( \mathbf{\check{T}}%
\right) \frac{\left\vert \mathbf{\Delta Z}_{1}\right\vert }{c}\right)
\right) ^{t}\exp \left( i\Upsilon _{p}\left( \mathbf{\check{T}}\right)
\theta \right)
\end{eqnarray*}

\subsection*{3.1 Averages computations and activated states}

\subsubsection*{3.1.1 Averages}

Now, considering (\ref{MFD}) for the group of shifted states, we rewrite the
action functional for this group by taking into account their particular
interactions. Given the activities, we can compute the average
connectivities. To do so, we replace in (\ref{TSN}) (see (\cite{GLt})):%
\begin{eqnarray*}
\left( \left( \left( Z-Z^{\prime }\right) \left( \nabla _{Z}+\nabla
_{Z}\omega \left( Z\right) \right) +\frac{\left\vert Z-Z^{\prime
}\right\vert }{c}\right) \Delta \omega \left( \theta ,Z,\left\vert \Psi
\right\vert ^{2}\right) \right) &\rightarrow &\frac{\left\vert Z-Z^{\prime
}\right\vert }{c}\Delta \omega \left( \theta ,Z,\left\vert \Psi \right\vert
^{2}\right) \\
&\rightarrow &g\left\vert Z-Z^{\prime }\right\vert \Delta \omega \left(
\theta ,Z,\left\vert \Psi \right\vert ^{2}\right)
\end{eqnarray*}%
assume $A\left( Z_{1}\right) $ proportional to activity on the group:%
\begin{eqnarray*}
&&\left( \left( Z-Z^{\prime }\right) \left( \nabla _{Z}+\nabla _{Z}\omega
_{0}\left( Z\right) \right) +\frac{\left\vert Z-Z^{\prime }\right\vert }{c}%
\right) \Delta \omega \left( \theta ,\mathbf{Z},\left\vert \Psi \right\vert
^{2}\right) \\
&=&\left\vert Z-Z^{\prime }\right\vert k\left\langle \left\vert \Psi
\right\vert ^{2}\right\rangle \left( 1,\left( 1-\mathbf{\check{T}}\exp
\left( -i\Upsilon _{p}\frac{\left\vert \mathbf{\Delta Z}\right\vert }{c}%
\right) \right) ^{-1}\check{T}\left( \mathbf{Z}\right) \exp \left(
-i\Upsilon _{p}\frac{\left\vert \mathbf{\Delta Z}_{1}\right\vert }{c}\right)
\right) ^{t}\exp \left( i\Upsilon _{p}\left( \mathbf{\check{T}}\right)
\theta \right)
\end{eqnarray*}%
In these equations, we decompose $\mathbf{T\rightarrow }\left\langle \mathbf{%
T}\right\rangle \mathbf{+\Delta T}$ where $\left\langle \mathbf{T}%
\right\rangle $ is defined by the background.

We thus consider the effective action resulting from the shift. It leads to
rewrite (\ref{MFD}):%
\begin{eqnarray}
&&-\Delta \Gamma ^{\dag }\left( T,\hat{T},\theta ,Z,Z^{\prime }\right) 
\label{NGD} \\
&&\times \left( \nabla _{T}\left( \sigma _{T}^{2}\nabla _{T}+\frac{\left(
\Delta T-\underline{\left\langle \Delta T\right\rangle }\right) -\lambda
\left( \Delta \hat{T}-\underline{\left\langle \Delta \hat{T}\right\rangle }%
\right) }{\tau \omega _{S}\left( Z\right) }\left\vert \Psi \left( \theta
,Z\right) \right\vert ^{2}\right) \right) \Delta \Gamma \left( T,\hat{T}%
,\theta ,Z,Z^{\prime }\right)   \notag \\
&&-\Delta \Gamma ^{\dag }\left( T,\hat{T},\theta ,Z,Z^{\prime }\right)
\nabla _{\hat{T}}\left( \sigma _{\hat{T}}^{2}\nabla _{\hat{T}}-\rho
\left\vert \bar{\Psi}_{0}\left( Z,Z^{\prime }\right) \right\vert ^{2}\left(
\Delta \hat{T}-\underline{\left\langle \Delta \hat{T}\right\rangle }\right)
\right.   \notag \\
&&\left. +\frac{\rho D\left( \theta \right) \left\langle \hat{T}%
\right\rangle \left\vert \Psi _{0}\left( Z^{\prime }\right) \right\vert ^{2}%
}{\omega _{S}\left( Z\right) }\left( \left( Z-Z^{\prime }\right) \left(
\nabla _{Z}+\nabla _{Z}\omega _{S}\left( Z\right) \right) +\frac{\left\vert
Z-Z^{\prime }\right\vert }{c}\right) \Delta \omega \left( \theta
,Z,\left\vert \Psi \right\vert ^{2}\right) \right) \Delta \Gamma \left( T,%
\hat{T},\theta ,Z,Z^{\prime }\right)   \notag
\end{eqnarray}%
At the scale considered, system's frequencies can be replaced by their
averages:%
\begin{equation*}
\Delta \omega \left( \theta ,Z,\left\vert \Psi \right\vert ^{2}\right)
\simeq \overline{\Delta \omega }\left( Z,\mathbf{\check{T}}\right) 
\end{equation*}%
and the second term of (\ref{NGD}) rewrites:%
\begin{eqnarray*}
&&\Delta \Gamma ^{\dag }\left( T,\hat{T},\theta ,Z,Z^{\prime }\right) \nabla
_{\hat{T}}\left( \rho \left\vert \Psi _{0}\left( Z\right) \right\vert
^{2}\left( \Delta \hat{T}-\underline{\left\langle \Delta \hat{T}%
\right\rangle }\right) \right.  \\
&&\left. +\frac{\rho }{\omega _{S}\left( Z\right) }\left( D\left( \theta
\right) \left\langle \hat{T}\right\rangle \left\vert \Psi _{0}\left(
Z^{\prime }\right) \right\vert ^{2}A\left\vert Z-Z^{\prime }\right\vert
\Delta \omega \left( \theta ,Z,\left\vert \Psi \right\vert ^{2}\right)
\right) \right) \Delta \Gamma \left( T,\hat{T},\theta ,Z,Z^{\prime }\right) 
\\
&=&\Delta \Gamma ^{\dag }\left( T,\hat{T},\theta ,Z,Z^{\prime }\right)  \\
&&\times \nabla _{\hat{T}}\left( \frac{\rho \left( C\left( \theta \right)
\left\vert \Psi _{0}\left( Z\right) \right\vert ^{2}\omega _{S}\left(
Z\right) +D\left( \theta \right) \hat{T}\left\vert \Psi _{0}\left( Z^{\prime
}\right) \right\vert ^{2}\omega _{S}\left( Z^{\prime }\right) \right) \left(
\Delta \hat{T}-\Delta ^{\prime }\hat{T}\right) }{\omega _{S}\left( Z\right) }%
\right) \Delta \Gamma \left( T,\hat{T},\theta ,Z,Z^{\prime }\right) 
\end{eqnarray*}%
wth:%
\begin{eqnarray*}
\Delta ^{\prime }\hat{T} &=&\underline{\left\langle \Delta \hat{T}%
\right\rangle }+\frac{\left( D\left( \theta \right) \left\langle \hat{T}%
\right\rangle \left\vert \Psi _{0}\left( Z^{\prime }\right) \right\vert
^{2}\left\vert Z-Z^{\prime }\right\vert g\overline{\Delta \omega }\left(
\theta ,Z,\left\vert \Psi \right\vert ^{2}\right) \right) }{\left( C\left(
\theta \right) \left\vert \Psi _{0}\left( Z\right) \right\vert ^{2}\omega
_{S}\left( Z\right) +D\left( \theta \right) \hat{T}\left\vert \Psi
_{0}\left( Z^{\prime }\right) \right\vert ^{2}\omega _{S}\left( Z^{\prime
}\right) \right) } \\
&\simeq &\underline{\left\langle \Delta \hat{T}\right\rangle }+\Delta
^{\omega }\hat{T}
\end{eqnarray*}%
whr:%
\begin{equation*}
\Delta ^{\omega }\hat{T}=\frac{\left( D\left( \theta \right) \left\langle 
\hat{T}\right\rangle \left\vert \Psi _{0}\left( Z^{\prime }\right)
\right\vert ^{2}\left\vert Z-Z^{\prime }\right\vert g\overline{\Delta \omega 
}\left( Z,\mathbf{T}\right) \right) }{\left( C\left( \theta \right)
\left\vert \Psi _{0}\left( Z\right) \right\vert ^{2}\omega _{S}\left(
Z\right) +D\left( \theta \right) \left\langle \hat{T}\right\rangle
\left\vert \Psi _{0}\left( Z^{\prime }\right) \right\vert ^{2}\omega
_{S}\left( Z^{\prime }\right) \right) }
\end{equation*}%
Dfn ls:%
\begin{equation*}
\Delta ^{\prime }T=\underline{\left\langle \Delta \hat{T}\right\rangle }%
+\lambda \Delta ^{\omega }\hat{T}
\end{equation*}

\subsubsection*{3.1.2 Change of variables and activated states}

Recall that the static part $\omega _{0}\left( Z\right) +\Delta \omega
\left( Z\right) $ is written $\omega _{S}\left( Z\right) $.

Performing the change of variable in (\ref{NGD}):%
\begin{eqnarray}
\Delta \Gamma \left( T,\hat{T},\theta ,Z,Z^{\prime }\right) &\rightarrow
&\exp \left( -\frac{\rho \left( \left( C\left( \theta \right) \left\vert
\Psi \left( Z\right) \right\vert ^{2}h_{C}+D\left( \theta \right) \left\vert
\Psi \left( Z^{\prime }\right) \right\vert ^{2}h_{D}\right) \left( \Delta 
\hat{T}-\Delta ^{\prime }\hat{T}\right) ^{2}\right) }{4\sigma _{\hat{T}%
}^{2}\omega \left( \theta ,Z,\left\vert \Psi \right\vert ^{2}\right) }\right)
\\
&&\times \exp \left( -\frac{\left( \left( \Delta T-\Delta ^{\prime }T\right)
^{2}-2\lambda \tau \left( \Delta \hat{T}-\Delta ^{\prime }\hat{T}\right)
\left( \Delta T-\Delta ^{\prime }T\right) \right) }{4\sigma _{T}^{2}\tau
\omega }\right) \Delta \Gamma \left( T,\hat{T},\theta ,Z,Z^{\prime }\right) 
\notag
\end{eqnarray}%
and:%
\begin{eqnarray}
\Delta \Gamma ^{\dag }\left( T,\hat{T},\theta ,Z,Z^{\prime }\right)
&\rightarrow &\exp \left( \frac{\rho \left( \left( C\left( \theta \right)
\left\vert \Psi \left( Z\right) \right\vert ^{2}h_{C}+D\left( \theta \right)
\left\vert \Psi \left( Z^{\prime }\right) \right\vert ^{2}h_{D}\right)
\left( \Delta \hat{T}-\Delta ^{\prime }\hat{T}\right) ^{2}\right) }{4\sigma
_{\hat{T}}^{2}\omega \left( \theta ,Z,\left\vert \Psi \right\vert
^{2}\right) }\right) \\
&&\times \exp \left( \frac{\left( \frac{\left( T-\left\langle T\right\rangle
\right) ^{2}}{\tau }-2\lambda \left( \Delta \hat{T}-\Delta ^{\prime }\hat{T}%
\right) \left( \Delta T-\Delta ^{\prime }T\right) \right) }{4\sigma
_{T}^{2}\tau \omega \left( \theta ,Z,\left\vert \Psi \right\vert ^{2}\right) 
}\right) \Delta \Gamma ^{\dag }\left( T,\hat{T},\theta ,Z,Z^{\prime }\right)
\notag
\end{eqnarray}%
We replace $\omega \left( \theta ,Z,\left\vert \Psi \right\vert ^{2}\right) $
with their static part $\omega _{S}\left( Z\right) $. It leads to the action
in first approximation in $\frac{\sigma _{\hat{T}}^{2}}{\sigma _{T}^{2}}$:%
\begin{eqnarray}
&&S\left( \left\{ \Delta \Gamma _{S_{\alpha }^{2}}\left( \left( \mathbf{T},%
\mathbf{\hat{T}},\mathbf{Z}\right) _{S_{\alpha }^{2}},\theta \right)
\right\} \right) \\
&=&-\Delta \Gamma ^{\dag }\left( T,\hat{T},\theta ,Z,Z^{\prime }\right)
\left( \sigma _{T}^{2}\nabla _{T}^{2}-\frac{1}{2\sigma _{T}^{2}}\left( \frac{%
\left( \Delta T-\Delta T^{\prime }\right) -\lambda \left( \Delta \hat{T}%
-\Delta ^{\prime }\hat{T}\right) }{\tau \omega _{S}\left( Z\right) }%
\left\vert \Psi _{0}\left( \theta ,Z\right) \right\vert ^{2}\right)
^{2}\right) \Delta \Gamma \left( T,\hat{T},\theta ,Z,Z^{\prime }\right) 
\notag \\
&&-\Delta \Gamma ^{\dag }\left( T,\hat{T},\theta ,Z,Z^{\prime }\right) 
\notag \\
&&\times \left( \sigma _{\hat{T}}^{2}\nabla _{\hat{T}}^{2}-\frac{1}{2\sigma
_{\hat{T}}^{2}}\left( \frac{\rho \left( C\left( \theta \right) \left\vert
\Psi _{0}\left( Z^{\prime }\right) \right\vert ^{2}\omega _{S}\left(
Z\right) +D\left( \theta \right) \left\vert \Psi _{0}\left( Z^{\prime
}\right) \right\vert ^{2}\omega _{S}\left( Z^{\prime }\right) \right) \left(
\Delta \hat{T}-\Delta ^{\prime }\hat{T}\right) }{\omega _{S}\left( Z\right) }%
\right) ^{2}\right) \Delta \Gamma \left( T,\hat{T},\theta ,Z,Z^{\prime
}\right)  \notag \\
&&+\left( \tau \omega _{S}\left( Z\right) +\frac{\rho \left( C\left( \theta
\right) \left\vert \Psi _{0}\left( Z^{\prime }\right) \right\vert ^{2}\omega
_{S}\left( Z\right) +D\left( \theta \right) \left\vert \Psi _{0}\left(
Z^{\prime }\right) \right\vert ^{2}\omega _{S}\left( Z^{\prime }\right)
\right) }{\omega _{S}\left( Z\right) }\right) \left\Vert \Delta \Gamma
\left( T,\hat{T},\theta ,Z,Z^{\prime }\right) \right\Vert ^{2}  \notag
\end{eqnarray}%
The average values satisfy:%
\begin{eqnarray}
\left\langle \Delta T\right\rangle &=&\left\langle \Delta ^{\prime
}T\right\rangle =\underline{\left\langle \Delta T\right\rangle }+\lambda 
\frac{\left( D\left( \theta \right) \left\langle \hat{T}\right\rangle
\left\vert \Psi _{0}\left( Z^{\prime }\right) \right\vert ^{2}\left\vert
Z-Z^{\prime }\right\vert g\overline{\Delta \omega }\left( Z,\left\langle
\Delta \mathbf{T}\right\rangle \right) \right) }{\left( C\left( \theta
\right) \left\vert \Psi _{0}\left( Z\right) \right\vert ^{2}\omega
_{S}\left( Z\right) +D\left( \theta \right) \left\langle \hat{T}%
\right\rangle \left\vert \Psi _{0}\left( Z^{\prime }\right) \right\vert
^{2}\omega _{S}\left( Z^{\prime }\right) \right) }  \label{vrb} \\
\left\langle \Delta \hat{T}\right\rangle &=&\left\langle \Delta ^{\prime }%
\hat{T}\right\rangle =\underline{\left\langle \Delta \hat{T}\right\rangle }+%
\frac{\left( D\left( \theta \right) \left\langle \hat{T}\right\rangle
\left\vert \Psi _{0}\left( Z^{\prime }\right) \right\vert ^{2}\left\vert
Z-Z^{\prime }\right\vert g\overline{\Delta \omega }\left( Z,\left\langle
\Delta \mathbf{T}\right\rangle \right) \right) }{\left( C\left( \theta
\right) \left\vert \Psi _{0}\left( Z\right) \right\vert ^{2}\omega
_{S}\left( Z\right) +D\left( \theta \right) \left\langle \hat{T}%
\right\rangle \left\vert \Psi _{0}\left( Z^{\prime }\right) \right\vert
^{2}\omega _{S}\left( Z^{\prime }\right) \right) }  \notag
\end{eqnarray}%
and the effective action wrts:%
\begin{eqnarray}
&&\hat{S}\left( \Delta \Gamma \left( T,\hat{T},\theta ,Z,Z^{\prime }\right)
\right)  \label{SH} \\
&=&-\Delta \Gamma ^{\dag }\left( T,\hat{T},\theta ,Z,Z^{\prime }\right) 
\notag \\
&&\left( \sigma _{T}^{2}\nabla _{T}^{2}-\frac{1}{2\sigma _{T}^{2}}\left( 
\frac{\left( \Delta T-\left\langle \Delta T\left( Z,Z^{\prime }\right)
\right\rangle _{p}^{\alpha }\right) -\lambda \left( \Delta \hat{T}%
-\left\langle \Delta \hat{T}\left( Z,Z^{\prime }\right) \right\rangle
_{p}^{\alpha }\right) }{\tau \omega _{S}\left( Z\right) }\left\vert \Psi
_{0}\left( \theta ,Z\right) \right\vert ^{2}\right) ^{2}\right) \Delta
\Gamma \left( T,\hat{T},\theta ,Z,Z^{\prime }\right)  \notag \\
&&-\Delta \Gamma ^{\dag }\left( T,\hat{T},\theta ,Z,Z^{\prime }\right) 
\notag \\
&&\times \left( \sigma _{\hat{T}}^{2}\nabla _{\hat{T}}^{2}-\frac{\left(
D\left( Z,Z^{\prime }\right) \left( \Delta \hat{T}-\left\langle \Delta \hat{T%
}\left( Z,Z^{\prime }\right) \right\rangle _{p}^{\alpha }\right) +\mathbf{M}%
^{\alpha }\left( Z,Z^{\prime }\right) \left( \Delta T-\left\langle \Delta
T\left( Z,Z^{\prime }\right) \right\rangle _{p}^{\alpha }\right) \right) ^{2}%
}{2\sigma _{\hat{T}}^{2}}\right) \Delta \Gamma \left( T,\hat{T},\theta
,Z,Z^{\prime }\right)  \notag \\
&&+C\left( Z,Z^{\prime }\right) \left\Vert \Delta \Gamma \left( T,\hat{T}%
,\theta ,Z,Z^{\prime }\right) \right\Vert ^{2}  \notag
\end{eqnarray}%
with:

\begin{equation*}
C\left( Z,Z^{\prime }\right) =\tau \omega _{S}\left( Z\right) +\frac{\rho
\left( C\left( \theta \right) \left\vert \Psi _{0}\left( Z^{\prime }\right)
\right\vert ^{2}\omega _{S}\left( Z\right) +D\left( \theta \right)
\left\vert \Psi _{0}\left( Z^{\prime }\right) \right\vert ^{2}\omega
_{S}\left( Z^{\prime }\right) \right) }{\omega _{S}\left( Z\right) }
\end{equation*}%
and:%
\begin{equation*}
\mathbf{M}^{\alpha }\left( Z,Z^{\prime }\right) =\left( \frac{\rho D\left(
\theta \right) \left\langle \hat{T}\right\rangle \left\vert \Psi _{0}\left(
Z^{\prime }\right) \right\vert ^{2}A\left\vert Z-Z^{\prime }\right\vert }{%
\omega _{S}\left( Z\right) }\left( \nabla _{\mathbf{\Delta T}_{\left(
Z_{1},Z_{1}^{\prime }\right) }}\left( \overline{\Delta \omega }\left(
Z,\left\langle \mathbf{\Delta T}\right\rangle \right) \right) _{\left(
\left\langle \Delta \mathbf{T}_{\left( Z_{1},Z_{1}^{\prime }\right)
}\right\rangle _{p}^{\alpha }\right) }\right) \right)
\end{equation*}%
and action (\ref{SH}) is rewritten as:%
\begin{eqnarray}
&&\hat{S}\left( \Delta \Gamma \left( T,\hat{T},\theta ,Z,Z^{\prime }\right)
\right) \\
&=&-\Delta \Gamma ^{\dag }\left( T,\hat{T},\theta ,Z,Z^{\prime }\right)
\left( \nabla _{T}^{2}+\nabla _{\hat{T}}^{2}-\frac{1}{2}\left( \mathbf{%
\Delta T-}\left\langle \mathbf{\Delta T}\right\rangle _{p}^{\alpha }\right)
^{t}\mathbf{A}_{p}^{\alpha }\left( \mathbf{\Delta T-}\left\langle \mathbf{%
\Delta T}\right\rangle _{p}^{\alpha }\right) \right) \Delta \Gamma \left( T,%
\hat{T},\theta ,Z,Z^{\prime }\right)  \notag \\
&&+C\left( Z,Z^{\prime }\right) \left\Vert \Delta \Gamma \left( T,\hat{T}%
,\theta ,Z,Z^{\prime }\right) \right\Vert ^{2}  \notag
\end{eqnarray}%
where the variables are:%
\begin{equation}
\mathbf{\Delta T-}\left\langle \mathbf{\Delta T}\right\rangle _{p}^{\alpha
}=\left( 
\begin{array}{c}
\Delta T-\left\langle \Delta T\right\rangle \\ 
\Delta \hat{T}-\left\langle \Delta \hat{T}\right\rangle%
\end{array}%
\right)
\end{equation}%
and:%
\begin{equation*}
\mathbf{A}_{p}^{\alpha }=\left( 
\begin{array}{cc}
\left( \frac{1}{\tau \omega _{S}\left( Z\right) }\right) ^{2}+\left( \mathbf{%
M}^{\alpha }\left( Z,Z^{\prime }\right) \right) ^{2} & -\lambda \left( \frac{%
1}{\tau \omega _{S}\left( Z\right) }\right) ^{2}+D\left( Z,Z^{\prime
}\right) \mathbf{M}^{\alpha }\left( Z,Z^{\prime }\right) \\ 
-\lambda \left( \frac{1}{\tau \omega _{S}\left( Z\right) }\right)
^{2}+D\left( Z,Z^{\prime }\right) \mathbf{M}^{\alpha }\left( Z,Z^{\prime
}\right) & \left( \frac{\lambda }{\tau \omega _{S}\left( Z\right) }\right)
^{2}+D^{2}\left( Z,Z^{\prime }\right)%
\end{array}%
\right)
\end{equation*}%
The matrix $\mathbf{\hat{A}}_{p}^{\alpha }$ is defined by:%
\begin{eqnarray*}
\mathbf{\hat{A}}_{p}^{\alpha } &=&\left( 
\begin{array}{cc}
\frac{1}{\sigma _{T}} & 0 \\ 
0 & \frac{1}{\sigma _{\hat{T}}}%
\end{array}%
\right) \mathbf{A}_{p}^{\alpha }\left( 
\begin{array}{cc}
\frac{1}{\sigma _{T}} & 0 \\ 
0 & \frac{1}{\sigma _{\hat{T}}}%
\end{array}%
\right) \\
&=&\left( 
\begin{array}{cc}
\frac{\left( \frac{1}{\tau \omega _{S}\left( Z\right) }\right) ^{2}+\left( 
\mathbf{M}^{\alpha }\left( Z,Z^{\prime }\right) \right) ^{2}}{\sigma _{T}^{2}%
} & \frac{-\lambda \left( \frac{1}{\tau \omega _{S}\left( Z\right) }\right)
^{2}+D\left( Z,Z^{\prime }\right) \mathbf{M}^{\alpha }\left( Z,Z^{\prime
}\right) }{\sigma _{T}\sigma _{\hat{T}}} \\ 
\frac{-\lambda \left( \frac{1}{\tau \omega _{S}\left( Z\right) }\right)
^{2}+D\left( Z,Z^{\prime }\right) \mathbf{M}^{\alpha }\left( Z,Z^{\prime
}\right) }{\sigma _{T}\sigma _{\hat{T}}} & \frac{\left( \frac{\lambda }{\tau
\omega _{S}\left( Z\right) }\right) ^{2}+D^{2}\left( Z,Z^{\prime }\right) }{%
\sigma _{\hat{T}}^{2}}%
\end{array}%
\right)
\end{eqnarray*}%
The minimization of $\hat{S}\left( \Delta \Gamma \left( T,\hat{T},\theta
,Z,Z^{\prime }\right) \right) $ is similar to (\ref{sdl}) and leads to the
solutions:

\begin{eqnarray}
\Delta \Gamma _{n,n^{\prime }}^{\alpha ,p}\left( T,\hat{T},\theta
,Z,Z^{\prime }\right) &=&\mathcal{N}\exp \left( -\frac{1}{2}\left( \mathbf{%
\Delta T-}\left\langle \mathbf{\Delta T}\right\rangle _{p}^{\alpha }\right)
^{t}\mathbf{\hat{A}}_{p}^{\alpha }\left( \mathbf{\Delta T-}\left\langle 
\mathbf{\Delta T}\right\rangle _{p}^{\alpha }\right) \right)  \label{STM} \\
&&\times H_{n}\left( \frac{\sigma _{T}}{2\sqrt{2}}\left( \left( \mathbf{%
\Delta T-}\left\langle \mathbf{\Delta T}\right\rangle _{p}^{\alpha }\right)
^{t}\mathbf{\hat{A}}_{p}^{\alpha }\left( \mathbf{\Delta T-}\left\langle 
\mathbf{\Delta T}\right\rangle _{p}^{\alpha }\right) \right) _{2}\right) 
\notag \\
&&\times H_{n^{\prime }}\left( \frac{\sigma _{\hat{T}}}{2\sqrt{2}}\left(
\left( \mathbf{\Delta T-}\left\langle \mathbf{\Delta T}\right\rangle
_{p}^{\alpha }\right) ^{t}\mathbf{\hat{A}}_{p}^{\alpha }\left( \mathbf{%
\Delta T-}\left\langle \mathbf{\Delta T}\right\rangle _{p}^{\alpha }\right)
\right) _{1}\right)  \notag
\end{eqnarray}%
and:%
\begin{eqnarray}
\Delta \Gamma _{n,n^{\prime }}^{\dagger \alpha ,p}\left( T,\hat{T},\theta
,Z,Z^{\prime }\right) &=&H_{n}\left( \frac{\sigma _{T}}{2\sqrt{2}}\left(
\left( \mathbf{\Delta T-}\left\langle \mathbf{\Delta T}\right\rangle
_{p}^{\alpha }\right) ^{t}\mathbf{\hat{A}}_{p}^{\alpha }\left( \mathbf{%
\Delta T-}\left\langle \mathbf{\Delta T}\right\rangle _{p}^{\alpha }\right)
\right) _{+}\right)  \label{STC} \\
&&\times H_{n^{\prime }}\left( \frac{\sigma _{\hat{T}}}{2\sqrt{2}}\left(
\left( \mathbf{\Delta T-}\left\langle \mathbf{\Delta T}\right\rangle
_{p}^{\alpha }\right) ^{t}\mathbf{\hat{A}}_{p}^{\alpha }\left( \mathbf{%
\Delta T-}\left\langle \mathbf{\Delta T}\right\rangle _{p}^{\alpha }\right)
\right) _{-}\right)  \notag
\end{eqnarray}%
where $H_{p}$ and $H_{p-\delta }$ are Hermite polynomials. Given the
diagonalization of $\mathbf{\hat{A}}_{p}^{\alpha }$%
\begin{equation*}
\mathbf{\hat{A}}_{p}^{\alpha }=PDP^{-1}
\end{equation*}%
\begin{eqnarray*}
\left( \left( \mathbf{\Delta T-}\left\langle \mathbf{\Delta T}\right\rangle
_{p}^{\alpha }\right) ^{t}\mathbf{\hat{A}}_{p}^{\alpha }\left( \mathbf{%
\Delta T-}\left\langle \mathbf{\Delta T}\right\rangle _{p}^{\alpha }\right)
\right) _{+} &=&P\left( 
\begin{array}{cc}
0 & 0 \\ 
0 & 1%
\end{array}%
\right) D\left( 
\begin{array}{cc}
0 & 0 \\ 
0 & 1%
\end{array}%
\right) P^{-1} \\
\left( \left( \mathbf{\Delta T-}\left\langle \mathbf{\Delta T}\right\rangle
_{p}^{\alpha }\right) ^{t}\mathbf{\hat{A}}_{p}^{\alpha }\left( \mathbf{%
\Delta T-}\left\langle \mathbf{\Delta T}\right\rangle _{p}^{\alpha }\right)
\right) _{\_} &=&P\left( 
\begin{array}{cc}
1 & 0 \\ 
0 & 0%
\end{array}%
\right) D\left( 
\begin{array}{cc}
1 & 0 \\ 
0 & 0%
\end{array}%
\right) P^{-1}
\end{eqnarray*}%
The constant $\mathcal{N}$ is the normalization factor. $\allowbreak $

The lowest eigenvalue state is:%
\begin{equation}
\Delta \Gamma _{0}^{\alpha ,p}\left( T,\hat{T},\theta ,Z,Z^{\prime }\right) =%
\mathcal{N}\exp \left( -\frac{1}{2}\left( \mathbf{\Delta T-}\left\langle 
\mathbf{\Delta T}\right\rangle _{p}^{\alpha }\right) ^{t}\mathbf{\hat{A}}%
_{p}^{\alpha }\left( \mathbf{\Delta T-}\left\langle \mathbf{\Delta T}%
\right\rangle _{p}^{\alpha }\right) \right)
\end{equation}%
and the state associated to the system is:%
\begin{equation*}
\prod\limits_{Z,Z^{\prime }}\Delta \Gamma _{\delta }^{\alpha ,p}\left( T,%
\hat{T},\theta ,Z,Z^{\prime }\right) \equiv \prod\limits_{Z,Z^{\prime
}}\left\vert \Delta T\left( Z,Z^{\prime }\right) ,\Delta \hat{T}\left(
Z,Z^{\prime }\right) ,\alpha \left( Z,Z^{\prime }\right) ,p\left(
Z,Z^{\prime }\right) ,S^{2}\right\rangle \equiv \left\vert \mathbf{\alpha },%
\mathbf{p},S^{2}\right\rangle
\end{equation*}%
To simplify the arguments, we assume as before that $\left\langle \Delta 
\hat{T}\right\rangle \simeq \frac{\left\langle \Delta T\right\rangle }{%
\lambda }$. This yields the equation for the average $\left\langle \Delta 
\hat{T}\right\rangle $ by writing:%
\begin{equation*}
\left\langle \Delta \hat{T}\right\rangle =\left\langle \Delta ^{\prime }\hat{%
T}\right\rangle
\end{equation*}%
which leads to the relation:There are several sets of solutions:%
\begin{equation*}
\left( \left\langle \Delta \mathbf{T}\right\rangle ^{\alpha },\left\langle
\Delta \mathbf{\hat{T}}\right\rangle ^{\alpha }=\frac{\left\langle \Delta 
\mathbf{T}\right\rangle ^{\alpha }}{\lambda }\right)
\end{equation*}%
For each of these solutions, a sequence of frequencies $\left( \Upsilon
_{p}^{\alpha }\right) $ are compatible, and the variable part of activities
is:%
\begin{equation*}
\Delta \omega _{p}^{\alpha }\left( \theta ,\mathbf{Z}\right) =A\left(
Z_{1}\right) \left( 1,\left( 1-\left( \left\langle \Delta \mathbf{T}%
\right\rangle _{p}^{\alpha }\right) \exp \left( -i\Upsilon _{p}\frac{%
\left\vert \mathbf{\Delta Z}\right\vert }{c}\right) \right)
^{-1}\left\langle \Delta \mathbf{T}\right\rangle _{p}^{\alpha }\exp \left(
-i\Upsilon _{p}\frac{\left\vert \mathbf{\Delta Z}\right\vert }{c}\right)
\right) ^{t}\exp \left( i\Upsilon _{p}^{\alpha }\left( \Delta \mathbf{T}%
_{p}^{\alpha }\right) \theta \right)
\end{equation*}

Ultimately, the effective action $S\left( \left\{ \underline{\Gamma }\left(
\Delta \mathbf{T},\mathbf{\alpha },\mathbf{p},S^{2},\theta \right) \right\}
\right) $ for the state writes:%
\begin{eqnarray}
&&S\left( \left\{ \underline{\Gamma }\left( \Delta \mathbf{T},\mathbf{\alpha 
},\mathbf{p},S^{2},\theta \right) \right\} \right)  \label{FCT} \\
&=&\underline{\Gamma }^{\dag }\left( \Delta \mathbf{T},\mathbf{\alpha },%
\mathbf{p},S^{2},\theta \right) \left( -\nabla _{\Delta \mathbf{T}}^{2}+%
\frac{1}{2}\left( \mathbf{A}\left( \Delta \mathbf{T}-\left\langle \Delta 
\mathbf{T}\right\rangle ^{\alpha }\right) \right) ^{2}+C\right) \underline{%
\Gamma }\left( \Delta \mathbf{T},\mathbf{\alpha },\mathbf{p},S^{2},\theta
\right) +U\left( \left\Vert \underline{\Gamma }\left( \Delta \mathbf{T},%
\mathbf{\alpha },\mathbf{p},S^{2},\theta \right) \right\Vert ^{2}\right) 
\notag
\end{eqnarray}%
where $\mathbf{D}$ is diagonal with elements:%
\begin{equation*}
\mathbf{D}\left( Z,Z^{\prime }\right) =D\left[ \frac{\rho \left( C\left(
\theta \right) \left\vert \Psi _{0}\left( Z\right) \right\vert ^{2}\omega
_{S}\left( Z\right) +D\left( \theta \right) \hat{T}\left\vert \Psi
_{0}\left( Z^{\prime }\right) \right\vert ^{2}\omega _{S}\left( Z^{\prime
}\right) \right) }{\omega _{S}\left( Z\right) }\right]
\end{equation*}%
The matrix $\mathbf{A}^{\alpha }$ and the constant $C^{\alpha }\left(
Z,Z^{\prime }\right) $ are defined by:%
\begin{equation*}
\mathbf{A}^{\alpha }=\sqrt{\mathbf{D}^{2}+\left( \mathbf{M}^{\alpha }\right)
^{t}\mathbf{M}^{\alpha }}
\end{equation*}%
\begin{equation*}
C=\sum_{\left( Z,Z^{\prime }\right) }\mathbf{C}\left( Z,Z^{\prime }\right)
\end{equation*}%
with:%
\begin{equation*}
\mathbf{C}\left( Z,Z^{\prime }\right) =\frac{\tau \omega _{S}\left( \mathbf{Z%
}_{1}\right) }{2}+\frac{\rho \left( C\left( \theta \right) \left\vert \Psi
_{0}\left( \mathbf{Z}_{1}\right) \right\vert ^{2}\omega _{S}\left( \mathbf{Z}%
_{1}\right) +D\left( \theta \right) \left\vert \Psi _{0}\left( \mathbf{Z}%
_{2}\right) \right\vert ^{2}\omega _{S}\left( \mathbf{Z}_{2}\right) \right) 
}{2\omega _{S}\left( \mathbf{Z}_{1}\right) }
\end{equation*}

\subsection*{3.2 Rewriting effective action for collective state}

Using a change of variable similar to (\cite{GLt}), we obtain:%
\begin{eqnarray}
&&S\left( \left\{ \Delta \Gamma _{S_{\alpha }^{2}}\left( \left( \mathbf{T},%
\mathbf{\hat{T}},\mathbf{Z}\right) _{S_{\alpha }^{2}},\theta \right)
\right\} \right)  \label{CNV} \\
&=&-\Delta \Gamma _{S_{\alpha }^{2}}^{\dag }\left( \left( \mathbf{T},\mathbf{%
\hat{T}},\mathbf{Z}\right) _{S_{\alpha }^{2}},\theta \right)  \notag \\
&&\left( \nabla _{\mathbf{T}}^{2}-\frac{1}{2}\left( \frac{\left( \Delta 
\mathbf{T}-\underline{\left\langle \Delta \mathbf{T}\left( \mathbf{Z}\right)
\right\rangle }\right) -\lambda \left( \Delta \mathbf{\hat{T}}-\underline{%
\left\langle \Delta \mathbf{\hat{T}}\left( \mathbf{Z}\right) \right\rangle }%
\right) }{\tau \omega _{S}\left( \mathbf{Z}_{1}\right) }\left\vert \Psi
_{0}\left( \theta ,\mathbf{Z}_{1}\right) \right\vert ^{2}\right) ^{2}\right)
\Delta \Gamma _{S_{\alpha }^{2}}\left( \left( \mathbf{T},\mathbf{\hat{T}},%
\mathbf{Z}\right) _{S_{\alpha }^{2}},\theta \right)  \notag \\
&&-\Delta \Gamma ^{\dag }\left( \left( \mathbf{T},\mathbf{\hat{T}},\mathbf{Z}%
\right) _{S_{\alpha }^{2}},\theta \right)  \notag \\
&&\times \left( \nabla _{\mathbf{\hat{T}}}^{2}-\frac{1}{2}\left( \frac{\rho
\left( C\left( \theta \right) \left\vert \Psi _{0}\left( \mathbf{Z}%
_{1}\right) \right\vert ^{2}\omega _{S}\left( \mathbf{Z}_{1}\right) +D\left(
\theta \right) \left\vert \Psi _{0}\left( \mathbf{Z}_{2}\right) \right\vert
^{2}\omega _{S}\left( \mathbf{Z}_{2}\right) \right) \left( \Delta \mathbf{%
\hat{T}}-\Delta \mathbf{\hat{T}}^{\prime }\left( \mathbf{Z}\right) \right) }{%
\omega _{S}\left( \mathbf{Z}_{1}\right) }\right) ^{2}\right) \Delta \Gamma
\left( \left( \mathbf{T},\mathbf{\hat{T}},\mathbf{Z}\right) _{S_{\alpha
}^{2}},\theta \right)  \notag \\
&&+\left( \tau \omega _{S}\left( \mathbf{Z}_{1}\right) +\frac{\rho \left(
C\left( \theta \right) \left\vert \Psi _{0}\left( \mathbf{Z}_{1}\right)
\right\vert ^{2}\omega _{S}\left( \mathbf{Z}_{1}\right) +D\left( \theta
\right) \left\vert \Psi _{0}\left( \mathbf{Z}_{2}\right) \right\vert
^{2}\omega _{S}\left( \mathbf{Z}_{2}\right) \right) }{\omega _{S}\left( 
\mathbf{Z}_{1}\right) }\right) \left\Vert \Delta \Gamma _{S_{\alpha
}^{2}}\left( \left( \mathbf{T},\mathbf{\hat{T}},\mathbf{Z}\right)
_{S_{\alpha }^{2}},\theta \right) \right\Vert ^{2}
\end{eqnarray}%
\bigskip

where:%
\begin{equation*}
\Delta \mathbf{\hat{T}}-\Delta \mathbf{\hat{T}}^{\prime }=\Delta \mathbf{%
\hat{T}}-\left\langle \Delta \mathbf{\hat{T}}\right\rangle +\frac{\left(
D\left( \theta \right) \left\langle \hat{T}\right\rangle \left\vert \Psi
_{0}\left( Z^{\prime }\right) \right\vert ^{2}\left\vert Z-Z^{\prime
}\right\vert g\Delta \omega \left( Z,\Delta \mathbf{T}\right) \right) }{%
\left( C\left( \theta \right) \left\vert \Psi _{0}\left( Z\right)
\right\vert ^{2}\omega _{S}\left( Z\right) +D\left( \theta \right)
\left\langle \hat{T}\right\rangle \left\vert \Psi _{0}\left( Z^{\prime
}\right) \right\vert ^{2}\omega _{S}\left( Z^{\prime }\right) \right) }
\end{equation*}%
and $\omega _{S}$ is the static part of the overall activity. We can rewrite
the differences in (\ref{CNV}). We define:

\begin{equation*}
\left[ \mathbf{N}_{p}^{\alpha }\right] _{\left( Z_{i},Z_{j}\right) }=\left(
\delta _{ij}-\left[ \mathbf{\Delta T}\right] _{\left( Z_{i},Z_{j}\right)
}\exp \left( -i\Upsilon _{p}\frac{\left\vert Z_{i}-Z_{j}\right\vert }{c}%
\right) \right) ^{-1}
\end{equation*}%
so that in first approximation we have:%
\begin{eqnarray*}
\Delta \mathbf{\hat{T}}-\Delta \mathbf{\hat{T}}^{\prime } &=&\Delta \mathbf{%
\hat{T}}-\left\langle \Delta \mathbf{\hat{T}}\right\rangle +\frac{\left(
D\left( \theta \right) \left\langle \hat{T}\right\rangle \left\vert \Psi
_{0}\left( Z^{\prime }\right) \right\vert ^{2}\left\vert Z-Z^{\prime
}\right\vert g\Delta \omega \left( Z,\Delta \mathbf{T}\right) \right) }{%
\left( C\left( \theta \right) \left\vert \Psi _{0}\left( Z\right)
\right\vert ^{2}\omega _{S}\left( Z\right) +D\left( \theta \right)
\left\langle \hat{T}\right\rangle \left\vert \Psi _{0}\left( Z^{\prime
}\right) \right\vert ^{2}\omega _{S}\left( Z^{\prime }\right) \right) } \\
&=&\Delta \mathbf{\hat{T}}-\left\langle \Delta \mathbf{\hat{T}}\right\rangle
+\left( \nabla _{\mathbf{\Delta T}_{\left( Z_{1},Z_{1}^{\prime }\right)
}}\left( \Delta \omega \left( Z,\left\langle \mathbf{\Delta T}\right\rangle
\right) \right) _{\left( \left\langle \Delta \mathbf{T}_{\left(
Z_{1},Z_{1}^{\prime }\right) }\right\rangle _{p}^{\alpha }\right) }\right)
\left( \Delta \mathbf{T}-\left\langle \Delta \mathbf{T}\right\rangle
^{\alpha }\right)
\end{eqnarray*}

and:%
\begin{equation*}
\frac{\left( \Delta \mathbf{T}-\underline{\left\langle \Delta \mathbf{T}%
\left( \mathbf{Z}\right) \right\rangle }\right) -\lambda \left( \Delta 
\mathbf{\hat{T}}-\underline{\left\langle \Delta \mathbf{\hat{T}}\left( 
\mathbf{Z}\right) \right\rangle }\right) }{\tau \omega _{S}\left( \mathbf{Z}%
_{1}\right) }\left\vert \Psi _{0}\left( \theta ,\mathbf{Z}_{1}\right)
\right\vert ^{2}=\frac{\left( \Delta \mathbf{T}-\left\langle \Delta \mathbf{T%
}\left( \mathbf{Z}\right) \right\rangle \right) -\lambda \left( \Delta 
\mathbf{\hat{T}}-\left\langle \Delta \mathbf{\hat{T}}\left( \mathbf{Z}%
\right) \right\rangle \right) }{\tau \omega _{S}\left( \mathbf{Z}_{1}\right) 
}\left\vert \Psi _{0}\left( \theta ,\mathbf{Z}_{1}\right) \right\vert ^{2}
\end{equation*}%
Then, we rewrite:%
\begin{eqnarray*}
&&\frac{\rho \left( C\left( \theta \right) \left\vert \Psi _{0}\left( 
\mathbf{Z}_{1}\right) \right\vert ^{2}\omega _{S}\left( \mathbf{Z}%
_{1}\right) +D\left( \theta \right) \left\vert \Psi _{0}\left( \mathbf{Z}%
_{2}\right) \right\vert ^{2}\omega _{S}\left( \mathbf{Z}_{2}\right) \right)
\left( \Delta \mathbf{\hat{T}}-\Delta \mathbf{\hat{T}}^{\prime }\left( 
\mathbf{Z}\right) \right) }{\omega _{S}\left( \mathbf{Z}_{1}\right) } \\
&=&\mathbf{D}\left( \Delta \mathbf{\hat{T}}-\left\langle \Delta \mathbf{\hat{%
T}}\right\rangle \right) +\mathbf{M}^{\alpha }\left( \Delta \mathbf{T}%
-\left\langle \Delta \mathbf{T}\right\rangle ^{\alpha }\right)
\end{eqnarray*}%
with:%
\begin{equation*}
\mathbf{D}\left( \Delta \mathbf{\hat{T}}-\left\langle \Delta \mathbf{\hat{T}}%
\right\rangle \right) =D\left[ \frac{\rho \left( C\left( \theta \right)
\left\vert \Psi _{0}\left( Z\right) \right\vert ^{2}\omega _{S}\left(
Z\right) +D\left( \theta \right) \hat{T}\left\vert \Psi _{0}\left( Z^{\prime
}\right) \right\vert ^{2}\omega _{S}\left( Z^{\prime }\right) \right) }{%
\omega _{S}\left( Z\right) }\right] \left( \Delta \mathbf{\hat{T}}%
-\left\langle \Delta \mathbf{\hat{T}}\right\rangle \right)
\end{equation*}%
and:%
\begin{eqnarray*}
&&\mathbf{M}^{\alpha }\left( \Delta \mathbf{T}-\left\langle \Delta \mathbf{T}%
\right\rangle ^{\alpha }\right) \\
&=&\frac{\rho }{\omega _{S}\left( Z\right) }\left( D\left( \theta \right)
\left\langle \hat{T}\right\rangle \left\vert \Psi _{0}\left( Z^{\prime
}\right) \right\vert ^{2}A\left\vert Z-Z^{\prime }\right\vert \left( \nabla
_{\mathbf{\Delta T}_{\left( Z_{1},Z_{1}^{\prime }\right) }}\left( \Delta
\omega \left( Z,\left\langle \mathbf{\Delta T}\right\rangle \right) \right)
_{\left( \left\langle \Delta \mathbf{T}_{\left( Z_{1},Z_{1}^{\prime }\right)
}\right\rangle _{p}^{\alpha }\right) }\right) \right) \left( \Delta \mathbf{T%
}-\left\langle \Delta \mathbf{T}\right\rangle ^{\alpha }\right)
\end{eqnarray*}%
For a given $\left\langle \Delta \mathbf{T}\right\rangle _{p}^{\alpha }$,
the action $S\left( \left\{ \Delta \Gamma _{S_{\alpha }^{2}}\left( \left( 
\mathbf{T},\mathbf{\hat{T}},\mathbf{Z}\right) _{S_{\alpha }^{2}},\theta
\right) \right\} \right) $ becomes in frst approximation:%
\begin{eqnarray}
&&S\left( \left\{ \Delta \Gamma _{S_{\alpha }^{2}}\left( \left( \mathbf{T},%
\mathbf{\hat{T}},\mathbf{Z}\right) _{S_{\alpha }^{2}},\theta \right)
\right\} \right) \\
&=&-\Delta \Gamma _{S_{\alpha }^{2}}^{\dag }\left( \left( \mathbf{T},\mathbf{%
\hat{T}},\mathbf{Z}\right) _{S_{\alpha }^{2}},\theta \right)  \notag \\
&&\times \left( \nabla _{\mathbf{T}}^{2}-\frac{1}{2}\left( \frac{\left(
\Delta \mathbf{T}-\left\langle \Delta \mathbf{T}\right\rangle ^{\alpha
}\right) -\lambda \left( \Delta \mathbf{\hat{T}}-\left\langle \Delta \mathbf{%
\hat{T}}\right\rangle \right) }{\tau \omega _{S}\left( \mathbf{Z}_{1}\right) 
}\left\vert \Psi _{0}\left( \theta ,\mathbf{Z}_{1}\right) \right\vert
^{2}\right) ^{2}\right) \Delta \Gamma _{S_{\alpha }^{2}}\left( \left( 
\mathbf{T},\mathbf{\hat{T}},\mathbf{Z}\right) _{S_{\alpha }^{2}},\theta
\right)  \notag \\
&&-\Delta \Gamma ^{\dag }\left( \left( \mathbf{T},\mathbf{\hat{T}},\mathbf{Z}%
\right) _{S_{\alpha }^{2}},\theta \right) \left( \nabla _{\mathbf{\hat{T}}%
}^{2}-\frac{1}{2}\left( \mathbf{D}\left( \Delta \mathbf{\hat{T}}%
-\left\langle \Delta \mathbf{\hat{T}}\right\rangle \right) +\mathbf{M}%
^{\alpha }\left( \Delta \mathbf{T}-\left\langle \Delta \mathbf{T}%
\right\rangle ^{\alpha }\right) \right) ^{2}\right) \Delta \Gamma \left(
\left( \mathbf{T},\mathbf{\hat{T}},\mathbf{Z}\right) _{S_{\alpha
}^{2}},\theta \right)  \notag
\end{eqnarray}%
Given our hypotheses, we have:%
\begin{equation*}
\left\Vert \Delta \mathbf{\hat{T}}-\left\langle \Delta \mathbf{\hat{T}}%
\right\rangle \right\Vert <<\left\Vert \Delta \mathbf{T}-\left\langle \Delta 
\mathbf{T}\right\rangle ^{\alpha }\right\Vert
\end{equation*}%
Ultimately, considering the potential, the field action simplifies in first
apprximation: 
\begin{eqnarray}
&&S\left( \left\{ \Delta \Gamma _{S_{\alpha }^{2}}\left( \left( \mathbf{T},%
\mathbf{\hat{T}},\mathbf{Z}\right) _{S_{\alpha }^{2}},\theta \right)
\right\} \right) \\
&\simeq &S\left( \left\{ \Delta \Gamma _{S_{\alpha }^{2}}\left( \left( 
\mathbf{T},\mathbf{Z}\right) _{S_{\alpha }^{2}},\theta \right) \right\}
\right)  \notag \\
&=&-\Delta \Gamma _{S_{\alpha }^{2}}^{\dag }\left( \left( \mathbf{T},\mathbf{%
Z}\right) _{S_{\alpha }^{2}},\theta \right) \left( \nabla _{\mathbf{T}}^{2}-%
\frac{1}{2}\left( \mathbf{A}\left( \Delta \mathbf{T}-\left\langle \Delta 
\mathbf{T}\right\rangle ^{\alpha }\right) \right) ^{2}-\mathbf{C}\right)
\Delta \Gamma _{S_{\alpha }^{2}}\left( \left( \mathbf{T},\mathbf{Z}\right)
_{S_{\alpha }^{2}},\theta \right)  \notag \\
&&+C\left\Vert \Delta \Gamma _{S_{\alpha }^{2}}\left( \left( \mathbf{T},%
\mathbf{Z}\right) _{S_{\alpha }^{2}},\theta \right) \right\Vert ^{2}+U\left(
\left\Vert \Gamma \left( \left( \mathbf{T},\mathbf{\hat{T}},\mathbf{Z}%
\right) _{S_{\alpha }^{2}},\theta \right) \right\Vert ^{2}\right)
\end{eqnarray}%
where:%
\begin{equation*}
\mathbf{A}^{\alpha }=\sqrt{\mathbf{D}^{2}+\left( \mathbf{M}^{\alpha }\right)
^{t}\mathbf{M}^{\alpha }}
\end{equation*}%
and:%
\begin{equation*}
\mathbf{C}\left( Z,Z^{\prime }\right) =\frac{\tau \omega _{S}\left( \mathbf{Z%
}_{1}\right) }{2}+\frac{\rho \left( C\left( \theta \right) \left\vert \Psi
_{0}\left( \mathbf{Z}_{1}\right) \right\vert ^{2}\omega _{S}\left( \mathbf{Z}%
_{1}\right) +D\left( \theta \right) \left\vert \Psi _{0}\left( \mathbf{Z}%
_{2}\right) \right\vert ^{2}\omega _{S}\left( \mathbf{Z}_{2}\right) \right) 
}{2\omega _{S}\left( \mathbf{Z}_{1}\right) }
\end{equation*}

\section*{Appendix 4. n interacting fields}

The Equations for activities are similar to the $n$ field case and are
defined by:%
\begin{eqnarray*}
&&\omega _{i}\left( J,\theta ,Z,\left\vert \Psi \right\vert ^{2}\right) \\
&=&G\left( \int \sum \frac{\kappa }{N}g^{ij}\frac{\omega _{j}\left( J,\theta
-\frac{\left\vert Z-Z_{1}\right\vert }{c},Z_{1},\Psi \right) T_{ij}\left(
Z,\theta ,Z_{1},\theta -\frac{\left\vert Z-Z_{1}\right\vert }{c}\right) }{%
\omega _{i}\left( J,\theta ,Z,\left\vert \Psi \right\vert ^{2}\right) }%
\left\vert \Psi _{j}\left( \theta -\frac{\left\vert Z-Z_{1}\right\vert }{c}%
,Z_{1}\right) \right\vert ^{2}dZ_{1}\right)
\end{eqnarray*}%
The $n\times n$ matrix $G$ has coefficients in the interval $\left[ -1,1%
\right] $. In the sequel, the sum over index $j$ is implicit. For instance,
if $n=2$, the matrix $g$:%
\begin{equation*}
G=\left( 
\begin{array}{cc}
1 & -g \\ 
-g & 1%
\end{array}%
\right)
\end{equation*}%
represents inhibitory interactions between two populations, with, similarly
to the one field case:

\begin{equation*}
T_{ij}\left( Z,Z_{1}\right) =\int T_{ij}\left\vert \Gamma _{ij}\left( T_{ij},%
\hat{T}_{ij},\theta ,Z,Z^{\prime },C_{ij},D_{ij}\right) \right\vert ^{2}
\end{equation*}%
The field action for connectivity is recalled in appendix 2 and decomposed
in four terms plus potential: 
\begin{equation*}
S_{\Gamma }^{\left( 1\right) }+S_{\Gamma }^{\left( 2\right) }+S_{\Gamma
}^{\left( 3\right) }+S_{\Gamma }^{\left( 4\right) }+U\left( \left\{
\left\vert \Gamma \left( \theta ,Z,Z^{\prime },C,D\right) \right\vert
^{2}\right\} \right)
\end{equation*}%
We also derive an expression for the average connectivity functions $T_{ij}$
and the equilibrium static activities as a function of the non interacting
frquencies, by assuming:%
\begin{equation*}
G^{ij}T_{ij}<<T_{ii}
\end{equation*}%
This is valid if structures $i$ and $j$ are distant so that we have:%
\begin{equation*}
G^{j}=\exp \left( -\frac{d}{\nu }\right) \bar{G}^{j}
\end{equation*}%
for $d$ the average distant between structures. In such case, we also have $%
T_{ij}<<T_{ii}$ for $i\neq j$.

The field action for connectivity flds is decomposed in four terms: 
\begin{equation*}
S_{\Gamma }^{\left( 1\right) }+S_{\Gamma }^{\left( 2\right) }+S_{\Gamma
}^{\left( 3\right) }+S_{\Gamma }^{\left( 4\right) }+U\left( \left\{
\left\vert \Gamma \left( \theta ,Z,Z^{\prime },C,D\right) \right\vert
^{2}\right\} \right)
\end{equation*}%
where: 
\begin{eqnarray}
S_{\Gamma }^{\left( 1\right) } &=&\int \Gamma _{ij}^{\dag }\left( T_{ij},%
\hat{T}_{ij},\theta ,Z,Z^{\prime },C_{ij},D_{ij}\right) \\
&&\times \nabla _{T_{ij}}\left( \frac{\sigma _{T}^{2}}{2}\nabla
_{T_{ij}}-\left( -\frac{1}{\tau \omega }T_{ij}+\frac{\lambda }{\omega }\hat{T%
}_{ij}\right) \left\vert \Psi _{i}\left( \theta ,Z\right) \right\vert
^{2}\right) \Gamma _{ij}\left( T_{ij},\hat{T}_{ij},\theta ,Z,Z^{\prime
},C_{ij},D_{ij}\right)  \notag
\end{eqnarray}%
\begin{eqnarray}
S_{\Gamma }^{\left( 2\right) } &=&\int \Gamma _{ij}^{\dag }\left( T_{ij},%
\hat{T}_{ij},\theta ,Z,Z^{\prime },C_{ij},D_{ij}\right) \\
&&\times \nabla _{\hat{T}_{ij}}\left( \frac{\sigma _{\hat{T}_{ij}}^{2}}{2}%
\nabla _{\hat{T}_{ij}}-\frac{\rho }{\omega _{i}\left( J,\theta ,Z,\left\vert
\Psi \right\vert ^{2}\right) }\left( \left( h_{ij}\left( Z,Z^{\prime
}\right) -\hat{T}_{ij}\right) C\left\vert \Psi \left( \theta ,Z\right)
\right\vert ^{2}h_{C}\left( \omega _{i}\left( J,\theta ,Z,\left\vert \Psi
\right\vert ^{2}\right) \right) \right. \right.  \notag \\
&&\left. \left. -D\hat{T}_{ij}\left\vert \Psi _{j}\left( \theta -\frac{%
\left\vert Z-Z^{\prime }\right\vert }{c},Z^{\prime }\right) \right\vert
^{2}h_{D}\left( \omega _{j}\left( J,\theta -\frac{\left\vert Z-Z^{\prime
}\right\vert }{c},Z^{\prime },\left\vert \Psi \right\vert ^{2}\right)
\right) \right) \right) \Gamma _{ij}\left( T_{ij},\hat{T}_{ij},\theta
,Z,Z^{\prime },C_{ij},D_{ij}\right)  \notag
\end{eqnarray}%
\begin{eqnarray}
S_{\Gamma }^{\left( 3\right) } &=&\Gamma _{ij}^{\dag }\left( T,\hat{T}%
,\theta ,Z,Z^{\prime },C_{ij},D_{ij}\right) \\
&&\times \nabla _{C_{ij}}\left( \frac{\sigma _{C_{ij}}^{2}}{2}\nabla
_{C}\right.  \notag \\
&&\left. +\left( \frac{C}{\tau _{C}\omega _{i}\left( J,\theta ,Z,\left\vert
\Psi \right\vert ^{2}\right) }-\alpha _{C}\left( 1-C\right) \frac{\omega
_{j}\left( J,\theta -\frac{\left\vert Z-Z^{\prime }\right\vert }{c}%
,Z^{\prime },\left\vert \Psi \right\vert ^{2}\right) \left\vert \Psi
_{j}\left( \theta -\frac{\left\vert Z-Z^{\prime }\right\vert }{c},Z^{\prime
}\right) \right\vert ^{2}}{\omega _{i}\left( J,\theta ,Z,\left\vert \Psi
\right\vert ^{2}\right) }\right) \left\vert \Psi _{i}\left( \theta ,Z\right)
\right\vert ^{2}\right)  \notag \\
&&\times \Gamma _{ij}\left( T,\hat{T},\theta ,Z,Z^{\prime },C,D\right) 
\notag
\end{eqnarray}%
and:%
\begin{eqnarray}
S_{\Gamma }^{\left( 4\right) } &=&\Gamma _{ij}^{\dag }\left( T,\hat{T}%
,\theta ,Z,Z^{\prime },C,D\right) \\
&&\times \nabla _{D}\left( \frac{\sigma _{D}^{2}}{2}\nabla _{D}+\left( \frac{%
D}{\tau _{D}\omega _{i}\left( J,\theta ,Z,\left\vert \Psi \right\vert
^{2}\right) }-\alpha _{D}\left( 1-D\right) \left\vert \Psi \left( \theta
,Z\right) \right\vert ^{2}\right) \right) \Gamma _{ij}\left( T,\hat{T}%
,\theta ,Z,Z^{\prime },C,D\right)  \notag
\end{eqnarray}%
In (\ref{flt}), we added a potential:%
\begin{equation*}
U\left( \left\{ \left\vert \Gamma \left( \theta ,Z,Z^{\prime },C,D\right)
\right\vert ^{2}\right\} \right) =U\left( \int T\left\vert \Gamma \left( T,%
\hat{T},\theta ,Z,Z^{\prime },C,D\right) \right\vert ^{2}dTd\hat{T}\right)
\end{equation*}%
that models the constraint about the number of active connections in the
system.

\subsection*{4.1 Equilibrium \ activities}

We saw in part III how to compute background activities. Writing activities
equations in the static approximation:%
\begin{equation}
\left( \omega _{S}\right) _{i}\left( Z\right) =G\left( \sum_{j}\int \frac{%
\kappa }{N}\frac{G^{ij}T_{ij}\left\vert \Gamma _{ij}\left( T,\hat{T}%
,Z,Z_{1}\right) \right\vert ^{2}\left( \omega _{S}\right) _{j}\left(
J,Z_{1}\right) }{\left( \omega _{S}\right) _{i}\left( Z\right) }\left( 
\mathcal{G}_{j0}+\left\vert \Psi _{j}\left( Z_{1}\right) \right\vert
^{2}\right) dZ_{1}\right)  \label{cnts}
\end{equation}%
For a given structures with specific points, we replace the integrals in the
previous formula (\ref{cnts}) by sums:%
\begin{equation}
\left( \omega _{S}\right) _{i}\left( Z_{a_{i}},\left\vert \Psi \right\vert
^{2}\right) =G\left( \sum_{j}\sum_{b_{j}}\frac{\kappa }{N}G^{ij}\frac{\left(
\omega _{S}\right) _{j}\left( Z_{b_{j}}\right) T_{ij}\left(
Z_{a_{i}},Z_{b_{j}}\right) }{\left( \omega _{S}\right) _{i}\left(
Z_{a_{i}},\left\vert \Psi \right\vert ^{2}\right) }\left( \mathcal{G}%
_{j0}+\left\vert \Psi _{j}\left( Z_{b_{j}}\right) \right\vert ^{2}\right)
\right)  \label{dsct}
\end{equation}%
Assuming:%
\begin{equation*}
G^{ij}T_{ij}<<T_{ii}
\end{equation*}%
we expand (\ref{dsct}) to the first order:%
\begin{eqnarray*}
\left( \omega _{S}\right) _{i}\left( Z_{a_{i}}\right) &\simeq &G\left(
\sum_{a_{i}}\frac{\kappa }{N}\frac{\left( \omega _{S}\right) _{i}\left(
Z_{\alpha _{j}}\right) T_{ii}\left( Z_{a_{i}},Z_{\alpha _{j}}\right) }{%
\left( \omega _{S}\right) _{i}\left( Z_{a_{i}},\left\vert \Psi \right\vert
^{2}\right) }\left( \mathcal{G}_{i0}+\left\vert \Psi _{i}\left(
Z_{a_{j}}\right) \right\vert ^{2}\right) \right) \\
&&+\sum_{j\neq i,\left\{ \left\{ b_{j}\right\} \right\} }G^{\prime }\left(
\sum_{a_{i}}\frac{\kappa }{N}\frac{\left( \omega _{S}\right) _{i}\left(
Z_{\alpha _{j}}\right) T_{ii}\left( Z_{a_{i}},Z_{b_{j}}\right) }{\left(
\omega _{S}\right) _{i}\left( Z_{a_{i}},\left\vert \Psi \right\vert
^{2}\right) }\left( \mathcal{G}_{i0}+\left\vert \Psi _{i}\left(
Z_{b_{j}}\right) \right\vert ^{2}\right) \right) \\
&&\times \frac{\kappa }{N}G^{ij}\frac{\left( \omega _{S}\right) _{j}\left(
Z_{b_{j}}\right) T_{ij}\left( Z_{a_{i}},Z_{b_{j}}\right) }{\left( \omega
_{S}\right) _{i}\left( Z_{a_{i}},\left\vert \Psi \right\vert ^{2}\right) }%
\left( \mathcal{G}_{j0}+\left\vert \Psi _{j}\left( Z_{b_{j}}\right)
\right\vert ^{2}\right)
\end{eqnarray*}%
where $G^{ii}=1$, and this becomes:%
\begin{eqnarray}
\left( \omega _{S}\right) _{i}\left( Z_{a_{i}}\right) &\simeq &\left( \omega
_{Sf}\right) _{i}\left( Z_{a_{i}}\right)  \label{FRS} \\
&&+G^{\prime }\left( G^{-1}\left( \left( \omega _{Sf}\right) _{i}\left(
Z_{a_{i}}\right) \right) \right)  \notag \\
&&\times \sum_{j\neq i,\left\{ \left\{ b_{j}\right\} \right\} }\frac{\kappa 
}{N}G^{ij}\frac{\left( \omega _{S}\right) _{j}\left( Z_{b_{j}}\right)
T_{ij}\left( Z_{a_{i}},Z_{b_{j}}\right) }{\left( \omega _{Sf}\right)
_{i}\left( Z_{a_{i}},\left\vert \Psi \right\vert ^{2}\right) }\left( 
\mathcal{G}_{j0}+\left\vert \Psi _{j}\left( Z_{b_{j}}\right) \right\vert
^{2}\right)  \notag
\end{eqnarray}%
where $\left( \omega _{Sf}\right) _{i}\left( Z\right) $ are static
activities without interactions (where $G^{ii}=1$):%
\begin{equation*}
\left( \omega _{Sf}\right) _{i}\left( Z_{a_{i}}\right) =G\left(
\sum_{a_{i}^{\prime }}\frac{\kappa }{N}\frac{\left( \omega _{Sf}\right)
_{j}\left( Z_{a_{i}^{\prime }}\right) T_{ii}\left(
Z_{a_{i}},Z_{a_{i}^{\prime }}\right) }{\left( \omega _{Sf}\right) _{i}\left(
Z_{a_{i}},\left\vert \Psi \right\vert ^{2}\right) }\left( \mathcal{G}%
_{i0}+\left\vert \Psi _{i}\left( Z_{a_{i}^{\prime }}\right) \right\vert
^{2}\right) \right)
\end{equation*}%
Equation (\ref{FRS}) has solutions:%
\begin{eqnarray}
&&\left( \omega _{S}\right) _{i}\left( Z_{a_{i}}\right)  \label{mdf} \\
&=&\sum_{\left\{ \left\{ b_{j}\right\} \right\} }\left( \delta _{\left(
i,a_{i}\right) \left( j,b_{j}\right) }\right.  \notag \\
&&\left. -G^{\prime }\left( G^{-1}\left( \left( \omega _{Sf}\right)
_{i}\left( Z_{a_{i}}\right) \right) \right) \left( \frac{\kappa }{N}\frac{%
G^{ij}\left( \omega _{Sf}\right) _{j}\left( Z_{b_{j}}\right) T_{ij}\left(
Z_{a_{i}},Z_{b_{j}}\right) }{\left( \omega _{Sf}\right) _{i}\left(
Z_{a_{i}}\right) }\left( \mathcal{G}_{j0}+\left\vert \Psi _{j}\left(
Z_{b_{j}}\right) \right\vert ^{2}\right) \right) _{j\neq i}\right)
_{ij}^{-1}\left( \omega _{Sf}\right) _{j}\left( Z_{b_{j}}\right)  \notag
\end{eqnarray}

Remark that for inhibitory interactions $\omega _{i}\left( Z\right) <\left(
\omega _{f}\right) _{i}\left( Z\right) $.

\section*{4.2 Connectivity functions}

In part 1 we obtained average connectivity functions:%
\begin{eqnarray}
T_{ij}\left( Z_{-},Z_{+}^{\prime }\right) &=&\frac{\lambda \tau \exp \left( -%
\frac{\left\vert Z-Z^{\prime }\right\vert }{\nu c}\right) \left( \frac{1}{%
\tau _{D}\alpha _{D}}+\frac{1}{bG^{i}\bar{T}\left\langle \bar{T}\left\vert 
\bar{\Psi}_{i0}\left( Z^{\prime }\right) \right\vert ^{2}\right\rangle _{Z}}%
\right) }{\frac{1}{\tau _{D}\alpha _{D}}+\frac{1}{\alpha _{C}\tau _{C}}+%
\frac{1}{bG^{i}\bar{T}\left\langle \bar{T}\left\vert \bar{\Psi}_{i0}\left(
Z^{\prime }\right) \right\vert ^{2}\right\rangle _{Z}}+\frac{\left( bG^{j}%
\bar{T}\right) \left( \bar{T}\left\langle \left\vert \bar{\Psi}_{j0}\left(
Z^{\prime }\right) \right\vert ^{2}\right\rangle _{Z^{\prime }}\right) ^{2}}{%
2}}\simeq 0  \label{cnv} \\
T_{ij}\left( Z_{+},Z_{+}^{\prime }\right) &=&\frac{\lambda \tau \exp \left( -%
\frac{\left\vert Z-Z^{\prime }\right\vert }{\nu c}\right) \left( \frac{1}{%
\tau _{D}\alpha _{D}}+\frac{\left( bG^{i}\bar{T}\right) \left( \bar{T}%
\left\langle \left\vert \bar{\Psi}_{i0}\left( Z^{\prime }\right) \right\vert
^{2}\right\rangle _{Z}\right) ^{2}}{2}\right) }{\frac{1}{\tau _{D}\alpha _{D}%
}+\frac{1}{\alpha _{C}\tau _{C}}+\frac{\left( bG^{i}\bar{T}\right) \left( 
\bar{T}\left\langle \left\vert \bar{\Psi}_{i0}\left( Z^{\prime }\right)
\right\vert ^{2}\right\rangle _{Z}\right) ^{2}}{2}+\frac{\left( bG^{j}\bar{T}%
\right) \left( \bar{T}\left\langle \left\vert \bar{\Psi}_{j0}\left(
Z^{\prime }\right) \right\vert ^{2}\right\rangle _{Z^{\prime }}\right) ^{2}}{%
2}}\simeq \frac{G^{i}\lambda \tau \exp \left( -\frac{\left\vert Z-Z^{\prime
}\right\vert }{\nu c}\right) }{\left( G^{i}+G^{j}\right) }  \notag \\
T_{ij}\left( Z_{+},Z_{-}^{\prime }\right) &=&\frac{\lambda \tau \exp \left( -%
\frac{\left\vert Z-Z^{\prime }\right\vert }{\nu c}\right) \left( \frac{1}{%
\tau _{D}\alpha _{D}}+\frac{\left( bG^{i}\bar{T}\right) \left( \bar{T}%
\left\langle \left\vert \bar{\Psi}_{i0}\left( Z^{\prime }\right) \right\vert
^{2}\right\rangle _{Z}\right) ^{2}}{2}\right) }{\frac{1}{\tau _{D}\alpha _{D}%
}+\frac{1}{\alpha _{C}\tau _{C}}+\frac{\left( bG^{i}\bar{T}\right) \left( 
\bar{T}\left\langle \left\vert \bar{\Psi}_{i0}\left( Z^{\prime }\right)
\right\vert ^{2}\right\rangle _{Z}\right) ^{2}}{2}+\frac{1}{bG^{j}\bar{T}%
\left\langle \bar{T}\left\vert \bar{\Psi}_{j0}\left( Z^{\prime }\right)
\right\vert ^{2}\right\rangle _{Z}}}\simeq \lambda \tau \exp \left( -\frac{%
\left\vert Z-Z^{\prime }\right\vert }{\nu c}\right)  \notag \\
T\left( Z_{-},Z_{-}^{\prime }\right) &\simeq &\frac{\lambda \tau \exp \left(
-\frac{\left\vert Z-Z^{\prime }\right\vert }{\nu c}\right) +\frac{1}{bG^{i}%
\bar{T}\left\langle \bar{T}\left\vert \bar{\Psi}_{i0}\left( Z^{\prime
}\right) \right\vert ^{2}\right\rangle _{Z}}}{1+\frac{\tau _{D}\alpha _{D}}{%
\alpha _{C}\tau _{C}}+\frac{1}{bG^{i}\bar{T}\left\langle \bar{T}\left\vert 
\bar{\Psi}_{i0}\left( Z^{\prime }\right) \right\vert ^{2}\right\rangle _{Z}}+%
\frac{1}{bG^{j}\bar{T}\left\langle \bar{T}\left\vert \bar{\Psi}_{j0}\left(
Z^{\prime }\right) \right\vert ^{2}\right\rangle _{Z}}}\simeq \frac{%
G^{j}\lambda \tau \exp \left( -\frac{\left\vert Z-Z^{\prime }\right\vert }{%
\nu c}\right) }{\left( G^{i}+G^{j}\right) }  \notag
\end{eqnarray}%
with $\bar{T}=\frac{\lambda \tau \nu cb}{2}$, $b$ a coefficient
characterizing the function $G$ in the linear approximation\footnote{$%
b\simeq G^{\prime }\left( 0\right) $}, and the coefficient $G^{i}$ measuring
the connectivity of the $i$-th field with the other types. The expressions $%
\left\langle \left\vert \Psi _{i0}\left( Z^{\prime }\right) \right\vert
^{2}\right\rangle _{Z}^{2}$, \ $\left\langle \left\vert \Psi _{i0}\left(
Z^{\prime }\right) \right\vert ^{2}\right\rangle _{Z}^{2}$ are some averaged
background fields for the $t$-th types of cells in regions surrounding $Z$
and $Z^{\prime }$ respctvl. They are determined by a potential describing
some average activity depending on the points. These results are derived
under the assumption of static fields $\Psi _{i0}\left( Z\right) $.

Here we also assumed $G^{j}<<G^{i}$. This is valid if structures $i$ and $j$
are distant so that we have:%
\begin{equation*}
G^{j}=\exp \left( -\frac{d}{\nu }\right) \bar{G}^{j}
\end{equation*}%
for $d$ the average distant between structures.

\section*{Appendix 5. Activities for collective states}

We look for activities equations for collective states. We write the
equations for activities:%
\begin{eqnarray}
&&\left( \omega _{i0}+\Delta \omega _{i}\right) ^{-1}\left( Z_{a_{i}},\theta
\right)   \label{FQN} \\
&=&G\left( \sum_{j}\sum_{b_{j}}\frac{\kappa }{N}g^{ij}\frac{\left( \omega
_{j0}+\Delta \omega _{j}\right) \left( \theta -\frac{\left\vert
Z_{a_{i}}-Z_{b_{j}}\right\vert }{c},Z_{b_{j}}\right) \left( \left\langle
T_{ij}\right\rangle +\left\langle \Delta T_{ij}\right\rangle \right) \left(
Z_{a_{i}},Z_{b_{j}},\theta -\frac{\left\vert Z_{a_{i}}-Z_{b_{j}}\right\vert 
}{c}\right) }{\left( \omega _{i0}+\Delta \omega _{i}\right) \left(
Z_{a_{i}},\theta \right) }\right.   \notag \\
&&\times \left. \left\vert \Psi _{j}\left( \theta -\frac{\left\vert
Z_{a_{i}}-Z_{b_{j}}\right\vert }{c},Z_{b_{j}}\right) \right\vert ^{2}\right) 
\notag
\end{eqnarray}%
where $\left\langle T_{ij}\right\rangle +\left\langle \Delta
T_{ij}\right\rangle $ are the average connectivity, plus its modification
due to nteractn.

Assuming:%
\begin{equation*}
G^{ij}\Delta T_{ij}<<\Delta T_{ii}
\end{equation*}%
leads to a first order expansion with respect to the activities of the
groups considered individually. The solutions will thus be described by
corrections to the individual groups.

\subsection*{5.1 Static part}

We define again:%
\begin{equation*}
\omega _{i}=\omega _{i0}+\Delta \omega _{i}
\end{equation*}%
where $\omega _{i0}$ is the background activt and the additional activity $%
\Delta \omega _{i}$ decomposes into static and dynamic part:%
\begin{equation*}
\Delta \omega _{i}=\overline{\Delta \omega _{i}}+\left( \Delta \omega
_{D}\right) _{i}
\end{equation*}%
The overall static activity is:%
\begin{equation*}
\left( \omega _{S}\right) _{i}=\omega _{i0}+\overline{\Delta \omega _{i}}
\end{equation*}

The static part is derived as for (\ref{mdf}). The static form of (\ref{FQN}%
) is:%
\begin{equation}
\left( \omega _{S}\right) _{i}^{-1}\left( Z_{a_{i}}\right) =G\left(
\sum_{j}\sum_{b_{j}}\frac{\kappa }{N}g^{ij}\frac{\left( \omega _{S}\right)
_{j}\left( Z_{b_{j}}\right) \Delta T_{ij}\left( Z_{a_{i}},Z_{b_{j}}\right) }{%
\left( \omega _{S}\right) _{i}\left( Z_{a_{i}},\theta \right) }\left\vert
\Psi _{j}\left( Z_{b_{j}}\right) \right\vert ^{2}\right)
\end{equation}

The expansion (\ref{FQN}) around non interacting solution:%
\begin{equation*}
\left( \omega _{Sf}\right) _{i}=\omega _{i0}+\left( \overline{\Delta \omega
_{f}}\right) _{i}
\end{equation*}%
that satisfy equation for $g^{ii}=1$:%
\begin{equation*}
\left( \omega _{0}+\left( \overline{\Delta \omega _{f}}\right) \right)
_{i}\left( Z_{a_{i}}\right) =G\left( \sum_{\beta _{i}}\frac{\kappa }{N}\frac{%
\omega _{i0}+\left( \overline{\Delta \omega _{f}}\right) _{i}\left( Z_{\beta
_{i}}\right) T_{ij}\left( Z_{a_{i}},Z_{b_{j}}\right) }{\omega _{i0}+\left( 
\overline{\Delta \omega _{f}}\right) _{i}\left( Z_{a_{i}},\theta \right) }%
\left\vert \Psi _{j}\left( Z_{b_{j}}\right) \right\vert ^{2}\right)
\end{equation*}%
yields an equation similar to (\ref{FRS}): 
\begin{eqnarray}
\left( \omega _{i0}+\overline{\Delta \omega }_{i}\right) \left(
Z_{a_{i}}\right) &\simeq &\left( \omega _{0}+\overline{\Delta \omega }%
_{f}\right) _{i}\left( Z_{a_{i}}\right) \\
&&+G^{\prime }\left( G^{-1}\left( \left( \omega _{Sf}\right) _{i}\left(
Z_{a_{i}}\right) \right) \right)  \notag \\
&&\times \sum_{j\neq i}\sum_{b_{j}}\frac{\kappa }{N}G^{ij}\frac{\left(
\omega _{S}\right) _{j}\left( Z_{b_{j}}\right) T_{ij}\left(
Z_{a_{i}},Z_{b_{j}}\right) }{\left( \omega _{S}\right) _{i}\left(
Z_{a_{i}},\left\vert \Psi \right\vert ^{2}\right) }\left( \mathcal{G}%
_{j0}+\left\vert \Psi _{j}\left( Z_{b_{j}}\right) \right\vert ^{2}\right) 
\notag
\end{eqnarray}%
with solution:%
\begin{eqnarray}
&&\overline{\Delta \omega }_{i}\left( Z_{a_{i}}\right)  \label{STCN} \\
&\simeq &\sum_{j,b_{j}}\left( \delta _{\left( i,a_{i}\right) \left(
j,b_{j}\right) }-G^{\prime }\left( G^{-1}\left( \left( \omega _{Sf}\right)
_{i}\left( Z\right) \right) \right) \right.  \notag \\
&&\left. \left( \frac{\kappa }{N}\frac{G^{ij}\left( \omega _{Sf}\right)
_{j}\left( Z_{b_{j}}\right) T_{ij}\left( Z_{a_{i}},Z_{b_{j}}\right) }{\left(
\omega _{Sf}\right) _{i}\left( Z_{a_{i}}\right) }\left( \mathcal{G}%
_{j0}+\left\vert \Psi _{j}\left( Z_{b_{j}}\right) \right\vert ^{2}\right)
\right) _{j\neq i}\right) _{ij}^{-1}\left( \overline{\Delta \omega }%
_{f}\right) _{j}\left( Z_{b_{j}}\right)  \notag
\end{eqnarray}%
The solutions (\ref{STCN}) mix the static activities of the several groups,
showing that composed structures and individual ones are binded by some
consistencies conditions.

\subsection*{5.2 Non static part}

\subsubsection*{5.2.1 Equation for non static frequencies}

Expand (\ref{frQ}) around (\ref{bsc}):%
\begin{eqnarray*}
&&\left( \Delta \omega _{D}\right) _{i}\left( Z_{a_{i}},\left\vert \Psi
\right\vert ^{2}\right) \\
&=&\Delta G\left( \sum_{j}\sum_{b_{j}}\frac{\kappa }{N}g^{ij}\frac{\left(
\omega _{S}\right) _{j}\left( \theta -\frac{\left\vert
Z_{a_{i}}-Z_{b_{j}}\right\vert }{c},Z_{b_{j}}\right) T_{ij}\left(
Z_{a_{i}},Z_{b_{j}},\theta -\frac{\left\vert Z_{a_{i}}-Z_{b_{j}}\right\vert 
}{c}\right) }{\left( \omega _{S}\right) _{i}\left( Z_{a_{i}},\theta \right) }%
\left\vert \Psi _{j}\left( \theta -\frac{\left\vert
Z_{a_{i}}-Z_{b_{j}}\right\vert }{c},Z_{b_{j}}\right) \right\vert ^{2}\right)
\end{eqnarray*}%
is given at the first order:

\begin{eqnarray*}
&&\left( \Delta \omega _{D}\right) _{i}\left( Z_{a_{i}},\theta \right) \\
&=&G^{\prime }\left( G^{-1}\left( \left( \omega _{S}\right) ^{-1}\left(
Z_{a_{i}},\left\vert \Psi \right\vert ^{2}\right) \right) \right) \\
&&\times \Delta \left( \sum_{j}\sum_{b_{j}}\frac{\kappa }{N}g^{ij}\frac{%
\left( \omega _{S}\right) _{j}\left( \theta -\frac{\left\vert
Z_{a_{i}}-Z_{b_{j}}\right\vert }{c},Z_{b_{j}}\right) T_{ij}\left(
Z_{a_{i}},Z_{b_{j}},\theta -\frac{\left\vert Z_{a_{i}}-Z_{b_{j}}\right\vert 
}{c}\right) }{\left( \omega _{S}\right) _{i}\left( Z_{a_{i}},\theta \right) }%
\left\vert \Psi _{j}\left( \theta -\frac{\left\vert
Z_{a_{i}}-Z_{b_{j}}\right\vert }{c},Z_{b_{j}}\right) \right\vert ^{2}\right)
\end{eqnarray*}

As for one field case, we can neglect:%
\begin{equation*}
\Delta \left\vert \Psi _{j}\left( \theta -\frac{\left\vert
Z_{a_{i}}-Z_{b_{j}}\right\vert }{c},Z_{b_{j}}\right) \right\vert ^{2}
\end{equation*}%
and we obtain:%
\begin{eqnarray*}
&&\left( \Delta \omega _{D}\right) _{i}\left( Z_{a_{i}},\theta \right) \\
&=&G^{\prime }\left( G^{-1}\left( \left( \omega _{S}\right) ^{-1}\left(
Z_{a_{i}},\left\vert \Psi \right\vert ^{2}\right) \right) \right) \\
&&\times \Delta \left( \sum_{j}\sum_{b_{j}}\frac{\kappa }{N}g^{ij}\frac{%
\left( \omega _{S}\right) _{j}\left( \theta -\frac{\left\vert
Z_{a_{i}}-Z_{b_{j}}\right\vert }{c},Z_{b_{j}},\Psi \right) T_{ij}\left(
Z_{a_{i}},Z_{b_{j}},\theta -\frac{\left\vert Z_{a_{i}}-Z_{b_{j}}\right\vert 
}{c}\right) }{\left( \omega _{S}\right) _{i}\left( Z_{a_{i}},\theta \right) }%
\left\vert \Psi _{j}\left( \theta -\frac{\left\vert
Z_{a_{i}}-Z_{b_{j}}\right\vert }{c},Z_{b_{j}}\right) \right\vert ^{2}\right)
\end{eqnarray*}%
The variation in the last equation is given by:

\begin{eqnarray*}
&&\Delta \left( \sum_{j}\sum_{b_{j}}\frac{\kappa }{N}g^{ij}\frac{\left(
\omega _{S}\right) _{j}\left( \theta -\frac{\left\vert
Z_{a_{i}}-Z_{b_{j}}\right\vert }{c},Z_{b_{j}},\Psi \right) T_{ij}\left(
Z_{a_{i}},Z_{b_{j}},\theta -\frac{\left\vert Z_{a_{i}}-Z_{b_{j}}\right\vert 
}{c}\right) }{\left( \omega _{S}\right) _{i}\left( Z_{a_{i}},\theta \right) }%
\left\vert \Psi _{j}\left( \theta -\frac{\left\vert
Z_{a_{i}}-Z_{b_{j}}\right\vert }{c},Z_{b_{j}}\right) \right\vert ^{2}\right) 
\\
&\simeq &\sum_{\left\{ \left\{ b_{j}\right\} \right\} }\frac{\kappa }{N}%
g^{ij}\frac{T_{ij}\left( Z_{a_{i}},Z_{b_{j}}\right) }{\left( \omega
_{S}\right) _{i}\left( Z_{a_{i}}\right) }\left\vert \Psi _{j}\left(
Z_{b_{j}}\right) \right\vert ^{2}\left( \Delta \omega _{D}\right) _{j}\left(
\theta -\frac{\left\vert Z_{a_{i}}-Z_{b_{j}}\right\vert }{c},Z_{b_{j}},\Psi
\right)  \\
&&-\sum_{\left\{ \left\{ b_{j}\right\} \right\} }\frac{\kappa }{N}g^{ij}%
\frac{\left( \omega _{S}\right) _{j}\left( Z_{b_{j}}\right) T_{ij}\left(
Z_{a_{i}},Z_{b_{j}}\right) }{\left( \omega _{S}\right) _{i}\left(
Z_{a_{i}}\right) }\left\vert \Psi _{j}\left( Z_{b_{j}}\right) \right\vert
^{2}\frac{\left( \Delta \omega _{D}\right) _{i}\left( Z_{a_{i}},\theta
\right) }{\left( \omega _{S}\right) _{i}\left( Z_{a_{i}}\right) } \\
&&+\sum_{\left\{ \left\{ b_{j}\right\} \right\} }\frac{\kappa }{N}g^{ij}%
\frac{\left( \omega _{S}\right) _{j}\left( Z_{b_{j}},\Psi \right)
T_{ij}\left( Z_{a_{i}},Z_{b_{j}}\right) }{\left( \omega _{S}\right)
_{i}\left( Z_{a_{i}},\theta \right) }\Delta \left\vert \Psi _{j}\left(
\theta -\frac{\left\vert Z_{a_{i}}-Z_{b_{j}}\right\vert }{c}%
,Z_{b_{j}}\right) \right\vert ^{2}
\end{eqnarray*}%
The frequencies $\left( \omega _{S}\right) _{j}^{-1}\left( Z_{b_{j}}\right) $
represent the static part of the activity:%
\begin{equation*}
\left( \omega _{S}\right) _{j}\left( Z_{b_{j}}\right) =\omega _{j0}\left(
Z_{b_{j}}\right) +\overline{\Delta \omega _{j}}\left( Z_{b_{j}}\right) 
\end{equation*}%
We neglect $\Delta \left\vert \Psi _{j}\left( \theta -\frac{\left\vert
Z_{a_{i}}-Z_{b_{j}}\right\vert }{c},Z_{b_{j}}\right) \right\vert ^{2}$ in
first approximation. Consequently, the activities equation writes:%
\begin{eqnarray*}
&&\left( \left( \left( \omega _{S}\right) _{i}\left( Z_{a_{i}}\right)
\right) ^{2}+\sum_{b_{j}}\frac{\kappa }{N}g^{ij}\left( \omega _{S}\right)
_{j}\left( Z_{b_{j}}\right) G^{\prime }\left( G^{-1}\left( \left( \omega
_{S}\right) ^{-1}\left( Z_{a_{i}},\left\vert \Psi \right\vert ^{2}\right)
\right) \right) \right) \left( \Delta \omega _{D}\right) _{i}\left(
Z_{a_{i}},\theta \right)  \\
&=&G^{\prime }\left( G^{-1}\left( \left( \omega _{S}\right) ^{-1}\left(
Z_{a_{i}},\left\vert \Psi \right\vert ^{2}\right) \right) \right)  \\
&&\times \sum_{j}\sum_{b_{j}}\frac{\kappa }{N}g^{ij}\left( \omega
_{S}\right) _{j}\left( Z_{b_{j}}\right) T_{ij}\left(
Z_{a_{i}},Z_{b_{j}}\right) \left\vert \Psi _{j}\left( Z_{b_{j}}\right)
\right\vert ^{2}\left( \Delta \omega _{D}\right) _{j}\left( \theta -\frac{%
\left\vert Z_{a_{i}}-Z_{b_{j}}\right\vert }{c},Z_{b_{j}},\Psi \right) 
\end{eqnarray*}%
with solution:%
\begin{equation}
\left( \Delta \omega _{D}\right) _{i}^{-1}\left( Z_{a_{i}},\theta \right)
=\sum_{j}\sum_{b_{j}}\check{T}_{ij}\left( Z_{a_{i}},Z_{b_{j}}\right) \left(
\Delta \omega _{D}\right) _{j}\left( \theta -\frac{\left\vert
Z_{a_{i}}-Z_{b_{j}}\right\vert }{c},Z_{b_{j}}\right)   \label{NSL}
\end{equation}

where:%
\begin{eqnarray*}
\check{T}_{ij}\left( Z_{a_{i}},Z_{b_{j}}\right) &=&G^{\prime }\left(
G^{-1}\left( \left( \omega _{S}\right) _{i}^{-1}\left( Z_{a_{i}},\left\vert
\Psi \right\vert ^{2}\right) \right) \right) \\
&&\times \frac{\kappa }{N}g^{ij}\frac{\left( \omega _{S}\right) _{j}\left(
Z_{b_{j}}\right) T_{ij}\left( Z_{a_{i}},Z_{b_{j}}\right) }{\sum_{b_{j}}\frac{%
\kappa }{N}g^{ij}\left( \omega _{S}\right) _{j}\left( Z_{b_{j}}\right)
G^{\prime }\left( G^{-1}\left( \left( \omega _{S}\right) _{i}^{-1}\left(
Z_{a_{i}},\left\vert \Psi \right\vert ^{2}\right) \right) \right) +\left(
\omega _{S}\right) _{i}^{2}\left( Z_{a_{i}}\right) }\left\vert \Psi
_{j}\left( Z_{b_{j}}\right) \right\vert ^{2}
\end{eqnarray*}

\subsubsection*{5.2.2 Equation for oscillating solutions}

Rewriting the solutions (\ref{NSL}) as a vector:%
\begin{equation*}
\left( \left( \Delta \omega \right) _{i}^{-1}\left( Z_{a_{i}},\theta \right)
\right) _{i}\equiv \left( \Delta \mathbf{\omega }_{D}\right) ^{-1}\left( Z_{%
\mathbf{\alpha }},\theta \right) 
\end{equation*}%
we look for oscillatory solutions:%
\begin{equation*}
\left( \Delta \mathbf{\omega }_{D}\right) ^{-1}\left( Z_{\mathbf{\alpha }%
},\theta \right) =\left( \Delta \mathbf{\omega }_{D}\right) ^{-1}\left( Z_{%
\mathbf{\alpha }}\right) \exp \left( i\Upsilon \theta \right) 
\end{equation*}%
which implies that for these solutions, (\ref{NSL}) rewrites as :%
\begin{equation*}
\left( \Delta \mathbf{\omega }_{D}\right) ^{-1}\left( Z_{\mathbf{\alpha }%
}\right) =M\left( \Delta \mathbf{\omega }_{D}\right) ^{-1}\left( Z_{\mathbf{%
\alpha }}\right) 
\end{equation*}%
with:%
\begin{equation*}
M_{\left( ia_{i}\right) ,\left( jb_{j}\right) }=\hat{T}_{ij}\left(
Z_{a_{i}},Z_{b_{j}}\right) \exp \left( -i\Upsilon \frac{\left\vert
Z_{a_{i}}-Z_{b_{j}}\right\vert }{c}\right) 
\end{equation*}%
We decompose the matrix $M_{\left( ia_{i}\right) ,\left( jb_{j}\right) }$
into a diagonal and a non-diagonal part: 
\begin{equation*}
M=\left( M_{\left( ia_{i}\right) ,\left( jb_{j}\right) }\right) +\left(
M_{\left( ia_{i}\right) ,\left( jb_{j}\right) }\right) _{i\neq j}
\end{equation*}%
so that $\left( M_{\left( ia_{i}\right) ,\left( jb_{j}\right) }\right)
_{i\neq j}$ can be studied perturbatively. The condition for oscillatory
solutions writes:%
\begin{equation}
\det \left( 1-M\right) =0  \label{DT}
\end{equation}%
Given our order of approximation, the elements of: 
\begin{equation*}
\left( M_{\left( ia_{i}\right) ,\left( jb_{j}\right) }\right) _{i\neq j}
\end{equation*}%
are of lower magnitude than that of:%
\begin{equation*}
1-\left( M_{\left( ia_{i}\right) ,\left( jb_{j}\right) }\right) 
\end{equation*}%
and we write (\ref{DT}) as:%
\begin{eqnarray}
\det \left( 1-M\right)  &=&\prod \det \left( 1-\left( M_{\left(
ia_{i}\right) ,\left( ia_{i}\right) }\right) \right)   \label{TD} \\
&&\times \exp \left( -\frac{1}{2}Tr\left( \left( 1-\left( M_{\left(
ia_{i}\right) ,\left( ia_{i}\right) }\right) \right) ^{-1}\left( M_{\left(
ia_{i}\right) ,\left( jb_{j}\right) }\right) _{i\neq j}\right) ^{2}\right)  
\notag \\
&=&\prod \det \left( 1-\left( M_{i,i}\right) \right) \left( 1-\frac{1}{2}%
Tr\left( \left( 1-M_{i,i}\right) ^{-1}\left( M_{i,j}\right) _{i\neq
j}\right) ^{2}\right)   \notag \\
&=&\prod \det \left( 1-\left( M_{i,i}\right) \right) \left( 1-\frac{1}{2}%
\sum_{i}\sum_{j\neq i}Tr\left( \left( 1-M_{j,j}\right) ^{-1}\left(
M_{j,i}\right) \left( 1-M_{i,i}\right) ^{-1}\left( M_{i,j}\right) \right)
\right)   \notag
\end{eqnarray}%
To solve further (\ref{TD}), we assume that $1-M$ can be diagonalized:%
\begin{equation*}
1-M_{i,i}=U_{i}\left( \left( 1-f_{l_{i}}\left( \gamma \right) \right)
\right) _{l_{i}}U_{i}^{-1}
\end{equation*}%
where:%
\begin{equation*}
\left( \left( 1-f_{l_{i}}\left( \gamma \right) \right) \right) _{l_{i}}
\end{equation*}%
is diagonal with elements:%
\begin{equation*}
1-f_{l_{i}}\left( \gamma \right) 
\end{equation*}%
Now, define the frequencies $\gamma _{l_{i}}$\ by the following equation:%
\begin{equation*}
f_{l_{i}}\left( \gamma _{l_{i}}\right) =1
\end{equation*}%
Then, $\det \left( 1-M\right) $ becomes:%
\begin{eqnarray*}
&&\det \left( 1-M\right)  \\
&=&\prod\limits_{i}\left( \prod\limits_{l_{i}}\left( 1-f_{l_{i}}\left(
\gamma \right) \right) \right) _{i}\left( 1-\frac{1}{2}Tr\left( \left(
1-f_{l_{j}}\left( \gamma \right) \right) _{j}\right)
^{-1}U_{j}^{-1}M_{j,i}U_{i}\left( \left( 1-f_{l_{i}}\left( \gamma \right)
\right) _{i}\right) ^{-1}U_{i}^{-1}M_{i,j}U_{j}\right)  \\
&=&\prod\limits_{i}\left( \prod\limits_{l_{i}}\left( 1-f_{l_{i}}\left(
\gamma \right) \right) \right) \left( 1-\frac{1}{2}\sum_{i}\sum_{j\neq
i}\sum_{k,l}\left( \left( 1-f_{j,l}\left( \gamma \right) \right) \right)
^{-1}\left[ \hat{M}_{j,i}\right] _{l,k}\left( 1-f_{i,k}\left( \gamma \right)
\right) ^{-1}\left[ \hat{M}_{i,j}\right] _{k,l}\right) 
\end{eqnarray*}%
where:%
\begin{eqnarray*}
\left[ \hat{M}_{j,i}\right]  &=&U_{j}^{-1}M_{j,i}U_{i} \\
\left[ \hat{M}_{i,j}\right]  &=&U_{i}^{-1}M_{i,j}U_{j}
\end{eqnarray*}%
Define similarly the frequencies $\gamma _{i,k}$\ by:%
\begin{equation*}
f_{i,k}\left( \gamma _{i,k}\right) =1
\end{equation*}%
and the equation for the frequencies:%
\begin{equation*}
\det \left( 1-M\right) =0
\end{equation*}%
writes ultimately:%
\begin{equation}
1-\frac{1}{2}\sum_{i}\sum_{j\neq i}\sum_{k,l}\left( \left( f_{j,l}\left(
\gamma _{j,l}\right) -f_{j,l}\left( \gamma \right) \right) \right) ^{-1}%
\left[ \hat{M}_{j,i}\right] _{l,k}\left( f_{i,k}\left( \gamma _{i,k}\right)
-f_{i,k}\left( \gamma \right) \right) ^{-1}\left[ \hat{M}_{i,j}\right]
_{k,l}=0  \label{fqc}
\end{equation}%
The $\gamma _{i,k}$ are the possible frequencies for structure $i$ without
interaction.

\subsubsection*{5.2.3 Solutions for interacting structures with close
frequencies}

If the structures before interaction are in states $\gamma _{j,l_{j}}$ and
these frequencies are relatively close from each other, that is, if we
assume:%
\begin{equation*}
\left\vert \gamma _{j,l_{j}}-\gamma _{j^{\prime },l_{j^{\prime
}}}\right\vert <<1
\end{equation*}%
we start by defining:%
\begin{equation*}
\bar{\gamma}=\frac{1}{m}\sum \gamma _{j,l_{j}}
\end{equation*}%
where $m$ is the number of structures involved in the interactions, we can
compute the expression arising in (\ref{fqc}):%
\begin{eqnarray}
&&\sum_{i}\sum_{j\neq i}\sum_{k,l}\left( \left( f_{j,l}\left( \gamma
_{j,l}\right) -f_{j,l}\left( \gamma \right) \right) \right) ^{-1}\left[ \hat{%
M}_{j,i}\right] _{l,k}\left( f_{i,k}\left( \gamma _{i,k}\right)
-f_{i,k}\left( \gamma \right) \right) ^{-1}\left[ \hat{M}_{i,j}\right]
\label{dvp} \\
&\simeq &\sum_{i}\sum_{j\neq i}\sum_{\substack{ k,l  \\ l\neq l_{j},k\neq
k_{i}}}\left( \left( f_{j,l}\left( \gamma _{j,l}\right) -f_{j,l}\left( \bar{%
\gamma}\right) \right) \right) ^{-1}\left[ \hat{M}_{j,i}\right] _{l,k}\left(
f_{i,k}\left( \gamma _{i,k}\right) -f_{i,k}\left( \bar{\gamma}\right)
\right) ^{-1}\left[ \hat{M}_{i,j}\right] _{k,l}  \notag \\
&&+\sum_{i}\sum_{j\neq i}\sum_{\substack{ k  \\ k\neq k_{i}}}\left( \left(
f_{j,l_{j}}\left( \gamma _{j,l_{j}}\right) -f_{j,l_{j}}\left( \gamma \right)
\right) \right) ^{-1}\left[ \hat{M}_{j,i}\right] _{l,k}\left( f_{i,k}\left(
\gamma _{i,k}\right) -f_{i,k}\left( \bar{\gamma}\right) \right) ^{-1}\left[ 
\hat{M}_{i,j}\right] _{k,l}  \notag \\
&&+\sum_{i}\sum_{j\neq i}\sum_{\substack{ l  \\ l\neq l_{j}}}\left( \left(
f_{j,l}\left( \gamma _{j,l}\right) -f_{j,l}\left( \bar{\gamma}\right)
\right) \right) ^{-1}\left[ \hat{M}_{j,i}\right] _{l,k}\left(
f_{i,k_{i}}\left( \gamma _{i,k_{i}}\right) -f_{i,k_{i}}\left( \gamma \right)
\right) ^{-1}\left[ \hat{M}_{i,j}\right] _{k,l}  \notag \\
&&+\sum_{i}\sum_{j\neq i}\left( f_{j,l_{j}}\left( \gamma _{j,l_{j}}\right)
-f_{j,l_{j}}\left( \gamma \right) \right) ^{-1}\left[ \hat{M}_{j,i}\right]
_{l_{j},k_{i}}\left( f_{i,k_{i}}\left( \gamma _{i,k_{i}}\right)
-f_{i,k_{i}}\left( \gamma \right) \right) ^{-1}\left[ \hat{M}_{i,j}\right]
_{k_{i},l_{j}}  \notag
\end{eqnarray}%
the last term in (\ref{dvp}) dominates and (\ref{fqc}) writes:%
\begin{equation}
1-\frac{1}{2}\sum_{i}\sum_{j\neq i}\left( f_{j,l_{j}}\left( \gamma
_{j,l_{j}}\right) -f_{j,l_{j}}\left( \gamma \right) \right) ^{-1}\left[ \hat{%
M}_{j,i}\right] _{l_{j},k_{i}}\left( f_{i,k_{i}}\left( \gamma
_{i,k_{i}}\right) -f_{i,k_{i}}\left( \gamma \right) \right) ^{-1}\left[ \hat{%
M}_{i,j}\right] _{k_{i},l_{j}}  \label{fqm}
\end{equation}%
To solve (\ref{fqm}), we consider each term of the sum individually, and
decompose (\ref{fqm}) in individual equations by wrtng: 
\begin{equation}
\left( f_{j,l_{j}}\left( \gamma _{j,l_{j}}\right) -f_{j,l_{j}}\left( \gamma
\right) \right) ^{-1}\left[ \hat{M}_{j,i}\right] _{l_{j},k_{i}}\left(
f_{i,k_{i}}\left( \gamma _{i,k_{i}}\right) -f_{i,k_{i}}\left( \gamma \right)
\right) ^{-1}\left[ \hat{M}_{i,j}\right] _{k_{i},l_{j}}=d_{i,j}  \label{fqn}
\end{equation}%
with:%
\begin{equation*}
\sum d_{i,j}=1
\end{equation*}%
Solving for the frequencies imply to find the $d_{i,j}$. This is performed
by writing in first approximation:%
\begin{equation*}
f_{j,l_{j}}\left( \gamma _{j,l_{j}}\right) -f_{j,l_{j}}\left( \gamma \right)
\simeq \left( \frac{\partial }{\partial \gamma }f_{j,l_{j}}\left( \gamma
\right) \right) _{\gamma _{j,l_{j}}}\left( \frac{\partial }{\partial \gamma }%
f_{i,k_{i}}\left( \gamma \right) \right) _{\gamma _{i,k_{i}}}\left( \gamma
_{j,l_{j}}-\gamma \right) \left( \gamma _{i,k_{i}}-\gamma \right)
\end{equation*}%
and (\ref{fqn}) becomes:%
\begin{equation}
\left( \gamma _{j,l_{j}}-\gamma \right) \left( \gamma _{i,k_{i}}-\gamma
\right) =\frac{\left[ \hat{M}_{j,i}\right] _{l_{j},k_{i}}\left[ \hat{M}_{i,j}%
\right] _{k_{i},l_{j}}}{d_{i,j}\left( \frac{\partial }{\partial \gamma }%
f_{j,l_{j}}\left( \gamma \right) \right) _{\gamma _{j,l_{j}}}\left( \frac{%
\partial }{\partial \gamma }f_{i,k_{i}}\left( \gamma \right) \right)
_{\gamma _{i,k_{i}}}}  \label{qnf}
\end{equation}%
There are thus $\frac{m\left( m-1\right) }{2}$ equations, $\frac{m\left(
m-1\right) }{2}-1$ coefficients, $\frac{m\left( m-1\right) }{2}$ variables
including the variable $\gamma $. We first start by solving for $\gamma $,
then the $d_{i,j}$, and ultimately for the variables $\gamma _{i,k_{i}}$.

Multiplying (\ref{qnf}) by $d_{i,j}$ and summing over $i$ nd $j\neq i$,
leads to: 
\begin{equation*}
\gamma ^{2}-\gamma \sum_{i,j\neq i}d_{i,j}\left( \gamma _{i,k_{i}}+\gamma
_{j,l_{j}}\right) +\sum_{i,j\neq i}d_{i,j}\gamma _{i,k_{i}}\gamma
_{j,l_{j}}-\sum_{i,j\neq i}\frac{\left[ \hat{M}_{j,i}\right] _{l_{j},k_{i}}%
\left[ \hat{M}_{i,j}\right] _{k_{i},l_{j}}}{\left( \frac{\partial }{\partial
\gamma }f_{j,l_{j}}\left( \gamma \right) \right) _{\gamma _{j,l_{j}}}\left( 
\frac{\partial }{\partial \gamma }f_{i,k_{i}}\left( \gamma \right) \right)
_{\gamma _{i,k_{i}}}}
\end{equation*}%
with solution:%
\begin{equation}
\gamma =\sum_{i,j\neq i}d_{i,j}\frac{\gamma _{i,k_{i}}+\gamma _{j,l_{j}}}{2}%
\pm \sqrt{\left( \sum_{i,j\neq i}d_{i,j}\frac{\gamma _{i,k_{i}}-\gamma
_{j,l_{j}}}{2}\right) ^{2}+\sum_{i,j\neq i}\frac{\left[ \hat{M}_{j,i}\right]
_{l_{j},k_{i}}\left[ \hat{M}_{i,j}\right] _{k_{i},l_{j}}}{\left( \frac{%
\partial }{\partial \gamma }f_{j,l_{j}}\left( \gamma \right) \right)
_{\gamma _{j,l_{j}}}\left( \frac{\partial }{\partial \gamma }%
f_{i,k_{i}}\left( \gamma \right) \right) _{\gamma _{i,k_{i}}}}}  \label{SL}
\end{equation}%
To complete the derivation, we have to find $d_{i,j}$, which will yield the $%
\gamma _{i,k_{i}}$. Using again (\ref{qnf}) leads to:%
\begin{equation}
C_{l_{j},k_{i}}d_{i,j}=\frac{\left[ \hat{M}_{j,i}\right] _{l_{j},k_{i}}\left[
\hat{M}_{i,j}\right] _{k_{i},l_{j}}}{\left( \frac{\partial }{\partial \gamma 
}f_{j,l_{j}}\left( \gamma \right) \right) _{\gamma _{j,l_{j}}}\left( \frac{%
\partial }{\partial \gamma }f_{i,k_{i}}\left( \gamma \right) \right)
_{\gamma _{i,k_{i}}}}  \label{NRR}
\end{equation}%
with:%
\begin{equation*}
C_{l_{j},k_{i}}=\left( \gamma _{j,l_{j}}-\gamma \right) \left( \gamma
_{i,k_{i}}-\gamma \right)
\end{equation*}%
Using our assumption:%
\begin{equation*}
\left\vert \gamma _{j,l_{j}}-\gamma _{j^{\prime },l_{j^{\prime
}}}\right\vert <<1
\end{equation*}%
we have:%
\begin{equation*}
C_{l_{j},k_{i}}\simeq C=\left\langle \left( \gamma _{j,l_{j}}-\gamma \right)
\left( \gamma _{i,k_{i}}-\gamma \right) \right\rangle
\end{equation*}%
and summing (\ref{NRR}) over $i$ nd $j\neq i$, leads to:%
\begin{equation*}
C=\sum_{i,j}\left[ \left[ M\right] \right] _{i,j}
\end{equation*}%
where:%
\begin{equation*}
\left[ \left[ M\right] \right] _{i,j}=\frac{\left[ \hat{M}_{j,i}\right]
_{l_{j},k_{i}}\left[ \hat{M}_{i,j}\right] _{k_{i},l_{j}}}{\left( \frac{%
\partial }{\partial \gamma }f_{j,l_{j}}\left( \gamma \right) \right)
_{\gamma _{j,l_{j}}}\left( \frac{\partial }{\partial \gamma }%
f_{i,k_{i}}\left( \gamma \right) \right) _{\gamma _{i,k_{i}}}}
\end{equation*}%
and as a consequence:%
\begin{equation*}
d_{i,j}\simeq \frac{\left[ \left[ M\right] \right] _{i,j}}{\sum_{i,j}\left[ %
\left[ M\right] \right] _{i,j}}
\end{equation*}%
so that, writing $\gamma _{\left( i,l_{i}\right) }$ to label the resulting
frequency for the new structure, we obtain:%
\begin{equation*}
\gamma _{\left( i,l_{i}\right) }=\sum \left[ \left[ M\right] \right] _{i,j}%
\frac{\gamma _{i,k_{i}}+\gamma _{j,l_{j}}}{2}\pm \sqrt{\left( \sum \frac{%
\left[ \left[ M\right] \right] _{i,j}}{\sum_{i,j}\left[ \left[ M\right] %
\right] _{i,j}}\frac{\gamma _{i,k_{i}}-\gamma _{j,l_{j}}}{2}\right)
^{2}+\sum \left[ \left[ M\right] \right] _{i,j}}
\end{equation*}%
Ultimately, the possible activities are:%
\begin{equation*}
\omega _{i0}\left( Z_{a_{i}}\right) +\overline{\Delta \omega }%
_{i}^{-1}\left( Z_{a_{i}}\right) +\left( \Delta \mathbf{\omega }_{D}\right)
^{-1}\left( Z_{\mathbf{\alpha }}\right) \exp \left( i\Upsilon _{\left\{
i,l_{i}\right\} }\theta \right)
\end{equation*}

\section*{Appendix 6}

We present the change of basis in the operator formalism. In a first step,
we derive the transformation that cancels the interactions with the
intermediate structure at the first order. Then we rewrite the transformed
operator at the second order. Interactions reappear as effective terms
between structure at the same time-scale in which the intermediate
structures action is hidden.

\subsection*{6.1 Finding $F$}

As explained in the text, we find operator $F$ by solving (\ref{CNT}):%
\begin{equation}
\left[ F,S_{0}\right] +I=0  \label{CN}
\end{equation}%
where:%
\begin{equation*}
S_{0}=\sum_{S\times S}\mathbf{\bar{D}}_{S^{2}}^{\alpha }\left( \mathbf{A}%
^{+}\left( \alpha ,p,S^{2}\right) \mathbf{A}^{-}\left( \alpha
,p,S^{2}\right) +\frac{1}{2}\right)
\end{equation*}%
and by postulating form of $F$ which is similar to that of $I$:%
\begin{eqnarray}
F &=&\sum_{n,n^{\prime }}\sum_{\substack{ k=1...n  \\ l=1,...,n^{\prime }}}%
\sum_{\left\{ S_{k},S_{l}\right\} _{\substack{ l=1,...,n^{\prime }  \\ %
k=1...n }}}\prod_{l=1}^{n^{\prime }}\prod\limits_{s=1}^{m_{l}^{\prime }}%
\mathbf{A}^{+}\left( \mathbf{\alpha }_{l}^{\prime },\mathbf{p}_{l}^{\prime
},S_{l}^{\prime 2}\right)  \label{SRN} \\
&&\times F_{n,n^{\prime }}\left( \left\{ \mathbf{\alpha }_{l}^{\prime },%
\mathbf{p}_{l}^{\prime },S_{l}^{\prime 2},m_{l}^{\prime }\right\} ,\left\{ 
\mathbf{\alpha }_{k},\mathbf{p}_{k},S_{k}^{2},m_{k}\right\} \right)
\prod_{k=1}^{n}\prod\limits_{s=1}^{m_{k}}\mathbf{A}^{-}\left( \mathbf{\alpha 
}_{k},\mathbf{p}_{k},S_{k}^{2}\right)  \notag
\end{eqnarray}%
To solve (\ref{CN}), we will use the commutation relations:%
\begin{equation*}
\left[ \mathbf{A}^{+}\left( \alpha ,p,S^{2}\right) \mathbf{A}^{-}\left(
\alpha ,p,S^{2}\right) ,\prod\limits_{s=1}^{m_{l}^{\prime }}\mathbf{A}%
^{+}\left( \mathbf{\alpha }_{l}^{\prime },\mathbf{p}_{l}^{\prime
},S_{l}^{\prime 2}\right) \right] =m_{l}^{\prime }\mathbf{\bar{D}}%
_{S_{l}^{\prime 2}}^{\mathbf{\alpha }_{l}^{\prime }}\delta \left( \left(
\alpha ,p,S^{2}\right) -\left( \mathbf{\alpha }_{l}^{\prime },\mathbf{p}%
_{l}^{\prime },S_{l}^{\prime 2}\right) \right)
\end{equation*}%
and:%
\begin{equation*}
\left[ \mathbf{A}^{+}\left( \alpha ,p,S^{2}\right) \mathbf{A}^{-}\left(
\alpha ,p,S^{2}\right) ,\prod\limits_{s=1}^{m_{k}}\mathbf{A}^{-}\left( 
\mathbf{\alpha }_{k},\mathbf{p}_{k},S_{k}^{2}\right) \right] =-m_{k}\mathbf{%
\bar{D}}_{S_{k}^{2}}^{\mathbf{\alpha }_{k}}\delta \left( \left( \alpha
,p,S^{2}\right) -\left( \mathbf{\alpha }_{k},\mathbf{p}_{k},S_{k}^{2}\right)
\right)
\end{equation*}%
Inserting these relations in (\ref{CN}) leads to:%
\begin{eqnarray*}
0 &=&\sum_{l=1}^{n^{\prime }}\sum_{k=1}^{n}\left( m_{l}^{\prime }\mathbf{%
\bar{D}}_{S_{l}^{\prime 2}}^{\mathbf{\alpha }_{l}^{\prime }}-m_{k}\mathbf{%
\bar{D}}_{S_{k}^{2}}^{\mathbf{\alpha }_{k}}\right)
F_{m_{1},...,m_{n},m_{1}^{\prime },...,m_{n^{\prime }}^{\prime }}\left(
\left\{ \mathbf{\alpha }_{l}^{\prime },\mathbf{p}_{l}^{\prime
},S_{l}^{\prime 2}\right\} ,\left\{ \mathbf{\alpha }_{k},\mathbf{p}%
_{k},S_{k}^{2}\right\} \right) \\
&&+V_{m_{1},...,m_{n},m_{1}^{\prime },...,m_{n^{\prime }}^{\prime }}\left(
\left\{ \mathbf{\alpha }_{l}^{\prime },\mathbf{p}_{l}^{\prime
},S_{l}^{\prime 2}\right\} ,\left\{ \mathbf{\alpha }_{k},\mathbf{p}%
_{k},S_{k}^{2}\right\} \right)
\end{eqnarray*}%
and this yields the matrices elements of the trnsfrmtn:%
\begin{equation}
F_{m_{1},...,m_{n},m_{1}^{\prime },...,m_{n^{\prime }}^{\prime }}\left(
\left\{ \mathbf{\alpha }_{l}^{\prime },\mathbf{p}_{l}^{\prime
},S_{l}^{\prime 2}\right\} ,\left\{ \mathbf{\alpha }_{k},\mathbf{p}%
_{k},S_{k}^{2}\right\} \right) =-\frac{V_{m_{1},...,m_{n},m_{1}^{\prime
},...,m_{n^{\prime }}^{\prime }}\left( \left\{ \mathbf{\alpha }_{l}^{\prime
},\mathbf{p}_{l}^{\prime },S_{l}^{\prime 2}\right\} ,\left\{ \mathbf{\alpha }%
_{k},\mathbf{p}_{k},S_{k}^{2}\right\} \right) }{\sum_{l=1}^{n^{\prime
}}m_{l}^{\prime }\mathbf{\bar{D}}_{S_{l}^{\prime 2}}^{\mathbf{\alpha }%
_{l}^{\prime }}-\sum_{k=1}^{n}m_{k}\mathbf{\bar{D}}_{S_{k}^{2}}^{\mathbf{%
\alpha }_{k}}}  \label{SLN}
\end{equation}

\subsection*{6.2 Computing transformed action}

Once the transformation matrix is found, we compute the transformed prtr $%
\left( S^{\left( O\right) }\right) ^{\prime }$:%
\begin{equation*}
\left( S^{\left( O\right) }\right) ^{\prime }=S_{0}+\frac{1}{2}\left[ I,F%
\right]
\end{equation*}%
by writing $\left[ I,F\right] $\ as a series expansion:%
\begin{eqnarray}
\left[ I,F\right] &=&\sum \prod_{L=1}^{n^{\prime }}\prod\limits_{s^{\prime
}=1}^{M_{L}^{\prime }}\mathbf{A}^{+}\left( \mathbf{\alpha }_{L}^{\prime },%
\mathbf{p}_{L}^{\prime },S_{L}^{\prime 2}\right)  \label{CFN} \\
&&\times \left[ I,F\right] _{M_{1},...,M_{n},M_{1}^{\prime
},...,M_{n^{\prime }}^{\prime }}\left( \left\{ \mathbf{\alpha }_{L}^{\prime
},\mathbf{p}_{L}^{\prime },S_{L}^{\prime 2}\right\} ,\left\{ \mathbf{\alpha }%
_{K},\mathbf{p}_{K},S_{K}^{2}\right\} \right)
\prod\limits_{K=1}^{n}\prod\limits_{s=1}^{M_{k}^{\prime }}\mathbf{A}%
^{-}\left( \mathbf{\alpha }_{K},\mathbf{p}_{K},S_{K}^{2}\right)  \notag
\end{eqnarray}%
Using the series (\ref{SRN}) for $F$ and (\ref{SNR}) for $I$. The commutator
involves contributions of the form:

\begin{equation}
A=\left[ \prod_{l=1}^{n^{\prime }}\prod\limits_{s=1}^{m_{l}^{\prime }}%
\mathbf{A}^{+}\left( \mathbf{\alpha }_{l}^{\prime },\mathbf{p}_{l}^{\prime
},S_{l}^{\prime 2}\right) \prod_{k=1}^{n}\prod\limits_{s=1}^{m_{k}}\mathbf{A}%
^{-}\left( \mathbf{\alpha }_{k},\mathbf{p}_{k},S_{k}^{2}\right)
,\prod_{q=1}^{n_{1}^{\prime }}\prod\limits_{s=1}^{m_{q}^{\prime }}\mathbf{A}%
^{+}\left( \mathbf{\alpha }_{q}^{\prime },\mathbf{p}_{q}^{\prime
},S_{q}^{\prime 2}\right) \prod_{p=1}^{n_{1}}\prod\limits_{s=1}^{m_{p}}%
\mathbf{A}^{-}\left( \mathbf{\alpha }_{p},\mathbf{p}_{p},S_{p}^{2}\right) %
\right]  \label{CF}
\end{equation}%
and $A$ is obtained by successive derivation of a given commutator. Actually:%
\begin{equation}
A=\prod_{l=1}^{n^{\prime }}\left[ \frac{\partial ^{m_{l}^{\prime }}}{%
\partial t_{l}^{m_{l}^{\prime }}}\right] _{t_{l}=0}\prod_{k=1}^{n}\left[ 
\frac{\partial ^{m_{k}}}{\partial t_{k}^{m_{k}}}\right] _{t_{k}=0}C
\label{CDR}
\end{equation}%
where $C$ is defined by:%
\begin{eqnarray}
C &=&\left[ \exp \left( \sum t_{k}\mathbf{A}^{-}\left( \mathbf{\alpha }_{k},%
\mathbf{p}_{k},S_{k}^{2}\right) \right) \exp \left( \sum t_{k}^{\prime }%
\mathbf{A}^{+}\left( \mathbf{\alpha }_{k},\mathbf{p}_{k},S_{k}^{2}\right)
\right) ,\right.  \label{CTR} \\
&&\left. \exp \left( \sum t_{l}\mathbf{A}^{-}\left( \mathbf{\alpha }_{l},%
\mathbf{p}_{l},S_{l}^{2}\right) \right) \exp \left( \sum t_{l}^{\prime }%
\mathbf{A}^{+}\left( \mathbf{\alpha }_{l},\mathbf{p}_{l},S_{l}^{2}\right)
\right) \right]
\end{eqnarray}%
The commutator (\ref{CTR}) is calculated by using the Campbell Hausdorf
formula:

\begin{eqnarray*}
&&\exp \left( \sum t_{k}\mathbf{A}^{-}\left( \mathbf{\alpha }_{k},\mathbf{p}%
_{k},S_{k}^{2}\right) \right) \exp \left( \sum t_{k}^{\prime }\mathbf{A}%
^{+}\left( \mathbf{\alpha }_{k},\mathbf{p}_{k},S_{k}^{2}\right) \right) \\
&&\times \exp \left( \sum t_{l}\mathbf{A}^{-}\left( \mathbf{\alpha }_{l},%
\mathbf{p}_{l},S_{l}^{2}\right) \right) \exp \left( \sum t_{l}^{\prime }%
\mathbf{A}^{+}\left( \mathbf{\alpha }_{l},\mathbf{p}_{l},S_{l}^{2}\right)
\right) \\
&=&\exp \left( \sum t_{k}\mathbf{A}^{-}\left( \mathbf{\alpha }_{k},\mathbf{p}%
_{k},S_{k}^{2}\right) +\sum t_{l}\mathbf{A}^{-}\left( \mathbf{\alpha }_{l},%
\mathbf{p}_{l},S_{l}^{2}\right) \right) \\
&&\times \exp \left( \sum t_{k}^{\prime }\mathbf{A}^{+}\left( \mathbf{\alpha 
}_{k},\mathbf{p}_{k},S_{k}^{2}\right) +\sum t_{l}^{\prime }\mathbf{A}%
^{+}\left( \mathbf{\alpha }_{l},\mathbf{p}_{l},S_{l}^{2}\right) \right) \exp
\left( \sum t_{k}^{\prime }t_{l}\delta \left( \left( \mathbf{\alpha }_{k},%
\mathbf{p}_{k},S_{k}^{2}\right) -\left( \mathbf{\alpha }_{l},\mathbf{p}%
_{l},S_{l}^{2}\right) \right) \right)
\end{eqnarray*}%
so that, we find for (\ref{CTR}):

\begin{eqnarray}
C &=&\exp \left( \sum t_{k}\mathbf{A}^{-}\left( \mathbf{\alpha }_{k},\mathbf{%
p}_{k},S_{k}^{2}\right) +\sum t_{l}\mathbf{A}^{-}\left( \mathbf{\alpha }_{l},%
\mathbf{p}_{l},S_{l}^{2}\right) \right)  \label{CPN} \\
&&\times \exp \left( \sum t_{k}^{\prime }\mathbf{A}^{+}\left( \mathbf{\alpha 
}_{k},\mathbf{p}_{k},S_{k}^{2}\right) +\sum t_{l}^{\prime }\mathbf{A}%
^{+}\left( \mathbf{\alpha }_{l},\mathbf{p}_{l},S_{l}^{2}\right) \right) 
\notag \\
&&\times \left( \exp \left( \sum t_{k}^{\prime }t_{l}\delta \left( \left( 
\mathbf{\alpha }_{k},\mathbf{p}_{k},S_{k}^{2}\right) -\left( \mathbf{\alpha }%
_{l},\mathbf{p}_{l},S_{l}^{2}\right) \right) \right) -\exp \left( \sum
t_{k}t_{l}^{\prime }\delta \left( \left( \mathbf{\alpha }_{k},\mathbf{p}%
_{k},S_{k}^{2}\right) -\left( \mathbf{\alpha }_{l},\mathbf{p}%
_{l},S_{l}^{2}\right) \right) \right) \right)  \notag
\end{eqnarray}%
Applying (\ref{CDR}) to (\ref{CPN}) leads to the expression of the
coefficient (\ref{CF}) involved in (\ref{CFN}). As a consequence, we find:

\begin{eqnarray}
&&\left[ I,F\right] _{M_{1},...,M_{n},M_{1}^{\prime },...,M_{n^{\prime
}}^{\prime }}\left( \left\{ \mathbf{\alpha }_{L}^{\prime },\mathbf{p}%
_{L}^{\prime },S_{L}^{\prime 2}\right\} ,\left\{ \mathbf{\alpha }_{K},%
\mathbf{p}_{K},S_{K}^{2}\right\} \right)  \label{MR} \\
=\sum_{P_{K},P_{L}} &&\sum_{\left\{ \epsilon _{d}^{\prime }\right\} ,\left\{
\epsilon _{c}\right\} }\sum_{\left\{ \delta _{k}\right\} ,\left\{ \delta
_{l}^{\prime }\right\} }\prod \left( \epsilon _{d}^{\prime }!\right)
^{2}\left( \epsilon _{c}!\right) ^{2}\prod\limits_{\bar{k}\bar{l}%
}\prod\limits_{kl}\left( -1\right) ^{\delta _{l}^{\prime }}C_{m_{l}^{\prime
}+\delta _{l}^{\prime }}^{\delta _{l}^{\prime }}C_{m_{k}+\delta
_{k}}^{\delta _{k}}\mathbf{\bar{D}}_{S_{k}^{2}}^{\mathbf{\alpha }_{k}}%
\mathbf{\bar{D}}_{S_{k}^{2}}^{\mathbf{\alpha }_{l}^{\prime }}\mathbf{\bar{D}}%
_{S_{c}^{2}}^{\mathbf{\alpha }_{c}}\mathbf{\bar{D}}_{S_{d}^{2}}^{\mathbf{%
\alpha }_{d}^{\prime }}  \notag \\
\times &&\delta \left( \left( \mathbf{\alpha }_{k},\mathbf{p}%
_{k},S_{k}^{2}\right) -\left( \mathbf{\bar{\alpha}}_{\bar{l}}^{\prime },%
\mathbf{\bar{p}}_{\bar{l}}^{\prime },\bar{S}_{\bar{l}}^{\prime 2}\right)
\right) \delta \left( \left( \mathbf{\alpha }_{l}^{\prime },\mathbf{p}%
_{l}^{\prime },S_{l}^{\prime 2}\right) -\left( \mathbf{\bar{\alpha}}_{\bar{k}%
},\mathbf{\bar{p}}_{\bar{k}},\bar{S}_{\bar{k}}^{2}\right) \right)  \notag \\
&&\delta \left( \left( \mathbf{\alpha }_{c},\mathbf{p}_{c},S_{c}^{2}\right)
-\left( \mathbf{\bar{\alpha}}_{\bar{d}}^{\prime },\mathbf{\bar{p}}_{\bar{d}%
}^{\prime },\bar{S}_{\bar{d}}^{\prime 2}\right) \right) \delta \left( \left( 
\mathbf{\alpha }_{d}^{\prime },\mathbf{p}_{d}^{\prime },S_{d}^{\prime
2}\right) -\mathbf{\bar{\alpha}}_{\bar{c}},\mathbf{\bar{p}}_{\bar{c}},\bar{S}%
_{\bar{c}}^{2}\right)  \notag \\
&&\times F\left( \left\{ \mathbf{\alpha }_{l}^{\prime },\mathbf{p}%
_{l}^{\prime },S_{l}^{\prime 2},m_{l}^{\prime }+\delta _{l}^{\prime
}\right\} \cup \left\{ \mathbf{\alpha }_{d}^{\prime },\mathbf{p}_{d}^{\prime
},S_{d}^{\prime 2},\epsilon _{d}^{\prime }\right\} ,\left\{ \mathbf{\alpha }%
_{k},\mathbf{p}_{k},S_{k}^{2},m_{l}+\delta _{l}\right\} \cup \left\{ \mathbf{%
\alpha }_{c},\mathbf{p}_{c},S_{c}^{2},\epsilon _{c}\right\} \right)  \notag
\\
&&\times V\left( \left\{ \mathbf{\bar{\alpha}}_{\bar{l}}^{\prime },\mathbf{%
\bar{p}}_{\bar{l}}^{\prime },\bar{S}_{\bar{l}}^{\prime 2},\bar{m}_{\bar{l}%
}^{\prime }+\delta _{k}\right\} \cup \left\{ \mathbf{\bar{\alpha}}_{\bar{d}%
}^{\prime },\mathbf{\bar{p}}_{\bar{d}}^{\prime },\bar{S}_{\bar{d}}^{\prime
2},\epsilon _{d}^{\prime }\right\} ,\left\{ \mathbf{\bar{\alpha}}_{\bar{k}},%
\mathbf{\bar{p}}_{\bar{k}},\bar{S}_{\bar{k}}^{2},\bar{m}_{\bar{k}}+\delta
_{l}^{\prime }\right\} \cup \left\{ \mathbf{\bar{\alpha}}_{\bar{c}},\mathbf{%
\bar{p}}_{\bar{c}},\bar{S}_{\bar{c}}^{2},\epsilon _{c}\right\} \right) 
\notag
\end{eqnarray}%
with $P_{K},P_{L}$ are partitions of $\left\{ \mathbf{\alpha }_{K},\mathbf{p}%
_{K},S_{K}^{2},M_{K}\right\} $ and $\left\{ \mathbf{\alpha }_{L}^{\prime },%
\mathbf{p}_{L}^{\prime },S_{L}^{\prime 2},M_{L}^{\prime }\right\} $:%
\begin{eqnarray*}
\left\{ \mathbf{\alpha }_{L}^{\prime },\mathbf{p}_{L}^{\prime
},S_{L}^{\prime 2},M_{L}^{\prime }\right\} &=&\left\{ \mathbf{\alpha }%
_{l}^{\prime },\mathbf{p}_{l}^{\prime },S_{l}^{\prime 2},\bar{m}_{l^{\prime
}}^{\prime }\right\} \cup \left\{ \mathbf{\bar{\alpha}}_{\bar{l}}^{\prime },%
\mathbf{\bar{p}}_{\bar{l}}^{\prime },\bar{S}_{\bar{l}}^{\prime 2},m_{\bar{l}%
}^{\prime }\right\} \\
\left\{ \mathbf{\alpha }_{K},\mathbf{p}_{K},S_{K}^{2},M_{K}\right\}
&=&\left\{ \mathbf{\alpha }_{k},\mathbf{p}_{k},S_{k}^{2},\bar{m}_{k}\right\}
\cup \left\{ \mathbf{\bar{\alpha}}_{\bar{k}},\mathbf{\bar{p}}_{\bar{k}},\bar{%
S}_{\bar{k}}^{2},m_{\bar{k}}\right\}
\end{eqnarray*}%
We can rewrite (\ref{MR}) by using the fact that the elmnts of $F$ are:%
\begin{eqnarray*}
&&F_{\left( m_{1}+\delta _{1},...,m_{n}+\delta _{n},\epsilon
_{1},,...,\epsilon _{p}\right) ,m_{1}^{\prime }+\delta _{1}^{\prime
},...,m_{n^{\prime }}^{\prime }+\delta _{n}^{\prime },\epsilon _{1}^{\prime
},,...,\epsilon _{p^{\prime }}^{\prime }}\left( \left\{ \mathbf{\alpha }%
_{l}^{\prime },\mathbf{p}_{l}^{\prime },S_{l}^{\prime 2}\right\} \cup
\left\{ \mathbf{\alpha }_{d}^{\prime },\mathbf{p}_{d}^{\prime
},S_{d}^{\prime 2}\right\} ,\left\{ \mathbf{\alpha }_{k},\mathbf{p}%
_{k},S_{k}^{2}\right\} \cup \left\{ \mathbf{\alpha }_{c},\mathbf{p}%
_{c},S_{c}^{2}\right\} \right) \\
&=&-\frac{V_{\left( m_{1}+\delta _{1},...,m_{n}+\delta _{n},\epsilon
_{1},,...,\epsilon _{p}\right) ,m_{1}^{\prime }+\delta _{1}^{\prime
},...,m_{n^{\prime }}^{\prime }+\delta _{n}^{\prime },\epsilon _{1}^{\prime
},,...,\epsilon _{p^{\prime }}^{\prime }}\left( \left\{ \mathbf{\alpha }%
_{l}^{\prime },\mathbf{p}_{l}^{\prime },S_{l}^{\prime 2}\right\} \cup
\left\{ \mathbf{\alpha }_{d}^{\prime },\mathbf{p}_{d}^{\prime
},S_{d}^{\prime 2}\right\} ,\left\{ \mathbf{\alpha }_{k},\mathbf{p}%
_{k},S_{k}^{2}\right\} \cup \left\{ \mathbf{\alpha }_{c},\mathbf{p}%
_{c},S_{c}^{2}\right\} \right) }{\sum_{l=1}^{n^{\prime }}\left(
m_{l}^{\prime }+\delta _{l^{\prime }}^{\prime }\right) \mathbf{\bar{D}}%
_{S_{l}^{\prime 2}}^{\mathbf{\alpha }_{l}^{\prime }}-\sum_{k=1}^{n}\left(
m_{k}+\delta _{k}\right) \mathbf{\bar{D}}_{S_{k}^{2}}^{\mathbf{\alpha }%
_{k}}+\sum_{d=1}^{p^{\prime }}\epsilon _{d}^{\prime }\mathbf{\bar{D}}%
_{S_{d}^{\prime 2}}^{\mathbf{\alpha }_{d}^{\prime }}-\sum_{c=1}^{p}\epsilon
_{c}\mathbf{\bar{D}}_{S_{c}^{2}}^{\mathbf{\alpha }_{c}}}
\end{eqnarray*}%
and this leads to the expression of the matrices elements for commutator (%
\ref{MR}):

\begin{eqnarray*}
&&\left[ I,F\right] _{M_{1},...,M_{n},M_{1}^{\prime },...,M_{n^{\prime
}}^{\prime }}\left( \left\{ \mathbf{\alpha }_{L}^{\prime },\mathbf{p}%
_{L}^{\prime },S_{L}^{\prime 2}\right\} ,\left\{ \mathbf{\alpha }_{K},%
\mathbf{p}_{K},S_{K}^{2}\right\} \right) \\
=-\sum_{P_{K},P_{L}} &&\sum_{\left\{ \epsilon _{d}^{\prime }\right\}
,\left\{ \epsilon _{c}\right\} }\sum_{\substack{ \left\{ \delta _{k}\right\}
,\left\{ \delta _{l}^{\prime }\right\}  \\ \left\{ \delta _{\bar{k}}\right\}
,\left\{ \delta _{\bar{l}}^{\prime }\right\} }}\prod \left( \epsilon
_{d}^{\prime }!\right) ^{2}\left( \epsilon _{c}!\right) ^{2}\prod\limits_{%
\bar{k}\bar{l}}\prod\limits_{kl}\left( -1\right) ^{\delta _{l}^{\prime
}}C_{m_{l}^{\prime }+\delta _{l}^{\prime }}^{\delta _{l}^{\prime
}}C_{m_{k}+\delta _{k}}^{\delta _{k}}\mathbf{\bar{D}}_{S_{k}^{2}}^{\mathbf{%
\alpha }_{k}}\mathbf{\bar{D}}_{S_{k}^{2}}^{\mathbf{\alpha }_{l}^{\prime }}%
\mathbf{\bar{D}}_{S_{c}^{2}}^{\mathbf{\alpha }_{0c}}\mathbf{\bar{D}}%
_{S_{d}^{2}}^{\mathbf{\alpha }_{0d}^{\prime }} \\
\times &&\delta \left( \left( \mathbf{\alpha }_{k},\mathbf{p}%
_{k},S_{k}^{2}\right) -\left( \mathbf{\bar{\alpha}}_{\bar{l}}^{\prime },%
\mathbf{\bar{p}}_{\bar{l}}^{\prime },\bar{S}_{\bar{l}}^{\prime 2}\right)
\right) \delta \left( \left( \mathbf{\alpha }_{l}^{\prime },\mathbf{p}%
_{l}^{\prime },S_{l}^{\prime 2}\right) -\left( \mathbf{\bar{\alpha}}_{\bar{k}%
},\mathbf{\bar{p}}_{\bar{k}},\bar{S}_{\bar{k}}^{2}\right) \right) \\
\times &&\delta \left( \left( \mathbf{\alpha }_{c},\mathbf{p}%
_{c},S_{c}^{2}\right) -\left( \mathbf{\bar{\alpha}}_{\bar{d}}^{\prime },%
\mathbf{\bar{p}}_{\bar{d}}^{\prime },\bar{S}_{\bar{d}}^{\prime 2}\right)
\right) \delta \left( \left( \mathbf{\alpha }_{d}^{\prime },\mathbf{p}%
_{d}^{\prime },S_{d}^{\prime 2}\right) -\mathbf{\bar{\alpha}}_{\bar{c}},%
\mathbf{\bar{p}}_{\bar{c}},\bar{S}_{\bar{c}}^{2}\right) \\
&&\times V_{1}\left( \left\{ \mathbf{\alpha }_{l}^{\prime },\mathbf{p}%
_{l}^{\prime },S_{l}^{\prime 2},m_{l}^{\prime }+\delta _{l}^{\prime
}\right\} \cup \left\{ \mathbf{\alpha }_{d}^{\prime },\mathbf{p}_{d}^{\prime
},S_{d}^{\prime 2},\epsilon _{d}^{\prime }\right\} ,\left\{ \mathbf{\alpha }%
_{k},\mathbf{p}_{k},S_{k}^{2}+\delta _{k}\right\} \cup \left\{ \mathbf{%
\alpha }_{c},\mathbf{p}_{c},S_{c}^{2},\epsilon _{c}\right\} \right) \\
&&\times V\left( \left\{ \mathbf{\bar{\alpha}}_{\bar{l}}^{\prime },\mathbf{%
\bar{p}}_{\bar{l}}^{\prime },\bar{S}_{\bar{l}}^{\prime 2},\bar{m}%
_{l}^{\prime }+\delta _{k}\right\} \cup \left\{ \mathbf{\bar{\alpha}}_{\bar{d%
}}^{\prime },\mathbf{\bar{p}}_{\bar{d}}^{\prime },\bar{S}_{\bar{d}}^{\prime
2},\epsilon _{c}\right\} ,\left\{ \mathbf{\bar{\alpha}}_{\bar{k}},\mathbf{%
\bar{p}}_{\bar{k}},\bar{S}_{\bar{k}}^{2},\bar{m}_{\bar{k}}+\delta
_{l}^{\prime }\right\} \cup \left\{ \mathbf{\bar{\alpha}}_{\bar{c}},\mathbf{%
\bar{p}}_{\bar{c}},\bar{S}_{\bar{c}}^{2},\epsilon _{d}^{\prime }\right\}
\right)
\end{eqnarray*}%
where we defined:%
\begin{eqnarray*}
&&V_{1}\left( \left\{ \mathbf{\alpha }_{l}^{\prime },\mathbf{p}_{l}^{\prime
},S_{l}^{\prime 2},m_{l}^{\prime }+\delta _{l}^{\prime }\right\} \cup
\left\{ \mathbf{\alpha }_{d}^{\prime },\mathbf{p}_{d}^{\prime
},S_{d}^{\prime 2},\epsilon _{d}^{\prime }\right\} ,\left\{ \mathbf{\alpha }%
_{k},\mathbf{p}_{k},S_{k}^{2}+\delta _{k}\right\} \cup \left\{ \mathbf{%
\alpha }_{c},\mathbf{p}_{c},S_{c}^{2},\epsilon _{c}\right\} \right) \\
&=&\frac{V_{\left( m_{1}+\delta _{1},...,m_{n}+\delta _{n},\epsilon
_{1},,...,\epsilon _{p}\right) ,m_{1}^{\prime }+\delta _{1}^{\prime
},...,m_{n^{\prime }}^{\prime }+\delta _{n}^{\prime },\epsilon _{1}^{\prime
},,...,\epsilon _{p^{\prime }}^{\prime }}\left( \left\{ \mathbf{\alpha }%
_{l}^{\prime },\mathbf{p}_{l}^{\prime },S_{l}^{\prime 2}\right\} \cup
\left\{ \mathbf{\alpha }_{d}^{\prime },\mathbf{p}_{d}^{\prime
},S_{d}^{\prime 2}\right\} ,\left\{ \mathbf{\alpha }_{k},\mathbf{p}%
_{k},S_{k}^{2}\right\} \cup \left\{ \mathbf{\alpha }_{c},\mathbf{p}%
_{c},S_{c}^{2}\right\} \right) }{\sum_{l=1}^{n^{\prime }}\left(
m_{l}^{\prime }+\delta _{l^{\prime }}^{\prime }\right) \mathbf{\bar{D}}%
_{S_{l}^{\prime 2}}^{\mathbf{\alpha }_{l}^{\prime }}-\sum_{k=1}^{n}\left(
m_{k}+\delta _{k}\right) \mathbf{\bar{D}}_{S_{k}^{2}}^{\mathbf{\alpha }%
_{k}}+\sum_{d=1}^{p^{\prime }}\epsilon _{d}^{\prime }\mathbf{\bar{D}}%
_{S_{d}^{\prime 2}}^{\mathbf{\alpha }_{d}^{\prime }}-\sum_{c=1}^{p}\epsilon
_{c}\mathbf{\bar{D}}_{S_{c}^{2}}^{\mathbf{\alpha }_{c}}}
\end{eqnarray*}%
This is the results presented in the text.

\subsection*{6.3 Exemple}

We present the previous computations for the example given in the text.

\subsubsection*{6.3.1 Computation of $F$}

In this case $F$ is defined by:%
\begin{equation*}
I+\left[ F,S_{0}\right] =0
\end{equation*}%
and this equation writes in expanded form%
\begin{eqnarray*}
0 &=&\left[ F,\mathbf{A}^{+}\left( \alpha _{i},\mathbf{p}_{i},S_{i}^{2}%
\right) \mathbf{A}^{-}\left( \mathbf{\alpha }_{i},\mathbf{p}%
_{i},S_{i}^{2}\right) \right] \\
&&+\sum_{i=1,2}I\left( \left( \mathbf{\alpha }_{i},\mathbf{p}_{i}\right)
,\left( \mathbf{\alpha }_{i}^{\prime },\mathbf{p}_{i}^{\prime }\right)
,\left( \mathbf{\alpha }_{0},\mathbf{p}_{0}\right)
,S_{i}^{2},S_{0}^{2}\right) \mathbf{A}^{+}\left( \mathbf{\alpha }_{i},%
\mathbf{p}_{i},S_{i}^{2}\right) \mathbf{A}^{-}\left( \mathbf{\alpha }%
_{i}^{\prime },\mathbf{p}_{i}^{\prime },S_{i}^{2}\right) \left( \mathbf{A}%
^{-}\left( \mathbf{\alpha }_{0},\mathbf{p}_{0},S_{0}^{2}\right) +\mathbf{A}%
^{+}\left( \mathbf{\alpha }_{0},\mathbf{p}_{0},S_{0}^{2}\right) \right)
\end{eqnarray*}%
which yields the matrices elements of the transformation. Considering the
interaction term:%
\begin{equation*}
I=\sum_{i}I\left( \left( \mathbf{\alpha }_{i},\mathbf{p}_{i}\right) ,\left( 
\mathbf{\alpha }_{i}^{\prime },\mathbf{p}_{i}^{\prime }\right) ,\left( 
\mathbf{\alpha }_{0},\mathbf{p}_{0}\right) ,S_{i}^{2},S_{0}^{2}\right) 
\mathbf{A}^{+}\left( \mathbf{\alpha }_{i},\mathbf{p}_{i},S_{i}^{2}\right) 
\mathbf{A}^{-}\left( \mathbf{\alpha }_{i}^{\prime },\mathbf{p}_{i}^{\prime
},S_{i}^{2}\right) \left( \mathbf{A}^{-}\left( \mathbf{\alpha }_{0},\mathbf{p%
}_{0},S_{0}^{2}\right) +\mathbf{A}^{+}\left( \mathbf{\alpha }_{0},\mathbf{p}%
_{0},S_{0}^{2}\right) \right)
\end{equation*}%
it can be written in a compact notation:%
\begin{eqnarray*}
I &=&I_{a,b,c,d,e,f}\left( \left( \mathbf{\alpha }_{1},\mathbf{p}_{1}\right)
,\left( \mathbf{\alpha }_{1}^{\prime },\mathbf{p}_{1}^{\prime }\right)
,\left( \mathbf{\alpha }_{0},\mathbf{p}_{0}\right)
,S_{1}^{2},S_{0}^{2}\right) \\
&&\left( \mathbf{A}^{+}\left( \mathbf{\alpha }_{i},\mathbf{p}%
_{i},S_{1}^{2}\right) \right) ^{a}\left( \mathbf{A}^{-}\left( \mathbf{\alpha 
}_{i},\mathbf{p}_{i},S_{1}^{2}\right) \right) ^{b}\left( \mathbf{A}%
^{+}\left( \mathbf{\alpha }_{i^{\prime }},\mathbf{p}_{i^{\prime
}},S_{i^{\prime }}^{2}\right) \right) ^{c}\left( \mathbf{A}^{-}\left( 
\mathbf{\alpha }_{i^{\prime }},\mathbf{p}_{i^{\prime }},S_{i^{\prime
}}^{2}\right) \right) ^{d} \\
&&\times \left( \mathbf{A}^{+}\left( \mathbf{\alpha }_{0},\mathbf{p}%
_{0},S_{0}^{2}\right) \right) ^{e}\left( \mathbf{A}^{-}\left( \mathbf{\alpha 
}_{0},\mathbf{p}_{0},S_{0}^{2}\right) \right) ^{f}
\end{eqnarray*}%
The only matrices elements arising in the interaction are:%
\begin{eqnarray*}
&&I_{1,1,0,0,1,0}\left( \left( \mathbf{\alpha }_{1},\mathbf{p}_{1}\right)
,\left( \mathbf{\alpha }_{1}^{\prime },\mathbf{p}_{1}^{\prime }\right)
,\left( \mathbf{\alpha }_{0},\mathbf{p}_{0}\right)
,S_{1}^{2},S_{0}^{2}\right) \\
&&I_{1,1,0,0,0,1}\left( \left( \mathbf{\alpha }_{1},\mathbf{p}_{1}\right)
,\left( \mathbf{\alpha }_{1}^{\prime },\mathbf{p}_{1}^{\prime }\right)
,\left( \mathbf{\alpha }_{0}^{\prime },\mathbf{p}_{0}^{\prime }\right)
,S_{1}^{2},S_{0}^{2}\right) \\
&&I_{0,0,1,1,1,0}\left( \left( \mathbf{\alpha }_{2},\mathbf{p}_{2}\right)
,\left( \mathbf{\alpha }_{2}^{\prime },\mathbf{p}_{2}^{\prime }\right)
,\left( \mathbf{\alpha }_{0},\mathbf{p}_{0}\right)
,S_{2}^{2},S_{0}^{2}\right) \\
&&I_{0,0,1,1,0,1}\left( \left( \mathbf{\alpha }_{2},\mathbf{p}_{2}\right)
,\left( \mathbf{\alpha }_{2}^{\prime },\mathbf{p}_{2}^{\prime }\right)
,\left( \mathbf{\alpha }_{0}^{\prime },\mathbf{p}_{0}^{\prime }\right)
,S_{2}^{2},S_{0}^{2}\right)
\end{eqnarray*}

\subsubsection*{6.3.2 Transformation matrix $F$}

With these notations, equation (\ref{CMT}):%
\begin{equation*}
F\left( \left\{ \mathbf{\alpha }_{l}^{\prime },\mathbf{p}_{l}^{\prime
},S_{l}^{\prime 2},m_{l}^{\prime }\right\} ,\left\{ \mathbf{\alpha }_{k},%
\mathbf{p}_{k},S_{k}^{2},m_{k}\right\} \right) =-\frac{%
V_{m_{1},...,m_{n},m_{1}^{\prime },...,m_{n^{\prime }}^{\prime }}\left(
\left\{ \mathbf{\alpha }_{l}^{\prime },\mathbf{p}_{l}^{\prime
},S_{l}^{\prime 2},m_{l}^{\prime }\right\} ,\left\{ \mathbf{\alpha }_{k},%
\mathbf{p}_{k},S_{k}^{2},m_{k}\right\} \right) }{\sum_{l=1}^{n^{\prime
}}m_{l}^{\prime }\mathbf{\bar{D}}_{S_{l}^{\prime 2}}^{\mathbf{\alpha }%
_{l}^{\prime }}-\sum_{k=1}^{n}m_{k}\mathbf{\bar{D}}_{S_{k}^{2}}^{\mathbf{%
\alpha }_{k}}}
\end{equation*}%
solves as:%
\begin{eqnarray*}
&&F_{1,1,0,0,1,0}\left( \left( \mathbf{\alpha }_{1},\mathbf{p}_{1}\right)
,\left( \mathbf{\alpha }_{1}^{\prime },\mathbf{p}_{1}^{\prime }\right)
,\left( \mathbf{\alpha }_{0},\mathbf{p}_{0}\right)
,S_{1}^{2},S_{0}^{2}\right) \\
&=&-\frac{I_{1,1,0,0,1,0}\left( \left( \mathbf{\alpha }_{1},\mathbf{p}%
_{1}\right) ,\left( \mathbf{\alpha }_{1}^{\prime },\mathbf{p}_{1}^{\prime
}\right) ,\left( \mathbf{\alpha }_{0},\mathbf{p}_{0}\right)
,S_{1}^{2},S_{0}^{2}\right) }{\left\{ \sqrt{\left( \mathbf{\bar{D}}_{\mathbf{%
\alpha }_{1},\mathbf{p}_{1},S_{1}^{2}}\right) ^{2}}-\sqrt{\left( \mathbf{%
\bar{D}}_{\mathbf{\alpha }_{1}^{\prime },\mathbf{p}_{1}^{\prime
},S_{1}^{2}}\right) ^{2}}+\sqrt{\left( \mathbf{\bar{D}}_{\mathbf{\alpha }%
_{0},\mathbf{p}_{0},S_{0}^{2}}\right) ^{2}}\right\} } \\
&&F_{1,1,0,0,1,0}\left( \left( \mathbf{\alpha }_{1},\mathbf{p}_{1}\right)
,\left( \mathbf{\alpha }_{1}^{\prime },\mathbf{p}_{1}^{\prime }\right)
,\left( \mathbf{\alpha }_{0}^{\prime },\mathbf{p}_{0}^{\prime }\right)
,S_{1}^{2},S_{0}^{2}\right) \\
&=&-\frac{I_{1,1,0,0,1,0}\left( \left( \mathbf{\alpha }_{1},\mathbf{p}%
_{1}\right) ,\left( \mathbf{\alpha }_{1}^{\prime },\mathbf{p}_{1}^{\prime
}\right) ,\left( \mathbf{\alpha }_{0},\mathbf{p}_{0}\right)
,S_{1}^{2},S_{0}^{2}\right) }{\left\{ \sqrt{\left( \mathbf{\bar{D}}_{\mathbf{%
\alpha }_{1},\mathbf{p}_{1},S_{1}^{2}}\right) ^{2}}-\sqrt{\left( \mathbf{%
\bar{D}}_{\mathbf{\alpha }_{1}^{\prime },\mathbf{p}_{1}^{\prime
},S_{1}^{2}}\right) ^{2}}+\sqrt{\left( \mathbf{\bar{D}}_{\mathbf{\alpha }%
_{0},\mathbf{p}_{0},S_{0}^{2}}\right) ^{2}}\right\} }
\end{eqnarray*}%
and:%
\begin{eqnarray*}
&&F_{0,0,1,1,1,0}\left( \left( \mathbf{\alpha }_{2},\mathbf{p}_{2}\right)
,\left( \mathbf{\alpha }_{2}^{\prime },\mathbf{p}_{2}^{\prime }\right)
,\left( \mathbf{\alpha }_{0},\mathbf{p}_{0}\right)
,S_{2}^{2},S_{0}^{2}\right) \\
&=&-\frac{I_{0,0,1,1,1,0}\left( \left( \mathbf{\alpha }_{2},\mathbf{p}%
_{2}\right) ,\left( \mathbf{\alpha }_{2}^{\prime },\mathbf{p}_{2}^{\prime
}\right) ,\left( \mathbf{\alpha }_{0},\mathbf{p}_{0}\right)
,S_{2}^{2},S_{0}^{2}\right) }{\left\{ \sqrt{\left( \mathbf{\bar{D}}_{\mathbf{%
\alpha }_{1},\mathbf{p}_{1},S_{1}^{2}}\right) ^{2}}-\sqrt{\left( \mathbf{%
\bar{D}}_{\mathbf{\alpha }_{1}^{\prime },\mathbf{p}_{1}^{\prime
},S_{1}^{2}}\right) ^{2}}+\sqrt{\left( \mathbf{\bar{D}}_{\mathbf{\alpha }%
_{0},\mathbf{p}_{0},S_{0}^{2}}\right) ^{2}}\right\} } \\
&&F_{0,0,1,1,1,0}\left( \left( \mathbf{\alpha }_{2},\mathbf{p}_{2}\right)
,\left( \mathbf{\alpha }_{2}^{\prime },\mathbf{p}_{2}^{\prime }\right)
,\left( \mathbf{\alpha }_{0}^{\prime },\mathbf{p}_{0}^{\prime }\right)
,S_{2}^{2},S_{0}^{2}\right) \\
&=&-\frac{I_{0,0,1,1,1,0}\left( \left( \mathbf{\alpha }_{2},\mathbf{p}%
_{2}\right) ,\left( \mathbf{\alpha }_{2}^{\prime },\mathbf{p}_{2}^{\prime
}\right) ,\left( \mathbf{\alpha }_{0}^{\prime },\mathbf{p}_{0}^{\prime
}\right) ,S_{2}^{2},S_{0}^{2}\right) }{\left\{ \sqrt{\left( \mathbf{\bar{D}}%
_{\mathbf{\alpha }_{1},\mathbf{p}_{1},S_{1}^{2}}\right) ^{2}}-\sqrt{\left( 
\mathbf{\bar{D}}_{\mathbf{\alpha }_{1}^{\prime },\mathbf{p}_{1}^{\prime
},S_{1}^{2}}\right) ^{2}}+\sqrt{\left( \mathbf{\bar{D}}_{\mathbf{\alpha }%
_{0},\mathbf{p}_{0},S_{0}^{2}}\right) ^{2}}\right\} }
\end{eqnarray*}%
Written in term of operator, this is:%
\begin{eqnarray}
F=- &&\frac{I\left( \left( \mathbf{\alpha }_{i},\mathbf{p}_{i}\right)
,\left( \mathbf{\alpha }_{i}^{\prime },\mathbf{p}_{i}^{\prime }\right)
,\left( \mathbf{\alpha }_{0},\mathbf{p}_{0}\right)
,S_{i}^{2},S_{0}^{2}\right) }{\left( \sqrt{\left( \mathbf{\bar{D}}_{\mathbf{%
\alpha }_{i},\mathbf{p}_{i},S_{i}^{2}}\right) ^{2}}-\sqrt{\left( \mathbf{%
\bar{D}}_{\mathbf{\alpha }_{i}^{\prime },\mathbf{p}_{i}^{\prime
},S_{i}^{2}}\right) ^{2}}\right) ^{2}-\left( \mathbf{\bar{D}}_{\mathbf{%
\alpha }_{0},\mathbf{p}_{0},S_{0}^{2}}\right) ^{2}}  \label{FML} \\
&&\times \left( \left( \sqrt{\left( \mathbf{\bar{D}}_{\mathbf{\alpha }_{i},%
\mathbf{p}_{i},S_{i}^{2}}\right) ^{2}}-\sqrt{\left( \mathbf{\bar{D}}_{%
\mathbf{\alpha }_{i}^{\prime },\mathbf{p}_{i}^{\prime },S_{i}^{2}}\right)
^{2}}+\sqrt{\left( \mathbf{\bar{D}}_{\mathbf{\alpha }_{0},\mathbf{p}%
_{0},S_{0}^{2}}\right) ^{2}}\right) \mathbf{A}^{-}\left( \mathbf{\alpha }%
_{0},\mathbf{p}_{0},S_{0}^{2}\right) \right.  \notag \\
&&\left. +\left( \sqrt{\left( \mathbf{\bar{D}}_{\mathbf{\alpha }_{i},\mathbf{%
p}_{i},S_{i}^{2}}\right) ^{2}}-\sqrt{\left( \mathbf{\bar{D}}_{\mathbf{\alpha 
}_{i}^{\prime },\mathbf{p}_{i}^{\prime },S_{i}^{2}}\right) ^{2}}-\sqrt{%
\left( \mathbf{\bar{D}}_{\mathbf{\alpha }_{0},\mathbf{p}_{0},S_{0}^{2}}%
\right) ^{2}}\right) \mathbf{A}^{+-}\left( \mathbf{\alpha }_{0},\mathbf{p}%
_{0},S_{0}^{2}\right) \right)  \notag
\end{eqnarray}

\subsubsection*{6.3.3 Remark}

Note that formula (\ref{FML}) could have been retrieved also by postulating
the form:%
\begin{eqnarray*}
F &=&\sum_{i}F\left( \left( \mathbf{\alpha }_{i},\mathbf{p}_{i}\right)
,\left( \mathbf{\alpha }_{i}^{\prime },\mathbf{p}_{i}^{\prime }\right)
,\left( \mathbf{\alpha }_{0},\mathbf{p}_{0}\right)
,S_{i}^{2},S_{0}^{2}\right) \mathbf{A}^{+}\left( \mathbf{\alpha }_{i},%
\mathbf{p}_{i},S_{i}^{2}\right) \mathbf{A}^{-}\left( \mathbf{\alpha }%
_{i}^{\prime },\mathbf{p}_{i}^{\prime },S_{i}^{2}\right) \\
&&\times \left( \varepsilon \mathbf{A}^{-}\left( \mathbf{\alpha }_{0},%
\mathbf{p}_{0},S_{0}^{2}\right) +\mathbf{A}^{+-}\left( \mathbf{\alpha }_{0},%
\mathbf{p}_{0},S_{0}^{2}\right) \right)
\end{eqnarray*}%
with $\varepsilon $ to be determined to ensure:%
\begin{equation*}
I+\left[ F,S_{0}\right] =0
\end{equation*}%
As a consequence:%
\begin{eqnarray*}
&&\left[ F,S_{0}\right] \\
&=&\left[ F\left( \left( \mathbf{\alpha }_{i},\mathbf{p}_{i}\right) ,\left( 
\mathbf{\alpha }_{i}^{\prime },\mathbf{p}_{i}^{\prime }\right) ,\left( 
\mathbf{\alpha }_{0},\mathbf{p}_{0}\right) ,S_{i}^{2},S_{0}^{2}\right) 
\mathbf{A}^{+}\left( \mathbf{\alpha }_{i},\mathbf{p}_{i},S_{i}^{2}\right) 
\mathbf{A}^{-}\left( \mathbf{\alpha }_{i}^{\prime },\mathbf{p}_{i}^{\prime
},S_{i}^{2}\right) \left( \mathbf{A}^{-}\left( \mathbf{\alpha }_{0},\mathbf{p%
}_{0},S_{0}^{2}\right) +\mathbf{A}^{+-}\left( \mathbf{\alpha }_{0},\mathbf{p}%
_{0},S_{0}^{2}\right) \right) ,S_{0}\right] \\
&=&\left( \sqrt{\left( \mathbf{\bar{D}}_{\mathbf{\alpha }_{i},\mathbf{p}%
_{i},S_{i}^{2}}\right) ^{2}}-\sqrt{\left( \mathbf{\bar{D}}_{\mathbf{\alpha }%
_{i}^{\prime },\mathbf{p}_{i}^{\prime },S_{i}^{2}}\right) ^{2}}\right) \\
&&\times F\left( \left( \mathbf{\alpha }_{i},\mathbf{p}_{i}\right) ,\left( 
\mathbf{\alpha }_{i}^{\prime },\mathbf{p}_{i}^{\prime }\right) ,\left( 
\mathbf{\alpha }_{0},\mathbf{p}_{0}\right) ,S_{i}^{2},S_{0}^{2}\right) 
\mathbf{A}^{+}\left( \mathbf{\alpha }_{i},\mathbf{p}_{i},S_{i}^{2}\right) 
\mathbf{A}^{-}\left( \mathbf{\alpha }_{i}^{\prime },\mathbf{p}_{i}^{\prime
},S_{i}^{2}\right) \left( \varepsilon \mathbf{A}^{-}\left( \mathbf{\alpha }%
_{0},\mathbf{p}_{0},S_{0}^{2}\right) +\mathbf{A}^{+-}\left( \mathbf{\alpha }%
_{0},\mathbf{p}_{0},S_{0}^{2}\right) \right) \\
&&+\sqrt{\left( \mathbf{\bar{D}}_{\mathbf{\alpha }_{0},\mathbf{p}%
_{0},S_{0}^{2}}\right) ^{2}}F\left( \left( \mathbf{\alpha }_{i},\mathbf{p}%
_{i}\right) ,\left( \mathbf{\alpha }_{i}^{\prime },\mathbf{p}_{i}^{\prime
}\right) ,\left( \mathbf{\alpha }_{0},\mathbf{p}_{0}\right)
,S_{i}^{2},S_{0}^{2}\right) \mathbf{A}^{+}\left( \mathbf{\alpha }_{i},%
\mathbf{p}_{i},S_{i}^{2}\right) \mathbf{A}^{-}\left( \mathbf{\alpha }%
_{i}^{\prime },\mathbf{p}_{i}^{\prime },S_{i}^{2}\right) \\
&&\times \left( -\varepsilon \mathbf{A}^{-}\left( \mathbf{\alpha }_{0},%
\mathbf{p}_{0},S_{0}^{2}\right) +\mathbf{A}^{+-}\left( \mathbf{\alpha }_{0},%
\mathbf{p}_{0},S_{0}^{2}\right) \right)
\end{eqnarray*}%
which leads to:%
\begin{equation*}
\left( \sqrt{\left( \mathbf{\bar{D}}_{\mathbf{\alpha }_{i},\mathbf{p}%
_{i},S_{i}^{2}}\right) ^{2}}-\sqrt{\left( \mathbf{\bar{D}}_{\mathbf{\alpha }%
_{i}^{\prime },\mathbf{p}_{i}^{\prime },S_{i}^{2}}\right) ^{2}}-\sqrt{\left( 
\mathbf{\bar{D}}_{\mathbf{\alpha }_{0},\mathbf{p}_{0},S_{0}^{2}}\right) ^{2}}%
\right) \varepsilon =\sqrt{\left( \mathbf{\bar{D}}_{\mathbf{\alpha }_{i},%
\mathbf{p}_{i},S_{i}^{2}}\right) ^{2}}-\sqrt{\left( \mathbf{\bar{D}}_{%
\mathbf{\alpha }_{i}^{\prime },\mathbf{p}_{i}^{\prime },S_{i}^{2}}\right)
^{2}}+\sqrt{\left( \mathbf{\bar{D}}_{\mathbf{\alpha }_{0},\mathbf{p}%
_{0},S_{0}^{2}}\right) ^{2}}
\end{equation*}%
and:%
\begin{eqnarray*}
&&\left( \left( \sqrt{\left( \mathbf{\bar{D}}_{\mathbf{\alpha }_{i},\mathbf{p%
}_{i},S_{i}^{2}}\right) ^{2}}-\sqrt{\left( \mathbf{\bar{D}}_{\mathbf{\alpha }%
_{i}^{\prime },\mathbf{p}_{i}^{\prime },S_{i}^{2}}\right) ^{2}}\right) +%
\sqrt{\left( \mathbf{\bar{D}}_{\mathbf{\alpha }_{0},\mathbf{p}%
_{0},S_{0}^{2}}\right) ^{2}}\right) F\left( \left( \mathbf{\alpha }_{i},%
\mathbf{p}_{i}\right) ,\left( \mathbf{\alpha }_{i}^{\prime },\mathbf{p}%
_{i}^{\prime }\right) ,\left( \mathbf{\alpha }_{0},\mathbf{p}_{0}\right)
,S_{i}^{2},S_{0}^{2}\right) \\
&=&-F\left( \left( \mathbf{\alpha }_{i},\mathbf{p}_{i}\right) ,\left( 
\mathbf{\alpha }_{i}^{\prime },\mathbf{p}_{i}^{\prime }\right) ,\left( 
\mathbf{\alpha }_{0},\mathbf{p}_{0}\right) ,S_{i}^{2},S_{0}^{2}\right)
\end{eqnarray*}%
These two equations lead to (\ref{FML}).

\subsection*{6.4 Bounded states}

We compute the eigenstates of $\left( S^{\left( O\right) }\right) $ that are
the stable collective states resulting from the effective interactions.

As stated in the text, those states are series expansion:%
\begin{equation}
\left\vert \prod\limits_{K}\left( \left( \mathbf{\alpha }_{K},\mathbf{p}%
_{K},S_{K}^{2},M_{K}\right) \right) \right\rangle =\sum_{\left( M_{K}\right)
}A\left( \left( \mathbf{\alpha }_{K},\mathbf{p}_{K},S_{K}^{2},M_{K}\right)
\right) \prod\limits_{K=1}^{n}\prod\limits_{s=1}^{M_{K}}\mathbf{A}^{+}\left( 
\mathbf{\alpha }_{K},\mathbf{p}_{K},S_{K}^{2}\right)
\prod\limits_{K}\left\vert Vac\right\rangle _{K}  \label{GS}
\end{equation}%
Our aim is first to find the lowest gnstts which have the form:%
\begin{equation}
\left\vert \prod\limits_{K}\left( \mathbf{\alpha }_{K},\mathbf{p}%
_{K},S_{K}^{2}\right) \right\rangle =\sum_{K}A\left( \left( \mathbf{\alpha }%
_{K},\mathbf{p}_{K},S_{K}^{2}\right) \right) \mathbf{A}^{+}\left( \mathbf{%
\alpha }_{K},\mathbf{p}_{K},S_{K}^{2}\right) \prod\limits_{K}\left\vert
Vac\right\rangle _{K}  \label{LG}
\end{equation}%
and then to generalize the result to obtain the full series expansion.

\subsubsection*{6.4.1 Lowest eigenvalues}

As in the text, we start with the action of $\left( S^{\left( O\right)
}\right) $ on (\ref{LG}):%
\begin{eqnarray}
&&\left( S^{\left( O\right) }\right) \left\vert \prod\limits_{K}\left( 
\mathbf{\alpha }_{K},\mathbf{p}_{K},S_{K}^{2}\right) \right\rangle
\label{SN} \\
&=&\sum_{K}A\left( \left( \mathbf{\alpha }_{K},\mathbf{p}_{K},S_{K}^{2}%
\right) \right) \left( \sum_{K}\mathbf{\bar{D}}_{S_{K^{\prime }}^{2}}^{%
\mathbf{\alpha }_{K^{\prime }}}+2\mathbf{\bar{D}}_{S_{K}^{2}}^{\mathbf{%
\alpha }_{K}}\right) \mathbf{A}^{+}\left( \mathbf{\alpha }_{K},\mathbf{p}%
_{K},S_{K}^{2}\right) \prod\limits_{K}\left\vert Vac\right\rangle _{K} 
\notag \\
&&+\sum_{K,L}A\left( \left( \mathbf{\alpha }_{K},\mathbf{p}%
_{K},S_{K}^{2}\right) \right) \left[ I,F\right] _{1,1,0,...,0}\left( \left\{ 
\mathbf{\alpha }_{L}^{\prime },\mathbf{p}_{L}^{\prime },S_{L}^{\prime
2}\right\} ,\left\{ \mathbf{\alpha }_{K},\mathbf{p}_{K},S_{K}^{2}\right\}
,\left( \mathbf{\alpha }_{P},\mathbf{p}_{P},S_{P}^{2}\right) \right) \mathbf{%
A}^{+}\left( \mathbf{\alpha }_{L}^{\prime },\mathbf{p}_{L}^{\prime
},S_{L}^{\prime 2}\right) \prod\limits_{K}\left\vert Vac\right\rangle _{K} 
\notag
\end{eqnarray}%
The condition for $\left\vert \prod\limits_{K}\left( \mathbf{\alpha }_{K},%
\mathbf{p}_{K},S_{K}^{2}\right) \right\rangle $ to be an eignstate for $\eta 
$\ writes:%
\begin{eqnarray}
\eta A\left( \left( \mathbf{\alpha }_{K},\mathbf{p}_{K},S_{K}^{2}\right)
\right) &=&A\left( \left( \mathbf{\alpha }_{K},\mathbf{p}_{K},S_{K}^{2}%
\right) \right) \left( \sum_{K^{\prime }}\mathbf{\bar{D}}_{S_{K^{\prime
}}^{2}}^{\mathbf{\alpha }_{K^{\prime }}}+2\mathbf{\bar{D}}_{S_{K}^{2}}^{%
\mathbf{\alpha }_{K}}\right)  \label{GN} \\
&&+\sum_{L}A\left( \left( \mathbf{\alpha }_{L}^{\prime },\mathbf{p}%
_{L}^{\prime },S_{L}^{\prime 2}\right) \right) \left[ I,F\right]
_{1,1,0,...,0}\left( \left\{ \mathbf{\alpha }_{K},\mathbf{p}%
_{K},S_{K}^{2}\right\} ,\left\{ \mathbf{\alpha }_{L}^{\prime },\mathbf{p}%
_{L}^{\prime },S_{L}^{\prime 2}\right\} ,\left( \mathbf{\alpha }_{P},\mathbf{%
p}_{P},S_{P}^{2}\right) \right)  \notag
\end{eqnarray}%
or, which is equivalent, in terms of determinant:

\begin{eqnarray}
0 &=&\det \left\{ \left[ I,F\right] _{1,1}\left( \left\{ \mathbf{\alpha }%
_{L}^{\prime },\mathbf{p}_{L}^{\prime },S_{L}^{\prime 2}\right\} ,\left\{ 
\mathbf{\alpha }_{K},\mathbf{p}_{K},S_{K}^{2}\right\} \right) \right.
\label{NV} \\
&&\left. +\left( \sum_{K^{\prime }}\mathbf{\bar{D}}_{S_{K^{\prime }}^{2}}^{%
\mathbf{\alpha }_{K^{\prime }}}+2\mathbf{\bar{D}}_{S_{K}^{2}}^{\mathbf{%
\alpha }_{K}}-\eta \right) \delta \left( \left( \mathbf{\alpha }_{L}^{\prime
},\mathbf{p}_{L}^{\prime },S_{L}^{\prime 2}\right) -\left( \mathbf{\alpha }%
_{K},\mathbf{p}_{K},S_{K}^{2}\right) \right) \right\}  \notag
\end{eqnarray}%
that is:%
\begin{equation*}
\prod\limits_{K}\left( \sum_{K^{\prime }}\mathbf{\bar{D}}_{S_{K^{\prime
}}^{2}}^{\mathbf{\alpha }_{K^{\prime }}}+2\mathbf{\bar{D}}_{S_{K}^{2}}^{%
\mathbf{\alpha }_{K}}-\eta \right) \det \left\{ 1+\frac{\left[ I,F\right]
_{1,1}\left( \left\{ \mathbf{\alpha }_{L}^{\prime },\mathbf{p}_{L}^{\prime
},S_{L}^{\prime 2}\right\} ,\left\{ \mathbf{\alpha }_{K},\mathbf{p}%
_{K},S_{K}^{2}\right\} \right) }{\sum_{K^{\prime }}\mathbf{\bar{D}}%
_{S_{K^{\prime }}^{2}}^{\mathbf{\alpha }_{K^{\prime }}}+2\mathbf{\bar{D}}%
_{S_{K}^{2}}^{\mathbf{\alpha }_{K}}-\eta }\right\} =0
\end{equation*}%
For relatively small interactions $\left[ I,F\right] _{\left( N_{K^{\prime
}}\right) ,M_{K}-M_{K^{\prime }}+N_{K^{\prime }}}<<1$ and the determinant
equation is in first approximation:%
\begin{eqnarray}
&&0=\prod\limits_{K}\left( \sum_{K^{\prime }}\mathbf{\bar{D}}_{S_{K^{\prime
}}^{2}}^{\mathbf{\alpha }_{K^{\prime }}}+2\mathbf{\bar{D}}_{S_{K}^{2}}^{%
\mathbf{\alpha }_{K}}-\eta \right)  \label{DTM} \\
&&\left( 1-\sum_{K,K^{\prime }}\frac{\left[ I,F\right] _{1,1}\left( \left( 
\mathbf{\alpha }_{K},\mathbf{p}_{K},S_{K}^{2}\right) ,\left( \mathbf{\alpha }%
_{K^{\prime }},\mathbf{p}_{K^{\prime }},S_{K^{\prime }}^{2}\right) \right) %
\left[ I,F\right] _{1,1}\left( \left( \mathbf{\alpha }_{K^{\prime }},\mathbf{%
p}_{K^{\prime }},S_{K^{\prime }}^{2}\right) ,\left( \mathbf{\alpha }_{K},%
\mathbf{p}_{K},S_{K}^{2}\right) \right) }{\left( \mathbf{\bar{D}}%
_{S_{K}^{2}}^{\mathbf{\alpha }_{K}}+2\mathbf{\bar{D}}_{S_{K}^{2}}^{\mathbf{%
\alpha }_{K}}-\eta \right) \left( \mathbf{\bar{D}}_{S_{K^{\prime }}^{2}}^{%
\mathbf{\alpha }_{K^{\prime }}}+2\mathbf{\bar{D}}_{S_{K^{\prime }}^{2}}^{%
\mathbf{\alpha }_{K^{\prime }}}-\eta \right) }\right)  \notag
\end{eqnarray}%
writing the eigenvalues solving (\ref{DTM}):%
\begin{equation}
\eta \left( \mathbf{\alpha }_{K},\mathbf{p}_{K},S_{K}^{2}\right)
=\sum_{K^{\prime }}\mathbf{\bar{D}}_{S_{K^{\prime }}^{2}}^{\mathbf{\alpha }%
_{K^{\prime }}}+2\mathbf{\bar{D}}_{S_{K}^{2}}^{\mathbf{\alpha }%
_{K}}-\varepsilon \left( \mathbf{\alpha }_{K},\mathbf{p}_{K},S_{K}^{2}\right)
\label{GV}
\end{equation}%
and inserting this expression in (\ref{DTM}) yields the value of $%
\varepsilon \left( \mathbf{\alpha }_{K},\mathbf{p}_{K},S_{K}^{2}\right) $:%
\begin{equation*}
\varepsilon \left( \mathbf{\alpha }_{K},\mathbf{p}_{K},S_{K}^{2}\right)
=\sum_{K^{\prime }}\frac{\left[ I,F\right] _{1,1}\left( \left( \mathbf{%
\alpha }_{K},\mathbf{p}_{K},S_{K}^{2}\right) ,\left( \mathbf{\alpha }%
_{K^{\prime }},\mathbf{p}_{K^{\prime }},S_{K^{\prime }}^{2}\right) \right) %
\left[ I,F\right] _{1,1}\left( \left( \mathbf{\alpha }_{K^{\prime }},\mathbf{%
p}_{K^{\prime }},S_{K^{\prime }}^{2}\right) ,\left( \mathbf{\alpha }_{K},%
\mathbf{p}_{K},S_{K}^{2}\right) \right) }{\left( \mathbf{\bar{D}}%
_{S_{K^{\prime }}^{2}}^{\mathbf{\alpha }_{K^{\prime }}}+2\mathbf{\bar{D}}%
_{S_{K^{\prime }}^{2}}^{\mathbf{\alpha }_{K^{\prime }}}-\eta \right) }
\end{equation*}%
Inserting the eigenvalue (\ref{GV}) in the eigenstate equation (\ref{DTM})
becomes:%
\begin{equation*}
0=\varepsilon \left( \mathbf{\alpha }_{K},\mathbf{p}_{K},S_{K}^{2}\right)
A\left( \mathbf{\alpha }_{K},\mathbf{p}_{K},S_{K}^{2}\right)
+\sum_{K^{\prime }}P\left( \left( \mathbf{\alpha }_{K},\mathbf{p}%
_{K},S_{K}^{2}\right) ,\left( \mathbf{\alpha }_{K^{\prime }},\mathbf{p}%
_{K^{\prime }},S_{K^{\prime }}^{2}\right) \right) A\left( \mathbf{\alpha }%
_{K^{\prime }},\mathbf{p}_{K^{\prime }},S_{K^{\prime }}^{2}\right)
\end{equation*}%
f $L\neq K$:%
\begin{equation*}
0\simeq \left( \mathbf{\bar{D}}_{S_{L}^{2}}^{\mathbf{\alpha }_{L}}+2\mathbf{%
\bar{D}}_{S_{L}^{2}}^{\mathbf{\alpha }_{L}}\right) A\left( \mathbf{\alpha }%
_{L},\mathbf{p}_{L},S_{L}^{2}\right) +\sum_{K^{\prime }M_{K^{\prime }}}\left[
I,F\right] _{1,1}\left( \left( \mathbf{\alpha }_{L},\mathbf{p}%
_{L},S_{L}^{2}\right) ,\left( \mathbf{\alpha }_{K^{\prime }},\mathbf{p}%
_{K^{\prime }},S_{K^{\prime }}^{2}\right) \right) A\left( \mathbf{\alpha }%
_{K^{\prime }},\mathbf{p}_{K^{\prime }},S_{K^{\prime }}^{2}\right)
\end{equation*}%
with solution:%
\begin{equation*}
A\left( \mathbf{\alpha }_{L},\mathbf{p}_{L},S_{L}^{2}\right) \simeq -\frac{%
\left[ I,F\right] _{1,1}\left( \left( \mathbf{\alpha }_{L},\mathbf{p}%
_{L},S_{L}^{2}\right) ,\left( \mathbf{\alpha }_{K^{\prime }},\mathbf{p}%
_{K^{\prime }},S_{K^{\prime }}^{2}\right) \right) }{\mathbf{\bar{D}}%
_{S_{L}^{2}}^{\mathbf{\alpha }_{L}}+2\mathbf{\bar{D}}_{S_{L}^{2}}^{\mathbf{%
\alpha }_{L}}}A\left( \mathbf{\alpha }_{K},\mathbf{p}_{K},S_{K}^{2}\right)
\end{equation*}%
In first approximation, using:%
\begin{equation*}
\eta _{K}=\sum_{K^{\prime }}\mathbf{\bar{D}}_{S_{K^{\prime }}^{2}}^{\mathbf{%
\alpha }_{K^{\prime }}}+2\mathbf{\bar{D}}_{S_{K}^{2}}^{\mathbf{\alpha }%
_{K}}+\varepsilon _{K}
\end{equation*}%
the $\varepsilon _{K}$ are solutions of: 
\begin{equation*}
\det \left\{ \left[ I,F\right] _{1,1}\left( \left\{ \mathbf{\alpha }%
_{L}^{\prime },\mathbf{p}_{L}^{\prime },S_{L}^{\prime 2}\right\} ,\left\{ 
\mathbf{\alpha }_{K},\mathbf{p}_{K},S_{K}^{2}\right\} \right) -\varepsilon
_{K}\delta \left( \left( \mathbf{\alpha }_{L}^{\prime },\mathbf{p}%
_{L}^{\prime },S_{L}^{\prime 2}\right) -\left( \mathbf{\alpha }_{K},\mathbf{p%
}_{K},S_{K}^{2}\right) \right) \right\} =0
\end{equation*}%
So that the $\varepsilon _{K}$ are eigenvalues f $\left[ I,F\right] _{1,1}$
and the coefficients $A\left( \left( \mathbf{\alpha }_{K},\mathbf{p}%
_{K},S_{K}^{2}\right) \right) $ are eigenvectors:%
\begin{equation*}
\varepsilon _{K}A\left( \left( \mathbf{\alpha }_{K},\mathbf{p}%
_{K},S_{K}^{2}\right) \right) =\sum_{L}A\left( \left( \mathbf{\alpha }%
_{L}^{\prime },\mathbf{p}_{L}^{\prime },S_{L}^{\prime 2}\right) \right) %
\left[ I,F\right] _{1,1,0,...,0}\left( \left\{ \mathbf{\alpha }_{K},\mathbf{p%
}_{K},S_{K}^{2}\right\} ,\left\{ \mathbf{\alpha }_{L}^{\prime },\mathbf{p}%
_{L}^{\prime },S_{L}^{\prime 2}\right\} ,\left( \mathbf{\alpha }_{P},\mathbf{%
p}_{P},S_{P}^{2}\right) \right)
\end{equation*}

As a consequence, the states arr:%
\begin{equation*}
\left\vert \prod\limits_{K}\left( \mathbf{\alpha }_{K},\mathbf{p}%
_{K},S_{K}^{2}\right) \right\rangle =\sum_{K}A\left( \left( \mathbf{\alpha }%
_{K},\mathbf{p}_{K},S_{K}^{2}\right) \right) \mathbf{A}^{+}\left( \mathbf{%
\alpha }_{K},\mathbf{p}_{K},S_{K}^{2}\right) \prod\limits_{K}\left\vert
Vac\right\rangle _{K}
\end{equation*}

\subsubsection*{6.4.3 Arbitrary eigenvalues and eigenstates. Full series
expansion}

To find the more general eigenstates, we consider a series expansion:%
\begin{equation}
\left\vert \prod\limits_{K}\left( \mathbf{\alpha }_{K},\mathbf{p}%
_{K},S_{K}^{2}\right) \right\rangle =\sum_{\left( M_{K}\right) }A\left(
\left( \mathbf{\alpha }_{K},\mathbf{p}_{K},S_{K}^{2},M_{K}\right) \right)
\prod\limits_{K=1}^{n}\prod\limits_{s=1}^{M_{K}}\mathbf{A}^{+}\left( \mathbf{%
\alpha }_{K},\mathbf{p}_{K},S_{K}^{2}\right) \prod\limits_{K}\left\vert
Vac\right\rangle _{K}  \label{CTN}
\end{equation}%
We first need to compute the correction to $\left( S^{\left( O\right)
}\right) \left\vert \prod\limits_{K}\left( \mathbf{\alpha }_{K},\mathbf{p}%
_{K},S_{K}^{2}\right) \right\rangle $ due to $\frac{1}{2}\left[ I,F\right] $%
, we thus consider the action of the full operator:%
\begin{equation*}
\left( S^{\left( O\right) }\right) \left\vert \prod\limits_{K}\left( \mathbf{%
\alpha }_{K},\mathbf{p}_{K},S_{K}^{2}\right) \right\rangle \rightarrow
\left( S^{\left( O\right) }\right) \left\vert \prod\limits_{K}\left( \mathbf{%
\alpha }_{K},\mathbf{p}_{K},S_{K}^{2}\right) \right\rangle +\frac{1}{2}\left[
I,F\right] \left\vert \prod\limits_{K}\left( \mathbf{\alpha }_{K},\mathbf{p}%
_{K},S_{K}^{2}\right) \right\rangle
\end{equation*}

\subsubsection*{6.4.4 Correction to eigenstates due to $\frac{1}{2}\left[ I,F%
\right] $}

This correction is computed by first considering the commutator:%
\begin{eqnarray}
C &=&\left[ \exp \left( \sum t_{k}\mathbf{A}^{-}\left( \mathbf{\alpha }_{k},%
\mathbf{p}_{k},S_{k}^{2}\right) \right) ,\exp \left( \sum t_{l}^{\prime }%
\mathbf{A}^{+}\left( \mathbf{\alpha }_{l},\mathbf{p}_{l},S_{l}^{2}\right)
\right) \right]  \label{CM} \\
&=&\exp \left( \sum t_{l}^{\prime }\mathbf{A}^{+}\left( \mathbf{\alpha }_{l},%
\mathbf{p}_{l},S_{l}^{2}\right) \right) \exp \left( \sum t_{k}\mathbf{A}%
^{-}\left( \mathbf{\alpha }_{k},\mathbf{p}_{k},S_{k}^{2}\right) \right) \exp
\left( \sum t_{k}t_{l}^{\prime }\delta \left( \left( \mathbf{\alpha }_{k},%
\mathbf{p}_{k},S_{k}^{2}\right) -\left( \mathbf{\alpha }_{l},\mathbf{p}%
_{l},S_{l}^{2}\right) \right) \right)  \notag
\end{eqnarray}%
From formula (\ref{CM}), we deduce that:%
\begin{eqnarray*}
&&\prod_{k=1}^{n}\prod\limits_{s=1}^{m_{k}}\mathbf{A}^{-}\left( \mathbf{%
\alpha }_{k},\mathbf{p}_{k},S_{k}^{2}\right) \prod_{q=1}^{n_{1}^{\prime
}}\prod\limits_{s=1}^{m_{q}^{\prime }}\mathbf{A}^{+}\left( \mathbf{\alpha }%
_{q}^{\prime },\mathbf{p}_{q}^{\prime },S_{q}^{\prime 2}\right)
\prod\limits_{K}\left\vert Vac\right\rangle _{K} \\
&=&\left[ \prod_{k=1}^{n}\prod\limits_{s=1}^{m_{k}}\mathbf{A}^{-}\left( 
\mathbf{\alpha }_{k},\mathbf{p}_{k},S_{k}^{2}\right)
,\prod_{q=1}^{n_{1}^{\prime }}\prod\limits_{s=1}^{m_{q}^{\prime }}\mathbf{A}%
^{+}\left( \mathbf{\alpha }_{q}^{\prime },\mathbf{p}_{q}^{\prime
},S_{q}^{\prime 2}\right) \right] \prod\limits_{K}\left\vert
Vac\right\rangle _{K}
\end{eqnarray*}%
This expression is different from $0$ if $\left\{ \left( \mathbf{\alpha }%
_{k},\mathbf{p}_{k},S_{k}^{2}\right) \right\} \subset \left\{ \left( \mathbf{%
\alpha }_{q}^{\prime },\mathbf{p}_{q}^{\prime },S_{q}^{\prime 2}\right)
\right\} $ and $m_{k}\leqslant m_{k}^{\prime }$. In this case, let:%
\begin{equation*}
\left\{ \left( \mathbf{\alpha }_{p}^{\prime },\mathbf{p}_{p}^{\prime
},S_{p}^{\prime 2},m_{p}^{\prime }\right) \right\} =\left\{ \left( \mathbf{%
\alpha }_{q}^{\prime },\mathbf{p}_{q}^{\prime },S_{q}^{\prime 2}\right)
\right\} \backslash \left\{ \left( \mathbf{\alpha }_{k},\mathbf{p}%
_{k},S_{k}^{2}\right) \right\}
\end{equation*}%
with $m_{p}^{\prime }=m_{q=p}^{\prime }-m_{k=p}$ and we have:%
\begin{eqnarray*}
&&\left[ \prod_{k=1}^{n}\prod\limits_{s=1}^{m_{k}}\mathbf{A}^{-}\left( 
\mathbf{\alpha }_{k},\mathbf{p}_{k},S_{k}^{2}\right)
,\prod_{q=1}^{n_{1}^{\prime }}\prod\limits_{s=1}^{m_{q}^{\prime }}\mathbf{A}%
^{+}\left( \mathbf{\alpha }_{q}^{\prime },\mathbf{p}_{q}^{\prime
},S_{q}^{\prime 2}\right) \right] \prod\limits_{K}\left\vert
Vac\right\rangle _{K} \\
&=&m_{q=p}^{\prime }!C_{m_{q=p}^{\prime
}}^{m_{k=p}}\prod_{p}\prod\limits_{s=1}^{m_{p}^{\prime }}\mathbf{A}%
^{+}\left( \mathbf{\alpha }_{q}^{\prime },\mathbf{p}_{q}^{\prime
},S_{q}^{\prime 2}\right) \prod\limits_{K}\left\vert Vac\right\rangle _{K}
\end{eqnarray*}%
As a consequence, using (\ref{CTN}), we have:%
\begin{eqnarray*}
&&\frac{1}{2}\left[ I,F\right] \left\vert \prod\limits_{K}\left( \mathbf{%
\alpha }_{K},\mathbf{p}_{K},S_{K}^{2}\right) \right\rangle \\
&=&\sum_{\left( M_{K}\right) }A\left( \left( \mathbf{\alpha }_{K},\mathbf{p}%
_{K},S_{K}^{2},M_{K}\right) \right) \prod_{L=1}^{n^{\prime
}}\prod\limits_{s^{\prime }=1}^{M_{L}^{\prime }}\mathbf{A}^{+}\left( \mathbf{%
\alpha }_{L}^{\prime },\mathbf{p}_{L}^{\prime },S_{L}^{\prime 2}\right) \\
&&\times \sum \left[ I,F\right] _{N_{1},...,N_{n},M_{1}^{\prime
},...,M_{n^{\prime }}^{\prime }}\left( \left\{ \mathbf{\alpha }_{L}^{\prime
},\mathbf{p}_{L}^{\prime },S_{L}^{\prime 2}\right\} ,\left\{ \mathbf{\alpha }%
_{K},\mathbf{p}_{K},S_{K}^{2}\right\} \right) \\
&&\times \left[ \prod\limits_{K=1}^{n}\prod\limits_{s=1}^{N_{K}\leqslant
M_{K}}\mathbf{A}^{-}\left( \mathbf{\alpha }_{K},\mathbf{p}%
_{K},S_{K}^{2}\right) ,\prod\limits_{K=1}^{n}\prod\limits_{s=1}^{M_{K}}%
\mathbf{A}^{+}\left( \mathbf{\alpha }_{K},\mathbf{p}_{K},S_{K}^{2}\right) %
\right] \prod\limits_{K}\left\vert Vac\right\rangle _{K}
\end{eqnarray*}%
and this is equal to:%
\begin{eqnarray*}
&&\frac{1}{2}\left[ I,F\right] \left\vert \prod\limits_{K}\left( \mathbf{%
\alpha }_{K},\mathbf{p}_{K},S_{K}^{2}\right) \right\rangle \\
&=&\sum_{\left( M_{K}\right) }A\left( \left( \mathbf{\alpha }_{K},\mathbf{p}%
_{K},S_{K}^{2},M_{K}\right) \right) \prod_{L=1}^{n^{\prime
}}\prod\limits_{s^{\prime }=1}^{M_{L}^{\prime }}\mathbf{A}^{+}\left( \mathbf{%
\alpha }_{L}^{\prime },\mathbf{p}_{L}^{\prime },S_{L}^{\prime 2}\right) \\
&&\times \sum \left[ I,F\right] _{N_{1},...,N_{n},M_{1}^{\prime
},...,M_{n^{\prime }}^{\prime }}\left( \left\{ \mathbf{\alpha }_{L}^{\prime
},\mathbf{p}_{L}^{\prime },S_{L}^{\prime 2}\right\} ,\left\{ \mathbf{\alpha }%
_{K},\mathbf{p}_{K},S_{K}^{2}\right\} \right) \\
&&\times \prod_{L=1}^{n^{\prime }}\prod\limits_{s^{\prime
}=1}^{M_{L}^{\prime }}\mathbf{A}^{+}\left( \mathbf{\alpha }_{L}^{\prime },%
\mathbf{p}_{L}^{\prime },S_{L}^{\prime 2}\right) \left(
\prod\limits_{K=1}^{n}N_{K}!C_{M_{K}}^{N_{K}}\right)
\prod\limits_{K=1}^{n}\prod\limits_{s=1}^{M_{K}-N_{K}}\mathbf{A}^{+}\left( 
\mathbf{\alpha }_{K},\mathbf{p}_{K},S_{K}^{2}\right)
\prod\limits_{K}\left\vert Vac\right\rangle _{K}
\end{eqnarray*}%
This expression can be reorganized as:%
\begin{equation*}
\frac{1}{2}\left[ I,F\right] \left\vert \prod\limits_{K}\left( \mathbf{%
\alpha }_{K},\mathbf{p}_{K},S_{K}^{2}\right) \right\rangle =\sum_{\left(
M_{K}\right) }\hat{A}\left( \left( \mathbf{\alpha }_{K},\mathbf{p}%
_{K},S_{K}^{2},M_{K}\right) \right)
\prod\limits_{K=1}^{n}\prod\limits_{s=1}^{M_{K}}\mathbf{A}^{+}\left( \mathbf{%
\alpha }_{K},\mathbf{p}_{K},S_{K}^{2}\right) \prod\limits_{K}\left\vert
Vac\right\rangle _{K}
\end{equation*}%
where:%
\begin{eqnarray}
&&\hat{A}\left( \left( \mathbf{\alpha }_{K},\mathbf{p}_{K},S_{K}^{2},M_{K}%
\right) \right)  \label{CFH} \\
&=&\sum_{\mathcal{P}}\sum_{\left( N_{K}^{\left( 2\right) }\right) }\left(
\prod\limits_{K=1}^{n_{2}}N_{K}^{\left( 2\right) }!C_{M_{K}^{\left( 2\right)
}+N_{K}^{\left( 2\right) }}^{N_{K}^{\left( 2\right) }}\right) A\left( \left( 
\mathbf{\alpha }_{K}^{\left( 2\right) },\mathbf{p}_{K}^{\left( 2\right)
},\left( S_{K}^{2}\right) ^{\left( 2\right) },M_{K}^{\left( 2\right)
}+N_{K}^{\left( 2\right) }\right) \right)  \notag \\
&&\times \left[ I,F\right] _{N_{1}^{\left( 2\right) },...,N_{n_{1}}^{\left(
2\right) },M_{1}^{\left( 1\right) },...,M_{n_{2}}^{\left( 1\right) }}\left(
\left\{ \mathbf{\alpha }_{K}^{\left( 1\right) },\mathbf{p}_{K}^{\left(
1\right) },\left( S_{K}^{2}\right) ^{\left( 1\right) },M_{K}^{\left(
1\right) }\right\} ,\left\{ \mathbf{\alpha }_{K}^{\left( 2\right) },\mathbf{p%
}_{K}^{\left( 2\right) },\left( S_{K}^{2}\right) ^{\left( 2\right)
},N_{K}^{\left( 2\right) }\right\} \right)  \notag
\end{eqnarray}%
and the sum is over partitions:%
\begin{equation*}
\mathcal{P}=\left( \mathbf{\alpha }_{K},\mathbf{p}_{K},S_{K}^{2},M_{K}%
\right) =\left( \mathbf{\alpha }_{K}^{\left( 1\right) },\mathbf{p}%
_{K}^{\left( 1\right) },\left( S_{K}^{2}\right) ^{\left( 1\right)
},M_{K}^{\left( 1\right) }\right) \cup \left( \mathbf{\alpha }_{K}^{\left(
2\right) },\mathbf{p}_{K}^{\left( 2\right) },\left( S_{K}^{2}\right)
^{\left( 2\right) },M_{K}^{\left( 2\right) }\right)
\end{equation*}%
and where $M_{K}=M_{K}^{\left( 1\right) }+M_{K}^{\left( 2\right) }$.

Note that (\ref{CFH}) can be rewritten matricially:%
\begin{equation*}
\hat{A}\left( \mathbf{\alpha }_{K},\mathbf{p}_{K},S_{K}^{2},M_{K}\right)
=\sum_{K^{\prime },\left( M_{K^{\prime }}\right) }P\left( \left( \mathbf{%
\alpha }_{K},\mathbf{p}_{K},S_{K}^{2},M_{K}\right) ,\left( \mathbf{\alpha }%
_{K^{\prime }},\mathbf{p}_{K^{\prime }},S_{K^{\prime }}^{2},M_{K^{\prime
}}\right) \right) A\left( \mathbf{\alpha }_{K^{\prime }},\mathbf{p}%
_{K^{\prime }},S_{K^{\prime }}^{2},M_{K^{\prime }}\right)
\end{equation*}%
with:%
\begin{eqnarray}
&&P\left( \left( \mathbf{\alpha }_{K},\mathbf{p}_{K},S_{K}^{2},M_{K}\right)
,\left( \mathbf{\alpha }_{K^{\prime }},\mathbf{p}_{K^{\prime }},S_{K^{\prime
}}^{2},M_{K^{\prime }}\right) \right)  \label{MTR} \\
&=&\sum_{N_{K^{\prime }}\leqslant M_{K^{\prime }}}\left(
\prod\limits_{K^{\prime }=1}^{n}N_{K^{\prime }}!C_{M_{K^{\prime
}}}^{N_{K^{\prime }}}\right) A\left( \left( \mathbf{\alpha }_{K^{\prime }},%
\mathbf{p}_{K^{\prime }},\left( S_{K^{\prime }}^{2}\right) ,M_{K^{\prime
}}\right) \right) \left[ I,F\right] _{\left( N_{K^{\prime }}\right)
,M_{K}-M_{K^{\prime }}+N_{K^{\prime }}}\left( \left\{ \mathbf{\alpha }_{K},%
\mathbf{p}_{K},\left( S_{K}^{2}\right) ,M_{K}\right\} \right)  \notag
\end{eqnarray}

\subsubsection{6.4.4 Eigenvalues and eigenstates}

The action of $\left( S^{\left( O\right) }\right) $ on $\left\vert
\prod\limits_{K}\left( \mathbf{\alpha }_{K},\mathbf{p}_{K},S_{K}^{2}\right)
\right\rangle $ writes:%
\begin{eqnarray*}
&&\left( S^{\left( O\right) }\right) \left\vert \prod\limits_{K}\left( 
\mathbf{\alpha }_{K},\mathbf{p}_{K},S_{K}^{2}\right) \right\rangle \\
&=&\sum_{\left( M_{K}\right) }A\left( \left( \mathbf{\alpha }_{K},\mathbf{p}%
_{K},S_{K}^{2},M_{K}\right) \right) \left( \sum_{K}\left( \mathbf{\bar{D}}%
_{S_{K}^{2}}^{\mathbf{\alpha }_{K}}+2M_{K}\mathbf{\bar{D}}_{S_{K}^{2}}^{%
\mathbf{\alpha }_{K}}\right) \right)
\prod\limits_{K=1}^{n}\prod\limits_{s=1}^{M_{K}}\mathbf{A}^{+}\left( \mathbf{%
\alpha }_{K},\mathbf{p}_{K},S_{K}^{2}\right) \prod\limits_{K}\left\vert
Vac\right\rangle _{K} \\
&&+\sum_{\left( M_{K}\right) }\hat{A}\left( \left( \mathbf{\alpha }_{K},%
\mathbf{p}_{K},S_{K}^{2},M_{K}\right) \right)
\prod\limits_{K=1}^{n}\prod\limits_{s=1}^{M_{K}}\mathbf{A}^{+}\left( \mathbf{%
\alpha }_{K},\mathbf{p}_{K},S_{K}^{2}\right) \prod\limits_{K}\left\vert
Vac\right\rangle _{K}
\end{eqnarray*}%
and we look for eigenstates:%
\begin{equation*}
\left( S^{\left( O\right) }\right) \left\vert \prod\limits_{K}\left( \mathbf{%
\alpha }_{K},\mathbf{p}_{K},S_{K}^{2}\right) \right\rangle =\eta _{\left(
M_{K}\right) }\left\vert \prod\limits_{K}\left( \mathbf{\alpha }_{K},\mathbf{%
p}_{K},S_{K}^{2}\right) \right\rangle
\end{equation*}%
and consider eigenvalues of the form:%
\begin{equation*}
\eta _{\left( M_{K}\right) }=\sum_{K}\left( \mathbf{\bar{D}}_{S_{K^{\prime
}}^{2}}^{\mathbf{\alpha }_{K^{\prime }}}+2M_{K}\mathbf{\bar{D}}_{S_{K}^{2}}^{%
\mathbf{\alpha }_{K}}\right) +\varepsilon _{\left( M_{K}\right) }
\end{equation*}%
corresponding to perturbations of eigenstates without interactions.

The eigenvalues equation becomes:%
\begin{eqnarray}
0 &=&\det \left( \left( \mathbf{\bar{D}}_{S_{K}^{2}}^{\mathbf{\alpha }%
_{K}}+2M_{K}\mathbf{\bar{D}}_{S_{K}^{2}}^{\mathbf{\alpha }_{K}}-\eta \right)
\delta \left( \left( \mathbf{\alpha }_{K},\mathbf{p}_{K},S_{K}^{2},M_{K}%
\right) -\mathbf{\alpha }_{K^{\prime }},\mathbf{p}_{K^{\prime
}},S_{K^{\prime }}^{2},M_{K^{\prime }}\right) \right.  \label{DM} \\
&&\left. +P\left( \left( \mathbf{\alpha }_{K},\mathbf{p}_{K},S_{K}^{2},M_{K}%
\right) ,\left( \mathbf{\alpha }_{K^{\prime }},\mathbf{p}_{K^{\prime
}},S_{K^{\prime }}^{2},M_{K^{\prime }}\right) \right) \right)  \notag
\end{eqnarray}%
where the matrix $P$ is defined in (\ref{MTR}). We factor the determinant
arising in (\ref{DM}) as:%
\begin{eqnarray}
&&\det \left( \left( \mathbf{\bar{D}}_{S_{K}^{2}}^{\mathbf{\alpha }%
_{K}}+2M_{K}\mathbf{\bar{D}}_{S_{K}^{2}}^{\mathbf{\alpha }_{K}}-\eta \right)
\delta \left( \left( \mathbf{\alpha }_{K},\mathbf{p}_{K},S_{K}^{2},M_{K}%
\right) -\mathbf{\alpha }_{K^{\prime }},\mathbf{p}_{K^{\prime
}},S_{K^{\prime }}^{2},M_{K^{\prime }}\right) \right.  \label{DR} \\
&&\left. +P\left( \left( \mathbf{\alpha }_{K},\mathbf{p}_{K},S_{K}^{2},M_{K}%
\right) ,\left( \mathbf{\alpha }_{K^{\prime }},\mathbf{p}_{K^{\prime
}},S_{K^{\prime }}^{2},M_{K^{\prime }}\right) \right) \right)  \notag \\
&=&\det \left( \left( \mathbf{\bar{D}}_{S_{K}^{2}}^{\mathbf{\alpha }%
_{K}}+2M_{K}\mathbf{\bar{D}}_{S_{K}^{2}}^{\mathbf{\alpha }_{K}}-\eta \right)
\delta \left( \left( \mathbf{\alpha }_{K},\mathbf{p}_{K},S_{K}^{2},M_{K}%
\right) -\mathbf{\alpha }_{K^{\prime }},\mathbf{p}_{K^{\prime
}},S_{K^{\prime }}^{2},M_{K^{\prime }}\right) \right)  \notag \\
&&\times \det \left( 1+\frac{P\left( \left( \mathbf{\alpha }_{K},\mathbf{p}%
_{K},S_{K}^{2},M_{K}\right) ,\left( \mathbf{\alpha }_{K^{\prime }},\mathbf{p}%
_{K^{\prime }},S_{K^{\prime }}^{2},M_{K^{\prime }}\right) \right) }{\left( 
\mathbf{\bar{D}}_{S_{K}^{2}}^{\mathbf{\alpha }_{K}}+2M_{K}\mathbf{\bar{D}}%
_{S_{K}^{2}}^{\mathbf{\alpha }_{K}}-\eta \right) }\right)  \notag
\end{eqnarray}%
For relatively small interactions:%
\begin{equation*}
\left[ I,F\right] _{\left( N_{K^{\prime }}\right) ,M_{K}-M_{K^{\prime
}}+N_{K^{\prime }}}<<1
\end{equation*}%
and in first approximation the determinant (\ref{DR}) write:%
\begin{eqnarray*}
&&\det \left( \left( \mathbf{\bar{D}}_{S_{K}^{2}}^{\mathbf{\alpha }%
_{K}}+2M_{K}\mathbf{\bar{D}}_{S_{K}^{2}}^{\mathbf{\alpha }_{K}}-\eta \right)
\delta \left( \left( \mathbf{\alpha }_{K},\mathbf{p}_{K},S_{K}^{2},M_{K}%
\right) -\mathbf{\alpha }_{K^{\prime }},\mathbf{p}_{K^{\prime
}},S_{K^{\prime }}^{2},M_{K^{\prime }}\right) \right) \\
&&\times \left( 1-\sum_{K,K^{\prime }}\frac{P\left( \left( \mathbf{\alpha }%
_{K},\mathbf{p}_{K},S_{K}^{2},M_{K}\right) ,\left( \mathbf{\alpha }%
_{K^{\prime }},\mathbf{p}_{K^{\prime }},S_{K^{\prime }}^{2},M_{K^{\prime
}}\right) \right) P\left( \left( \mathbf{\alpha }_{K^{\prime }},\mathbf{p}%
_{K^{\prime }},S_{K^{\prime }}^{2},M_{K^{\prime }}\right) ,\left( \mathbf{%
\alpha }_{K},\mathbf{p}_{K},S_{K}^{2},M_{K}\right) ,\right) }{\left( \mathbf{%
\bar{D}}_{S_{K}^{2}}^{\mathbf{\alpha }_{K}}+2M_{K}\mathbf{\bar{D}}%
_{S_{K}^{2}}^{\mathbf{\alpha }_{K}}-\eta \right) \left( \mathbf{\bar{D}}%
_{S_{K^{\prime }}^{2}}^{\mathbf{\alpha }_{K^{\prime }}}+2M_{K^{\prime }}%
\mathbf{\bar{D}}_{S_{K^{\prime }}^{2}}^{\mathbf{\alpha }_{K^{\prime }}}-\eta
\right) }\right)
\end{eqnarray*}%
As in the previous paragraph, we replace the eigenvalue by:%
\begin{equation}
\eta \left( \mathbf{\alpha }_{K},\mathbf{p}_{K},S_{K}^{2},M_{K}\right) =%
\mathbf{\bar{D}}_{S_{K}^{2}}^{\mathbf{\alpha }_{K}}+2M_{K}\mathbf{\bar{D}}%
_{S_{K}^{2}}^{\mathbf{\alpha }_{K}}-\varepsilon \left( \mathbf{\alpha }_{K},%
\mathbf{p}_{K},S_{K}^{2},M_{K}\right)  \label{GNV}
\end{equation}%
where $\varepsilon \left( \mathbf{\alpha }_{K},\mathbf{p}%
_{K},S_{K}^{2},M_{K}\right) $ is given in first approximation by equation:%
\begin{equation*}
1-\sum_{K^{\prime }}\frac{P\left( \left( \mathbf{\alpha }_{K},\mathbf{p}%
_{K},S_{K}^{2},M_{K}\right) ,\left( \mathbf{\alpha }_{K^{\prime }},\mathbf{p}%
_{K^{\prime }},S_{K^{\prime }}^{2},M_{K^{\prime }}\right) \right) P\left(
\left( \mathbf{\alpha }_{K^{\prime }},\mathbf{p}_{K^{\prime }},S_{K^{\prime
}}^{2},M_{K^{\prime }}\right) ,\left( \mathbf{\alpha }_{K},\mathbf{p}%
_{K},S_{K}^{2},M_{K}\right) ,\right) }{\varepsilon \left( \mathbf{\alpha }%
_{K},\mathbf{p}_{K},S_{K}^{2},M_{K}\right) \left( \mathbf{\bar{D}}%
_{S_{K^{\prime }}^{2}}^{\mathbf{\alpha }_{K^{\prime }}}+2M_{K^{\prime }}%
\mathbf{\bar{D}}_{S_{K^{\prime }}^{2}}^{\mathbf{\alpha }_{K^{\prime
}}}\right) }\simeq 0
\end{equation*}%
with solution: 
\begin{eqnarray*}
&&\varepsilon \left( \mathbf{\alpha }_{K},\mathbf{p}_{K},S_{K}^{2},M_{K}%
\right) \\
&\simeq &\sum_{K^{\prime }}\frac{P\left( \left( \mathbf{\alpha }_{K},\mathbf{%
p}_{K},S_{K}^{2},M_{K}\right) ,\left( \mathbf{\alpha }_{K^{\prime }},\mathbf{%
p}_{K^{\prime }},S_{K^{\prime }}^{2},M_{K^{\prime }}\right) \right) P\left(
\left( \mathbf{\alpha }_{K^{\prime }},\mathbf{p}_{K^{\prime }},S_{K^{\prime
}}^{2},M_{K^{\prime }}\right) ,\left( \mathbf{\alpha }_{K},\mathbf{p}%
_{K},S_{K}^{2},M_{K}\right) ,\right) }{\left( \mathbf{\bar{D}}_{S_{K^{\prime
}}^{2}}^{\mathbf{\alpha }_{K^{\prime }}}+2M_{K^{\prime }}\mathbf{\bar{D}}%
_{S_{K^{\prime }}^{2}}^{\mathbf{\alpha }_{K^{\prime }}}\right) }
\end{eqnarray*}%
Once the eigenvalue is found, we can come back to the eigenstate equation:%
\begin{eqnarray*}
0 &=&\left( \mathbf{\bar{D}}_{S_{K}^{2}}^{\mathbf{\alpha }_{K}}+2M_{K}%
\mathbf{\bar{D}}_{S_{K}^{2}}^{\mathbf{\alpha }_{K}}-\eta \right) A\left( 
\mathbf{\alpha }_{K},\mathbf{p}_{K},S_{K}^{2},M_{K}\right) \\
&&+\sum_{K^{\prime }M_{K^{\prime }}}P\left( \left( \mathbf{\alpha }_{K},%
\mathbf{p}_{K},S_{K}^{2},M_{K}\right) ,\left( \mathbf{\alpha }_{K^{\prime }},%
\mathbf{p}_{K^{\prime }},S_{K^{\prime }}^{2},M_{K^{\prime }}\right) \right)
A\left( \mathbf{\alpha }_{K^{\prime }},\mathbf{p}_{K^{\prime }},S_{K^{\prime
}}^{2},M_{K^{\prime }}\right)
\end{eqnarray*}%
for:%
\begin{equation*}
\eta =\eta \left( \mathbf{\alpha }_{K},\mathbf{p}_{K},S_{K}^{2},M_{K}\right)
\end{equation*}%
Using (\ref{GNV}) this eigenstate equation is:%
\begin{eqnarray*}
0 &=&\varepsilon \left( \mathbf{\alpha }_{K},\mathbf{p}_{K},S_{K}^{2},M_{K}%
\right) A\left( \mathbf{\alpha }_{K},\mathbf{p}_{K},S_{K}^{2},M_{K^{\prime
}}\right) \\
&&+\sum_{K^{\prime }M_{K^{\prime }}}P\left( \left( \mathbf{\alpha }_{K},%
\mathbf{p}_{K},S_{K}^{2},M_{K}\right) ,\left( \mathbf{\alpha }_{K^{\prime }},%
\mathbf{p}_{K^{\prime }},S_{K^{\prime }}^{2},M_{K^{\prime }}\right) \right)
A\left( \mathbf{\alpha }_{K^{\prime }},\mathbf{p}_{K^{\prime }},S_{K^{\prime
}}^{2},M_{K^{\prime }}\right)
\end{eqnarray*}%
Now, if $L\neq K$ it writes:%
\begin{equation*}
0\simeq \left( \mathbf{\bar{D}}_{S_{L}^{2}}^{\mathbf{\alpha }_{L}}+2M_{L}%
\mathbf{\bar{D}}_{S_{L}^{2}}^{\mathbf{\alpha }_{L}}\right) A\left( \mathbf{%
\alpha }_{L},\mathbf{p}_{L},S_{L}^{2},M_{L}\right) +\sum_{K^{\prime
}M_{K^{\prime }}}P\left( \left( \mathbf{\alpha }_{L},\mathbf{p}%
_{L},S_{L}^{2},M_{L}\right) ,\left( \mathbf{\alpha }_{K^{\prime }},\mathbf{p}%
_{K^{\prime }},S_{K^{\prime }}^{2},M_{K^{\prime }}\right) \right) A\left( 
\mathbf{\alpha }_{K^{\prime }},\mathbf{p}_{K^{\prime }},S_{K^{\prime
}}^{2},M_{K^{\prime }}\right)
\end{equation*}%
and leads to:%
\begin{equation*}
A\left( \mathbf{\alpha }_{L},\mathbf{p}_{L},S_{L}^{2},M_{L}\right) \simeq -%
\frac{P\left( \left( \mathbf{\alpha }_{L},\mathbf{p}_{L},S_{L}^{2},M_{L}%
\right) ,\left( \mathbf{\alpha }_{K},\mathbf{p}_{K},S_{K}^{2},M_{K}\right)
\right) }{\mathbf{\bar{D}}_{S_{L}^{2}}^{\mathbf{\alpha }_{L}}+2M_{L}\mathbf{%
\bar{D}}_{S_{L}^{2}}^{\mathbf{\alpha }_{L}}}A\left( \mathbf{\alpha }_{K},%
\mathbf{p}_{K},S_{K}^{2},M_{K}\right)
\end{equation*}

\section*{Appendix 6 variable localisation}

\subsection*{6. 1 Interaction term}

To compute the interaction term:%
\begin{equation*}
I=\sum_{\delta S^{2}}\underline{\Gamma }^{\dag }\left( \mathbf{T},\mathbf{%
\alpha }+\delta \mathbf{\alpha },\mathbf{p+}\delta \mathbf{p},S^{2}+\delta
S^{2}\right) U\left( S^{2},S^{2}+\delta S^{2}\right) \underline{\Gamma }%
\left( \mathbf{T},\mathbf{\alpha },\mathbf{p},S^{2}\right)
\end{equation*}%
we choose the following form for the potential $U\left( S^{2},S^{2}+\delta
S^{2}\right) $:%
\begin{eqnarray*}
U\left( S^{2},S^{2}+\delta S^{2}\right) &=&\int_{\delta S^{2}}d^{+/-}\left(
Z,Z^{\prime }\right) U\left( \delta \mathbf{\bar{T}}_{p}^{\alpha }\left(
Z,Z^{\prime }\right) \right) \\
&=&\int_{S^{2}+\delta S^{2}/S^{2}}d\left( Z,Z^{\prime }\right) U\left(
\delta \mathbf{\bar{T}}_{p}^{\alpha }\left( Z,Z^{\prime }\right) \right)
-\int_{S^{2}/S^{2}+\delta S^{2}}d\left( Z,Z^{\prime }\right) U\left( \delta 
\mathbf{\bar{T}}_{p}^{\alpha }\left( Z,Z^{\prime }\right) \right)
\end{eqnarray*}%
where:%
\begin{eqnarray*}
S^{2}+\delta S^{2}/S^{2} &=&\left( S^{2}+\delta S^{2}\right) \backslash
S^{2}\cap \left( S^{2}+\delta S^{2}\right) \\
S^{2}/S^{2}+\delta S^{2} &=&\left( S^{2}\right) \backslash S^{2}\cap \left(
S^{2}+\delta S^{2}\right)
\end{eqnarray*}%
and $\delta \mathbf{\bar{T}}_{p}^{\alpha }\left( Z,Z^{\prime }\right) $ is
the variation defined by:%
\begin{equation*}
\delta \mathbf{\bar{T}}_{p}^{\alpha }\left( Z,Z^{\prime }\right) =\mathbf{%
\bar{T}}_{p}^{\alpha }\left( \left( Z,Z^{\prime }\right) +\delta \left(
Z,Z^{\prime }\right) \right) -\mathbf{\bar{T}}_{p}^{\alpha }\left(
Z,Z^{\prime }\right)
\end{equation*}

In a continuous approximation, this corresponds to:%
\begin{equation*}
U\left( S^{2},S^{2}+\delta S^{2}\right) =\int_{\delta S^{2}}U\left( \delta 
\mathbf{\bar{T}}_{p}^{\alpha }\left( Z,Z^{\prime }\right) \right) \left(
n\left( \left( Z,Z^{\prime }\right) \right) .\frac{\delta \left( Z,Z^{\prime
}\right) }{\left\Vert \delta \left( Z,Z^{\prime }\right) \right\Vert }%
\right) d\left( Z,Z^{\prime }\right)
\end{equation*}%
with $n\left( \left( Z,Z^{\prime }\right) \right) $ the normal to $S^{2}$.

A second order expansion rewrites the potential: 
\begin{eqnarray}
I &=&\sum_{\delta S^{2}}\underline{\Gamma }^{\dag }\left( \mathbf{T},\mathbf{%
\alpha }+\delta \mathbf{\alpha },\mathbf{p+}\delta \mathbf{p},S^{2}+\delta
S^{2}\right) U\left( S^{2},S^{2}+\delta S^{2}\right) \underline{\Gamma }%
\left( \mathbf{T},\mathbf{\alpha },\mathbf{p},S^{2}\right)  \label{NR} \\
&=&\sum_{\delta S^{2}}\underline{\Gamma }^{\dag }\left( \mathbf{T},\mathbf{%
\alpha },\mathbf{p},S^{2}\right) U\left( S^{2},S^{2}+\delta S^{2}\right) 
\underline{\Gamma }\left( \mathbf{T},\mathbf{\alpha },\mathbf{p},S^{2}\right)
\notag \\
&&-\sum_{\delta S^{2}}\underline{\Gamma }^{\dag }\left( \mathbf{T},\mathbf{%
\alpha },\mathbf{p},S^{2}\right) U\left( S^{2},S^{2}+\delta S^{2}\right)
\delta S^{2}\frac{\delta }{\delta S^{2}}\underline{\Gamma }\left( \mathbf{T},%
\mathbf{\alpha },\mathbf{p},S^{2}\right)  \notag
\end{eqnarray}%
Here, $\frac{\delta \underline{\Gamma }\left( \mathbf{T},\mathbf{\alpha },%
\mathbf{p},S^{2}\right) }{\delta S^{2}}$ should be considered as a full
derivatives, but, to focus on nonlocalization, we neglect variations in $%
\mathbf{\alpha },\mathbf{p}$. We have defined:%
\begin{equation*}
\delta S^{2}\frac{\delta }{\delta S^{2}}\underline{\Gamma }\left( \mathbf{T},%
\mathbf{\alpha },\mathbf{p},S^{2}\right) =\int_{S^{2}}\delta \left(
Z,Z^{\prime }\right) \nabla _{\left( Z,Z^{\prime }\right) }\underline{\Gamma 
}\left( \mathbf{T},\mathbf{\alpha },\mathbf{p},S^{2}\right)
\end{equation*}

In (\ref{NR}), the first term can be neglected. Actually: 
\begin{eqnarray*}
&&\sum_{\delta S^{2}}U\left( S^{2},S^{2}+\delta S^{2}\right) \\
&=&\sum_{\delta S^{2}}\left( \int d\left( S^{2}+\delta S^{2}/S^{2}\right)
U\left( \mathbf{\bar{T}}_{p}^{\alpha }\left( Z,Z^{\prime }\right) \right)
-\int d\left( S^{2}/S^{2}+\delta S^{2}\right) U\left( \mathbf{\bar{T}}%
_{p}^{\alpha }\left( Z,Z^{\prime }\right) \right) \right) \rightarrow 0
\end{eqnarray*}

The second term in (\ref{NR}) can be rewriten as:%
\begin{equation*}
\sum_{\delta S^{2}}U\left( S^{2},S^{2}+\delta S^{2}\right)
\int_{S^{2}}\delta \left( Z,Z^{\prime }\right) \nabla _{\left( Z,Z^{\prime
}\right) }\underline{\Gamma }\left( \mathbf{T},\mathbf{\alpha },\mathbf{p}%
,S^{2}\right) =\int_{S^{2}}\bar{U}\left( S^{2},Z,Z^{\prime }\right) \nabla
_{\left( Z,Z^{\prime }\right) }\underline{\Gamma }\left( \mathbf{T},\mathbf{%
\alpha },\mathbf{p},S^{2}\right)
\end{equation*}%
where:%
\begin{equation*}
\bar{U}\left( S^{2},Z,Z^{\prime }\right) =\sum_{\delta S^{2}}U\left(
S^{2},S^{2}+\delta S^{2}\right) \delta \left( Z,Z^{\prime }\right)
\end{equation*}%
As a consequence, the potential (\ref{PTL}) writes:%
\begin{eqnarray*}
I &=&\sum_{\delta S^{2}}\underline{\Gamma }^{\dag }\left( \mathbf{T},\mathbf{%
\alpha }+\delta \mathbf{\alpha },\mathbf{p+}\delta \mathbf{p},S^{2}+\delta
S^{2}\right) U\left( S^{2},S^{2}+\delta S^{2}\right) \underline{\Gamma }%
\left( \mathbf{T},\mathbf{\alpha },\mathbf{p},S^{2}\right) \\
&=&\int_{S^{2}}\bar{U}\left( S^{2},Z,Z^{\prime }\right) \nabla _{\left(
Z,Z^{\prime }\right) }\underline{\Gamma }\left( \mathbf{T},\mathbf{\alpha },%
\mathbf{p},S^{2}\right) \\
&\equiv &-\bar{U}\left( S^{2}\right) \frac{\delta }{\delta S^{2}}\underline{%
\Gamma }\left( \mathbf{T},\mathbf{\alpha },\mathbf{p},S^{2}\right)
\end{eqnarray*}

\subsection*{6.2 Computation of the Green function}

The Green function between two states $\left\vert \mathbf{T}_{p}^{\alpha
},S^{2}\right\rangle $ and $\left\vert \mathbf{T}_{p}^{\prime \alpha
},S^{\prime 2}\right\rangle $ is computed by changing the variables in the
action:%
\begin{equation*}
\underline{\Gamma }^{\dag }\left( \mathbf{T},\mathbf{\alpha },\mathbf{p}%
,S^{2}\right) \left( -\frac{1}{2}\frac{\delta ^{2}}{\delta \left(
S^{2}\right) ^{2}}-\bar{U}\left( S^{2}\right) \frac{\delta }{\delta S^{2}}-%
\frac{1}{2}\nabla _{\left( \mathbf{T}\right) _{S^{2}}}^{2}+\frac{1}{2}\left(
\Delta \mathbf{T}_{p}^{\alpha }\right) ^{t}\left( \mathbf{A}_{p}^{\alpha
}\right) _{S^{2}}\Delta \mathbf{T}_{p}^{\alpha }+\mathbf{C}\left(
S^{2}\right) \right) \underline{\Gamma }\left( \mathbf{T},\mathbf{\alpha },%
\mathbf{p},S^{2}\right)
\end{equation*}%
Replacing:%
\begin{eqnarray*}
\underline{\Gamma }\left( \mathbf{T},\mathbf{\alpha },\mathbf{p}%
,S^{2}\right) &\rightarrow &\exp \left( \int^{S^{2}}\bar{U}\left(
S^{2}\right) dS^{2}\right) \underline{\Gamma }\left( \mathbf{T},\mathbf{%
\alpha },\mathbf{p},S^{2}\right) \\
\underline{\Gamma }^{\dag }\left( \mathbf{T},\mathbf{\alpha },\mathbf{p}%
,S^{2}\right) &\rightarrow &\exp \left( -\int^{S^{2}}\bar{U}\left(
S^{2}\right) dS^{2}\right) \underline{\Gamma }^{\dag }\left( \mathbf{T},%
\mathbf{\alpha },\mathbf{p},S^{2}\right)
\end{eqnarray*}%
the action becomes:%
\begin{eqnarray}
&&\underline{\Gamma }^{\dag }\left( \mathbf{T},\mathbf{\alpha },\mathbf{p}%
,S^{2}\right)  \label{TR} \\
&&\times \left( -\frac{1}{2}\frac{\delta ^{2}}{\delta \left( S^{2}\right)
^{2}}+\frac{1}{2}\bar{U}^{2}\left( S^{2}\right) +\frac{1}{2}\frac{\delta }{%
\delta S^{2}}\bar{U}\left( S^{2}\right) -\frac{1}{2}\nabla _{\left( \mathbf{T%
}\right) _{S^{2}}}^{2}+\frac{1}{2}\left( \Delta \mathbf{T}_{p}^{\alpha
}\right) ^{t}\left( \mathbf{A}_{p}^{\alpha }\right) _{S^{2}}\Delta \mathbf{T}%
_{p}^{\alpha }+\mathbf{C}\left( S^{2}\right) \right) \underline{\Gamma }%
\left( \mathbf{T},\mathbf{\alpha },\mathbf{p},S^{2}\right)  \notag
\end{eqnarray}%
with:%
\begin{equation*}
\frac{\delta }{\delta S^{2}}\bar{U}\left( S^{2}\right) =\int \nabla _{\left(
Z,Z^{\prime }\right) }\bar{U}\left( S^{2},Z,Z^{\prime }\right) d\left(
Z,Z^{\prime }\right)
\end{equation*}%
For $\bar{U}^{2}\left( S^{2},Z,Z^{\prime }\right) $ quadratic in $\left(
Z,Z^{\prime }\right) $:%
\begin{eqnarray*}
\frac{1}{2}\bar{U}\left( S^{2},Z,Z^{\prime }\right) +\frac{1}{2}\frac{\delta 
}{\delta S^{2}}\bar{U}\left( S^{2}\right) &\simeq &\int_{S^{2}}\frac{1}{2}%
A\left( Z,Z^{\prime }\right) \left( Z^{2}+Z^{\prime 2}\right) \\
&\simeq &\int_{S^{2}}\frac{1}{2}\left\langle A\right\rangle \left(
Z^{2}+Z^{\prime 2}\right)
\end{eqnarray*}%
with $\left\langle A\right\rangle $ is the average of $A\left( Z,Z^{\prime
}\right) $ on $S^{2}$.

We also expand $\mathbf{C}\left( S^{2}\right) $ with respect to some $%
\mathbf{C}\left( S_{0}^{2}\right) $. In first approximation:%
\begin{equation*}
\mathbf{C}\left( S^{2}\right) \simeq \mathbf{C}\left( S_{0}^{2}\right)
+\int_{S^{2}}\left( \left( Z-Z_{0}\right) +\left( Z^{\prime }-Z_{0}^{\prime
}\right) \right) \mathbf{\bar{C}}\left( S^{2}\right)
\end{equation*}%
so that, up to come constant:%
\begin{equation*}
\frac{1}{2}\bar{U}\left( S^{2},Z,Z^{\prime }\right) +\frac{1}{2}\frac{\delta 
}{\delta S^{2}}\bar{U}\left( S^{2}\right) \simeq \frac{1}{2}%
\int_{S^{2}}\left\langle A\right\rangle \left( \left( Z-\left\langle
A\right\rangle ^{-1}\mathbf{\bar{C}}\left( S^{2}\right) \right) ^{2}+\left(
Z^{\prime }-\left\langle A\right\rangle ^{-1}\mathbf{\bar{C}}\left(
S^{2}\right) \right) ^{2}\right)
\end{equation*}

The Green function of:%
\begin{eqnarray*}
O &=&-\frac{1}{2}\frac{\delta ^{2}}{\delta \left( S^{2}\right) ^{2}}+\frac{1%
}{2}\int \left\langle A\right\rangle \left( \left( Z-\left\langle
A\right\rangle ^{-1}\mathbf{\bar{C}}\left( S^{2}\right) \right) ^{2}+\left(
Z^{\prime }-\left\langle A\right\rangle ^{-1}\mathbf{\bar{C}}\left(
S^{2}\right) \right) ^{2}\right) \\
&&-\frac{1}{2}\nabla _{\left( \mathbf{T}\right) _{S^{2}}}^{2}+\frac{1}{2}%
\left( \Delta \mathbf{T}_{p}^{\alpha }\right) ^{t}\left( \mathbf{A}%
_{p}^{\alpha }\right) _{S^{2}}\Delta \mathbf{T}_{p}^{\alpha }+\mathbf{C}%
\left( S^{2}\right)
\end{eqnarray*}%
is written:%
\begin{equation*}
G\left( S^{\prime 2},\mathbf{T}_{p}^{\prime \alpha },S^{2},\mathbf{T}%
_{p}^{\alpha }\right)
\end{equation*}%
between two close structures $S^{2}$ and $S^{2}+\delta S^{2}$ is
approximatively:%
\begin{eqnarray}
G\left( S^{\prime 2},\mathbf{T}_{p}^{\prime \alpha },S^{2},\mathbf{T}%
_{p}^{\alpha }\right) &\simeq &\frac{\exp \left( -\left( \mathbf{X}^{\prime
}\right) ^{t}\left\langle \mathbf{A}\right\rangle _{S^{2}+\delta S^{2}}%
\mathbf{X}^{\prime }-\left( \mathbf{X}\right) ^{t}\left\langle \mathbf{A}%
\right\rangle _{S^{2}}\mathbf{X-}\left( \mathbf{C}\left( S^{\prime 2}\right)
-\mathbf{C}\left( S^{2}\right) \right) \right) }{\sqrt{\det \left\langle 
\mathbf{A}\right\rangle }} \\
&&\times \frac{\exp \left( -\left( \mathbf{T}_{p}^{\prime \alpha }\right)
^{t}\mathbf{A}_{S^{2}}^{\alpha }\mathbf{T}_{p}^{\prime \alpha }-\left( 
\mathbf{T}_{p}^{\alpha }\right) ^{t}\mathbf{A}_{S^{2}}^{\alpha }\mathbf{T}%
_{p}^{\alpha }\right) }{\sqrt{\det \left( \mathbf{A}_{S^{2}}^{\alpha
}\right) }}  \notag
\end{eqnarray}%
where 
\begin{equation*}
\mathbf{X}^{\prime }=\left( Z-\left\langle A\right\rangle ^{-1}\mathbf{\bar{C%
}}\left( S^{2}\right) ,Z^{\prime }-\left\langle A\right\rangle ^{-1}\mathbf{%
\bar{C}}\left( S^{2}\right) \right) _{\left( Z,Z^{\prime }\right) \in
S^{\prime 2}}
\end{equation*}%
and: 
\begin{equation*}
\mathbf{X}=\left( Z-\left\langle A\right\rangle ^{-1}\mathbf{\bar{C}}\left(
S^{2}\right) ,Z^{\prime }-\left\langle A\right\rangle ^{-1}\mathbf{\bar{C}}%
\left( S^{2}\right) \right) _{\left( Z,Z^{\prime }\right) \in S^{2}}
\end{equation*}

We have set $\left\langle \mathbf{A}\right\rangle _{S^{2}}$ for the average $%
A\left( Z,Z^{\prime }\right) $ in $S^{2}$ and $\left\langle \mathbf{A}%
\right\rangle _{S^{2}+\delta S^{2}}$ for the average $A\left( Z,Z^{\prime
}\right) $ in $S^{2}+\delta S^{2}$. As a consequence, successive
convolutions yield in first approximation the Green function for the system
defined in (\ref{TR}):

\begin{eqnarray}
G\left( S^{\prime 2},\mathbf{T}_{p}^{\prime \alpha },S^{2},\mathbf{T}%
_{p}^{\alpha }\right) &\simeq &\frac{\exp \left( -\left( \mathbf{X}^{\prime
}\right) ^{t}\left\langle \mathbf{A}\right\rangle _{S^{\prime 2}}\mathbf{X}%
^{\prime }-\left( \mathbf{X}\right) ^{t}\left\langle \mathbf{A}\right\rangle
_{S^{2}}\mathbf{X-}\left( \mathbf{C}\left( S^{\prime 2}\right) -\mathbf{C}%
\left( S^{2}\right) \right) \right) }{\left( \left\langle \mathbf{A}%
\right\rangle _{S^{\prime 2}}\left\langle \mathbf{A}\right\rangle
_{S^{2}}\right) ^{\frac{1}{4}}}  \label{TS} \\
&&\times \frac{\exp \left( -\left( \mathbf{T}_{p}^{\prime \alpha }\right)
^{t}\mathbf{A}_{S^{\prime 2}}^{\alpha }\mathbf{T}_{p}^{\prime \alpha
}-\left( \mathbf{T}_{p}^{\alpha }\right) ^{t}\mathbf{A}_{S^{2}}^{\alpha }%
\mathbf{T}_{p}^{\alpha }\right) }{\sqrt{\det \left( \mathbf{A}%
_{S^{2}}^{\alpha }\right) }}  \notag
\end{eqnarray}%
Expanding $\mathbf{X}$ and $\mathbf{X}^{\prime }$ leads to:%
\begin{eqnarray}
&&G\left( S^{\prime 2},\mathbf{T}_{p}^{\prime \alpha },S^{2},\mathbf{T}%
_{p}^{\alpha }\right) \\
&\simeq &\frac{\exp \left( -\left( \left( \mathbf{Z,Z}^{\prime }\right)
^{\prime }\right) ^{t}\left\langle \mathbf{A}\right\rangle _{S^{\prime
2}}\left( \mathbf{Z,Z}^{\prime }\right) ^{\prime }-\left( \mathbf{Z,Z}%
^{\prime }\right) ^{t}\left\langle \mathbf{A}\right\rangle _{S^{2}}\left( 
\mathbf{Z,Z}^{\prime }\right) ^{\prime }\mathbf{-}\left( \mathbf{C}\left(
S^{\prime 2}\right) -\mathbf{C}\left( S^{2}\right) \right) \right) }{\sqrt{%
\det \left\langle \mathbf{A}\right\rangle }}  \notag \\
&&\times \frac{\exp \left( -\left( \mathbf{T}_{p}^{\prime \alpha }\right)
^{t}\mathbf{A}_{S^{\prime 2}}^{\alpha }\mathbf{T}_{p}^{\prime \alpha
}-\left( \mathbf{T}_{p}^{\alpha }\right) ^{t}\mathbf{A}_{S^{2}}^{\alpha }%
\mathbf{T}_{p}^{\alpha }\right) }{\sqrt{\det \left( \mathbf{A}%
_{S^{2}}^{\alpha }\right) }}  \notag
\end{eqnarray}%
where $\left( \mathbf{Z,Z}^{\prime }\right) ^{\prime }=S^{\prime 2}$ and $%
\left( \mathbf{Z,Z}^{\prime }\right) =S^{2}$.

Coming back to the initial variable amounts to add a factor in (\ref{TS}):%
\begin{equation}
\exp \left( \frac{\delta }{\delta S^{2}}\bar{U}\left( S^{2}\right) -\frac{%
\delta }{\delta S^{\prime 2}}\bar{U}\left( S^{\prime 2}\right) \right)
\end{equation}%
which yields the result in the text.

\end{document}